\documentclass[aps,prc,superscriptaddress,showpacs,floatfix,nobibnotes]{revtex4}
\usepackage{xspace}
\usepackage{longtable}
\usepackage{graphicx}
\usepackage{afterpage}
\usepackage{placeins}
\newcommand{\MeV}{\ensuremath{\mbox{MeV}}\xspace}
\newcommand{\GeVc}{\ensuremath{\mbox{GeV}/c}\xspace}
\newcommand{\MeVc}{\ensuremath{\mbox{MeV}/c}\xspace}
\newcommand{\T}{\ensuremath{\mbox{T}}\xspace}

\newcommand{\mm}{\ensuremath{\mbox{mm}}\xspace}

\newcommand{\mrad}{\ensuremath{\mbox{mrad}}\xspace}
\newcommand{\rad}{\ensuremath{\mbox{rad}}\xspace}

\newcommand{\dedx}{\ensuremath{\mbox{d}E/\mbox{d}x}\xspace}
\newcommand{\pip}{\ensuremath{\pi^+}\xspace}
\newcommand{\pim}{\ensuremath{\pi^-}\xspace}
\newcommand{\piz}{\ensuremath{\pi^0}\xspace}
\newcommand{\bfpip}{\ensuremath{\mathbf {\pi^+}}\xspace}
\newcommand{\bfpim}{\ensuremath{\mathbf {\pi^-}}\xspace}

\newcommand{\evtspill}{\ensuremath{N_{\mathrm{evt}}}\xspace}
\newcommand{\pt}{\ensuremath{p_{\mathrm{T}}}\xspace}

\newcommand{\bgr}{\ensuremath{\mbox{barn}/(\rad \ \GeVc)}\xspace}

\def\be{\begin{equation}}
\def\ee{\end{equation}}
\def\bea{\begin{eqnarray}}
\def\eea{\end{eqnarray}}

\newcommand{\bfGeVc}{\ensuremath{\mathbf {\mbox{\bf GeV}/c}}\xspace}
\begin{document}
\title{Large-angle production of  charged pions 
       with 3-12.9 \GeVc incident protons on nuclear targets}

\begin{abstract}
  Measurements of the  double-differential $\pi^{\pm}$ production
  cross-section    in the range of momentum $100~\MeVc \leq p \le 800~\MeVc$ 
  and angle $0.35~\rad \leq \theta  \le 2.15~\rad$
  in proton--beryllium, proton--carbon, proton--aluminium,  proton--copper,
  proton--tin, proton--tantalum and proton--lead collisions are presented. 
  The data were taken  with the large acceptance HARP detector in the T9 beam
  line of the CERN PS.
  The pions were produced by proton beams in a momentum range from
  3~\GeVc to  12.9~\GeVc hitting a target with a thickness of
  5\% of a nuclear interaction length.  
  The tracking and identification of the
  produced particles was performed using a small-radius
  cylindrical time projection chamber (TPC) placed inside a solenoidal
  magnet. 
  Incident particles were identified by an elaborate system of beam
  detectors.
  Results are obtained for the double-differential cross-sections 
  $
  {{\mathrm{d}^2 \sigma}}/{{\mathrm{d}p\mathrm{d}\theta }}
  $
  at six incident proton beam momenta (3~\GeVc, 5~\GeVc, 8~\GeVc, 
  8.9~\GeVc (Be only), 12~\GeVc and 12.9 \GeVc (Al only)).
  They are based on a complete correction of static and dynamic distortions
  of tracks in the HARP TPC which allows the complete statistics of
  collected data set to be used.
  The results include and supersede our previously published results and
 are compatible with these. 
  Results are compared with the  GEANT4 and MARS MonteCarlo
  simulation.
\end{abstract}

\author{M.G.~Catanesi}
\author{E.~Radicioni} 
\affiliation{Sezione INFN, Bari, Italy}
\author{R.~Edgecock}
\author{ M.~Ellis}
\altaffiliation{Now at FNAL, Batavia, Illinois, USA.}
\author{F.J.P.~Soler}
\affiliation{Rutherford Appleton Laboratory, Chilton, Didcot, UK} 
\author{C.~G\"{o}\ss ling}
\affiliation{ Institut f\"{u}r Physik, Universit\"{a}t Dortmund, Germany}
\author{S.~Bunyatov}
\author{A.~Krasnoperov}
\author{B.~Popov}
\altaffiliation{Also supported by LPNHE, Paris, France.}
\author{V.~Serdiouk}
\author{ V.~Tereschenko }
\affiliation{Joint Institute for Nuclear Research, JINR Dubna, Russia} 
\author{E.~Di~Capua}
\author{G.~Vidal--Sitjes}
\altaffiliation{Now at Imperial College, University of London, UK.}
\affiliation{Universit\`{a} degli Studi e Sezione INFN, Ferrara, Italy} 
\author{A.~Artamonov}
\altaffiliation{ITEP, Moscow, Russian Federation.}
\author{S.~Giani}
\author{S.~Gilardoni}  
\author{P.~Gorbunov}  
\altaffiliation{ITEP, Moscow, Russian Federation.}
\author{A.~Grant}  
\author{A.~Grossheim} 
\altaffiliation{Now at TRIUMF, Vancouver, Canada.}
\author{A.~Ivanchenko}
\altaffiliation{ On leave from Novosibirsk University,  Russia.}  
\author{V.~Ivanchenko}  
\altaffiliation{On leave  from Ecoanalitica, Moscow State University,
Moscow, Russia.}
\author{A.~Kayis-Topaksu}
\altaffiliation{Now at \c{C}ukurova University, Adana, Turkey.}
\author{J.~Panman} 
\author{I.~Papadopoulos}  
\author{E.~Tcherniaev} 
\author{I.~Tsukerman} 
\altaffiliation{ITEP, Moscow, Russian Federation.}
\author{R.~Veenhof} 
\author{C.~Wiebusch}  
\altaffiliation{Now at III Phys. Inst. B, RWTH Aachen, Germany.}
\author{P.~Zucchelli}
\altaffiliation{Now at SpinX Technologies, Geneva, Switzerland; \\
on leave  from INFN, Sezione di Ferrara, Italy.}
\affiliation{ CERN, Geneva, Switzerland} 
\author{A.~Blondel} 
\author{S.~Borghi}
\author{M.C.~Morone} 
\altaffiliation{Now at University of Rome Tor Vergata, Italy.}
\author{G.~Prior} 
\altaffiliation{Now at LBL, Berkeley, California, USA.}
\author{R.~Schroeter}
\affiliation{Section de Physique, Universit\'{e} de Gen\`{e}ve, Switzerland} 
\author{C.~Meurer}
\affiliation{Institut f\"{u}r Physik, Forschungszentrum Karlsruhe, Germany}
\author{U.~Gastaldi}, 
\affiliation{Laboratori Nazionali di Legnaro dell' INFN, Legnaro, Italy} 
\author{G.~B.~Mills}  
\altaffiliation{MiniBooNE Collaboration.}
\affiliation{ Los Alamos National Laboratory, Los Alamos, USA}  
\author{J.S.~Graulich}
\altaffiliation{Now at Section de Physique, Universit\'{e} de Gen\`{e}ve, Switzerland.}
\author{G.~Gr\'{e}goire} 
\affiliation{Institut de Physique Nucl\'{e}aire, UCL, Louvain-la-Neuve,
  Belgium} 
\author{M.~Bonesini}
\email{maurizio.bonesini@mib.infn.it}
\author{ F.~Ferri  }
\affiliation{Sezione INFN Milano Bicocca, Milano, Italy} 
\author{M.~Kirsanov}
\affiliation{Institute for Nuclear Research, Moscow, Russia}
\author{A. Bagulya}
\author{V.~Grichine}
\author{N.~Polukhina} 
\affiliation{P. N. Lebedev Institute of Physics (FIAN), Russian Academy of
Sciences, Moscow, Russia}
\author{V.~Palladino}
\affiliation{Universit\`{a} ``Federico II'' e Sezione INFN, Napoli, Italy}  
\author{L.~Coney}
\altaffiliation{MiniBooNE Collaboration.}
\author{ D.~Schmitz}
\altaffiliation{MiniBooNE Collaboration.}
\affiliation{Columbia University, New York, USA} 
\author{G.~Barr} 
\author{A.~De~Santo}
\affiliation{Nuclear and Astrophysics Laboratory, University of Oxford, UK} 
\author{F.~Bobisut$^{a,b}$}
\author{D.~Gibin$^{a,b}$}
\author{A.~Guglielmi$^{b}$}
\author{M.~Mezzetto$^{b}$}
\affiliation{Universit\`{a} degli Studi$^a$ e Sezione INFN$^b$, Padova, Italy} 
\author{J.~Dumarchez} 
\affiliation{ LPNHE, Universit\'{e}s de Paris VI et VII, Paris, France} 
\author{U.~Dore}
\affiliation{ Universit\`{a} ``La Sapienza'' e Sezione INFN Roma I, Roma,
  Italy} 
\author{D.~Orestano$^{c,d}$}
\author{F.~Pastore$^{c,d}$}
\author{A.~Tonazzo$^{c,d}$}
\author{L.~Tortora$^{d}$}
\affiliation{Universit\`{a} degli Studi$^{c}$ e Sezione INFN Roma Tre$^{d}$, Roma, Italy}
\author{C.~Booth}
\author{L.~Howlett}
\author{G.~Skoro}
\affiliation{ Dept. of Physics, University of Sheffield, UK} 
\author{M.~Bogomilov}
\author{M.~Chizhov}
\author{D.~Kolev}
\author{R.~Tsenov}
\affiliation{ Faculty of Physics, St. Kliment Ohridski University, Sofia,
  Bulgaria} 
\author{ S.~Piperov}
\author{P.~Temnikov}
\affiliation{ Institute for Nuclear Research and Nuclear Energy, 
Academy of Sciences, Sofia, Bulgaria} 
\author{M.~Apollonio}
\author{P.~Chimenti}
\author{G.~Giannini} 
\affiliation{Universit\`{a} degli Studi e Sezione INFN, Trieste, Italy}
\author{J.~Burguet--Castell}
\author{A.~Cervera--Villanueva}
\author{J.J.~G\'{o}mez--Cadenas} 
\author{J. Mart\'{i}n--Albo}
\author{P.~Novella}
\author{M.~Sorel}
\author{A.~Tornero}
\affiliation{Instituto de F\'{i}sica Corpuscular, IFIC, CSIC and Universidad de Valencia,
Spain} 
\collaboration{\bf HARP Collaboration}
\noaffiliation
\date{\today}

\maketitle
\section{Introduction}

The HARP experiment~\cite{harp-prop} 
makes use of a large-acceptance spectrometer for
a systematic study of hadron
production on a large range of target nuclei for beam momenta from 1.5~\GeVc to 15~\GeVc. 
This corresponds to a proton momentum region of great interest for neutrino beams and  far from coverage by earlier dedicated hadroproduction experiments
\cite{SPY}, \cite{Atherton}.

The main motivations are to measure pion yields for a quantitative
design of the proton driver of a future neutrino factory~\cite{ref:nufact}, 
to provide measurements to allow substantially improved calculations of
the atmospheric neutrino
flux~\cite{Battistoni,Stanev,Gaisser,Engel,Honda} to be made
and to measure particle yields as input for the flux
calculation of accelerator neutrino experiments~\cite{ref:physrep}, 
such as K2K~\cite{ref:k2k,ref:k2kfinal},
MiniBooNE~\cite{ref:miniboone} and SciBooNE~\cite{ref:sciboone}. 


This paper presents our final  
measurements of the double-differential cross-section, 
$
{{\mathrm{d}^2 \sigma^{\pi}}}/{{\mathrm{d}p\mathrm{d}\theta }}
$
for $\pi^{\pm}$ production at large angles by
protons of 3~\GeVc, 5~\GeVc, 8~\GeVc, 8.9~\GeVc (Be only),  12~\GeVc 
and 12.9~\GeVc (Al only) momentum impinging
on a thin beryllium, carbon, aluminium, copper, tin, tantalum  or lead target of 5\% 
nuclear interaction length.
\noindent
A first set of results on the production of pions at large angles
have been published by the HARP Collaboration in
references \cite{ref:harp:tantalum,ref:harp:cacotin,
ref:harp:bealpb}, based on the analysis of the data in the beginning of
each accelerator spill. 
The reduction of the data set was necessary to avoid problems in the TPC
detector responsible for dynamic distortions to the image of the
particle trajectories as the ion charge was building up during each
spill.   
Corrections that allow the use of the full statistics to be made, 
correcting for
such distortions, have been developed in \cite{ref:tpcmom} and are fully
applied in this analysis. The obtained results are fully
compatible within the statistical errors and differential systematic
uncertainties with those previously published. 
The increase of statistics is particularly useful in the 3 \GeVc data
sets. Comparisons with MonteCarlo models are then shown for a light and
a heavy target. 

This paper, covering an extended range of solid targets in
the same experiment, makes it possible to perform systematic
comparison of hadron production models with measurements 
at different incoming beam momenta over a large range 
of target atomic number $A$.
Results for pion production in the forward direction 
are the subject of
other HARP publications \cite{ref:alPaper,ref:CaFW,ref:bePaper}.

Pion production data at low momenta ($\simeq 200$ MeV/c) are extremely scarce and HARP
is the first experiment to provide a large data set, taken with many 
different targets, full particle identification and large detector 
acceptance.
In addition, the acceptance of the large-angle detector of the 
HARP experiment matches well the required phase space region of
pions relevant to the production of $\mu$'s in a neutrino factory
(NF). It covers the large majority of the pions accepted in
the focussing system of a typical design. 
For the optimization of the NF target, data were taken with high-$Z$ 
nuclear targets such as tantalum and lead.
%

Data were taken in the T9 beam of the CERN PS. The collected statistics,
for the different nuclear targets, are reported in table \ref{tab:events}. 

The analysis proceeds by selecting tracks in the Time Projection
Chamber (TPC), after corrections for static and dynamic distortions
(see later for details),
 in events with incident beam of protons.  
Momentum and polar angle measurements and particle identification are
based on the measurements of track position and energy deposition in
the TPC.
An unfolding method is used to correct for experimental resolution,
efficiency and acceptance and to obtain the double-differential pion
production cross-sections.  The method allows a full error evaluation to
be made.
A comparison with available data is 
presented. 
The analysis follows the same methods as used for the determination of
$\pi^{\pm}$ production by
protons on a tantalum target which are described 
in Ref.~\cite{ref:harp:tantalum} and will be only briefly outlined here.

\section{Experimental apparatus and data selection.}
\label{sec:apparatus}
\begin{figure*}[tbp]
  \begin{center}
    \includegraphics[width=12cm]{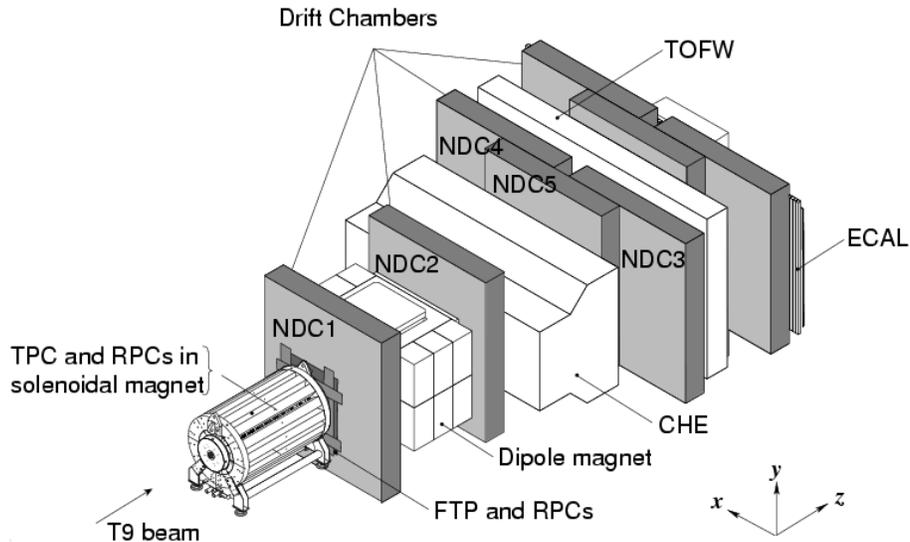}
  \end{center}
\caption{Schematic layout of the HARP detector. 
The convention for the coordinate system is shown in the lower-right
corner. 
The three most downstream (unlabelled) drift chamber modules are only partly
equipped with electronics and are not used for tracking. 
The detector covers a total length of 13.5 m along the beam axis 
and has a maximum width of 6.5 m perpendicular to the beam.
}
\label{fig:harp}
\end{figure*}

 The HARP detector is shown in Fig.~\ref{fig:harp} and is fully
 described in reference \cite{ref:harpTech}.
The forward spectrometer, mainly used in the analysis for the conventional
neutrino beams and atmospheric neutrino flux, comprises of a dipole magnet,
 large planar drift chambers 
(NDC)~\cite{NOMAD_NIM_DC} , a time-of-flight wall (TOFW) \cite{ref:tofPaper}, 
a threshold Cerenkov counter
(CHE) and an electromagnetic calorimer (ECAL).
%
In the large-angle region a cylindrical TPC with a radius of 408~\mm
is  positioned inside a solenoidal magnet with a field of 0.7~\T. 
The TPC detector was designed to measure and identify tracks in the
angular region from 0.25~\rad to 2.5~\rad with respect to the beam axis.
The target is placed inside the inner field cage (IFC) of the TPC such that,
in addition to particles produced in the forward direction, 
backward-going tracks can be measured.
The TPC is used
for tracking, momentum determination and measurement of the
energy deposition \dedx for particle identification~\cite{ref:tpc:ieee}.
A set of resistive plate chambers (RPC) form a barrel inside the solenoid 
around the TPC to measure the arrival time of the secondary
particles~\cite{ref:rpc}. 
Charged particle identification (PID) can be achieved by measuring the 
ionization per unit length in the gas (\dedx) as a function of the total
momentum of the particle. 
Additional PID can be performed through a time-of-flight 
measurement with the RPCs.

%
 
The momentum of the T9 beam is known with a precision of
the order of 1\%~\cite{ref:t9}. 
The absolute normalization of the number of incident protons was
performed using a total of 1,148,120 incident proton triggers. 
These are triggers where the same selection on the beam particle was
applied but no selection on the interaction was performed.
The rate of this trigger was down-scaled by a factor 64.
A cross-check of the absolute normalization was provided by counting
tracks in the forward spectrometer.


 Beam instrumentation provides identification of the incoming
 particle, the determination of the time when it hits the target, 
 and the impact point and direction of the beam particle
 on the target. 
 It is based on a set of four multi-wire proportional chambers (MWPC)
 to measure position and direction of the incoming beam particles 
 and time-of-flight detectors and 
 N$_2$-filled Cherenkov counters to identify incoming particles.  
 Several trigger detectors are installed to select events with an
 interaction and to define the normalization.

Besides the usual need for calibration of the detector, a number of
hardware shortfalls, discovered mainly after the end of data-taking,
had to be overcome to use the TPC data reliably in the analysis.
The TPC is affected by a relatively large number of dead or noisy 
pads and static and dynamic distortions of the reconstructed trajectories.
The applied corrections are briefly resumed in the next section.


The beam of positive particles used for this measurement contains mainly 
positrons, pions and protons, with small components of kaons and
deuterons and heavier ions.
Its composition depends on the selected beam momentum.
The proton fraction in the incoming beam varies from 35\% at
3 GeV/c to 92\% at 12 GeV/c.

At momenta higher than 5~\GeVc protons are selected by rejecting
particles with a measured signal in either of the beam Cherenkov
detectors.  
At 3~\GeVc the time-of-flight measurement allows the selection of
pions from protons to be made at more than 5{$\sigma$}.
Deuterons (and heavier ions) are removed by TOF measurements.  
The selection of protons for the beam momenta with the Cherenkov
detectors has been described in detail in Ref~\cite{ref:alPaper}.
More details on the beam particle selection can be found in 
Ref.~\cite{ref:harpTech}.
The purity of the selection of protons is higher than 99\% at all
momenta. 
A set of MWPCs is used to select events with only one
beam particle for which the trajectory extrapolates to the target.
An identical beam particle selection was performed for events
triggered with the incident-proton trigger in order to provide an
absolute normalization of the incoming protons.
This trigger selected every 64$^{th}$ beam particle coincidence
outside the dead-time of the data acquisition system.

The length of the accelerator spill is 400~ms with a typical intensity
of 15~000 beam particles per spill.
The average number of events recorded by the data acquisition ranges
from 300 to 350 per spill for the different beam momenta.
The analysis proceeds by first selecting a beam proton hitting the
target, not accompanied by other tracks. 
Then an event is required to give a large angle interaction (LAI) trigger  to be
retained. 
After the event selection the sample of tracks to be used for analysis
is defined.
Table~\ref{tab:events} shows the number of events and the number of $\pi^{\pm}$ 
selected in the analysis.
The large difference between the first and second set of rows (``total events
taken by the DAQ'' and ``Accepted protons with Large Angle
Interaction'') is due to the relatively large fraction of pions in the
beam and to the larger number of triggers taken for the measurements
with the forward dipole spectrometer.
 
\begin{table*}[htbp!] 
\caption{Total number of events and tracks used in the various nuclear 
  5\%~$\lambda_{\mathrm{I}}$ target data sets and the number of
  protons on target as calculated from the pre-scaled incident proton triggers.} 
\label{tab:events}
{\small
\begin{center}
\begin{tabular}{llrrrrrr} \hline
\bf{Data set (\bfGeVc)}          &         &\bf{3}&\bf{5}&\bf{8}&\bf{8.9} &\bf{12} &\bf{12.9}\\ \hline
    Total DAQ events     & (Be)     & 1399714 & 1473815      & 1102415    &  7236396        &  1211220        &   --            \\
                         &  (C)     & 1345461 & 2628362      & 1878590    &  --             &  1855615        &   --              \\
                         & (Al)     & 1586331 & 1787620      & 1706919    &   --            &  619021         &  5401701        \\
                         & (Cu)     & 623965  & 2089292      & 2613229    &   --            &  748443         &  --              \\
                         & (Sn)     & 1652751 & 2827934      & 2422110    &   --            &  1803035        &  --     \\
                         & (Ta)     & 2202760 & 2094286      & 2045631    &   --            &  886307         &  -- \\   
                         & (Pb)     & 1299264 & 2110904      & 2314552    &   --            &  486875         &  -- \\   
  Acc. protons with LAI  &  (Be)    & 76694   &  157625      &  200352    &  1267418        &  282272         &   --             \\
                         & (C)     & 58421   &  228490      &  337150    &   --            &  504945         &  --         \\
                         &  (Al)    & 69794   &  195912      & 341687    &   --            &  169151         &  1391159              \\
                         &  (Cu)    & 38290   &  229316      &  544615    &   --            &  226245         &  --              \\
                         &  (Sn)    & 84330   &  304949      &  523432    &   --            &  558306         &  --              \\
                         &  (Ta)    & 97732   &  218293      &  442625    &   --            &  269927         &  --              \\
                         &  (Pb)    & 79188   &  194064      &  491672    &   --            &  145843         &  --              \\
  Fraction of triggers used   & (Be)  & 79\%  &  75\%       &   83\%      &  94\%         &    79\%         &   --              \\
                              & (C)   &  95\% &  90\%       &   83\%      &  --           &    84\%         &  --            \\   
                              & (Al)  & 78\%  &  80\%       &   63\%      &  --             &  96\%        &   72\%             \\
                              & (Cu)  & 91\%  &  76\%       &   66\%      &  --             &   76\%        &   --             \\
                              & (Sn)  & 97\%  &  73\%       &   67\%      &  --             &   76\%        &   --             \\
                              & (Ta)  & 86\%  &  81\%       &   69\%      &  --             &   76\%        &   --             \\
                              & (Pb)  & 74\%  &  56\%       &   69\%      &  --             &   50\%        &   --             \\
  \bf{$\bfpim$ selected with PID} & (Be)  & 6553    &   20020      &  32078       &  231278         &  47608    &   --              \\
                                  & (C)   & 4831    &   33436      &  52105       &  --             &  72307    &   --          \\
                                  & (Al)  & 5496    &   26502      &  45442       &  --             &  37812    &   250037               \\ 
                                  & (Cu)  & 3065    &   28395      &  79497       &  --             &  46153    &   --               \\
                                  & (Sn)  & 7146    &   38337      &  89799       &  --             &  124925   &   --               \\
                                  & (Ta)  & 6758    &   27767      &  74977       &  --             &  63349    &   --               \\
                                  & (Pb)  & 4408    &   17766      &  81821       &  --             &  25050    &   --               \\
  \bf{$\bfpip$ selected with PID} & (Be)  & 11245   &   27796      &  41683       &  294594         &  58882    &   --              \\
                                  & (C)   & 9944    &   52633      &  73157       &  --             &  95151    &   --          \\
                                  & (Al)  & 9519    &   38657      &  59345       &  --             &  47609    &   314552               \\
                                  & (Cu)  & 4976    &   39823      &  102797       &  --            &  56665    &   --               \\
                                  & (Sn)  & 10179   &   48820      &  104239       &  --            &  145923    &   --               \\
                                  & (Ta)  & 9270    &   33985      &  87226       &  --             &  72275    &   --               \\
                                  & (Pb)  & 6160    &   21074      &  92913       &  --             &  27085         &   --               \\

\end{tabular}
\end{center}
}
\end{table*}

\subsection{Corrections for distortion of tracks in the TPC}
\label{sec;dyn}
The TPC contains a relatively large number of dead or noisy pads.
Noisy pads were considered equivalent to dead channels in this analysis.
The large number of dead pads ($\simeq 15 \%$) in the experimental
data taking required a day-by-day determination of the dead channel map.
The same map was used in the simulation to provide a description of
the TPC performances on a short term scale.
A method based on tracks accumulated during the data taking was used
to measure the gain variations of each pad, see Ref.~\cite{ref:harpTech}
for details. It is used to reduce the fluctuations in response
between different pads down to a $3 \%$ level.  

Static distortions on reconstructed tracks are caused by the
inhomogeneity of the electric field, 
due to an accidental mismatch between the inner and outer field cage 
(powered by two distinct HV supplies).
The day--by--day variations of this mismatch are consistent with the
specifications of the stability and reproducibility of the power
supplies.   
A specific calibration for each setting has been made.

\begin{figure*}
\begin{center}
\includegraphics[width=0.40\textwidth]{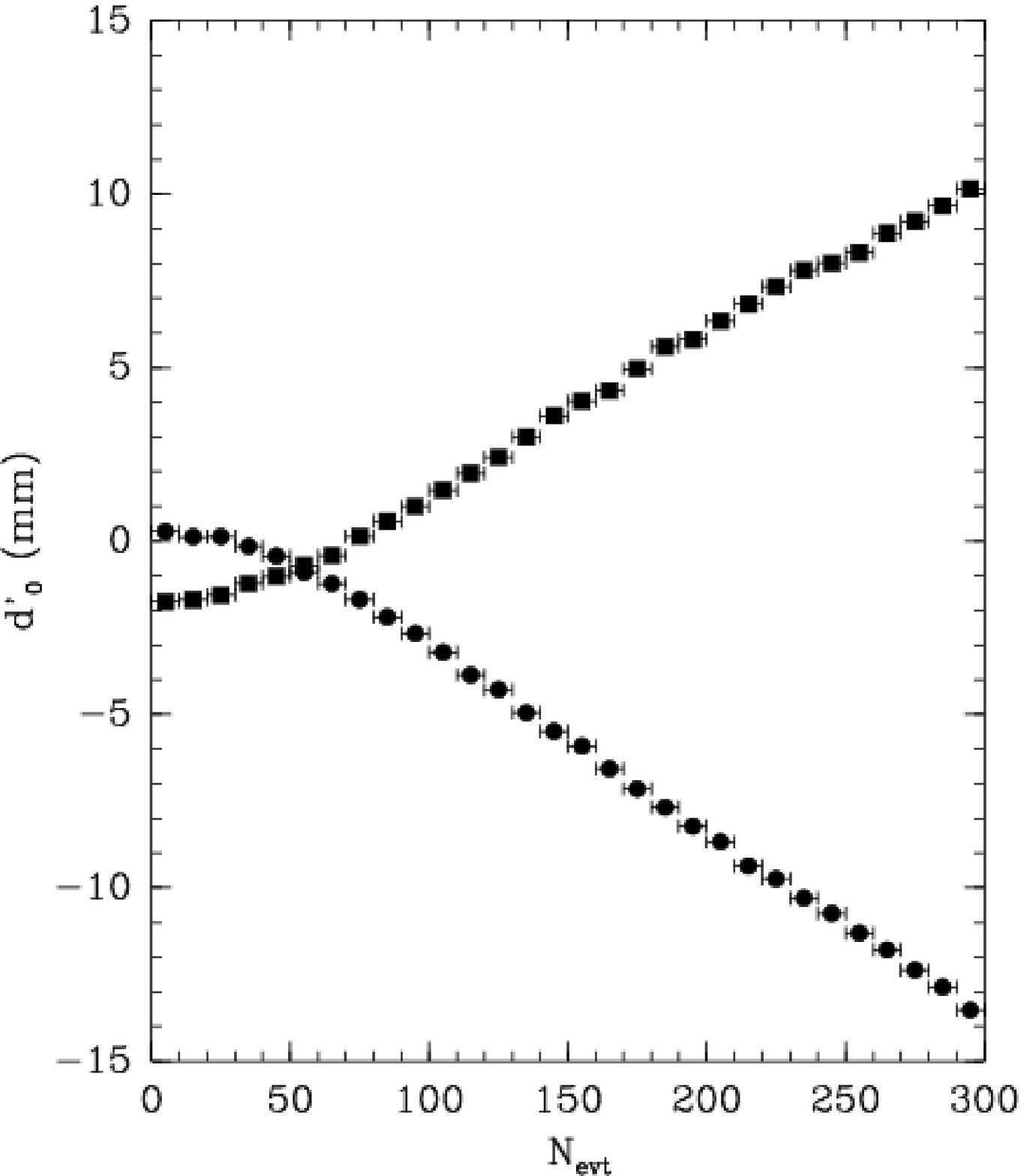}
~
\includegraphics[width=0.40\textwidth]{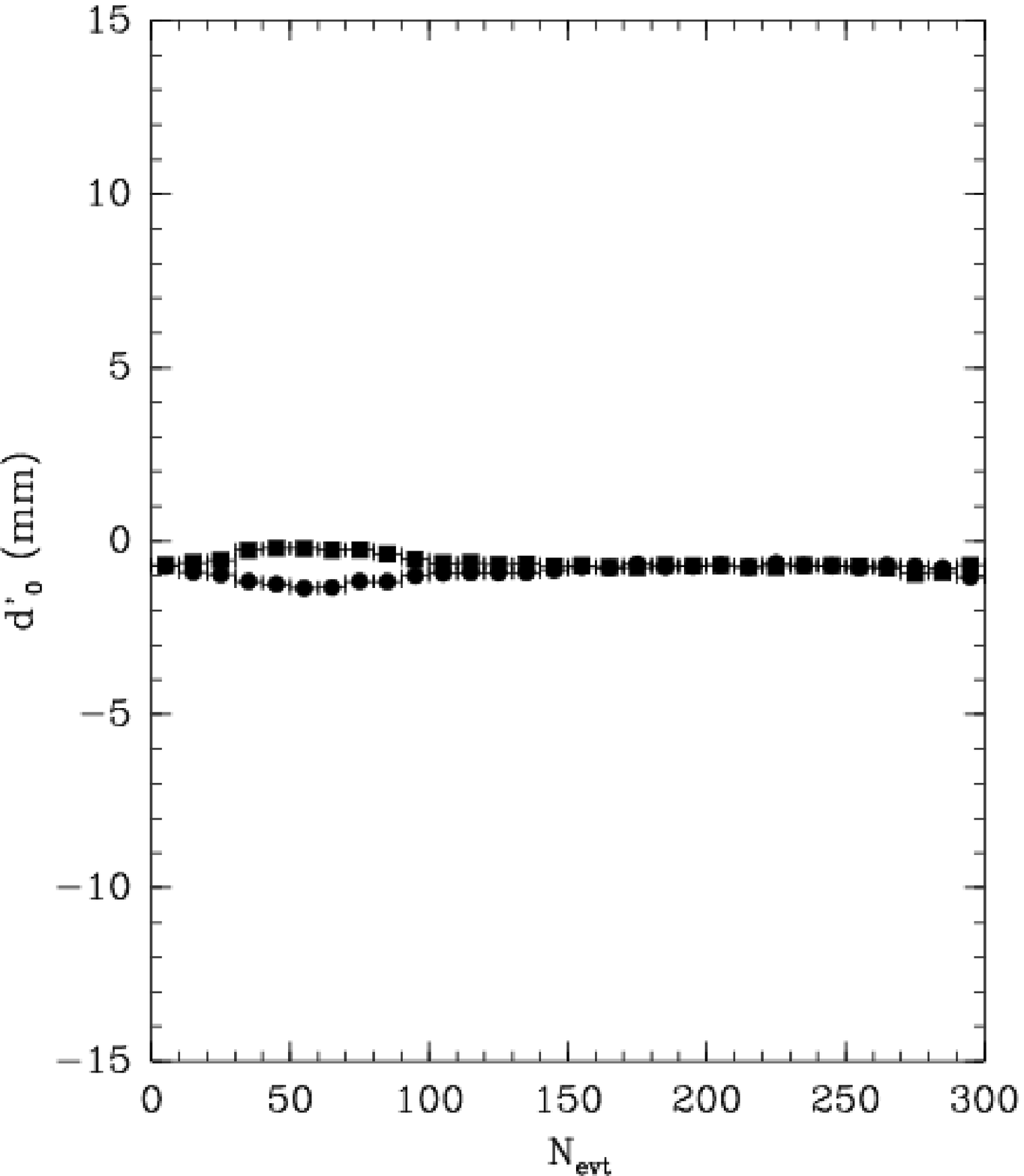}
\end{center}
\caption{Effect of dynamic distortions on a track in the HARP TPC for a
p--Be interactions sample at 5 GeV/c, as a 
function of  the event number in the spill. Left panel: before corrections, 
right panel: after dynamic distortion corrections. The {\it symbols} show the
average extrapolated distance from the incoming beam particle trajectory for
$\pi^{-} (filled \ squares)$, $\pi^{+} (filled \  circles)$}
\label{fig:dyn}
\end{figure*}

Dynamic distortions are caused instead by the
build-up of ion-charge density in the drift volume during the 400~ms 
long beam spill. 
All these effects were fully studied and available corrections are described 
in detail in Ref.~\cite{ref:harp:tantalum,ref:tpcmom}. In our earlier
published analyses a practical approach has been
followed. Only the events corresponding to the
early part of the spill, where the effects of the dynamic distortions are
still small, have been used (this translates into a cut on the maximum 
number of events ($N_{evt} \simeq 100$) to be retained). 
The time interval between spills 
is large enough to drain all charges in the TPC related to the effect of the beam.
The combined effect of the distortions on the kinematic quantities used
in the analysis has been studied in detail and only that part of the data
for which the systematic errors can be assessed with physical 
benchmarks was used, as fully explained in \cite{ref:harp:tantalum}. 
More than $30-40 \%$ of the
recorded data were thus used in the published analyses.
The influence of distortions was monitored taking the average value of 
the extrapolated minimum distance of the secondary tracks from the 
incoming beam particle trajectory $< d'_{0} >$.
An example of the result of the corrections for one setting is shown in
Fig.~\ref{fig:dyn}. 

In the presented analysis, instead, the TPC track dynamic distortions are
corrected on an event--by--event basis, as outlined in
\cite{ref:tpcmom}.
A direct measurement of the distortions as a function of the radius $R$ and
time--in--spill was performed using the prediction of the trajectory of
the recoil proton in elastic scattering events on a liquid hydrogen
target. 
The measurement of the direction of the  forward scattered proton
determined the kinematics and predicts the trajectory of the recoil
proton. 
The actual measurements in the chamber are then compared with the
prediction as a funtion of time.
In addition to this direct measurement also a model of ion charges and
their effect was developed. 

The strength of the effect depends on many parameters such as the beam
intensity, the momentum and the target.
An iterative procedure is applied to find the value of the strength
parameter until $< d'_{0} >$ is flattened down to 
$\pm 2$ mm, by using the empirical model described below to shape the
corrections.  
Taking into account the beam intensity, the data acquisition rate and the 
target thickness, the HARP experiment was operated in conditions of
$\approx 95 \%$
deadtime. The electrons are normally multiplied near the TPC plane with a 
multiplication factor $\sim 10^{5}$, producing an equivalent number of Ar 
ions. Any inefficiency of the gating grid at the $10^{-4}-10^{-5}$ level 
would give an overwhelming number of ions drifting in the TPC gas volume.
The phenomenological model is based on the fact that the field which is
responsible for the force acting on each drift electron is equivalent to:
\begin{itemize}
\item{} a field system where ions, in a given angular section at $R$ values
        internal to the drift electron position contribute to attract the drift 
        electrons inwards;
\item{} a field system where ions, in a given angular section at $R$ values
        external to the drift electron position, contribute to attract the drift
        electrons outwards.
\end{itemize}

This model makes it possible to understand all the peculiar features of the TPC dynamic 
distortions:

\begin{itemize}
\item{} the dependence of the distortion on the event number in the spill,
\item{} the dependence of the distortion on tracks generated at
        different longitudinal coordinate $Z$
        in the TPC. In particular tracks produced at increasingly larger $Z$
        exhibit the distortion saturation at increasingly later times and 
        the distortions tend to zero at $Z$ values already passed by the ion packet;
\item{} the dependence of the distortion of cosmic-ray tracks
        collected out of the spill
        as a function of time and $Z$, with the non-trivial fact that cosmic-rays just
        after the spill are more distorted than the cosmic-rays taken later;
\item{} the $R-{\phi}$ dependence as measured with elastic scattering. 
        The distortions have a 
        peculiar behaviour as a function of the TPC pad rows: from the $E \times B$ 
        calculation it follows that the inner rows are distorted by a radial electric
        field pointing inwards; there exists a pad row around the middle of the chamber
        where the radial electric field vanishes.
\end{itemize}

\begin{figure*}
\begin{center}
\includegraphics[width=0.40\textwidth]{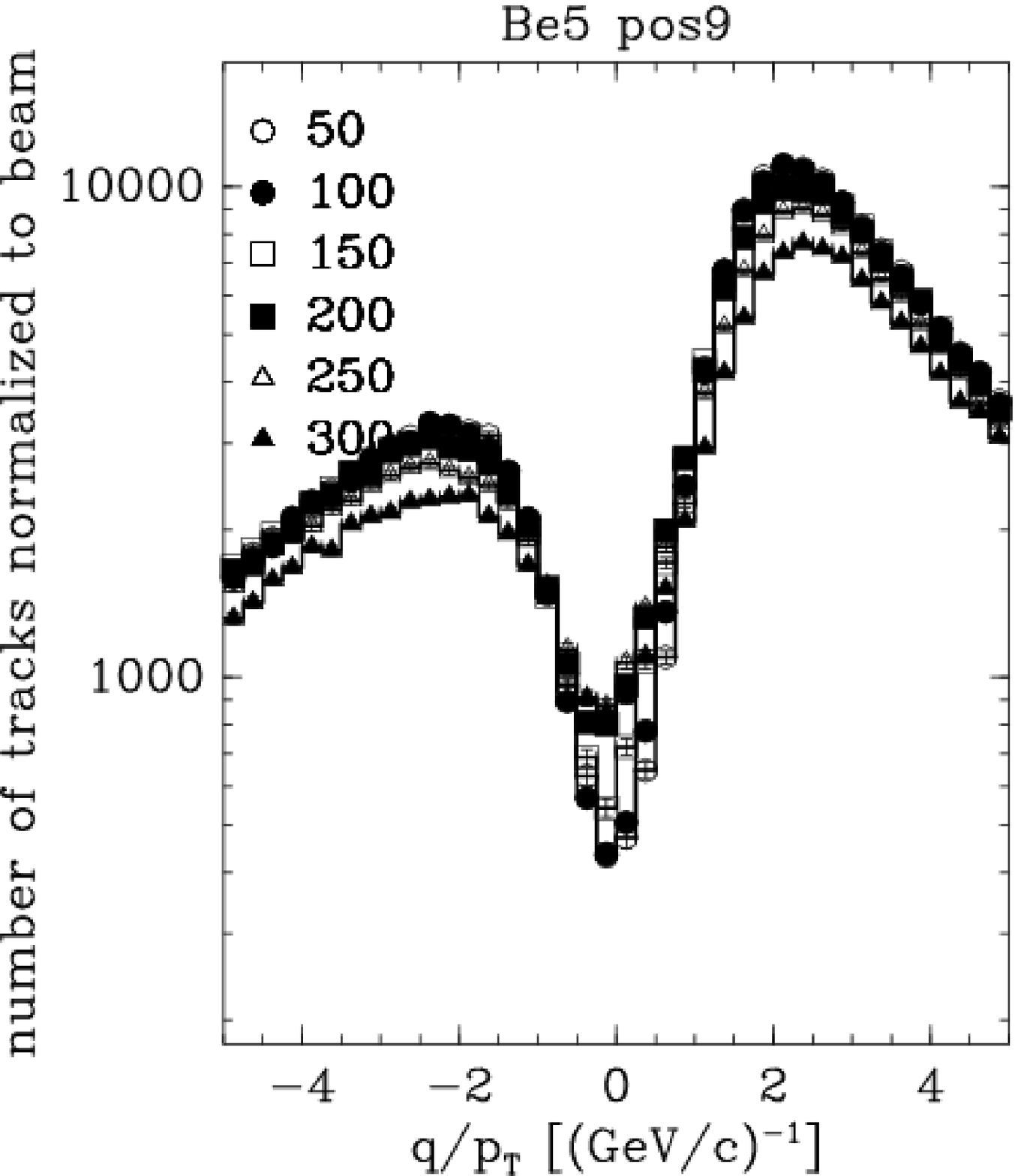}
\includegraphics[width=0.40\textwidth]{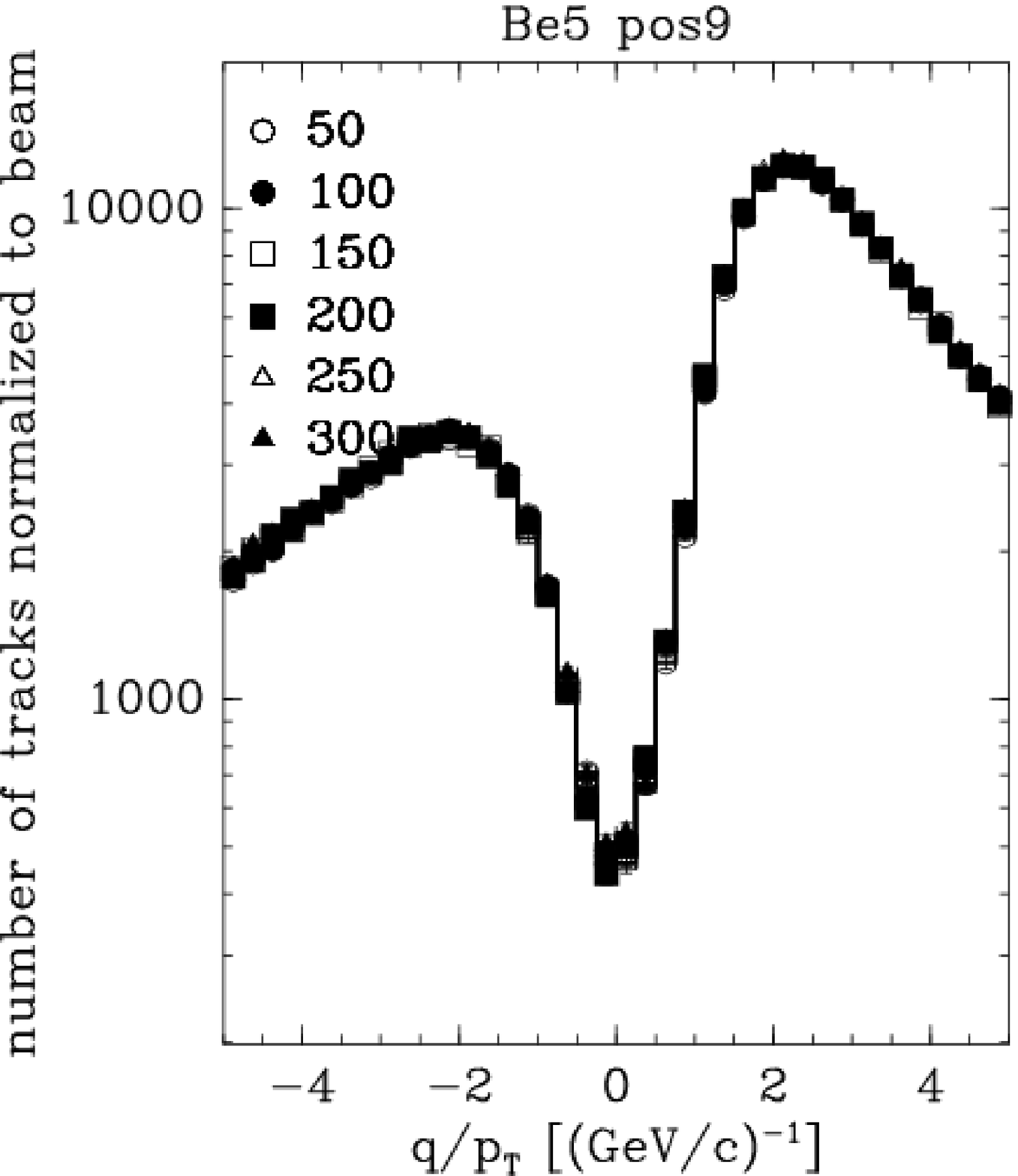}
\end{center}
\caption{Full spill analysis of $Q/p_T$ for the highest statistics
data sample: p--Be at 8.9 GeV/c. Left panel: before corrections, 
right panel: after corrections. Six curves are drawn, each for the next 50
events in the spill.
One notices that the distributions in the right panel are not distinguishable.
}
\label{fig:model}
\end{figure*}

The corrections, provided by the model, calibrated on a run-by-run basis allow a full
control of the TPC response along the spill. In figure \ref{fig:model} the $Q/p_{T}$ spectrum 
for p--Be interactions at 8.9 GeV/c is shown separately for the first 50 events in the spill,
the next 50 and so on up to the last 50, before and after the dynamic distortion 
correction.  After the correction, the curves are compatible within the statistical
errors, indicating that the correction is adequate.

\begin{figure}
\begin{center}
\includegraphics[width=0.45\textwidth,angle=0]{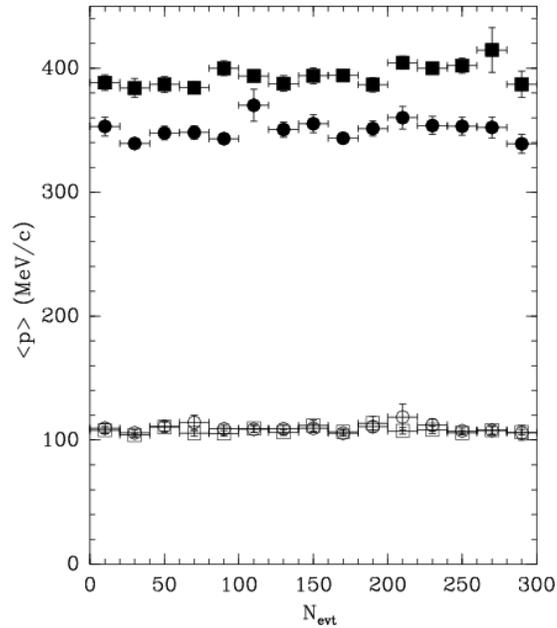}
\end{center}
\caption{
As a momentum benchmark, after the dynamic distortion correction,
the closed box shows the average momentum observed for protons selected
 using their range (reaching the second RPC) and \dedx. Closed circles
 show protons selected within a high \dedx region; open circles: \pim
 selected with \dedx; open boxes: \pip selected with \dedx.
The angle of the particles is restricted in a range with $\sin \theta
 \approx 0.9$.
The variation in the uncorrected sample was 
 $\approx 5\%$ for the high \pt samples.
The corrected data stay stable well within 3\%.
 The low \pt data remain stable with or without correction.
}
\label{fig:be:momentum}
\end{figure}

The main point, using beginning of spill data or full spill data,
is the presence of possible residual momentum bias in the TPC measurement 
due to the dynamic distortions. A dedicated paper~\cite{ref:tpcmom}
addresses this point and shows that our estimation of momentum bias
is below $3\%$, although the systematics can in principle be different in 
the uncorrected begin-of-spill data and the fully corrected data.
From the studies made we conclude that the data of the full spill can be
used for the analysis once the corrections for dynamic distortions have
been applied. 
In a small number of data sets the distortions in the last part of the
spill are too large to be reliably corrected.
These are mainly high $A$, high beam momentum settings, where sufficient
statistics are available even without this part of the spill.
The reliability of the correction has been checked by observing the
stability as a function of the event number in the spill of the average
momentum of protons in a small window of large \dedx. 
An additional benchmark assesses the stability of the momentum
measurement inside a spill 
after the dynamic distortion correction for the TPC tracks.
The dependence of the average momentum for four different track samples
as a function of \evtspill is shown in Fig.~\ref{fig:be:momentum}.
The range of \pt tested by this benchmark covers nearly the full range
used in the analysis.

\begin{figure*}[tbp]
\begin{center}
\includegraphics[width=0.45\textwidth]{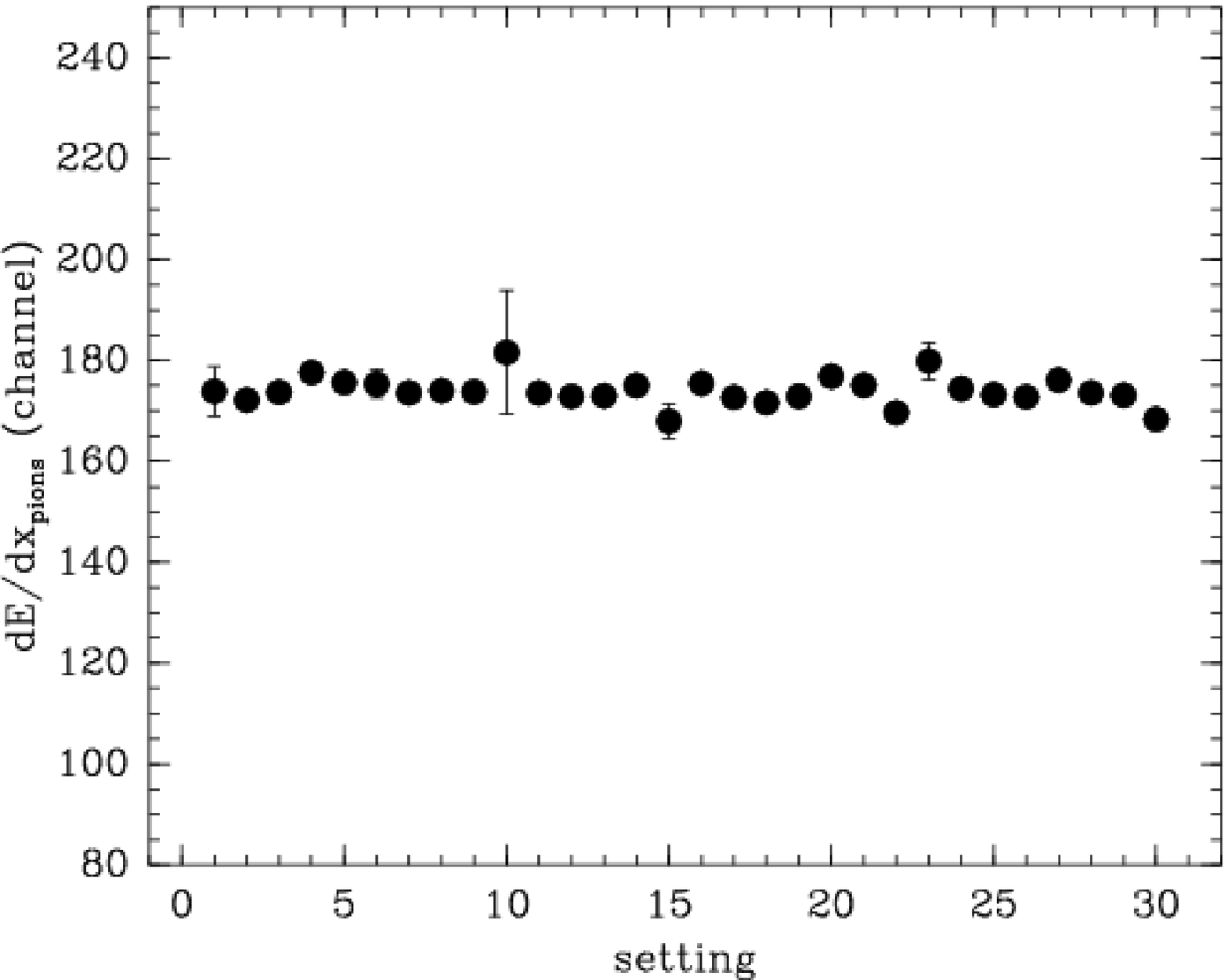}
\includegraphics[width=0.45\textwidth]{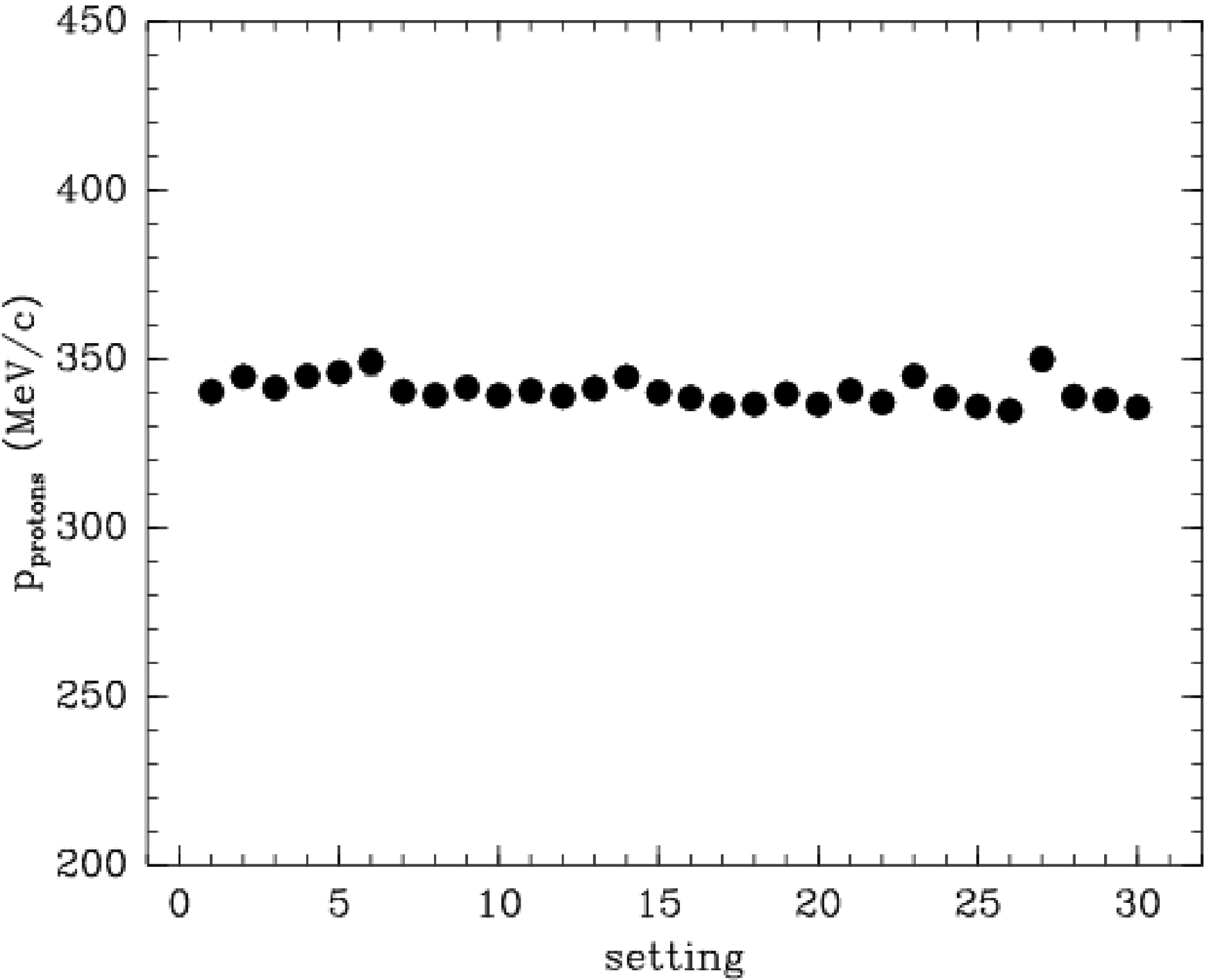}

\caption{
Stability of the TPC calibration. Left panel: \dedx versus settings for 
pions with $300 \leq p \leq 500$ MeV/c. Right panel: mean momentum for 
protons with a \dedx between 7 and 8 MIP versus setting, Settings go from
low $A$ to high $A$ including beam momenta from 3 \GeVc to 12.9 \GeVc.}
\label{fig:trend}
\end{center}
\end{figure*}

The stability of the TPC \dedx and momentum calibration over the
collected data sets is shown in Fig.~\ref{fig:trend}. 
The $X$-axis runs over all settings starting at  low $A$ (Be) up to high
$A$ such (Pb) and all used beam momenta from 3~\GeVc to 12.9~\GeVc.

\section{Data analysis}
\label{sec:analysis}

Only a short outline of the data analysis procedure is presented here, for further details see Ref. \cite{ref:harp:tantalum}. The most relevant difference is
the use of the full spill statistics by means of a correction of the 
dynamic distortions in the TPC tracks, as outlined before.

The double-differential cross-section for the production of a particle of 
type $\alpha$ can be expressed in the laboratory system as:

\begin{equation}
{\frac{{\mathrm{d}^2 \sigma_{\alpha}}}{{\mathrm{d}p_i \mathrm{d}\theta_j }}} =
\frac{1}{{N_{\mathrm{pot}} }}\frac{A}{{N_A \rho t}}
\frac{1}{\Delta p_j \Delta \theta_j}
 \sum_{i',j',\alpha'} M_{ij\alpha i'j' \alpha'}^{-1} \cdot
{N_{i'j'}^{\alpha'} } 
\ 
\label{eq:cross}
\end{equation}

where $\frac{{\mathrm{d}^2 \sigma_{\alpha}}}{{\mathrm{d}p_i \mathrm{d}\theta_j }}$
is expressed in bins of true momentum ($p_i$), angle ($\theta_j$) and
particle type ($\alpha$).
$\Delta p_J$ and $\Delta \theta_j$ are the bin sizes in momentum and and
angle, respectively.



The `raw yield' $N_{i'j'}^{\alpha'}$ 
is the number of particles of observed type $\alpha'$ in bins of reconstructed
momentum ($p_{i'}$) and  angle ($\theta_{j'}$). 
These particles must satisfy the event, track and PID 
selection criteria.
Although, owing to the stringent PID selection,  the background from
misidentified protons in the pion sample is small, the pion and proton
raw yields ($N_{i'j'}^{\alpha'}$, for 
$\alpha'=\pim, \pip, \mathrm{p}$) have been measured simultaneously. 
It is thus possible to correct for the small remaining proton
background in the pion data without prior assumptions concerning the
proton production cross-section.

The matrix $ M_{ij\alpha i'j' \alpha'}^{-1}$ 
corrects for the  efficiency and the resolution of the detector. 
It unfolds the true variables $ij\alpha$ from the reconstructed
variables $i'j'\alpha'$  with a Bayesian technique~\cite{dagostini} 
and corrects  
the observed number of particles to take into account effects such as 
trigger efficiency, reconstruction efficiency, acceptance, absorption,
pion decay, tertiary production, 
PID efficiency, PID misidentification and electron background. 
The method used to correct for the various effects is  described in
more detail in Ref.~\cite{ref:harp:tantalum}.

In order to predict the population of the migration matrix element 
$M_{ij\alpha i'j'\alpha'}$, the resolution, efficiency
and acceptance of the detector are obtained from the MonteCarlo.
This is accurate provided the MonteCarlo
simulation describes these quantities correctly. 
Where some deviations
from the control samples measured from the data are found, 
the data are used to introduce (small) {\em ad hoc} corrections to the
MonteCarlo. 
Using the unfolding approach, possible known biases in the measurements
are taken into account automatically as long as they are described by
the MonteCarlo.
In the experiment simulation, which is based on the GEANT4
toolkit~\cite{ref:geant4}, the materials in the beam-line and the 
detector are accurately described as well as
the relevant features of the detector response and 
the digitization process.
In general, the MonteCarlo
simulation compares well with the data, as shown in
Ref.~\cite{ref:harp:tantalum}. 
For all important issues physical benchmarkes have been used to validate
the analysis.
The absolute efficiency and the measurement of the angle and momentum
was determined with elastic scattering. 
The momentum and angular resolution was determined exploiting the two
halves of cosmic-ray tracks crossing the TPC volume.
The efficiency of the particle identification was checked using two
independent detector systems.
Only the latter needs a small {\em ad hoc} correction compared to the
simulation.  

The factor  $\frac{A}{{N_A \rho t}}$ in Eq.~\ref{eq:cross}
is the inverse of the number of target nuclei per unit area
($A$ is the atomic mass,
$N_A$ is the Avogadro number, $\rho$ and $t$ are the target density
and thickness)~\footnote{We do not make a correction for the attenuation
of the proton beam in the target, so that strictly speaking the
cross-sections are valid for a $\lambda_{\mathrm{I}}=5\%$ target.}. 
The result is normalized to the number of incident protons on the target
$N_{\mathrm{pot}}$. 
The absolute normalization of the result is calculated in the first
instance relative to the number of incident beam particles accepted by
the selection. 
After unfolding, the factor  $\frac{A}{{N_A \rho t}}$ is applied.
The beam normalization using down-scaled incident proton triggers 
has uncertainties smaller than 2\%
  for all beam momentum settings.

The background due to interactions of the primary
protons outside the target (called `Empty target background') is
measured using data taken without the target mounted in the target
holder.
Owing to the selection criteria which only accept events from the
target region and the good definition of the interaction point this
background is negligible ($< 10^{-5}$).

The effects of the systematic uncertainties on the final results are estimated
by repeating the analysis with the relevant input modified within the
estimated uncertainty intervals.
In many cases this procedure requires the construction of a set of
different migration matrices.
The correlations of the variations between the cross-section bins are
evaluated and expressed in the covariance matrix.
Each systematic error source is represented by its own covariance
matrix.
The sum of these matrices describes the total systematic error.
The magnitude of the systematic errors and their dependence on momentum
and angle will be shown in Section \ref{sec:results}.

\section{Experimental results}
\label{sec:results}

\begin{figure*}[htbp]
\begin{center}
\includegraphics[width=0.48\textwidth]{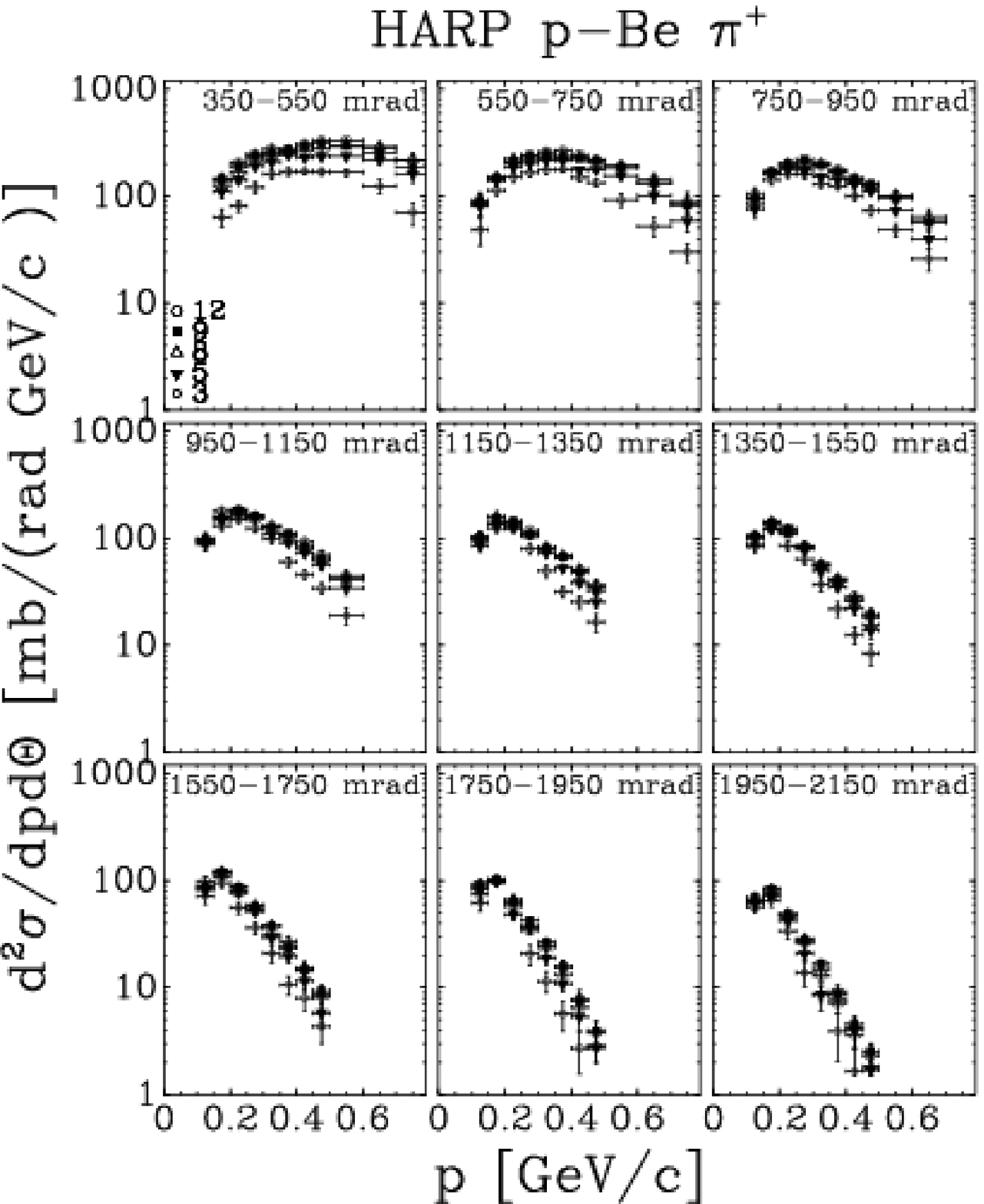}
\includegraphics[width=0.48\textwidth]{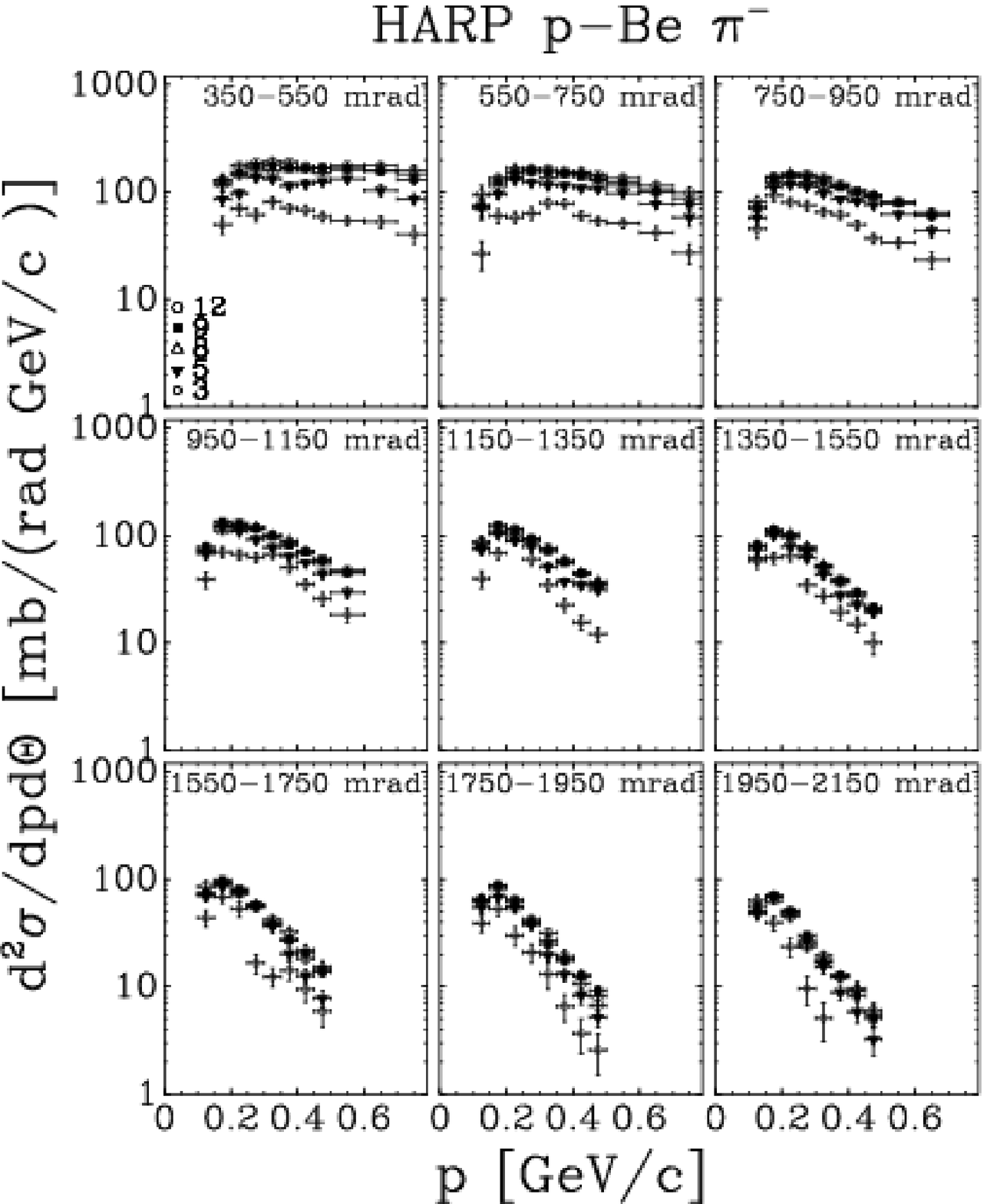}

\caption{
Double-differential cross-sections for \pip production (left) and  \pim
 production (right) in
p--Be interactions as a function of momentum displayed in different
angular bins (shown in \mrad in the panels).
In the figure, the symbol legend 9 refers to 8.9~\GeVc nominal
beam momentum.
The error bars represent the combination of statistical and systematic
 uncertainties. 
}
\label{fig:xs-p-th-pbeam-be}
\end{center}
\end{figure*}

\begin{figure*}[tbp]
\begin{center}
\includegraphics[width=0.48\textwidth]{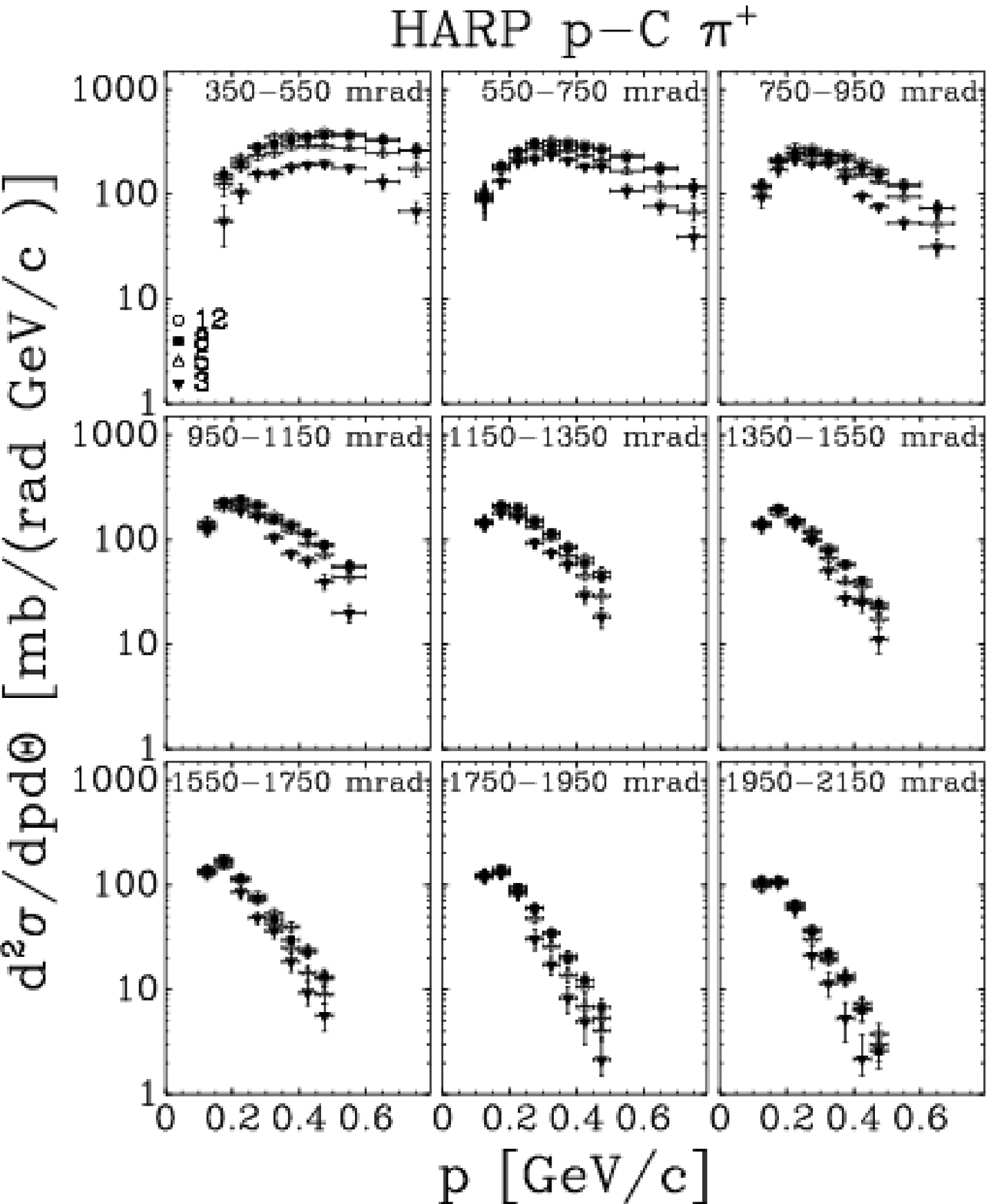}
\includegraphics[width=0.48\textwidth]{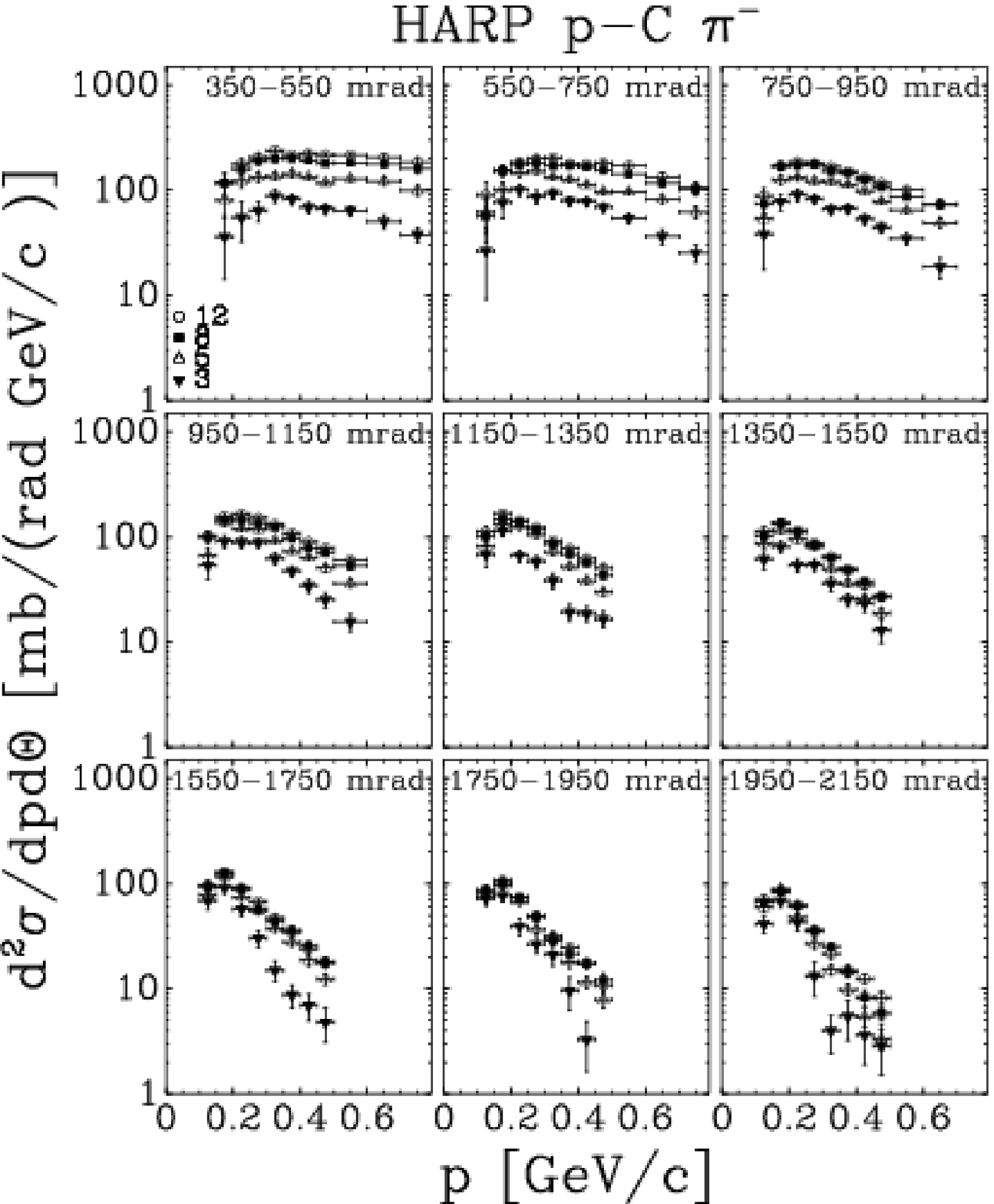}
\caption{
Double-differential cross-sections for \pip production (left) and  \pim
 production (right) in
p--C interactions as a function of momentum displayed in different
angular bins (shown in \mrad in the panels).
The error bars represent the combination of statistical and systematic
 uncertainties. 
}
\label{fig:xs-p-th-pbeam-c}
\end{center}
\end{figure*}

\begin{figure*}[tbp]
\begin{center}
\includegraphics[width=0.48\textwidth]{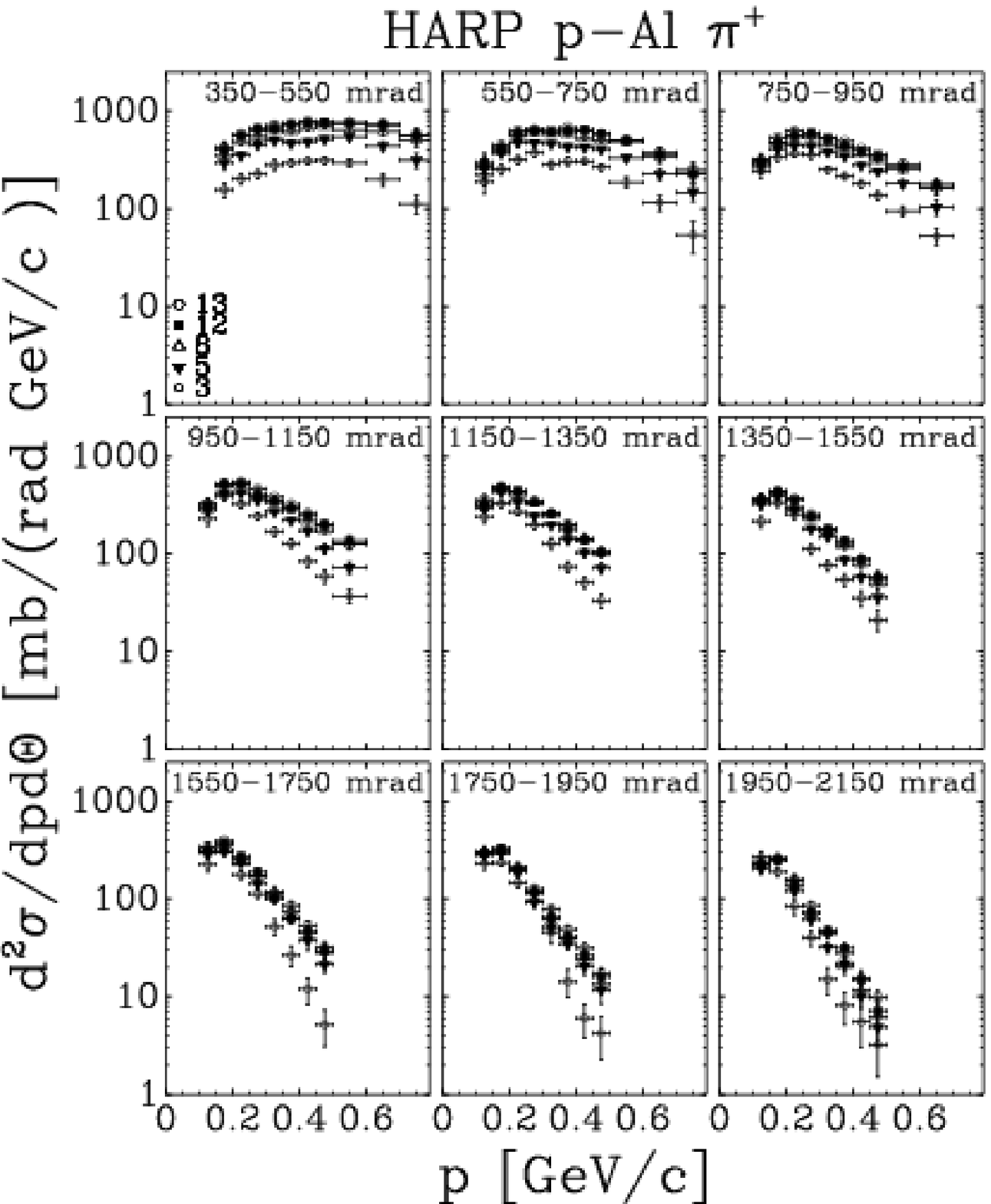}
\includegraphics[width=0.48\textwidth]{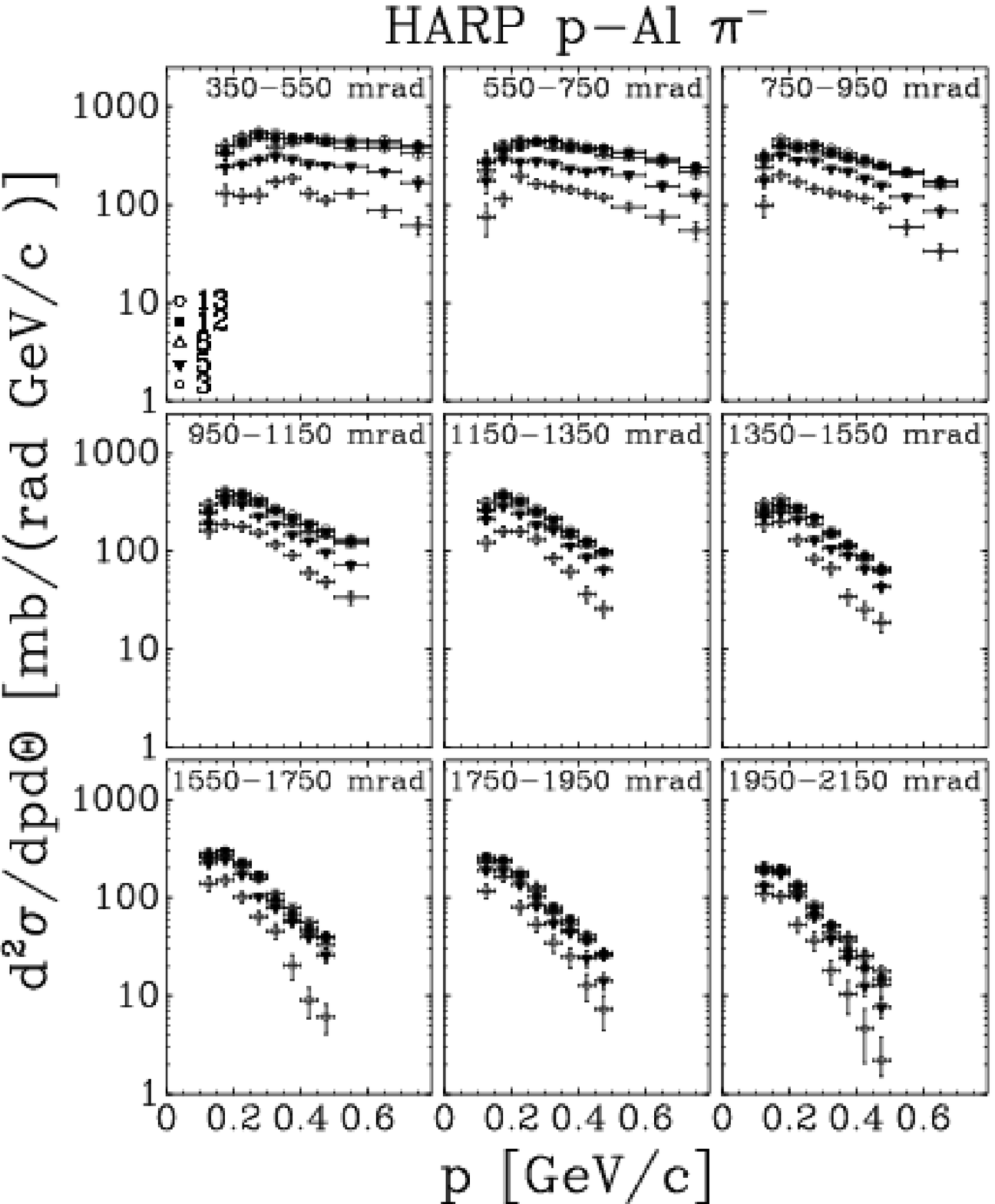}
\caption{
Double-differential cross-sections for \pip production (left) and  \pim
 production (right) in
p--Al interactions as a function of momentum displayed in different
angular bins (shown in \mrad in the panels).
In the figure, the symbol legend 13 refers to 12.9~\GeVc nominal
beam momentum.
The error bars represent the combination of statistical and systematic
 uncertainties. 
}
\label{fig:xs-p-th-pbeam-al}
\end{center}
\end{figure*}

\begin{figure*}[tbp]
\begin{center}
\includegraphics[width=0.48\textwidth]{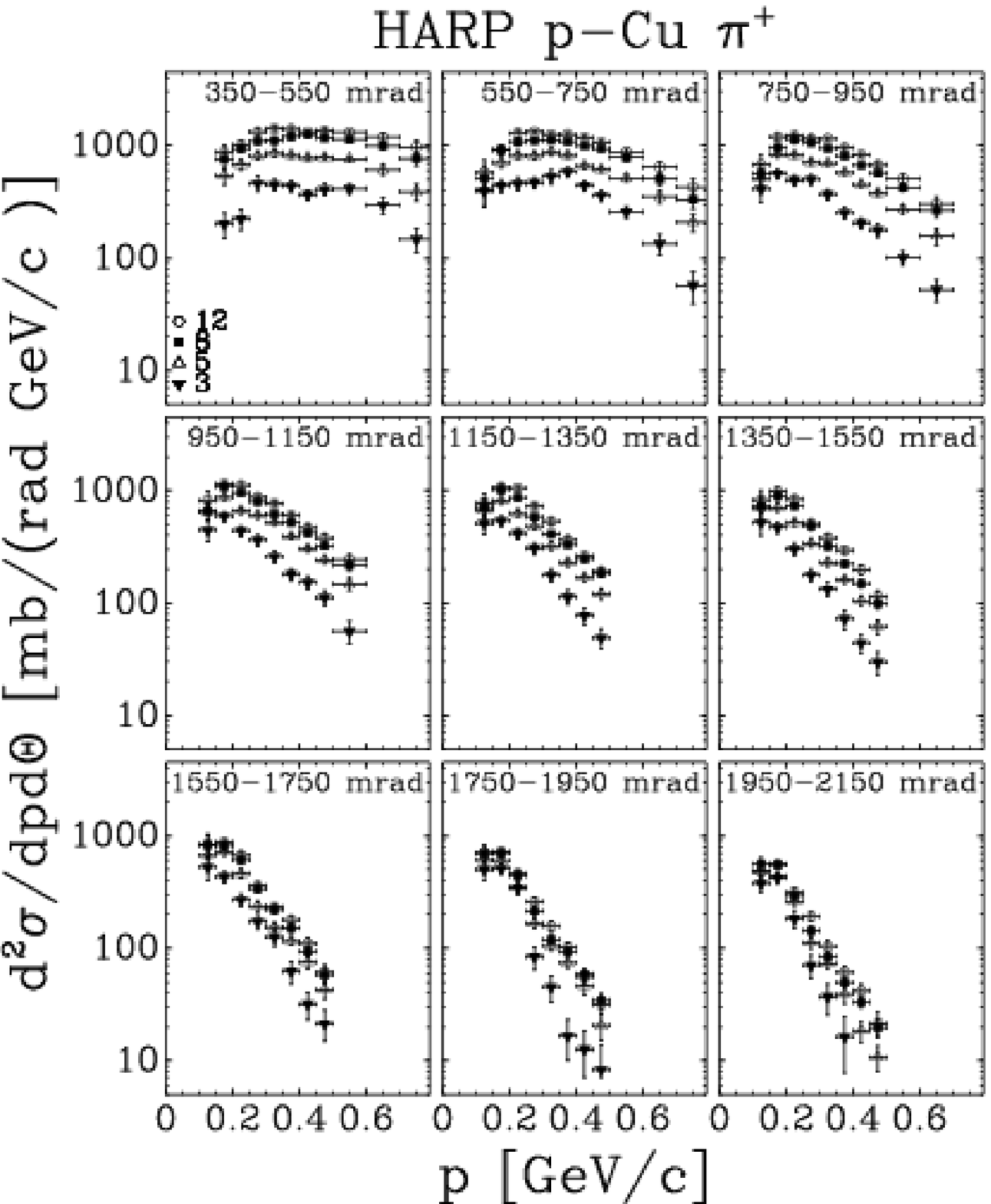}
\includegraphics[width=0.48\textwidth]{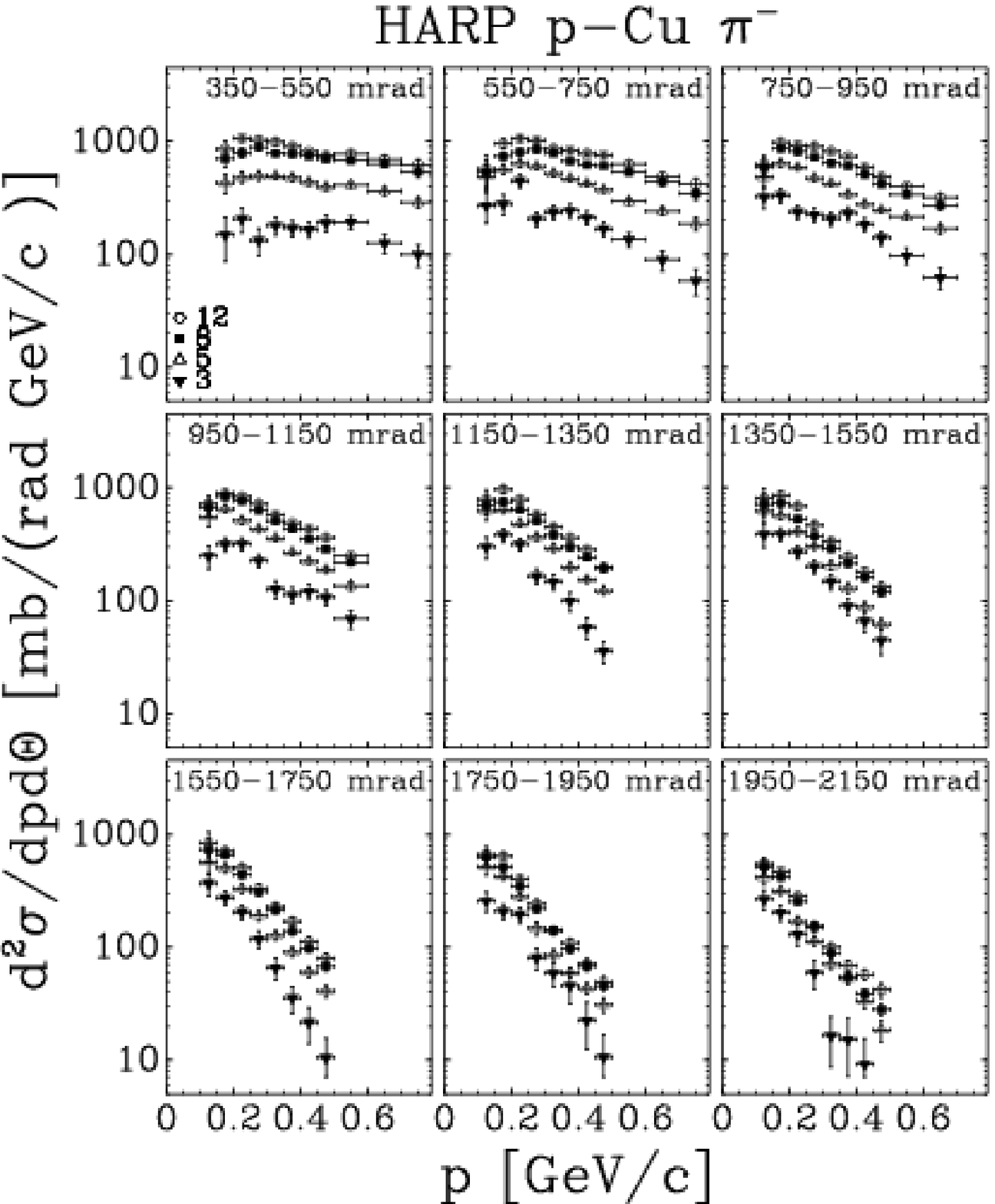}
\caption{
Double-differential cross-sections for \pip production (left) and  \pim
 production (right) in
p--Cu interactions as a function of momentum displayed in different
angular bins (shown in \mrad in the panels).
The error bars represent the combination of statistical and systematic
 uncertainties. 
}
\label{fig:xs-p-th-pbeam-cu}
\end{center}
\end{figure*}

\begin{figure*}[tbp]
\begin{center}
\includegraphics[width=0.48\textwidth]{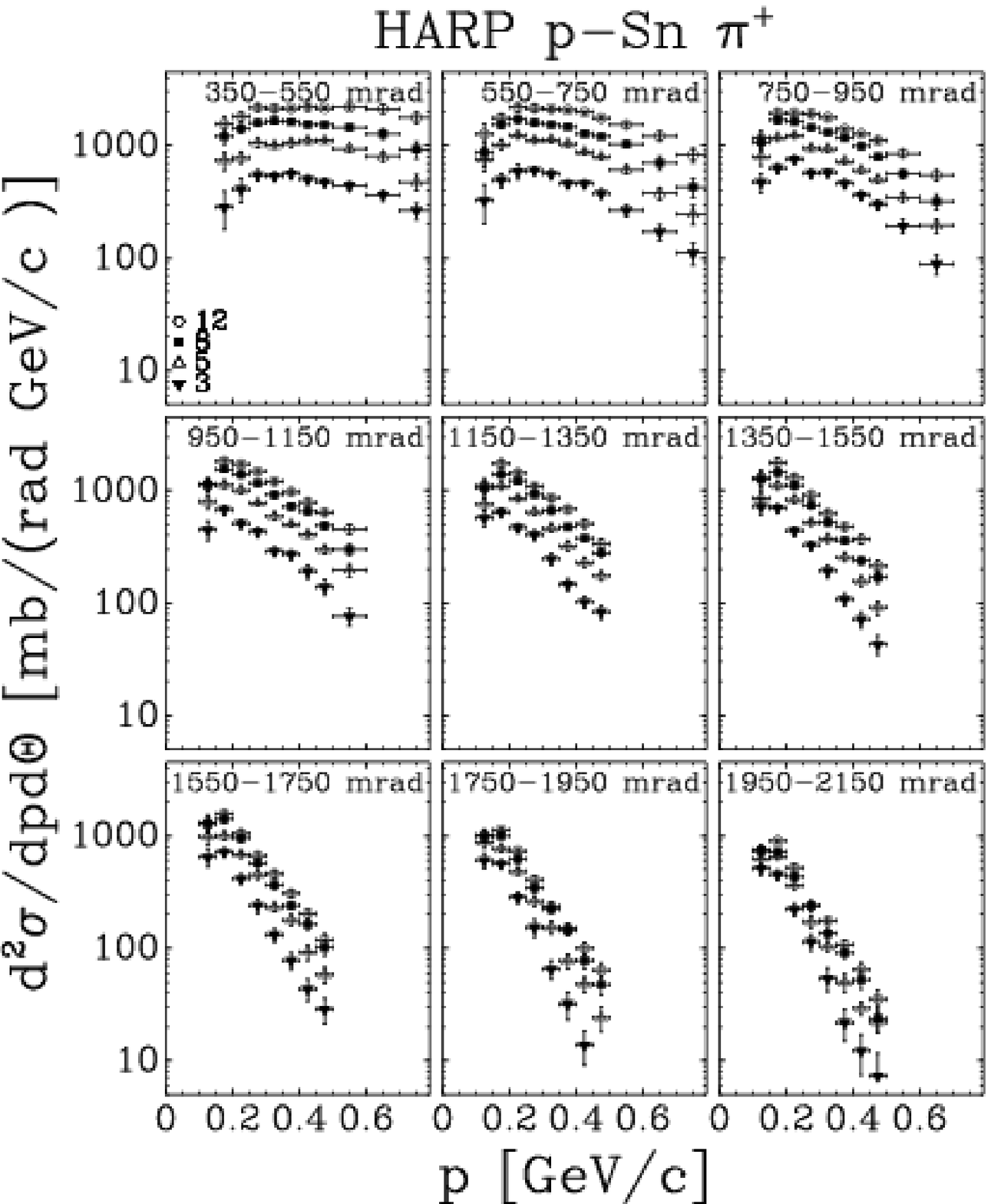}
\includegraphics[width=0.48\textwidth]{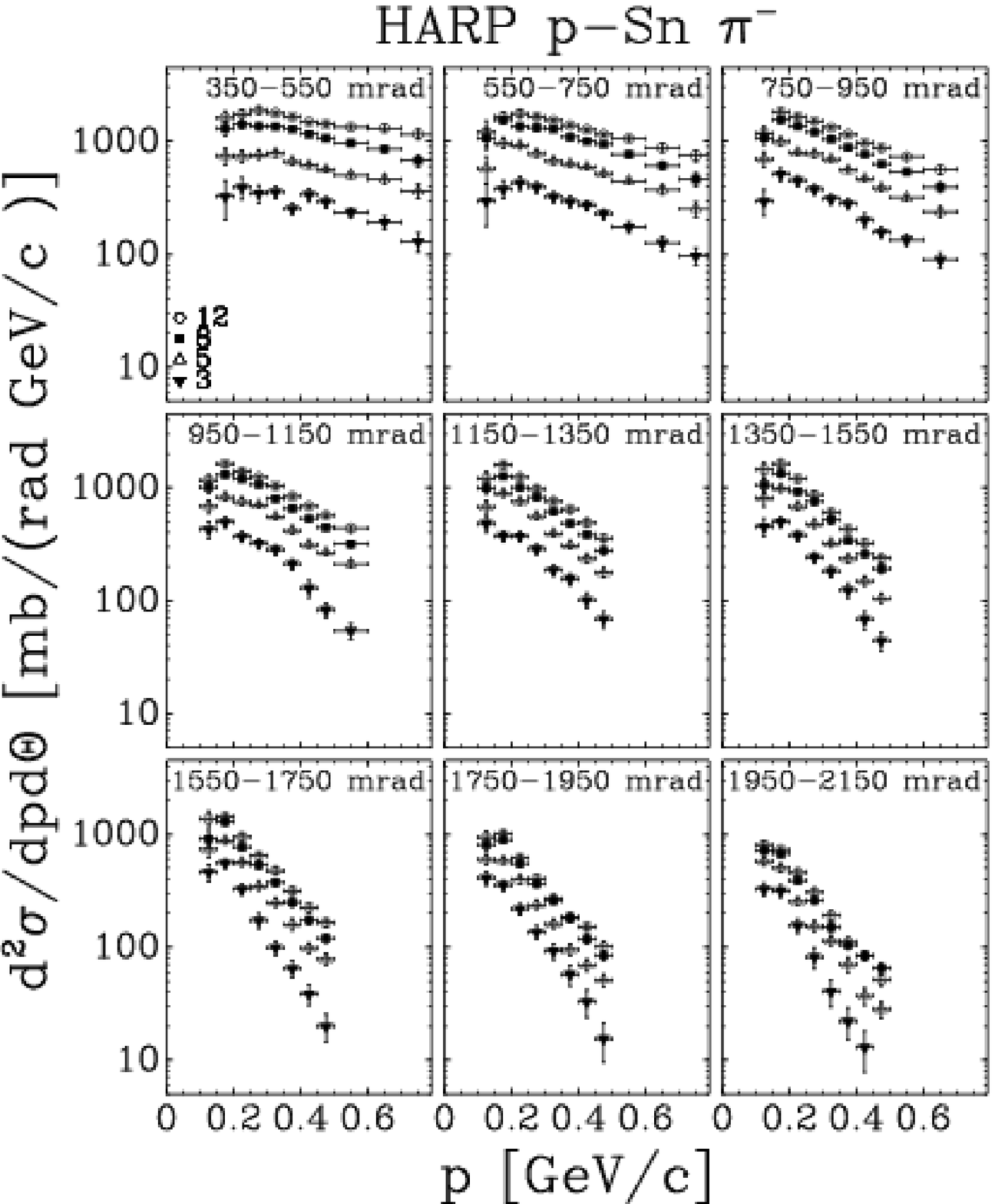}
\caption{
Double-differential cross-sections for \pip production (left) and  \pim
 production (right) in
p--Sn interactions as a function of momentum displayed in different
angular bins (shown in \mrad in the panels).
The error bars represent the combination of statistical and systematic
 uncertainties. 
}
\label{fig:xs-p-th-pbeam-sn}
\end{center}
\end{figure*}

\begin{figure*}[tbp]
\begin{center}
\includegraphics[width=0.48\textwidth]{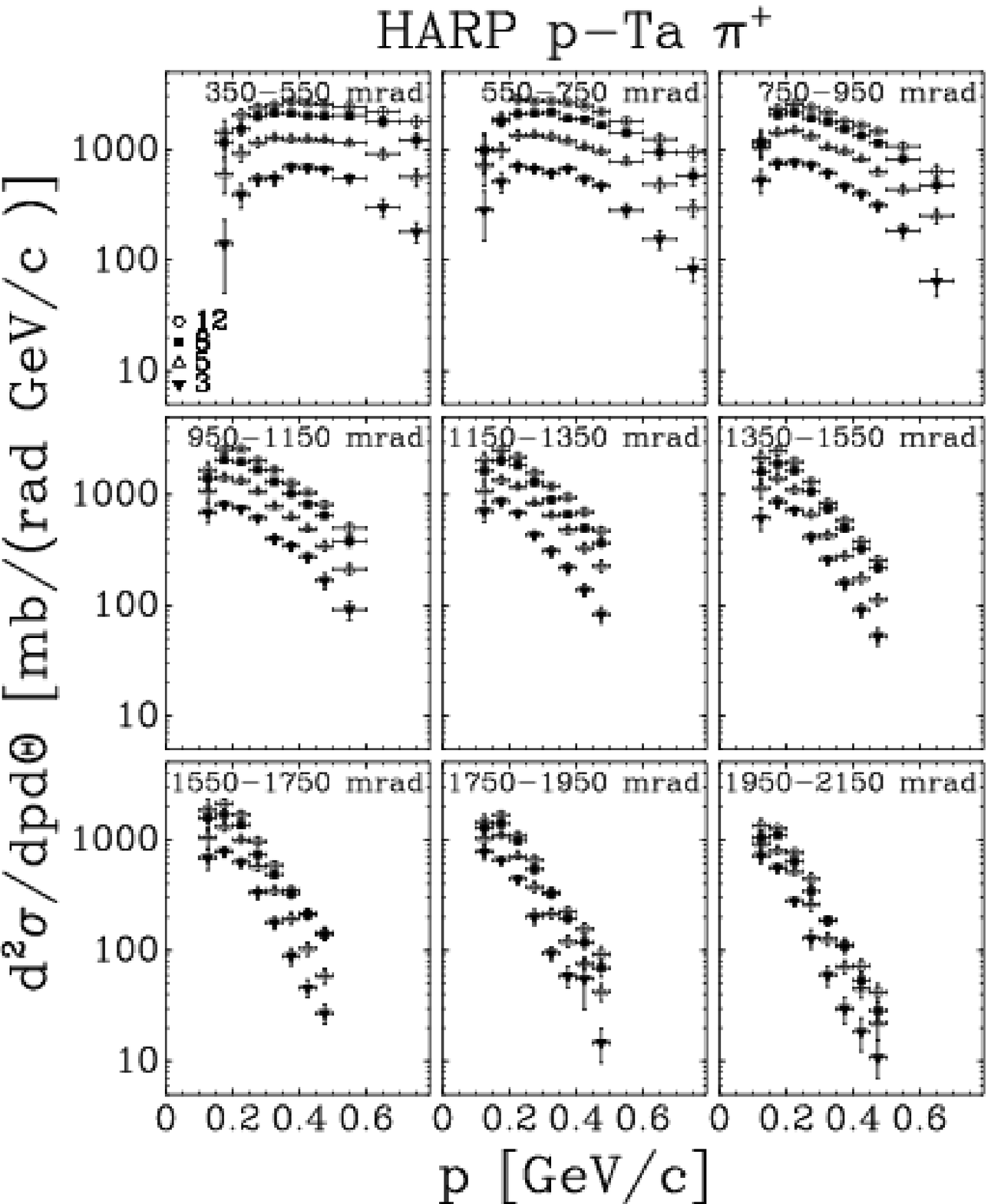}
\includegraphics[width=0.48\textwidth]{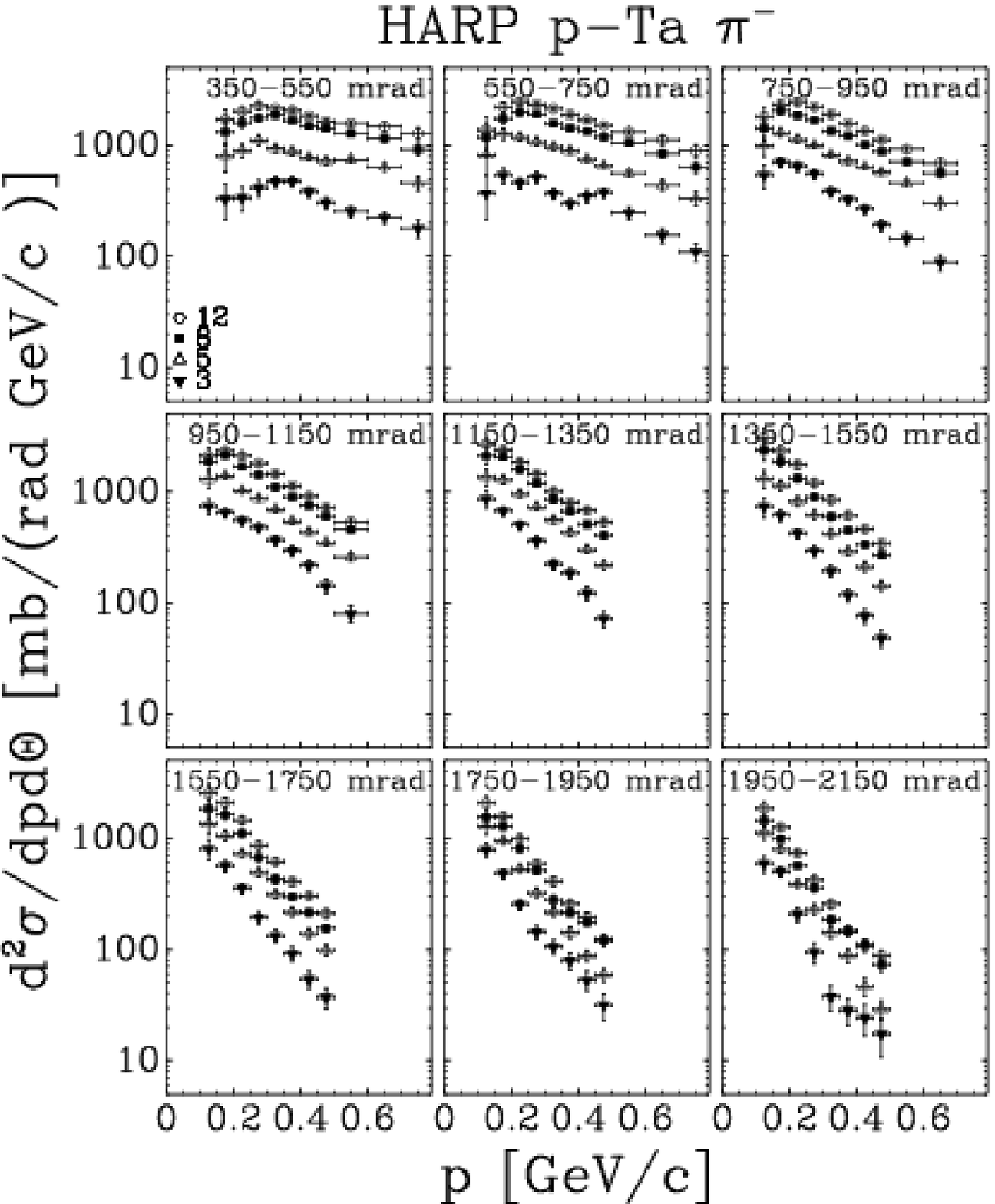}
\caption{
Double-differential cross-sections for \pip production (left) and  \pim
 production (right) in
p--Ta interactions as a function of momentum displayed in different
angular bins (shown in \mrad in the panels).
The error bars represent the combination of statistical and systematic
 uncertainties. 
}
\label{fig:xs-p-th-pbeam-ta}
\end{center}
\end{figure*}

\begin{figure*}[tbp]
\begin{center}
\includegraphics[width=0.48\textwidth]{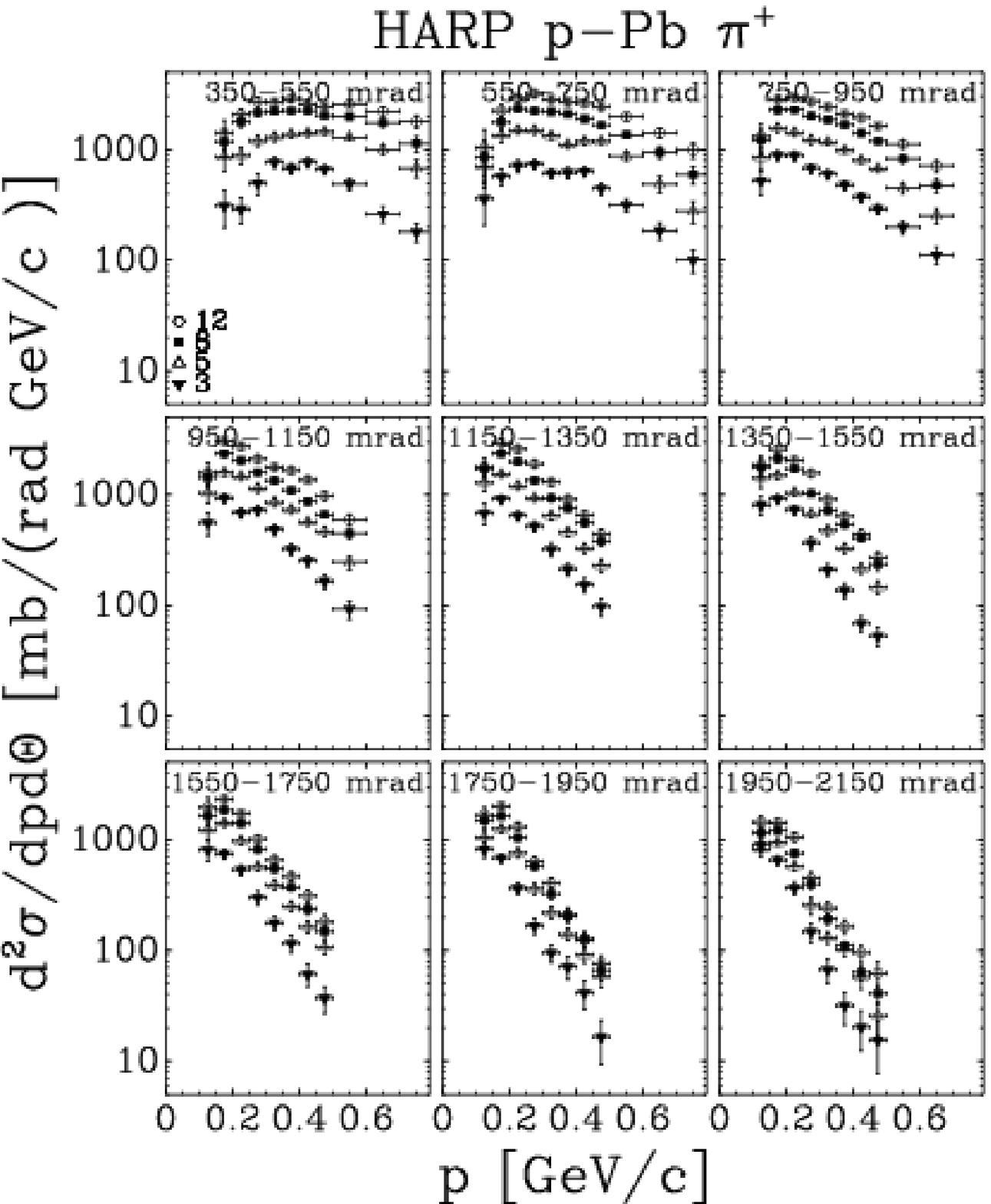}
\includegraphics[width=0.48\textwidth]{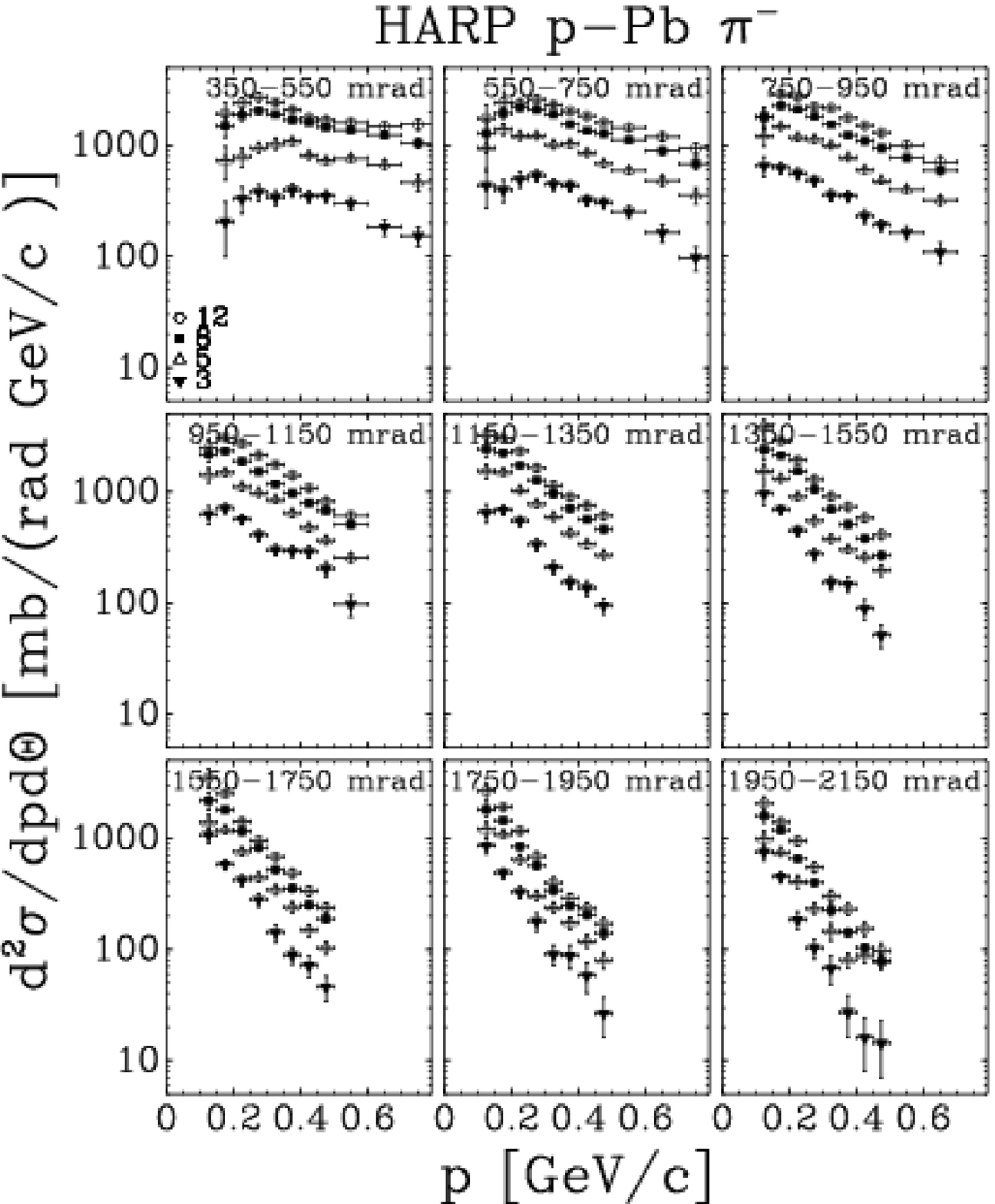}
\caption{
Double-differential cross-sections for \pip production (left) and  \pim
 production (right) in
p--Pb interactions as a function of momentum displayed in different
angular bins (shown in \mrad in the panels).
The error bars represent the combination of statistical and systematic
 uncertainties. 
}
\label{fig:xs-p-th-pbeam-pb}
\end{center}
\end{figure*}

The measured double-differential cross-sections for the 
production of \pip and \pim in the laboratory system as a function of
the momentum and the polar angle for each incident beam momentum are
shown in Figures \ref{fig:xs-p-th-pbeam-be} to \ref{fig:xs-p-th-pbeam-pb} for 
targets from Be to Pb.
The error bars  shown are the
square-roots of the diagonal elements in the covariance matrix,
where statistical and systematic uncertainties are combined
in quadrature.
The correlation of the statistical errors (introduced by the unfolding
procedure) are typically smaller than 20\% for adjacent momentum bins and
even smaller for adjacent angular bins.
The correlations of the systematic errors are larger, typically 80\% for
adjacent bins.
The overall scale error ($< 2\%$) is not shown.
The results of this analysis are also tabulated in Appendix A.

\begin{figure*}[tbp]
\begin{center}
\includegraphics[width=0.45\textwidth]{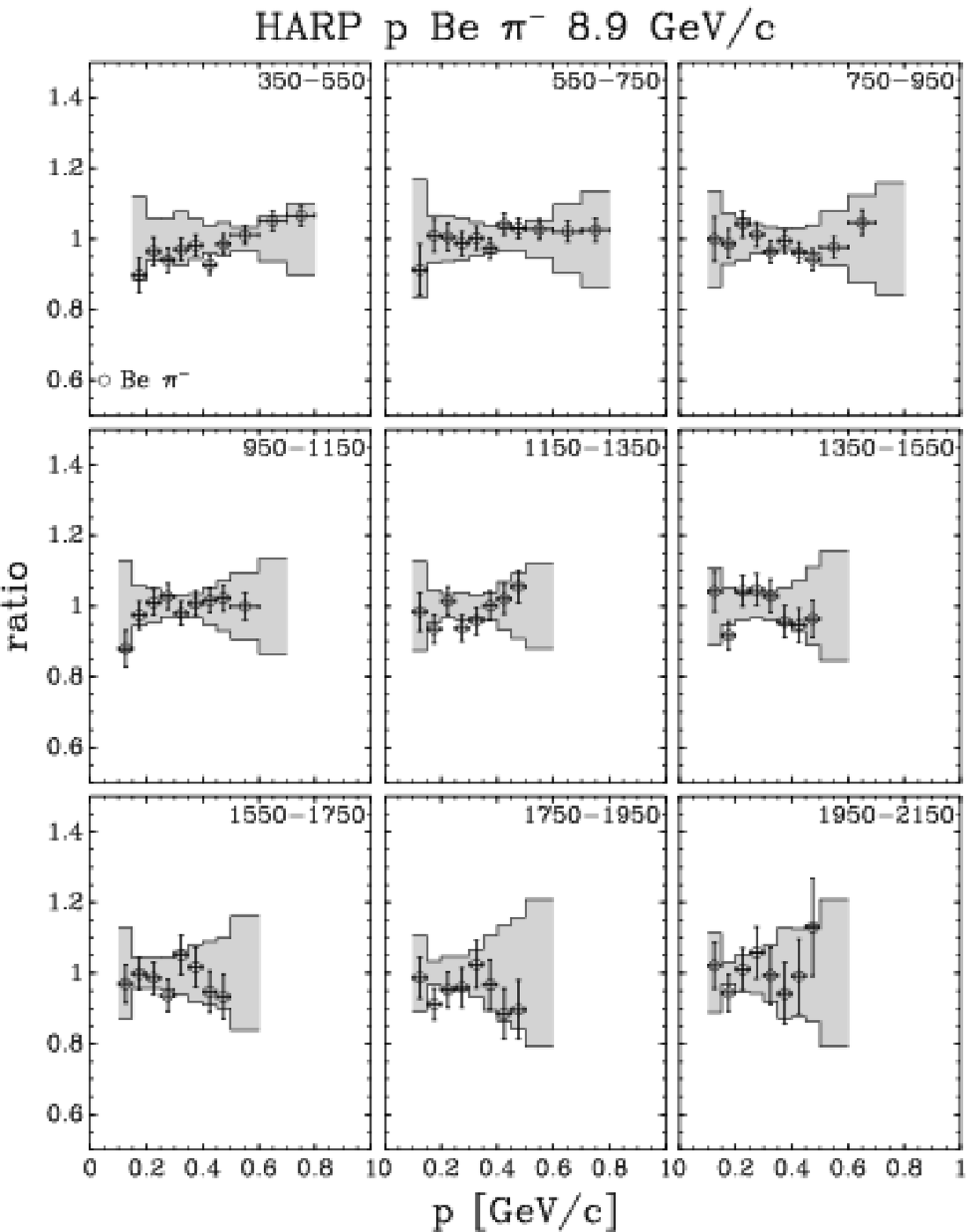}
\caption{Ratio of the \pim production cross-sections measured without and
with corrections for dynamic distortions in p--Be interactions at 8.9 GeV/c,
as a function of momentum for  different angular bins (shown in mrad
in the panels). The error band in the ratio takes into account momentum
error and the error on the efficiency, the other errors being correlated.
The errors on the data points are statistical.
}
\label{fig:becomp1}
\end{center}
\end{figure*}

\begin{figure*}[tbp]
\begin{center}
\includegraphics[width=0.45\textwidth]{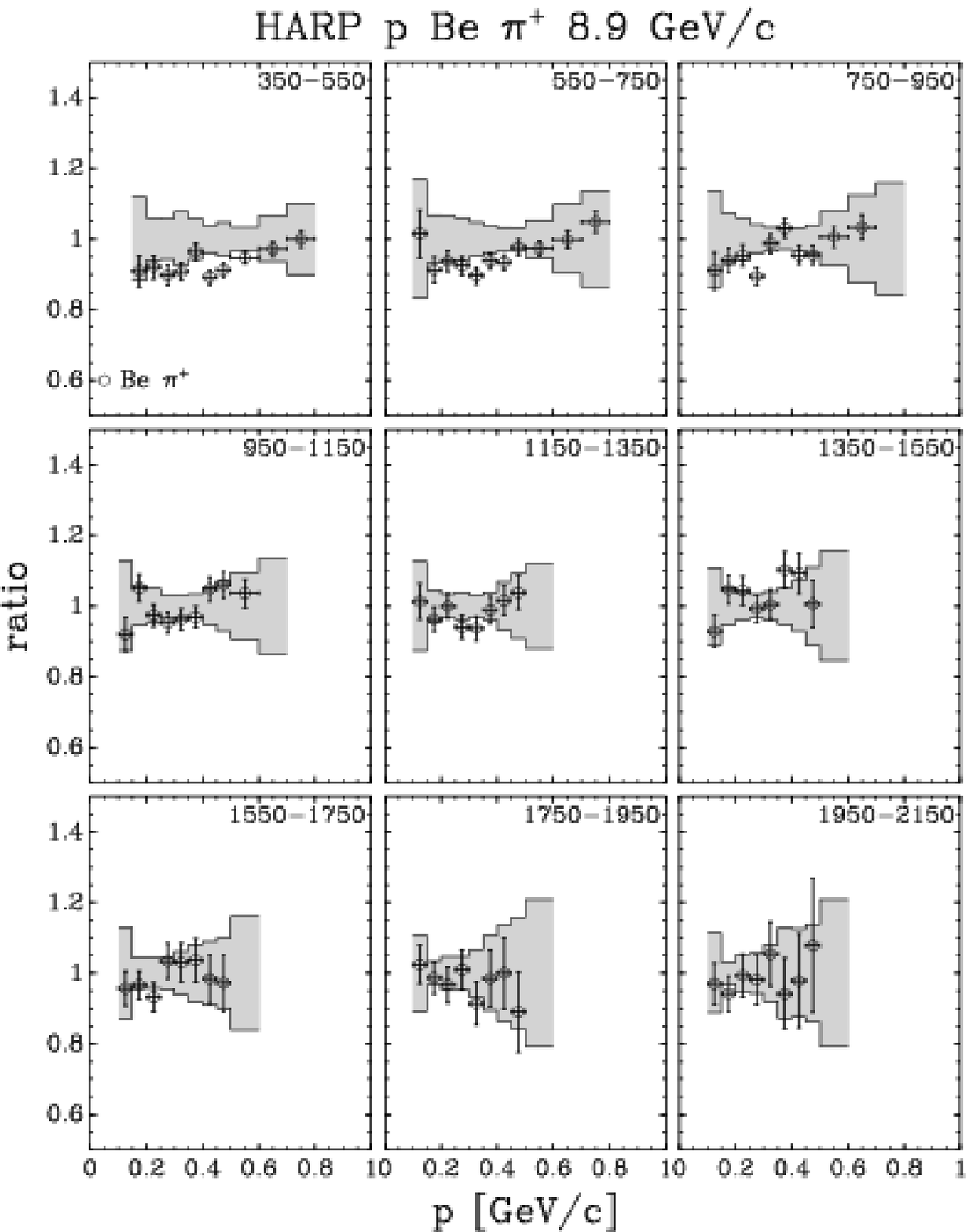}
\caption{Ratio of the \pip production cross-sections measured without and
with corrections for dynamic distortions in p--Be interactions at 8.9 GeV/c,
as a function of momentum for different angular bins (shown in mrad
in the panels). The error band in the ratio takes into account momentum
error and the error on the efficiency, the other errors being correlated.
The errors on the data points are statistical.
}
\label{fig:becomp2}
\end{center}
\end{figure*}

\begin{figure*}[tbp]
\begin{center}
\includegraphics[width=0.45\textwidth]{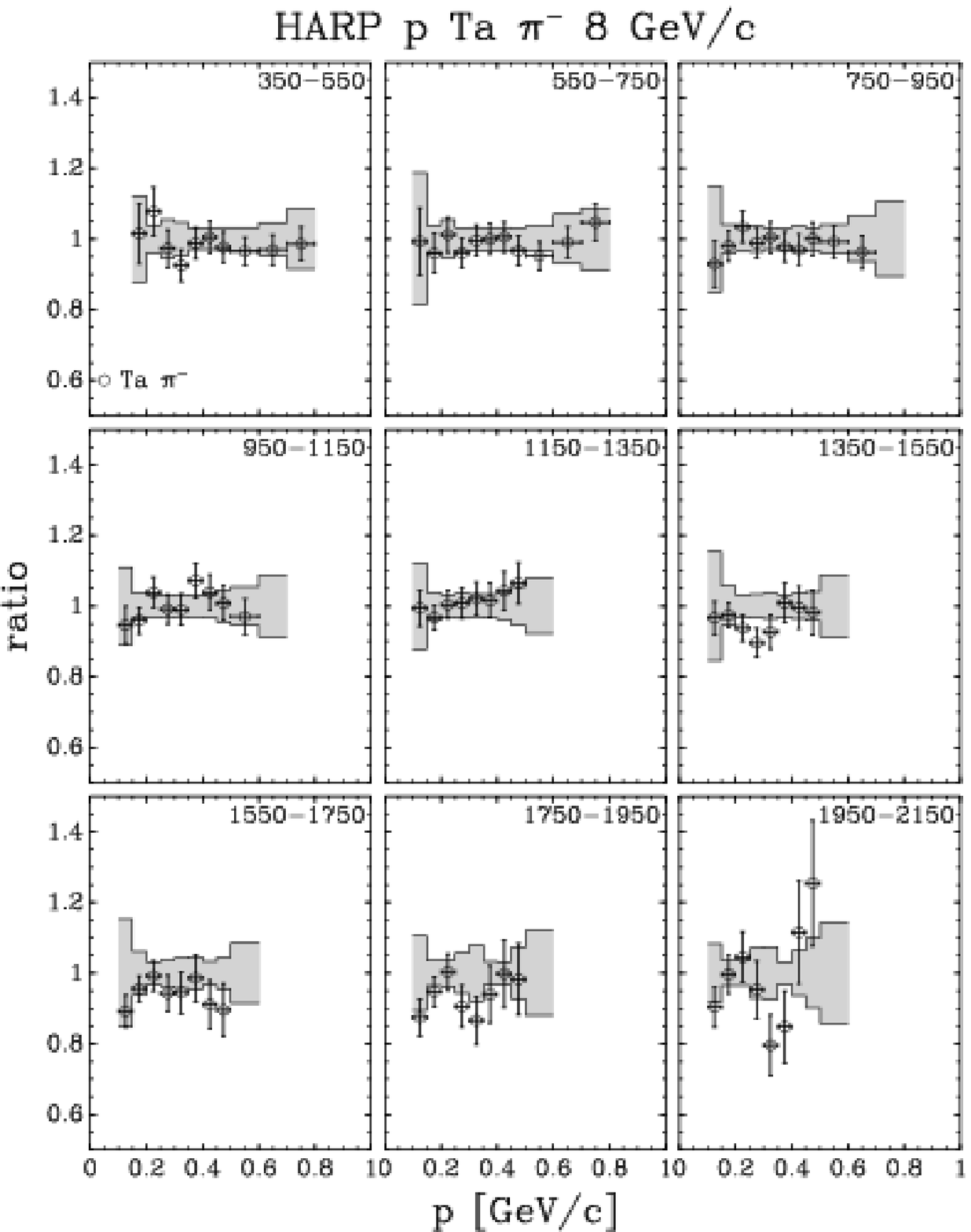}
\caption{Ratio of the \pim production cross-sections measured without and
with corrections for dynamic distortions in p--Ta interactions at 8 GeV/c,
as a function of momentum for different angular bins (shown in mrad
in the panels). The error band in the ratio takes into account momentum
error and the error on the efficiency, the other errors being correlated.
The errors on the data points are statistical.
}
\label{fig:becomp3}
\end{center}
\end{figure*}

\begin{figure*}[tbp]
\begin{center}
\includegraphics[width=0.45\textwidth]{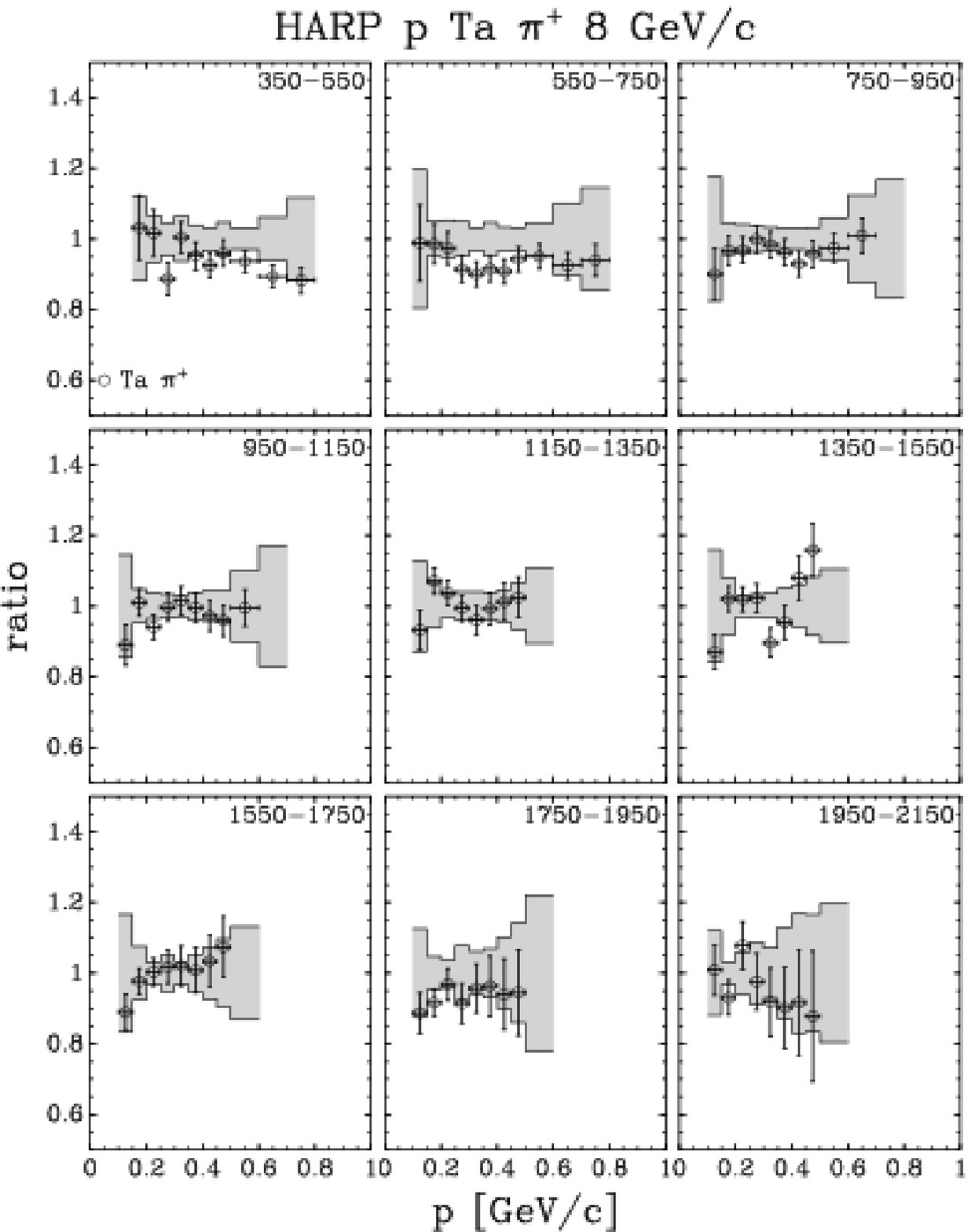}
\caption{Ratio of the \pip production cross-sections measured without and
with corrections for dynamic distortions in p--Ta interactions at 8 GeV/c,
as a function of momentum for different angular bins (shown in mrad
in the panels). The error band in the ratio takes into account momentum
error and the error on the efficiency, the other errors being correlated.
The errors on the data points are statistical.
}
\label{fig:becomp4}
\end{center}
\end{figure*}

These results are in agreement with what was previously found using only
the first part of the spill and using no dynamic distortions 
corrections. Figures~\ref{fig:becomp1} to \ref{fig:becomp4} show the ratio of
the cross sections without and with the correction factor for dynamic
distortions in 8.9 GeV/c Beryllium data (as an example of light target), where the statistics is bigger
and in 8 GeV/c Tantalum data (as an example of heavy target).
The error band in the ratio takes into account the usual estimate of
momentum error and the error on efficiency, the other errors are 
correlated. The agreement is within $1 \sigma$ for most of the points.

\begin{figure*}[tbp]
\begin{center}
  \includegraphics[width=0.325\textwidth]{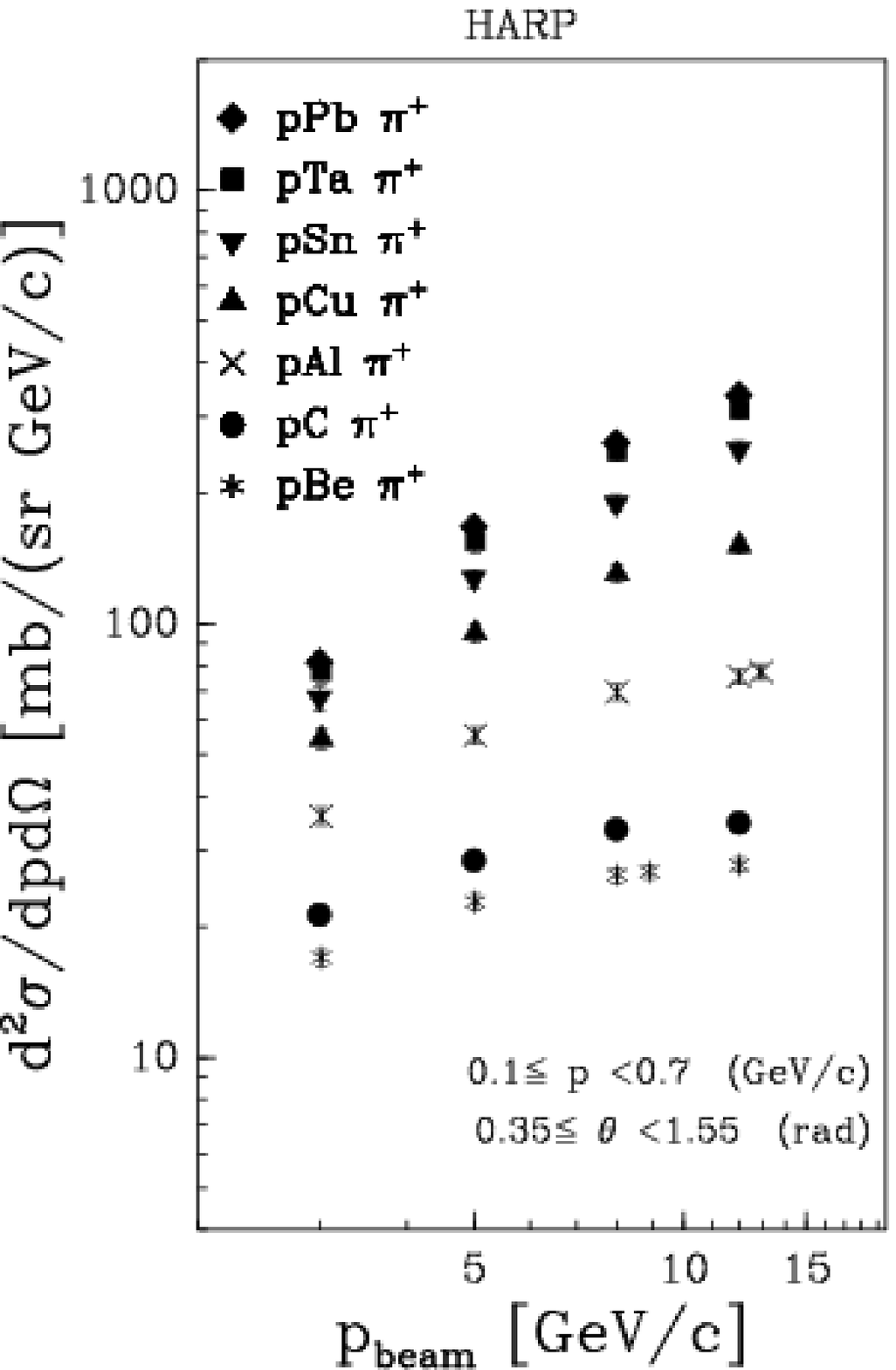}
  \includegraphics[width=0.325\textwidth]{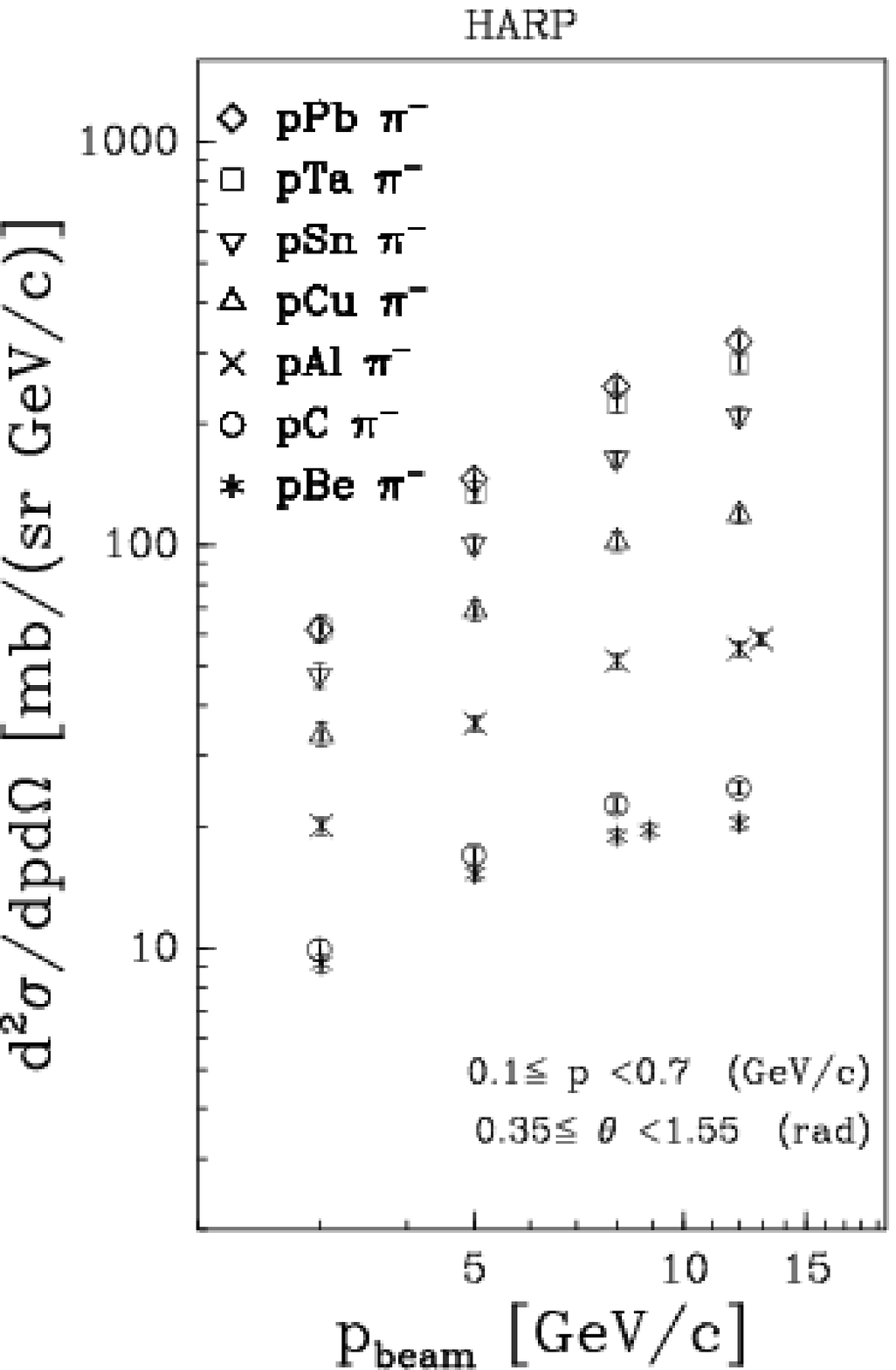}
\end{center}
\caption{
 The dependence on the beam momentum of the \pim (right) and \pip (left) 
  production yields
 in p--Be, p--C, p--Al, p--Cu, p--Sn, p--Ta, p--Pb
 interactions averaged over the forward angular region 
 ($0.350~\rad \leq \theta < 1.550~\rad$) 
 and momentum region $100~\MeVc \leq p < 700~\MeVc$.
 The results are given in arbitrary units, with a consistent scale
 between the left and right panel.
Data points for different target nuclei and equal momenta are slightly
 shifted horizontally with respect to each other to increase the visibility.
}
\label{fig:xs-trend}
\end{figure*}

The dependence of the averaged pion yields on the incident beam
momentum is shown in Fig.~\ref{fig:xs-trend}. 
The \pip and \pim yields are averaged over the region 
$0.35~\rad \leq \theta < 1.55~\rad$ and $100~\MeVc \leq p < 700~\MeVc$
(pions produced in the forward direction only).
Whereas the beam energy dependence of the yields in the
 p--Be, p--C data differs clearly from the dependence in the p--Ta,
p--Pb data one can
observe that the p--Al, p--Cu and p--Sn data display a
smooth transition between them.
The dependence in the p--Be, p--C data is much more flat with a saturation of
the yield between 8~\GeVc and 12~\GeVc
with the p--Al, p--Cu and p--Sn showing an intermediate behaviour.

\begin{figure*}[tbp]
\begin{center}
  \includegraphics[width=0.27\textwidth]{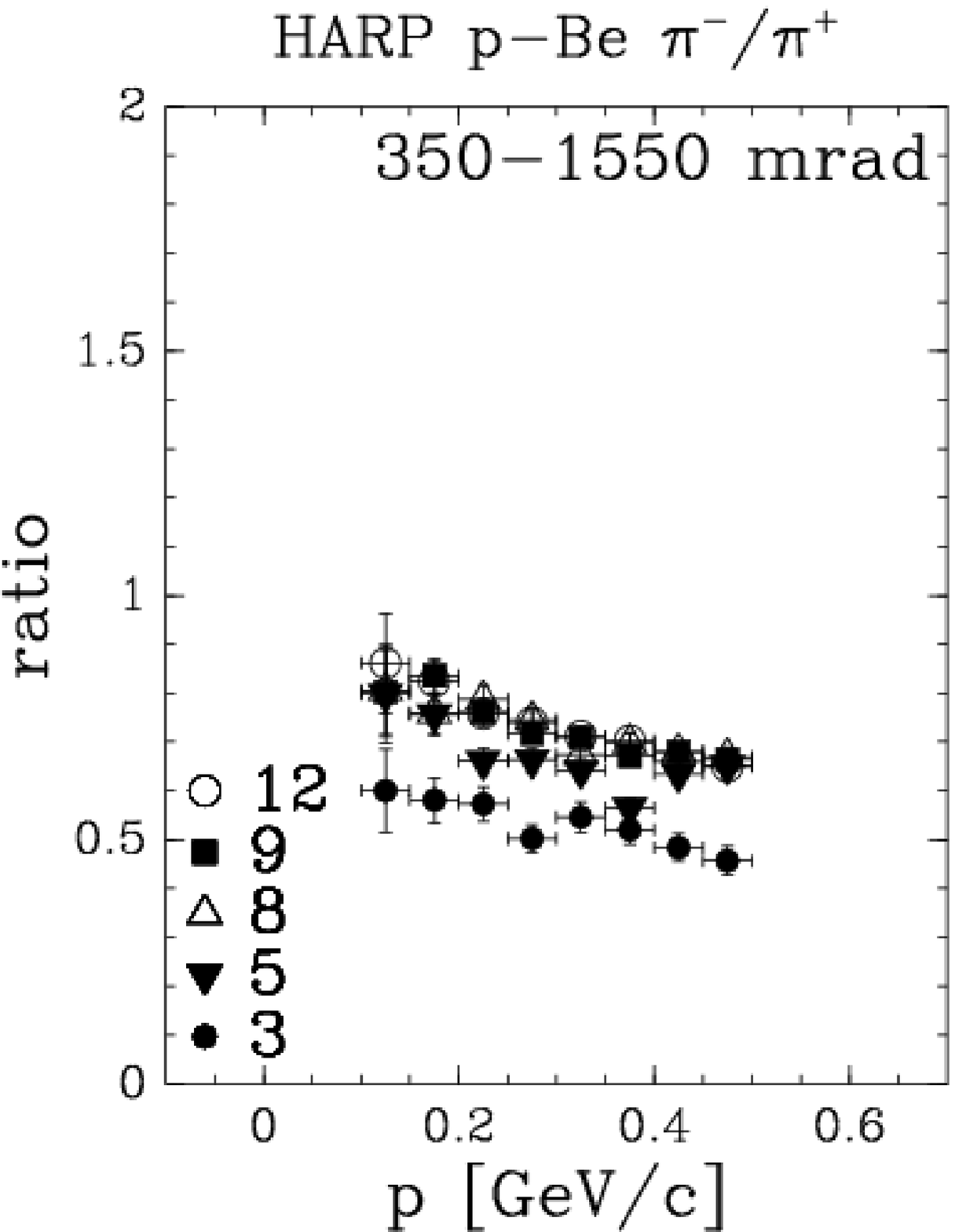}
  \includegraphics[width=0.27\textwidth]{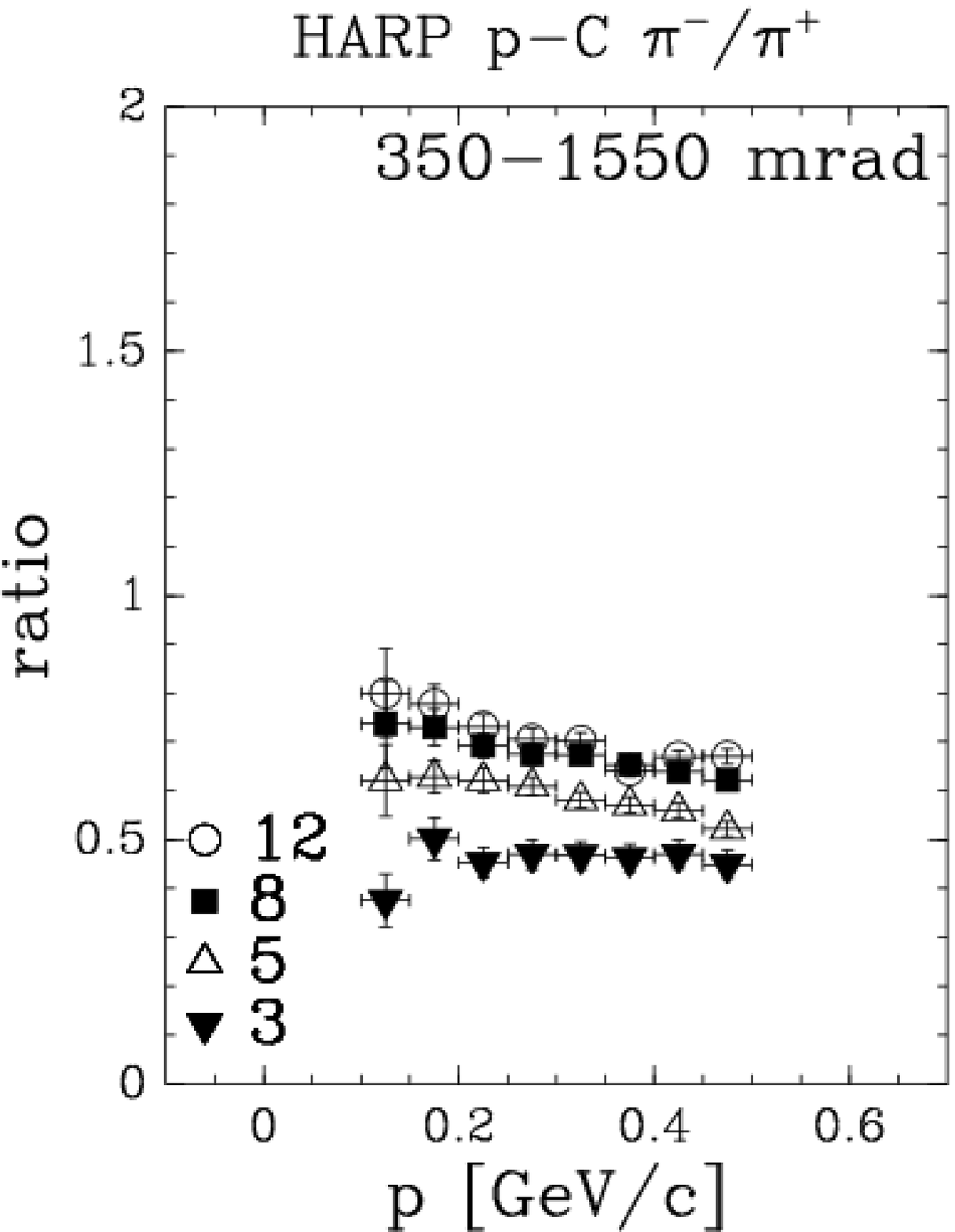}
  \includegraphics[width=0.27\textwidth]{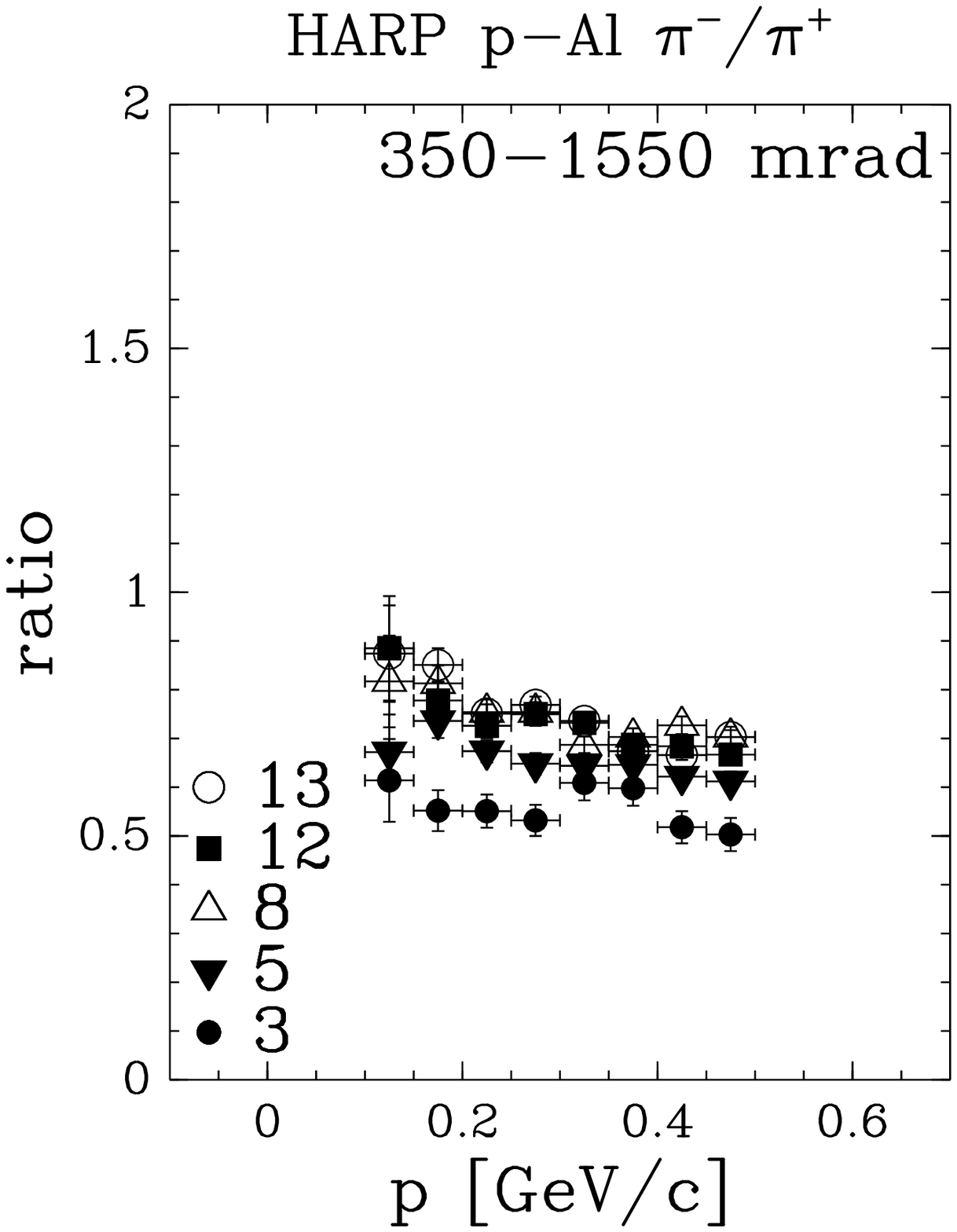}
  \includegraphics[width=0.27\textwidth]{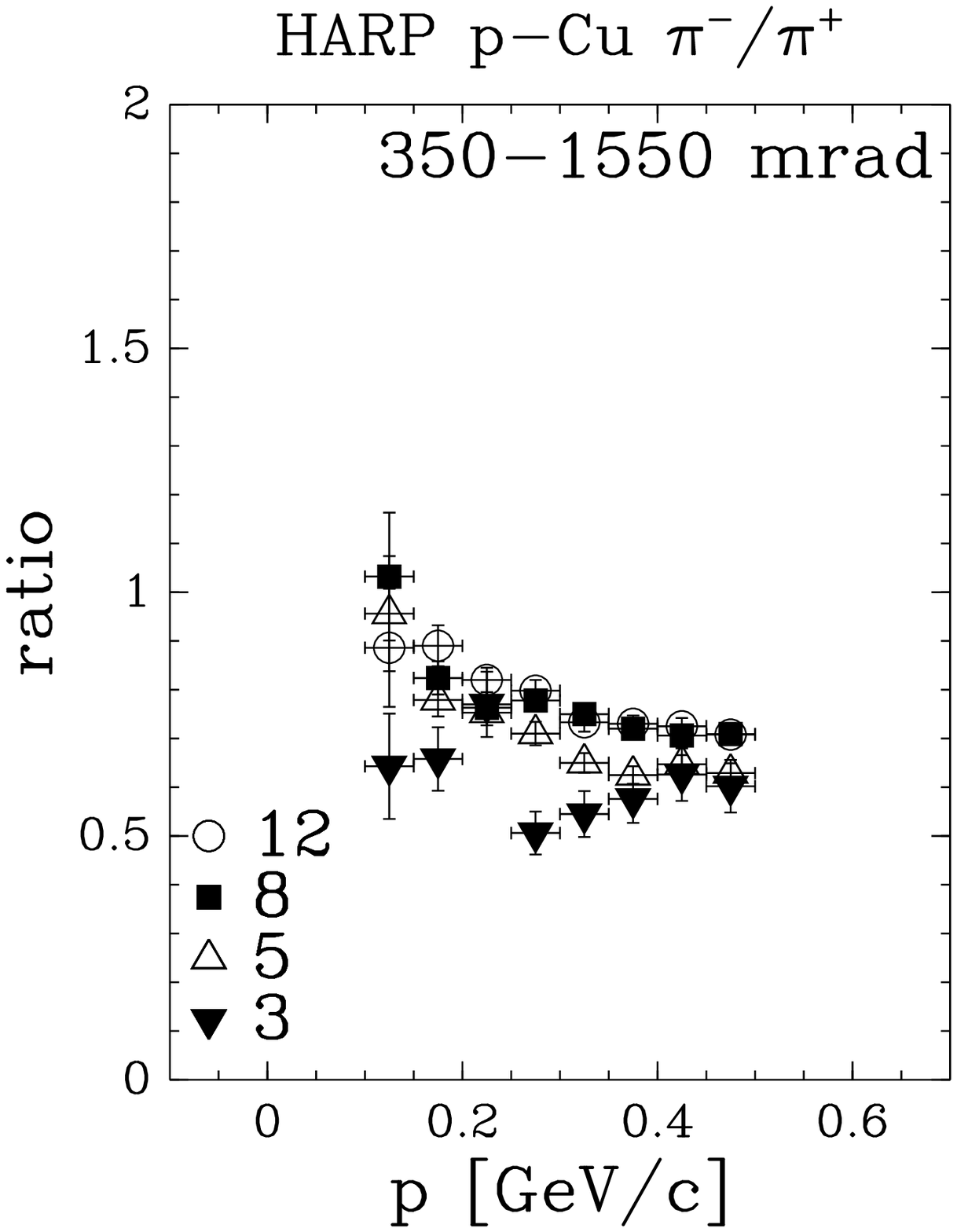}
  \includegraphics[width=0.27\textwidth]{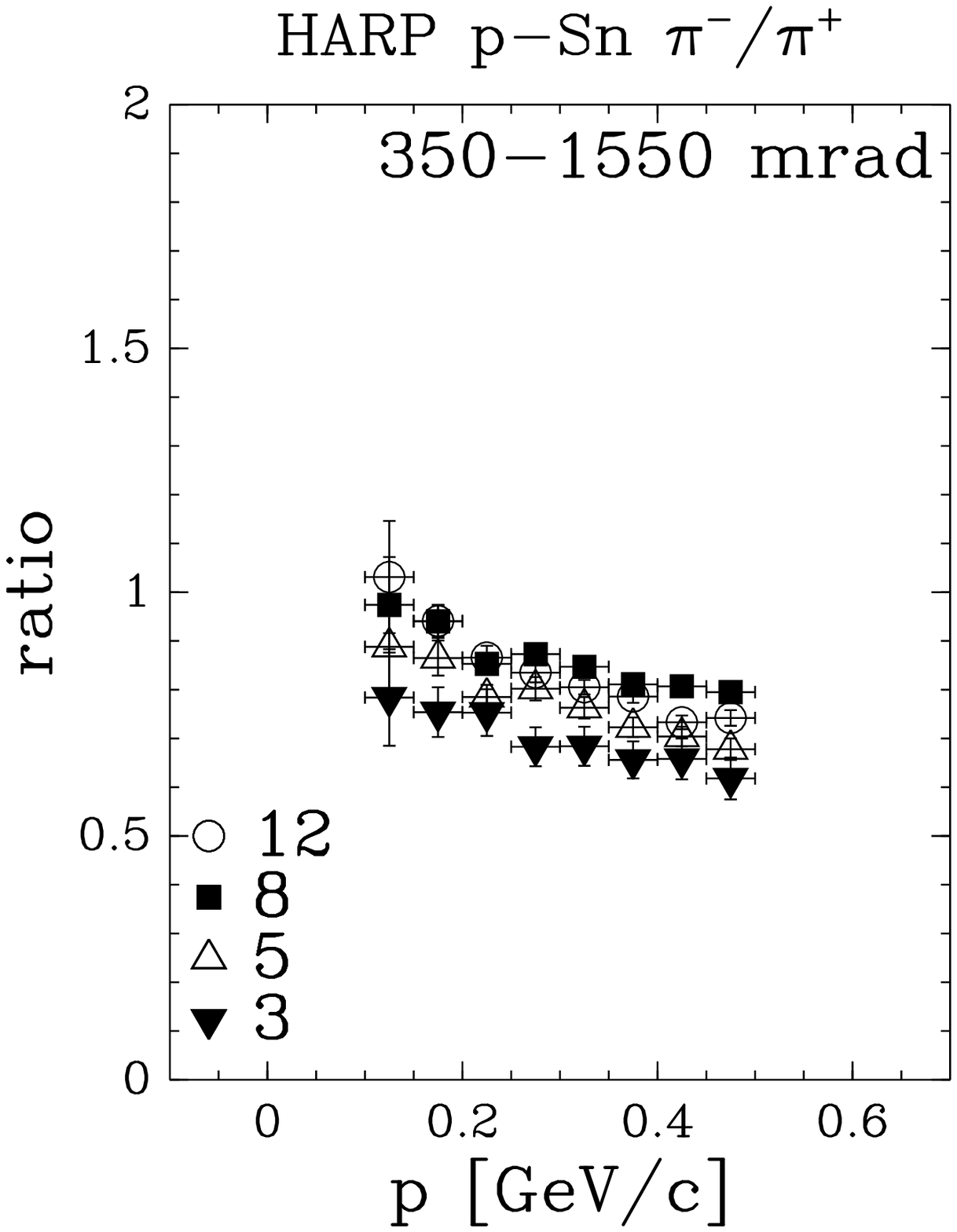}
  \includegraphics[width=0.27\textwidth]{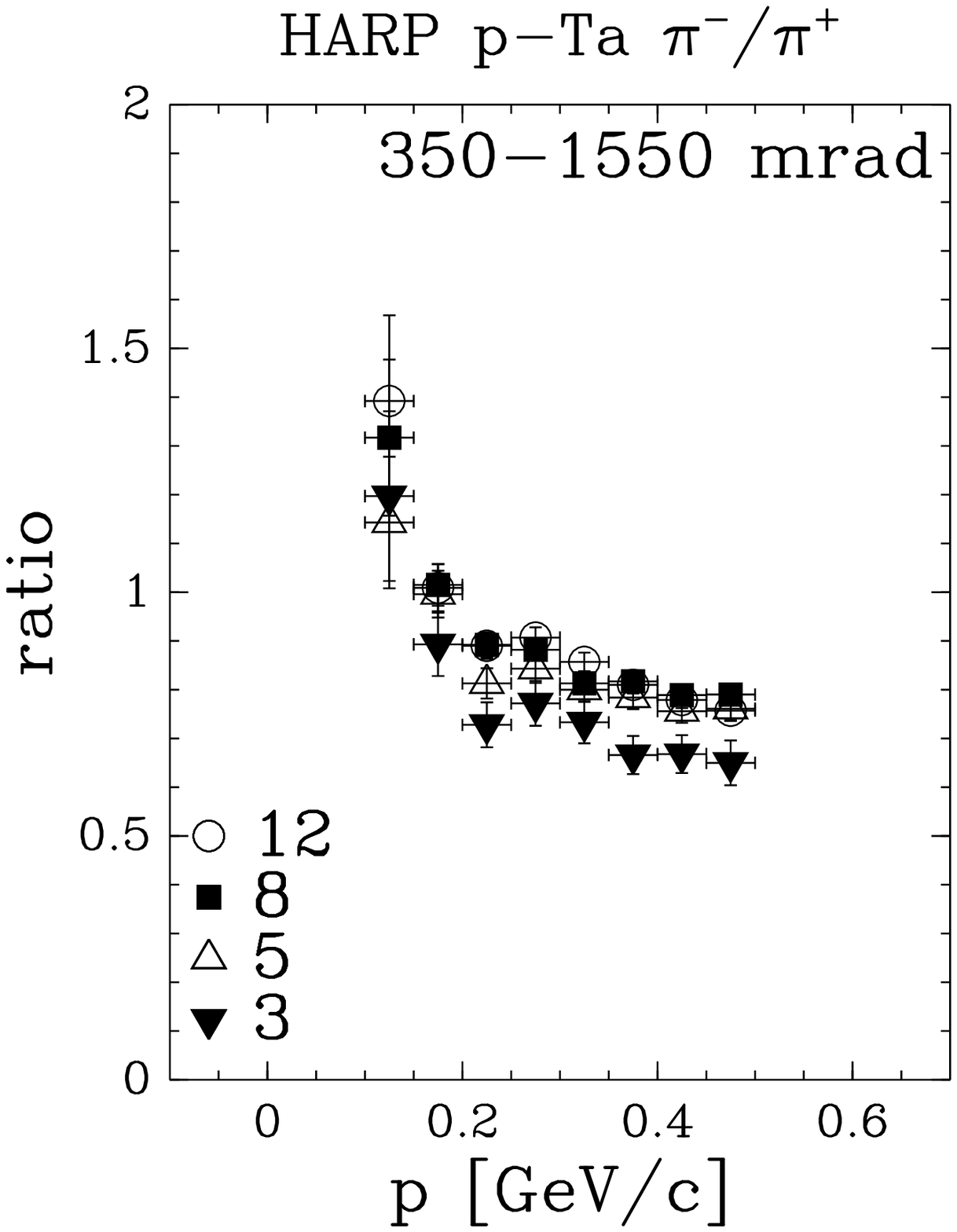}
  \includegraphics[width=0.27\textwidth]{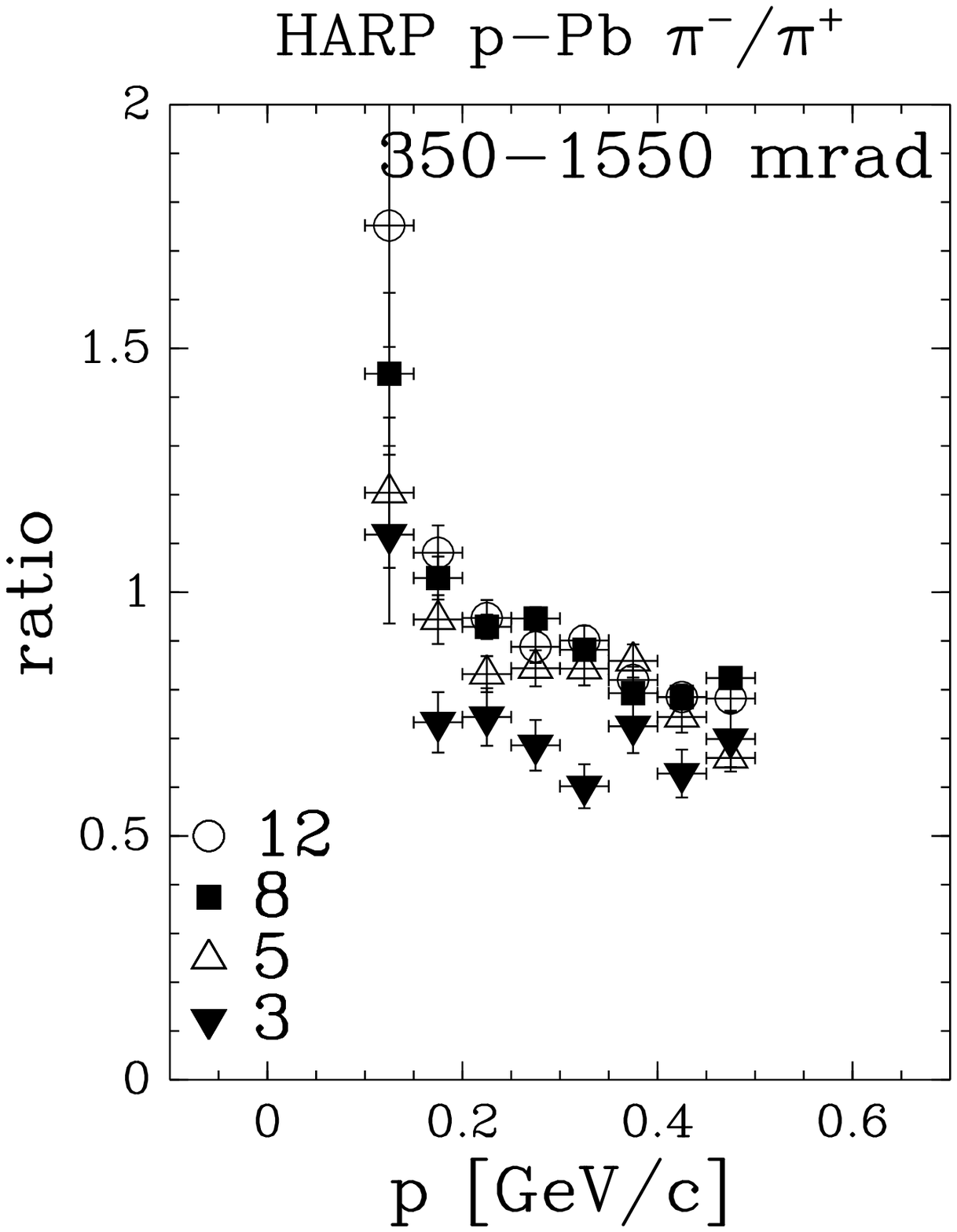}
\end{center}
\caption{
From top-left panel to bottom-right panel, the ratio of the differential cross-sections for \pim and \pip
 production in
p--Be, p--C, p--Al, p--Cu, p--Sn, p--Ta   and p--Pb   interactions as a function of 
the secondary momentum integrated over the
forward angular region (shown in mrad).
In the figure, the symbol legends 13 and 9 refer to 12.9 and 8.9~\GeVc nominal
beam momentum, respectively.
}
\label{fig:xs-ratio}
\end{figure*}

The integrated \pim/\pip ratio in the forward direction is displayed in
Fig.~\ref{fig:xs-ratio} as a function of the secondary momentum. 
In the covered  part of the momentum range in most bins more \pip's are 
produced than \pim's.
In the p--Ta and p--Pb data the ratio is closer to unity than 
for the p--Be, p--C and p--Al data.
The \pim/\pip ratio is larger for higher incoming beam momenta than for
lower momenta.


In the tantalum and  lead data, the number of \pip's produced
is smaller than the number of \pim's in the lowest momentum bin
(100~\MeVc--150~\MeVc) for the 8~\GeVc and 12~\GeVc incoming beam
momenta.
A similar effect was seen by E910 in their p--Au data~\cite{ref:E910}.
Lighter targets do not show this behaviour.

\begin{figure*}[tbp]
\begin{center}
  \includegraphics[width=0.30\textwidth]{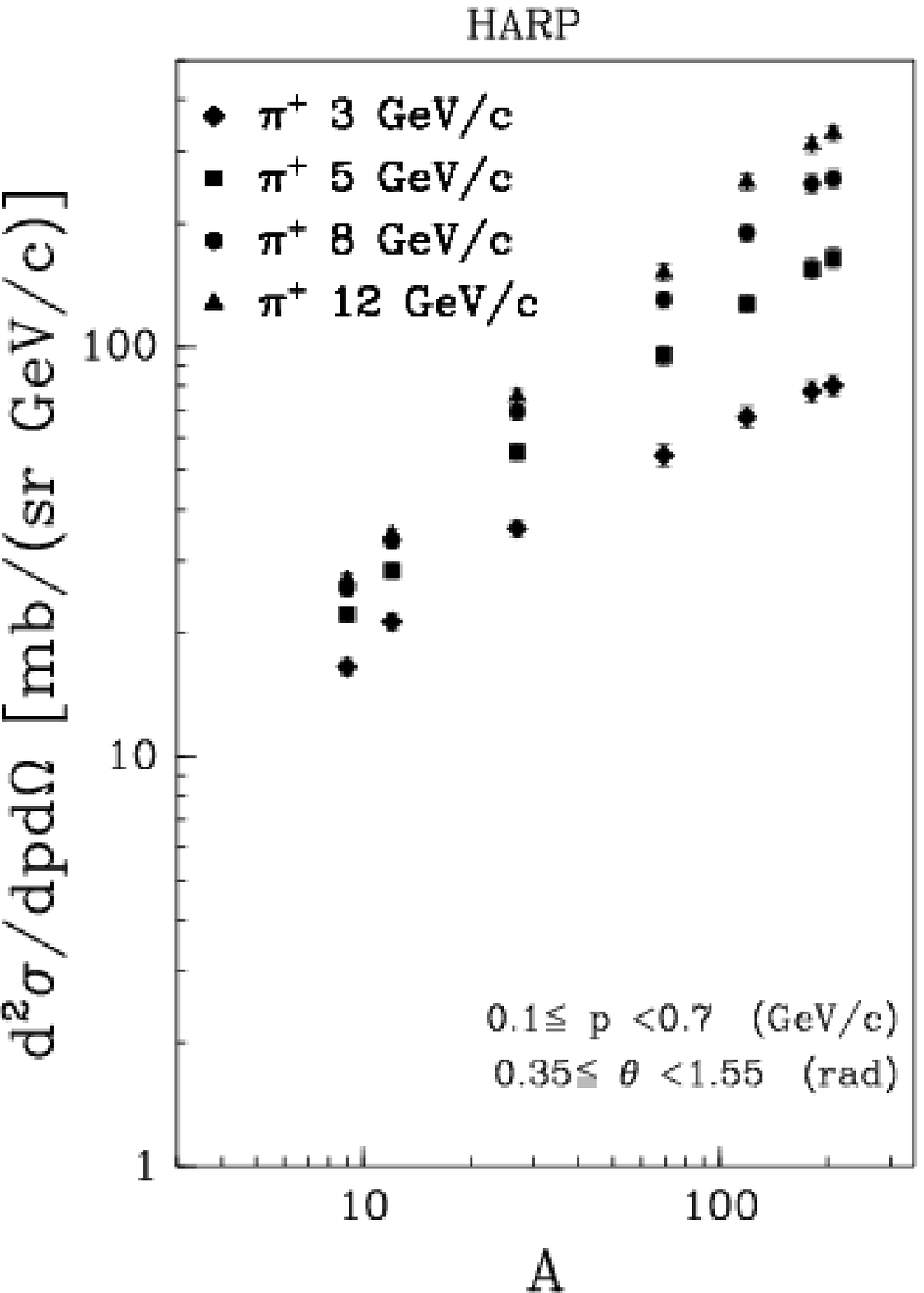}
  \includegraphics[width=0.30\textwidth]{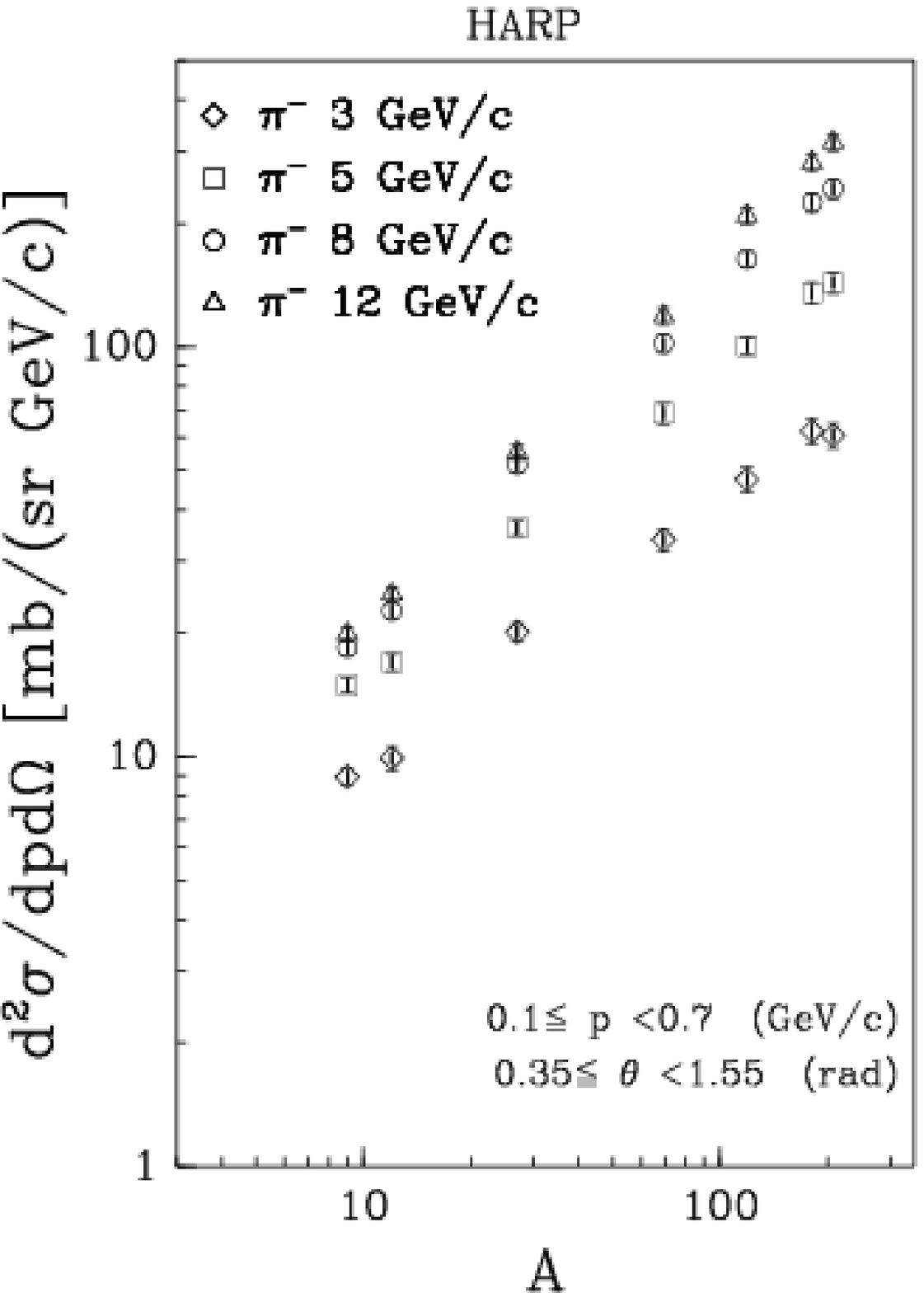}
\end{center}
\caption{
 The dependence on the atomic number $A$ of the pion production yields
 in p--Be, p--Al, p--C, p--Cu, p--Sn, p--Ta, p--Pb
 interactions averaged over the forward angular region 
 ($0.35~\rad \leq \theta < 1.55~\rad$) 
 and momentum region $100~\MeVc \leq p < 700~\MeV/c$.
 The results are given in arbitrary units, with a consistent scale
 between the left and right panel.
 The vertical scale used in this figure is consistent with the one in
 Fig.~\ref{fig:xs-trend}. 
}
\label{fig:xs-a-dep}
\end{figure*}

The dependence of the averaged pion yields on the atomic number $A$ is
shown in Fig.~\ref{fig:xs-a-dep}.  
The \pip yields averaged over the region 
$0.350~\rad \leq \theta < 1.550~\rad$ and $100~\MeVc \leq p < 700~\MeVc$ are
shown in the left panel and the \pim data averaged over the same region
in the right panel for four different beam momenta.
One observes a smooth behaviour of the averaged yields.
The $A$-dependence is slightly different for \pim and \pip production,
the latter saturating earlier towards higher $A$, especially at lower
beam momenta.

%
%
The analysis reported here (for p--Ta and p--Pb interactions) covers the major part of pions produced in
the target and accepted by the focusing system of the input stage of a
neutrino factory~\cite{ref:iss}.
The effective coverage of the kinematic range can be defined as the
fraction of the number of muons transported by the input stage of a
neutrino factory design originating from decays for which the pion
production cross-section is within the kinematic range measured by the
present experiment.
As an example, this effective coverage was evaluated for 
the ``International scoping study of a Neutrino Factory and
super-beam facility'' (ISS) input
stage~\cite{ref:iss} to be 69\% for \pip and 72\% for \pim,
respectively~\cite{ref:fernow}, using a particular model for pion
production at an incoming beam momentum of 10.9~\GeVc~\cite{ref:brooks}.
%
Since the data covers already a large fraction of the relevant
phase-space, one would expect that the extrapolation to the full region
with hadronic production models can be done reliably, once these models
are adjusted to reproduce this data set in the region covered.
Such tuning of models can also profit from the additional data
provided with the HARP forward spectrometer.

\begin{figure*}[tbp]
  \begin{center}
  \includegraphics[width=0.48\textwidth]{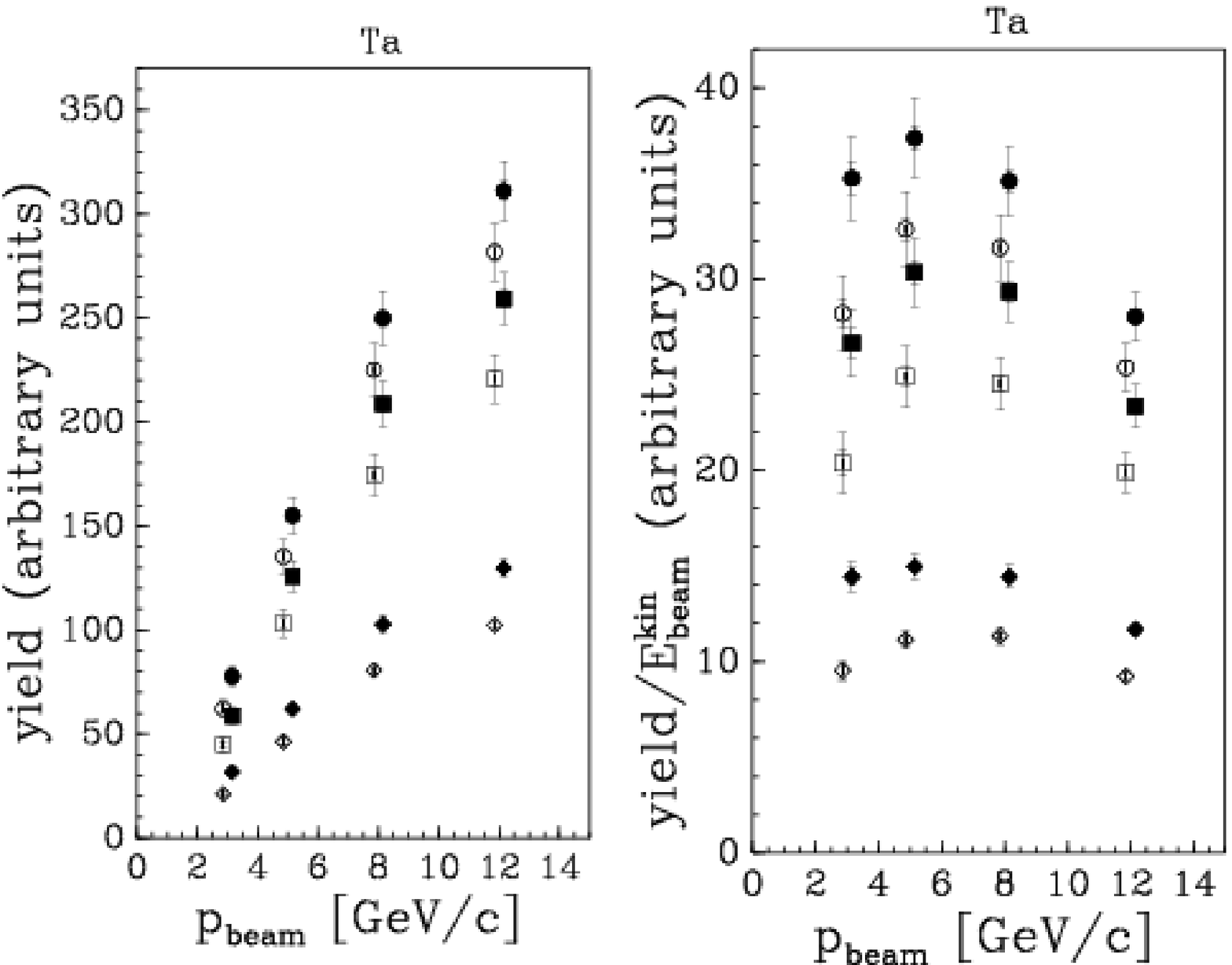}
  \includegraphics[width=0.48\textwidth]{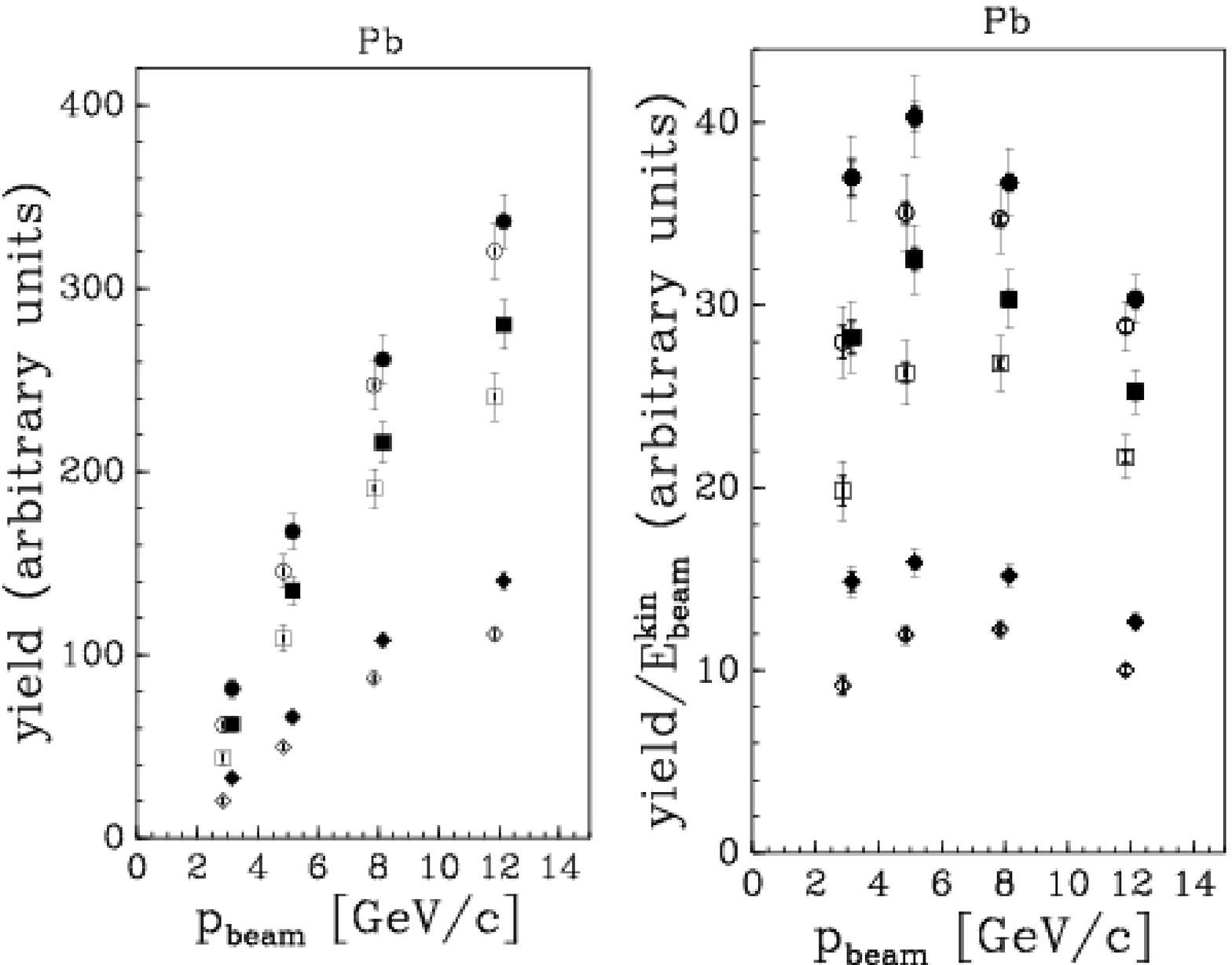}
  \end{center}
\caption{
Predictions of the $\pi^{+}$ (closed symbols) and
$\pi^{-}$ (open symbols) yields for different design of the NF
focussing stage.
Integrated yields (left) and the integrated yields
normalized to the kinetic energy of the proton (right) for 
p--Ta  and p--Pb  interactions.
The circles indicate the integral over the full HARP acceptance
($100~\MeVc<p<700~\MeVc$ and $0.35 \ \rad <\theta < 1.55 \ \rad$), the
squares are integrated over $0.35 \ \rad <\theta< 0.95 \ \rad$, while the diamonds
are calculated for the smaller angular range and
$250~\MeVc<p<500~\MeVc$.
Although the units are indicated as ``arbitrary'',
for the largest region the yield is expressed as
${{\mathrm{d}^2 \sigma}}/{{\mathrm{d}p\mathrm{d}\Omega }}$ in
mb/(\GeVc~sr).
For the
other regions the same normalization is chosen, but now scaled with the
relative bin size to show visually the correct ratio of number of pions
produced in these kinematic regions.
The full error bar shows the overall (systematic and statistical)
error, while the inner error bar shows the error relevant for the
point--to-point comparison.
For the latter error only the uncorrelated systematic uncertainties
were added to the statistical error.
}
\label{fig:nufact-yield}
\end{figure*}

As an indication of the overall pion yield as a function of incoming
beam momentum, the \pip and \pim production cross-sections were
averaged over the full HARP kinematic range in the forward hemisphere
($100~\MeVc<p<700~\MeVc$ and $0.35 <\theta< 1.55$).
The results are shown in Fig.~\ref{fig:nufact-yield}.
The integrated yields are shown in the left panel and the integrated
yields normalized to the kinetic energy of the incoming beam particles
are shown in the right panel.
The outer error bars indicate the total statistical and systematic errors.
If one compares the \pip and \pim rates for a given beam momentum or
if one compares the rates at a different beam momentum the relative
systematic error is reduced by about a factor of two.
The relative uncertainties are shown as inner error bar.
It is shown that the pion yield increases with
momentum and that in our kinematic coverage the optimum yield is
between 5~\GeVc and 8~\GeVc.

However, these calculations should be completed with more realistic
kinematical cuts in the integration.
To show the trend the rates within restricted ranges are also given: a
restricted angular range ($0.35 <\theta< 0.95$)  and a range further
restricted in momentum ($250~\MeVc<p<500~\MeVc$).
The latter range may be most representative for the neutrino factory.

Of course this analysis only gives a simplified picture of the results.
One should note that the best result can be obtained by using the
full information of the double-differential cross-section and
by developing designs optimized specifically for each single beam
momentum.
Then these optimized designs can be compared.
                                                                                                                           
\begin{figure*}[tbp]
  \begin{center}
\includegraphics[width=0.90\textwidth]{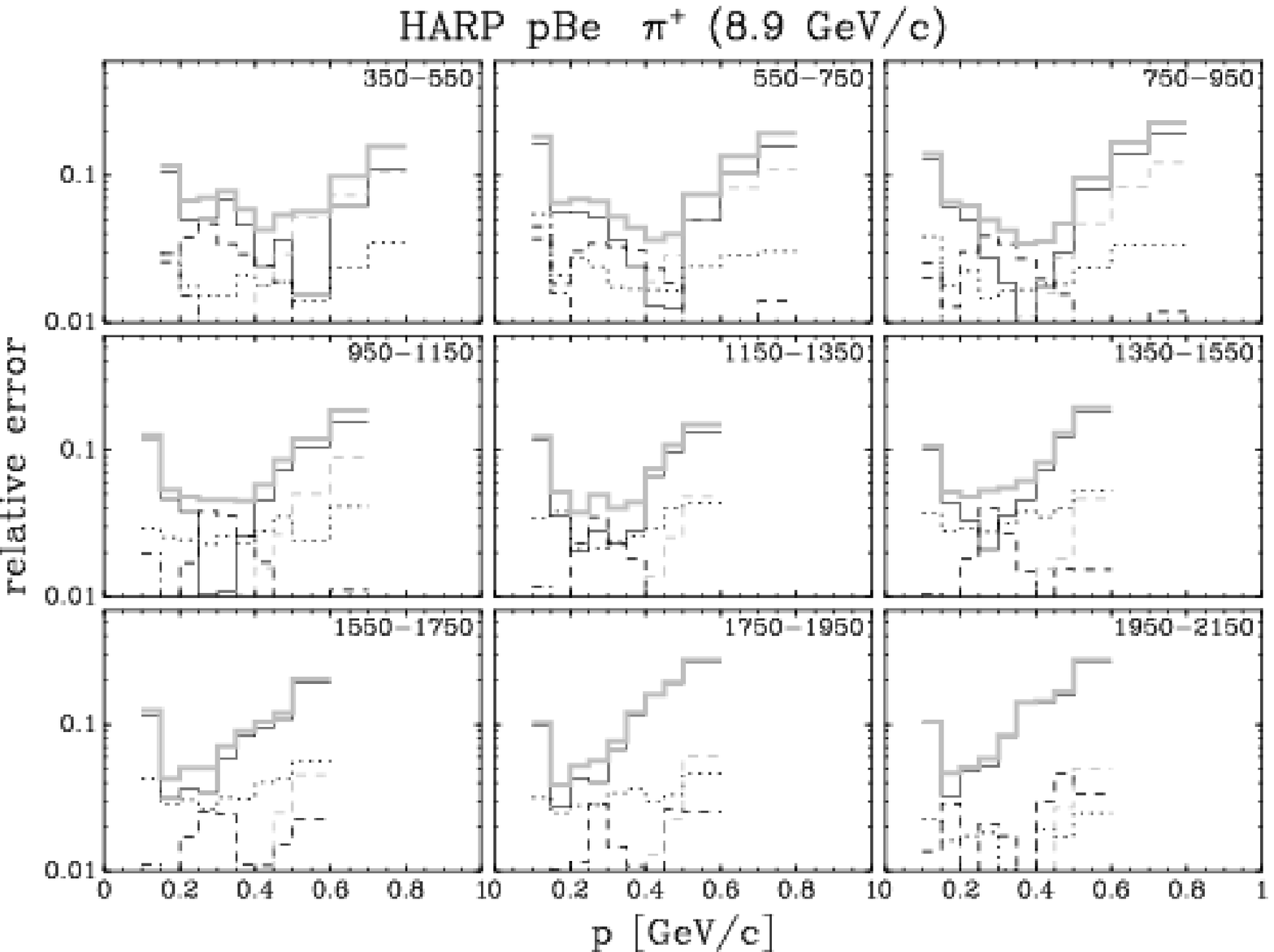}
\includegraphics[width=0.90\textwidth]{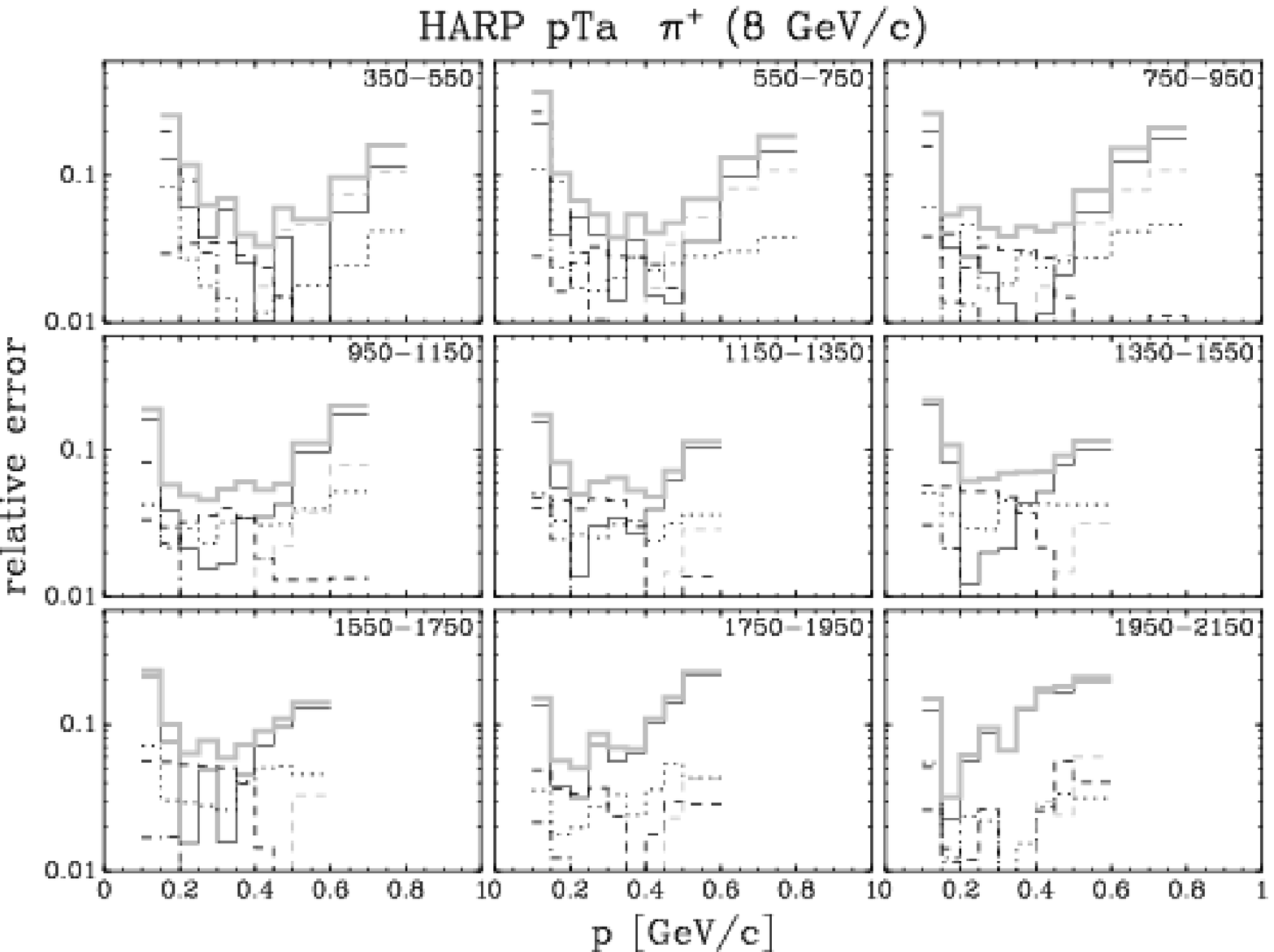}
\end{center}
\caption{Total systematic error (grey solid line) and main components: 
black short-dashed line for absorption+tertiares interactions,
black dotted line for track efficiency and target pointing efficiency,
black dot-dashed line for $\pi^{0}$ subtraction, black solid line for 
momentum scale+resolution and angle scale, grey short-dashed line 
for PID for two typical targets (Be and Ta)
}
\label{fig:syst}
\end{figure*}

The experimental uncertainties are summarized for $\pi^{+}$  in
Table~\ref{tab:errors-3} 
 for all used targets and shown for two typical targets
(one light and one heavy) in Figure~\ref{fig:syst}. 
They are very similar for $\pi^{-}$ and at the other beam energies:
3, 5, 8, 12 and 12.9 GeV/c. Going from lighter (Be, C) to heavier targets (Ta, Pb)
the corrections for $\pi^{0}$ (conversion) and absorption/tertiares 
are bigger. 
   
One observes that only for the 3~\GeVc beam is the statistical error 
similar in magnitude to the systematic error, while the statistical
error is negligible for the 8~\GeVc and 12~\GeVc beam settings.
The statistical error is calculated by error propagation as part of the
unfolding procedure. 
It takes into account that the unfolding matrix is obtained from the
data themselves~\footnote{The migration matrix is calculated without
prior knowledge of the cross-sections, while the unfolding procedure
determined the unfolding matrix from the migration matrix and the
distributions found in the data.} and hence contributes also to the
statistical error. 
This procedure almost doubles the statistical error, but avoids an important 
systematic error which would otherwise be introduced by assuming a
cross-section model {\em a priori} to calculate the corrections. 

The largest systematic error corresponds to the uncertainty in the
absolute momentum scale, which was estimated to be around 3\% using elastic
scattering~\cite{ref:harp:tantalum}.
Although the corrections for the dynamic distortions have been applied
in this analysis, contrary to the previously published ones, the
systematic error has not been reduced.
This is due to the fact that now the full
statistics has been used introducing data with larger distortions.
The benefit of the introduction of the larger statistics is
evident mainly at low incident beam momenta (3 GeV/c), in the
bins with higher secondary momenta and larger scattering angle. 
At low momentum in the relatively small angle forward
direction the uncertainty in the subtraction of the electron and
positron background due to \piz production is dominant ($\sim 6 - 10 \%$).
This uncertainty is split between the variation in the shape of the
\piz spectrum and the normalization using the recognized electrons. 
The target region definition  and
the uncertainty in the PID efficiency and background from tertiaries
(particles produced in secondary interactions)
are of similar size and are not negligible ($\sim 2-3 \%$).
Relatively small errors are introduced by the uncertainties in
the absorption correction, absolute knowledge of the angular and the
momentum resolution.
The correction for tertiaries is relatively large at low momenta and large 
angles ($\sim 3-5 \%$). 
As expected, this region is most affected by this component.
The errors are quoted for the positive pion data.
Owing to the similarity of the spectra the errors are very similar
for the negative pions.

As already mentioned above, the overall normalization has an uncertainty
of 2\%, and is not reported in the table.
It is mainly due to the uncertainty in the efficiency that beam protons
counted in the normalization actually hit the target, with smaller
components from the target density and beam particle counting procedure.

%
\begin{table*}[htbp!] 
\small{
\begin{center}
\caption{Experimental uncertainties for the analysis of the data taken
 with beryllium, carbon, aluminium, copper, tin  and lead targets at
3, 5, 8, 8.9, 12 and 12.9  GeV/c. The
 numbers represent the uncertainty in percent of 
  the cross-section integrated over the angle and momentum region indicated. 
} 
\label{tab:errors-3}
\vspace{2mm}
\begin{tabular}{ l l rrr | rrr | rr} \hline
\bf{p (\GeVc) }&\multicolumn{4}{c|}{0.1 -- 0.3}
                            &\multicolumn{3}{c|}{0.3 -- 0.5}
                            &\multicolumn{2}{c}{0.5 -- 0.7} \\
\hline
\bf{Angle (\mrad)}& &350--950&950--1550&1550--2150&350--950&950--1550
&1550--2150&350--950&950--1550 \\
\hline
\bf{3 \GeVc }&&&&&&&&\\
\hline
\bf{Total syst.} & (Be)    & 8.2 & 4.7  & 3.5 & 3.9 & 6.3  & 8.9  &9.5 &14.9 \\
                 & (C)     & 13.2& 5.2  & 3.2 & 3.8 & 6.9  & 10.1 &9.5 &16.5     \\                 
                 & (Al)    &9.3 &  4.8  & 3.5 & 3.8 & 7.2  & 13.1  &10.0&13.6   \\                 
                 & (Cu)    &10.0 &  7.7  & 6.7 & 3.7 & 5.4  & 9.0  &9.0&12.6   \\                 
                 & (Sn)    &14.2 &  6.7  & 5.1 & 3.4 & 6.5  & 9.8  &8.1 &14.2   \\                 
                 & (Ta)    &13.7 &  8.1  & 7.5 & 3.8 & 6.0  & 6.0  &9.9&13.9   \\                 
                 & (Pb)    &  13.2 &  7.5  &  6.7 &  3.7 &  5.7 & 7.8 & 9.5 & 14.1  \\
                          
\bf{Statistics}  & (Be)        & 2.8 & 2.6 & 3.6 & 2.2  & 3.7 & 9.0 & 2.9 & 6.8 \\
                 & (C)        & 3.2 & 2.6 & 3.3 & 2.4  & 3.8 & 8.9&   3.1&   7.5  \\
                 & (Al)        & 3.0 & 2.5 & 3.2 & 2.4  & 3.9 & 8.2&   3.2&   7.0  \\
                 & (Cu)        & 4.4 & 3.8 & 4.6 & 3.6  & 5.4 & 11.2&   4.7&   9.7  \\
                 & (Sn)        & 3.1 & 2.5 & 2.9 & 2.5  & 3.6 & 6.9&   3.3&   6.3  \\
                 & (Ta)        & 3.3 & 2.6 & 3.1 & 2.5  & 3.5 & 6.5&   3.4&   6.2  \\
                 & (Pb)        & 4.2      &  3.1     &  3.7     & 3.2       &  4.5     & 8.5      &   4.4    & 8.2  \\ 
\hline
\bf{5 \GeVc }&&&&&&&&\\
\hline
\bf{Total syst.}  & (Be)   &9.0  & 4.8  & 3.2  & 4.2  & 5.1  & 10.0  & 7.9  &13.8 \\
                  & (C)   &11.1   & 5.0  & 3.2  & 3.9  & 5.3  & 8.0  & 6.9  &11.7 \\    
                  & (Al)   &9.5   & 5.0  & 3.3  & 3.9  & 4.8  & 8.9  & 7.8  &12.7 \\    
                  & (Cu)   &10.6   & 8.1  & 7.0  & 3.7  & 5.2  & 6.7  & 7.1  &12.8 \\    
                  & (Sn)   &10.1   & 6.2  & 5.1  & 3.5  & 5.2  & 8.0  & 7.3  &11.6 \\    
                  & (Ta)   &12.5   & 8.1  & 7.7  & 3.7  & 5.2  & 7.7  & 7.1  &10.7 \\    
                  & (Pb)   & 12.6       & 7.6       & 6.6       & 3.7       & 5.6       &  6.5      &  7.3      & 11.5 \\ 
\bf{Statistics        } & (Be)  & 1.7  & 1.6  & 2.1  & 1.3  & 2.1  & 4.3  & 1.5  & 3.3  \\
                        & (C)  & 1.6  & 1.3  & 1.6  & 1.0  & 1.5  & 2.9  & 1.1  & 2.3  \\
                        & (Al)  & 1.5  & 1.3  & 1.7  & 1.1  & 1.7  & 3.2  & 1.4  & 2.7  \\
                        & (Cu)  & 1.6  & 1.4  & 1.7  & 1.2  & 1.7  & 3.1  & 1.4  & 2.6  \\
                        & (Sn)  & 1.5  & 1.3  & 1.5  & 1.1  & 1.6  & 2.8  & 1.4  & 2.4  \\
                        & (Ta)  & 1.7  & 1.4  & 1.7  & 1.3  & 1.7  & 3.0  & 1.5  & 2.5 \\
                        & (Pb)  &  2.2      &  1.8      &    2.1    &   1.7     &  2.2      & 3.9       & 2.0       &  3.4       \\ 
\hline
\bf{8 \GeVc }&&&&&&&&\\
\hline
\bf{Total syst.}    & (Be) &8.6  & 4.7  & 3.1  & 4.0  & 4.5  & 7.5  & 7.2  &12.0 \\
                    & (C) & 10.4   &4.8   &3.1  & 3.7  & 3.8  & 7.0  & 6.6  &11.8  \\
                    & (Al) & 9.4   &5.0   &3.5  & 3.8  & 4.3  & 7.5  & 7.2  &11.2  \\
                    & (Cu) & 10.1   &7.8   &7.1  & 3.6  & 4.3  & 6.3  & 6.5  &10.8  \\
                    & (Sn) &  9.1      &    6.2    &  5.3      &   3.3     &  5.0      &  6.9      &  7.1      & 11.2      \\ 
                    & (Ta) &  11.5      &    8.2    &  7.6      &   3.7     &  5.0      &  5.9      &  7.1      & 10.4      \\ 
                    & (Pb) &  11.2      &    7.5    &  6.5      &   3.5     &  4.9      &  6.9      &  6.4      & 9.9      \\ 
\bf{Statistics}     & (Be)      & 1.4  & 1.3  & 1.7  & 1.0  & 1.6  & 3.1 & 1.2 & 2.4 \\
                    & (C)      & 1.3  & 1.0  & 1.3  & 0.7  & 1.1  & 2.1  & 0.8 &  1.6 \\
                    & (Al)      & 1.2  & 1.1  & 1.4  & 0.9  & 1.3  & 2.5  & 1.0 &  1.9 \\
                    & (Cu)      & 1.0  & 0.9 & 1.1  & 0.7  & 1.0  & 1.8  & 0.8 &  1.4 \\
                    & (Sn)      & 1.0  & 0.9  & 1.1  & 0.7  & 1.0  & 1.8  & 0.9 &  1.5 \\
                    & (Ta)      & 1.1  & 0.9  & 1.2  & 0.8  & 1.1  & 1.9  & 0.9 &  1.6 \\
                    & (Pb)      &   1.1     &   0.9     &   1.0     &   0.7     &   1.0     &  1.8      &  0.9      & 1.5 \\          
\end{tabular}
\end{center}
}
\end{table*}
\begin{table*}[h!]
\small{

\begin{center}
\vspace{2mm}
\begin{tabular}{ l l rrr | rrr | rr} \hline
\bf{p (\GeVc) }&\multicolumn{4}{c|}{0.1 -- 0.3}
                            &\multicolumn{3}{c|}{0.3 -- 0.5}
                            &\multicolumn{2}{c}{0.5 -- 0.7} \\
\hline
\bf{Angle (\mrad)}& &350--950&950--1550&1550--2150&350--950&950--1550&
1550--2150&350--950&950--1550 \\

\hline
\bf{8.9 \GeVc }&&&&&&&&\\
\hline
\bf{Total syst.}   & (Be) & 8.8 & 4.7  & 3.1  & 4.2  & 4.4  & 7.7  & 7.9  & 12.4 \\
\bf{Statistics}    & (Be)  & 0.5 & 0.5  & 0.6  & 0.4  & 0.6  & 1.1  & 0.4  & 0.9  \\
\hline
\bf{12 \GeVc }&&&&&&&&\\
\hline
\bf{Total syst.}   & (Be) &9.0 & 4.8  & 3.3  & 3.8 & 4.1  & 7.3  & 7.3  &12.1 \\
                   & (C) &9.8  & 5.8  & 4.1  & 3.9 &  4.5 &  6.8 &  7.5 & 10.8  \\
                   & (Al) &9.6  & 5.4  & 3.9  & 3.5 &  4.0 &  7.6 &  7.6 & 11.7  \\
                   & (Cu) &10.1  & 7.8  & 7.1  & 2.9 &  4.3 &  6.3 &  6.7 & 11.2  \\
                   & (Sn) &10.0  & 6.7  & 5.7  & 2.9 &  4.7 &  6.6 &  6.1 & 9.6  \\
                   & (Ta) &11.4  & 7.2  & 6.8  & 2.3 &  4.4 &  5.9 &  6.6 & 9.4  \\
                   & (Pb) & 10.8      &  7.1      &   6.7     &    2.9   &    4.4    &  5.6      &  7.1      &  9.1 \\  
\bf{Statistics}    & (Be)  & 1.1  & 1.1  & 1.5  & 0.8  & 1.3  & 2.6  & 0.9  & 2.0 \\
                   & (C)  & 1.9  & 1.9  & 2.4  & 1.5  & 2.2  & 4.4  & 1.6  & 3.2  \\
                   & (Al)  & 1.3  & 1.2  & 1.5  & 0.9  & 1.5  & 2.8  & 1.1  & 2.1  \\
                   & (Cu)  & 1.2  & 1.2  & 1.5  & 0.9  & 1.3  & 2.5  & 1.0  & 1.9  \\
                   & (Sn)  & 0.8  & 0.8  & 0.9  & 0.6  & 0.9  & 1.5  & 0.7  & 1.2  \\
                   & (Ta)  & 1.1  & 1.0  & 1.2  & 0.8  & 1.2  & 2.1  & 1.0  & 1.7  \\
                   & (Pb)  & 1.8  & 1.6  & 2.0  & 1.4  & 1.9  & 3.5  & 1.6  & 2.7  \\
\hline
\bf{12.9 \GeVc }&&&&&&&&\\
\hline
\bf{Total syst.}   & (Al) &9.6 &  5.2 &  3.7 &  3.5 &  4.1 &  6.8 &  7.4 & 11.7 \\
\bf{Statistics}    & (Al)  & 0.5 &  0.5 &  0.6 &  0.4 &  0.6 &  1.1 &  0.4 &  0.8 \\
\end{tabular}
\end{center}
}
\end{table*}
\FloatBarrier
\subsection{Comparisons with MC predictions}
\label{sec:compare}

As our final results, obtained with a full correction of the
distortions of TPC tracks (static+dynamic),
are compatible with our preliminary ones published in 
\cite{ref:harp:tantalum, ref:harp:cacotin, ref:harp:bealpb},
we refer to these papers for a comparison with published data. 
We only stress here that previous data sets are scarce, with big total errors
and cover only a limited region of the phase space covered by the
HARP experiment. 

\begin{figure*}[htbp]
\begin{center}
  \includegraphics[width=.47\textwidth]{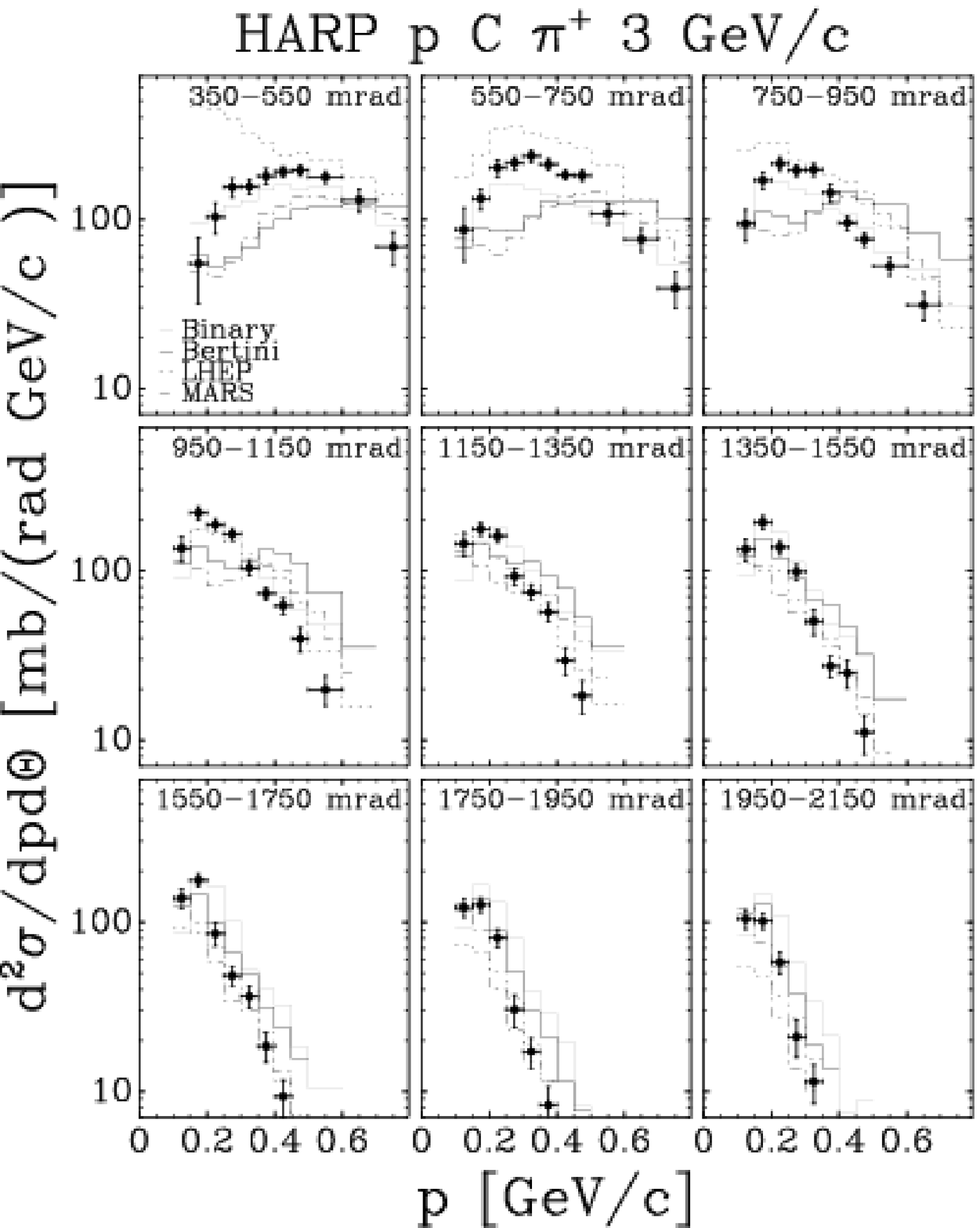}
  \includegraphics[width=0.47\textwidth]{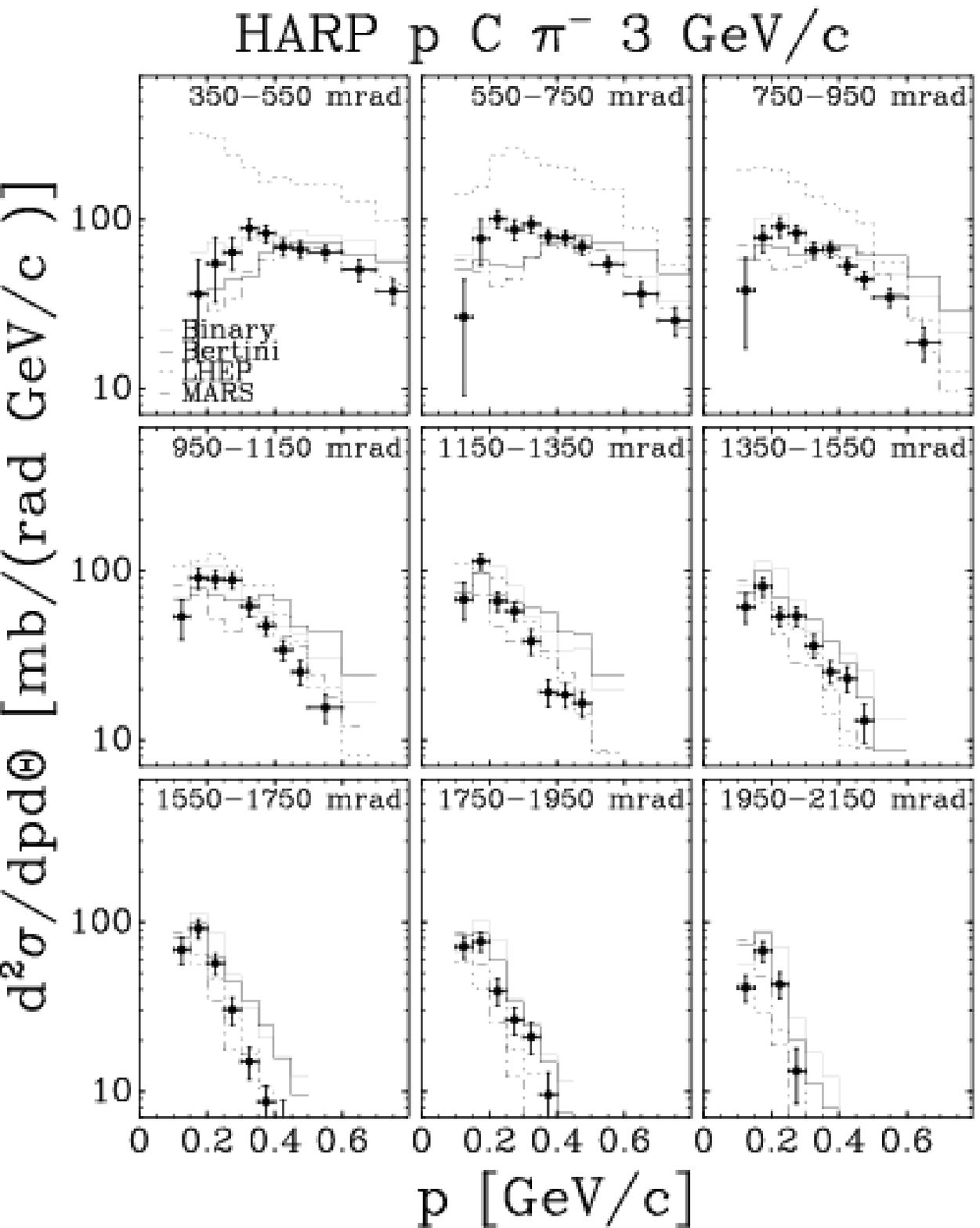}
\end{center}
\caption{
 Comparison of HARP double-differential $\pi^{+}$ ($\pi^{-}$) cross sections for p--C at 3 GeV/c with
 GEANT4 and MARS MC predictions, using several generator models (see text for details).
}
\label{fig:G43a}
\end{figure*}

\begin{figure*}[htbp]
\begin{center}
  \includegraphics[width=0.47\textwidth]{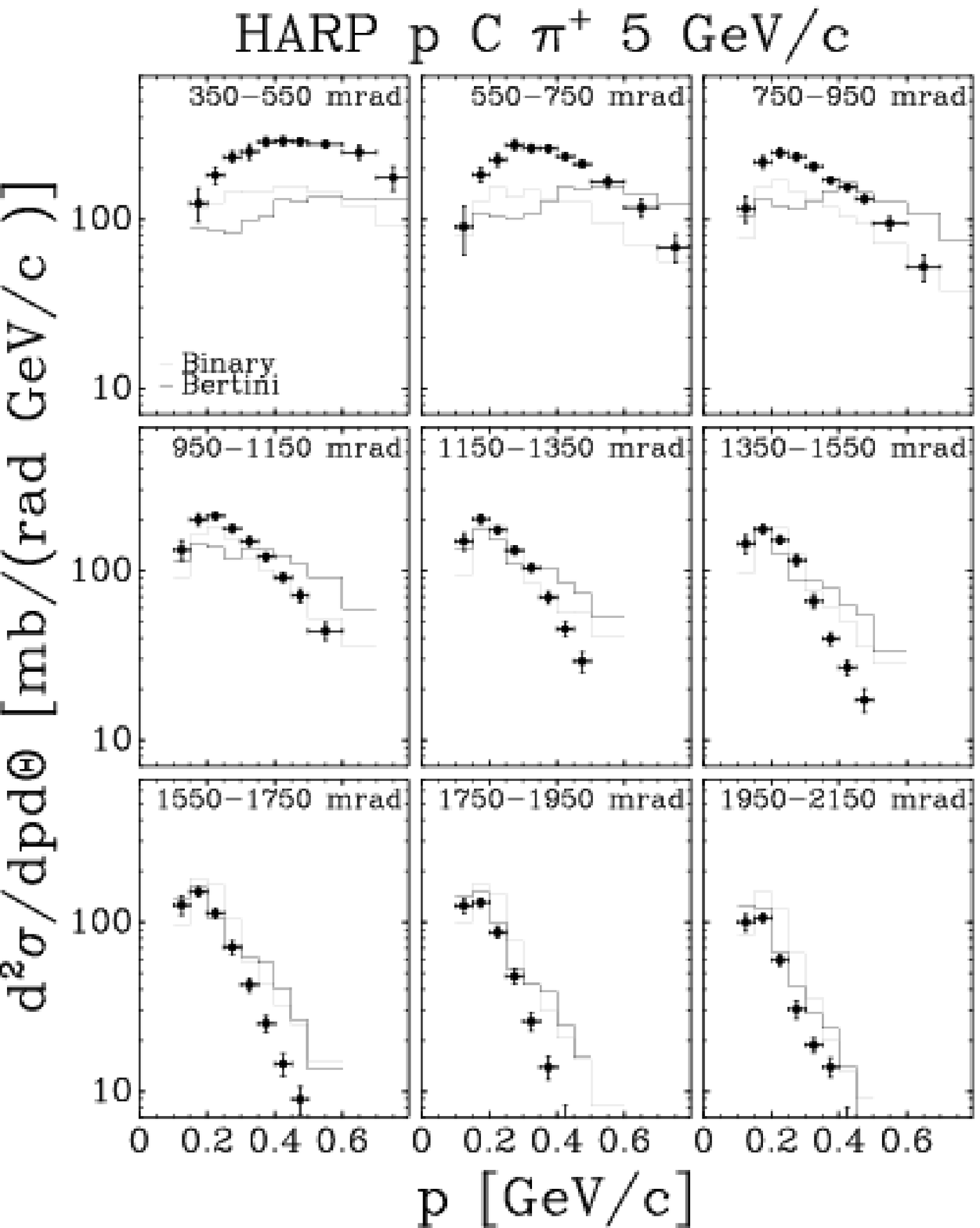}
  \includegraphics[width=0.47\textwidth]{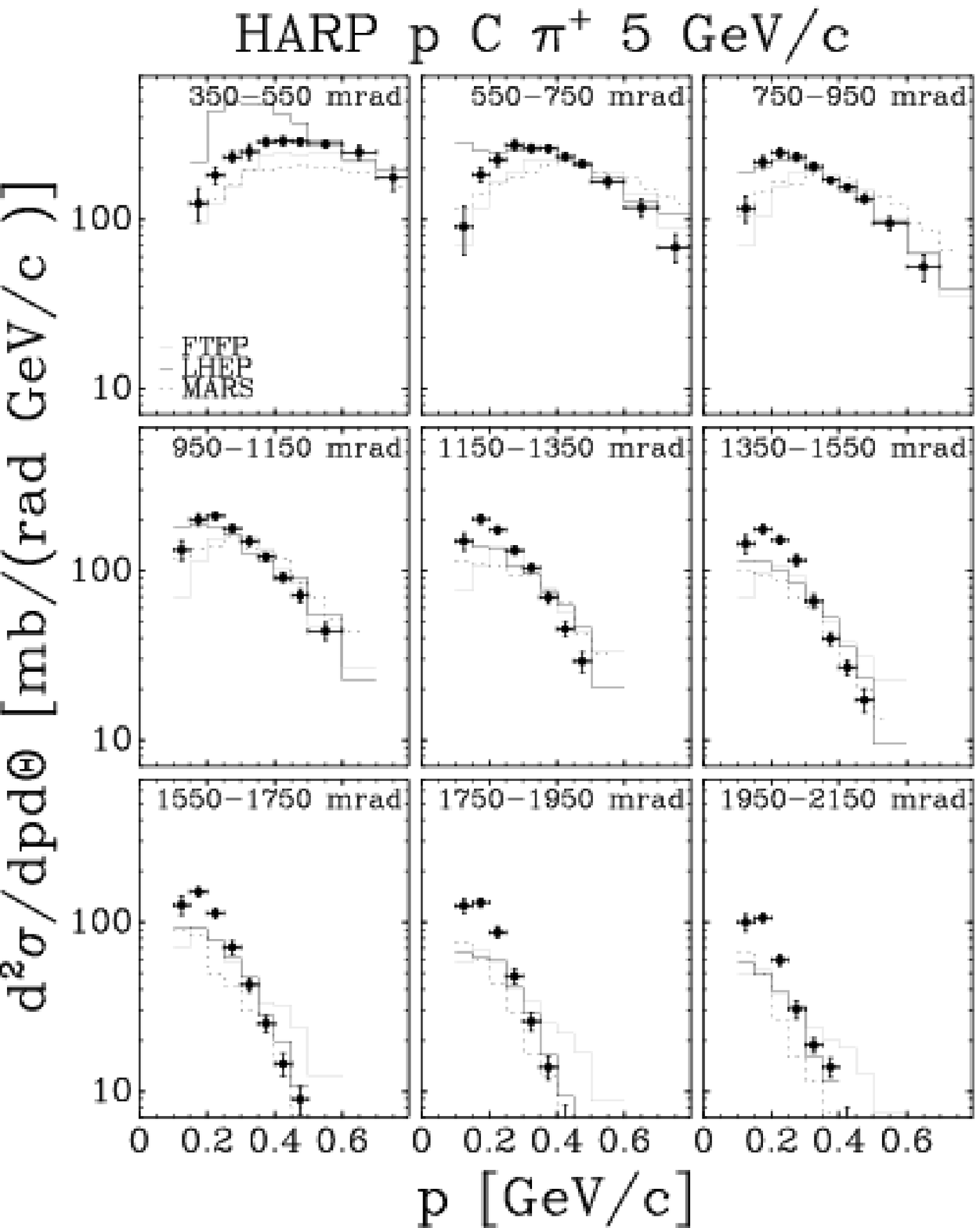}
\end{center}
\caption{
 Comparison of HARP double-differential $\pi^{+}$ cross sections for p--C at 5 GeV/c with
 GEANT4 and MARS MC predictions, using several generator models (see text for details).
}
\label{fig:G44a}
\end{figure*}

\begin{figure*}[htbp]
\begin{center}
  \includegraphics[width=.47\textwidth]{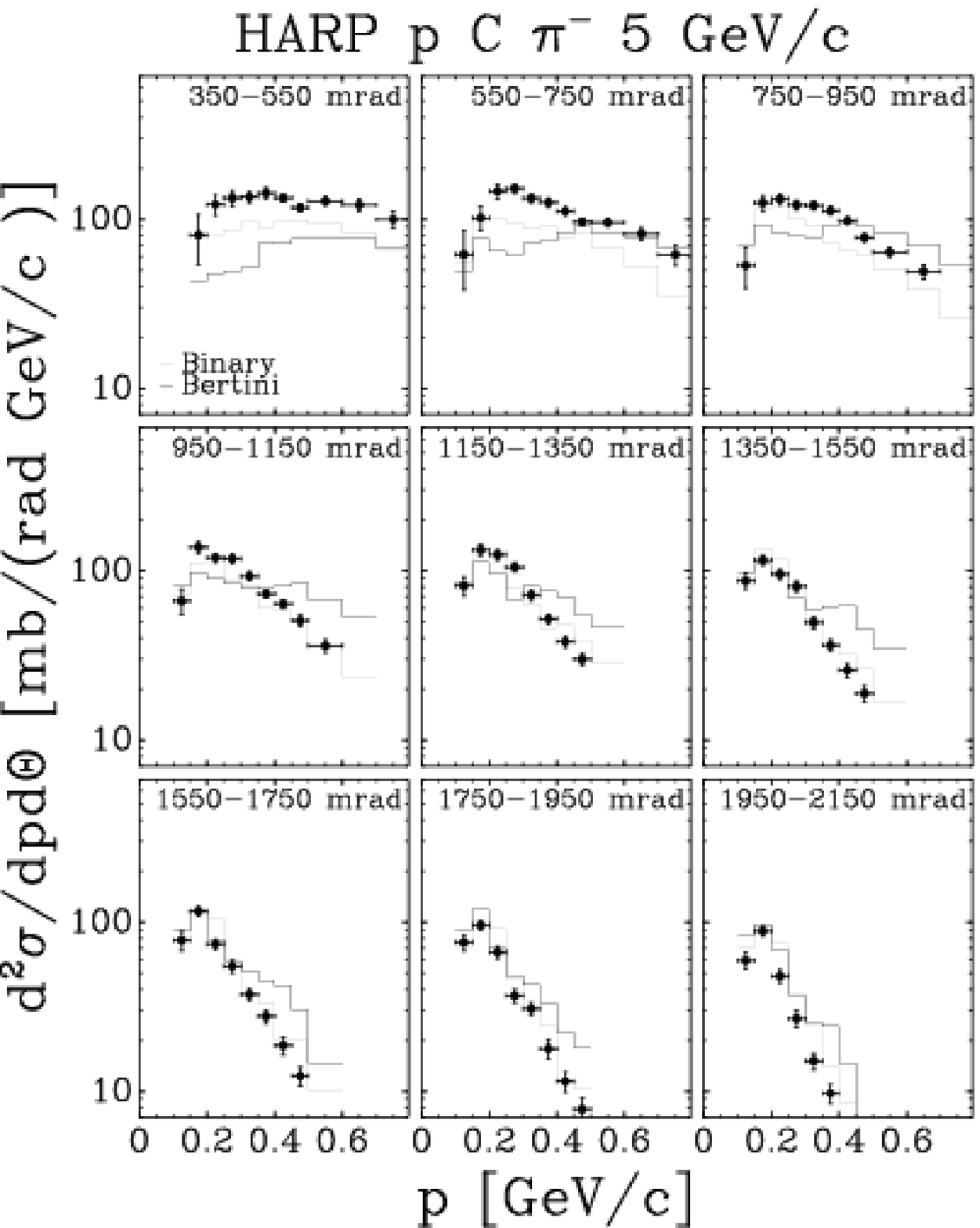}
  \includegraphics[width=.47\textwidth]{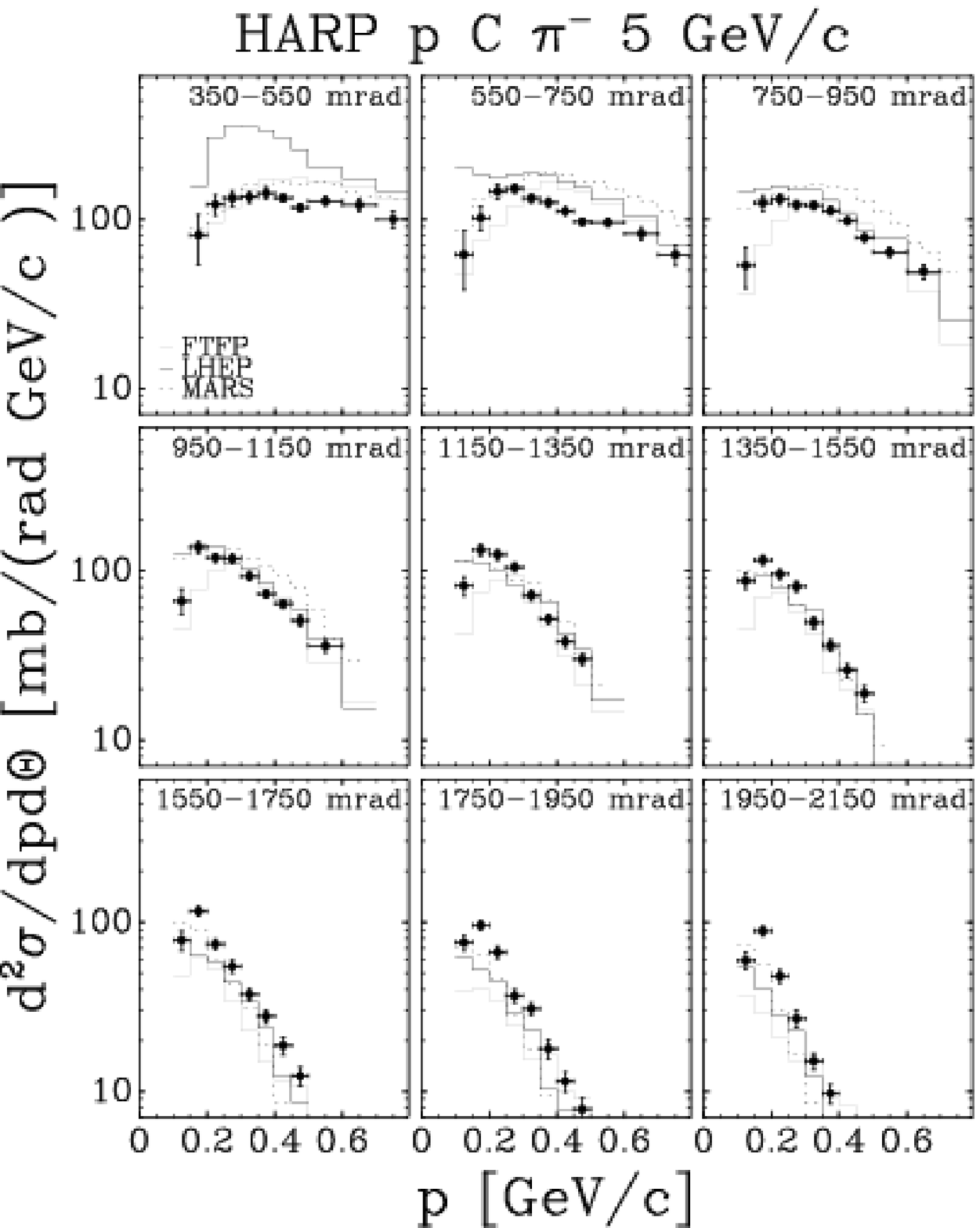}
\end{center}
\caption{
 Comparison of HARP double-differential $\pi^{-}$ cross sections for p--C at 5 GeV/c with
 GEANT4 MC predictions, using several generator models (see text for details).
}
\label{fig:G44b}
\end{figure*}

\begin{figure*}[htbp]
\begin{center}
  \includegraphics[width=0.47\textwidth]{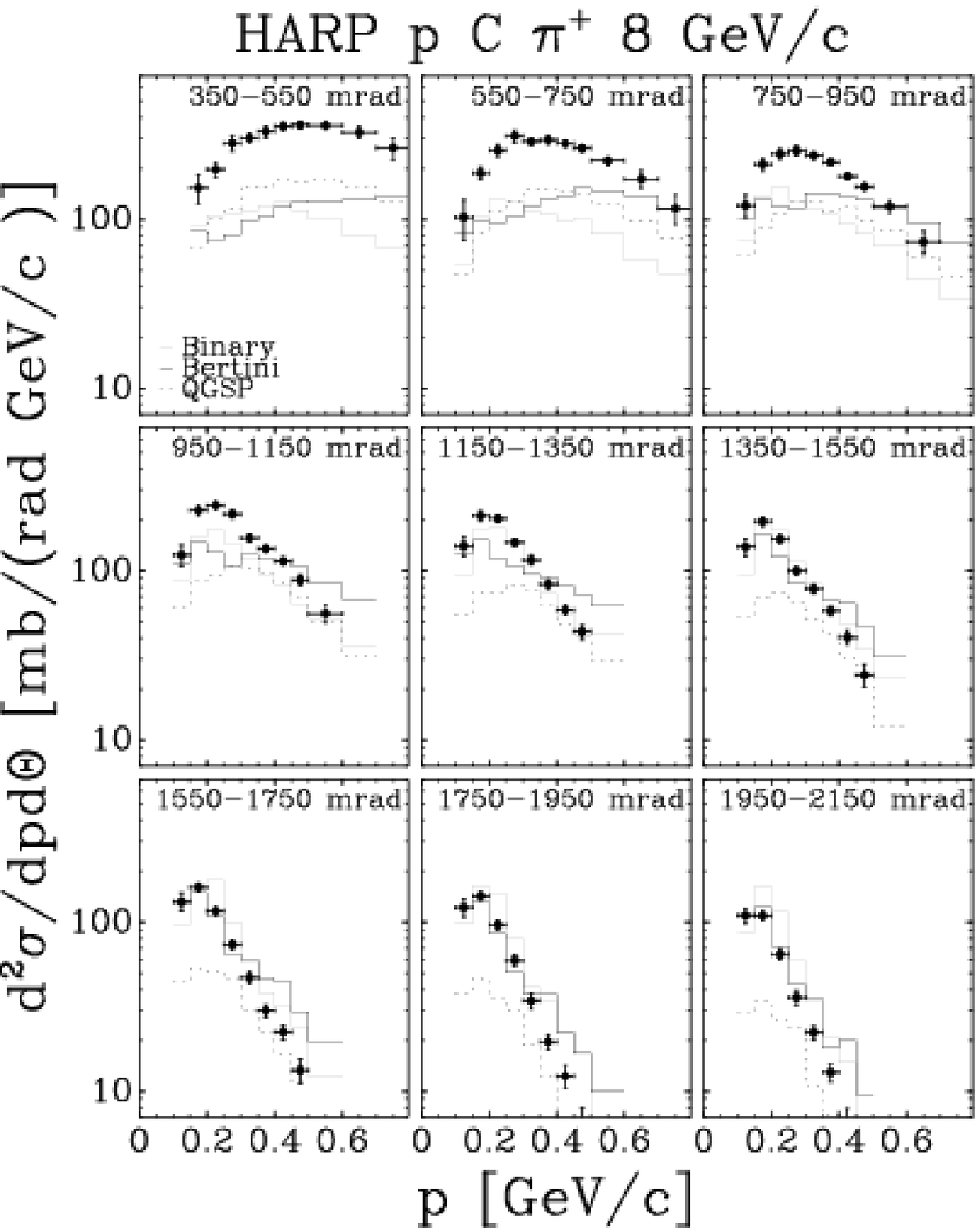}
  \includegraphics[width=0.47\textwidth]{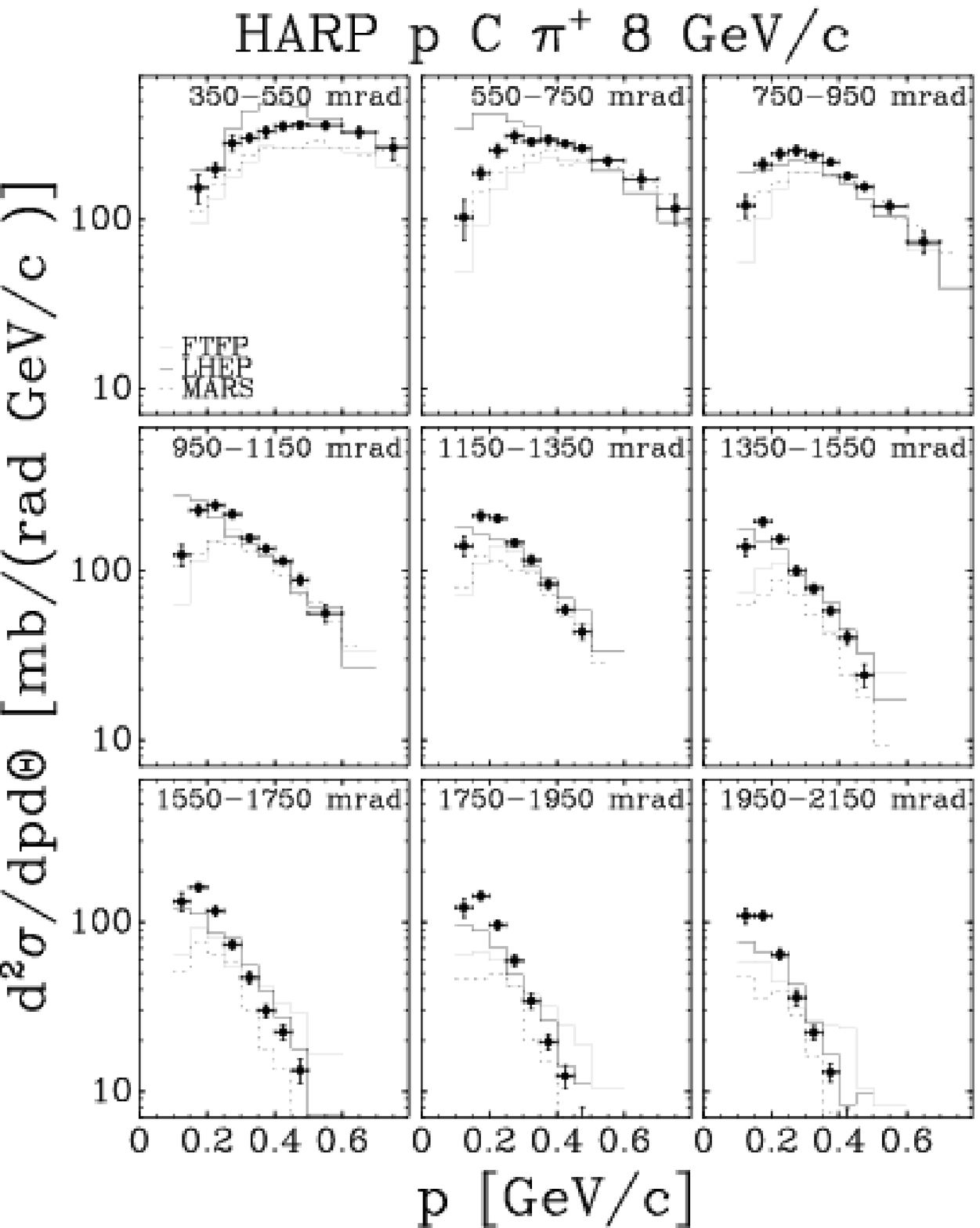}
\end{center}
\caption{
 Comparison of HARP double-differential $\pi^{+}$ cross sections for p--C at 8 GeV/c with
 GEANT4 and MARS MC predictions, using several generator models (see text for details).
}
\label{fig:G45a}
\end{figure*}

\begin{figure*}[htbp]
\begin{center}
  \includegraphics[width=0.47\textwidth]{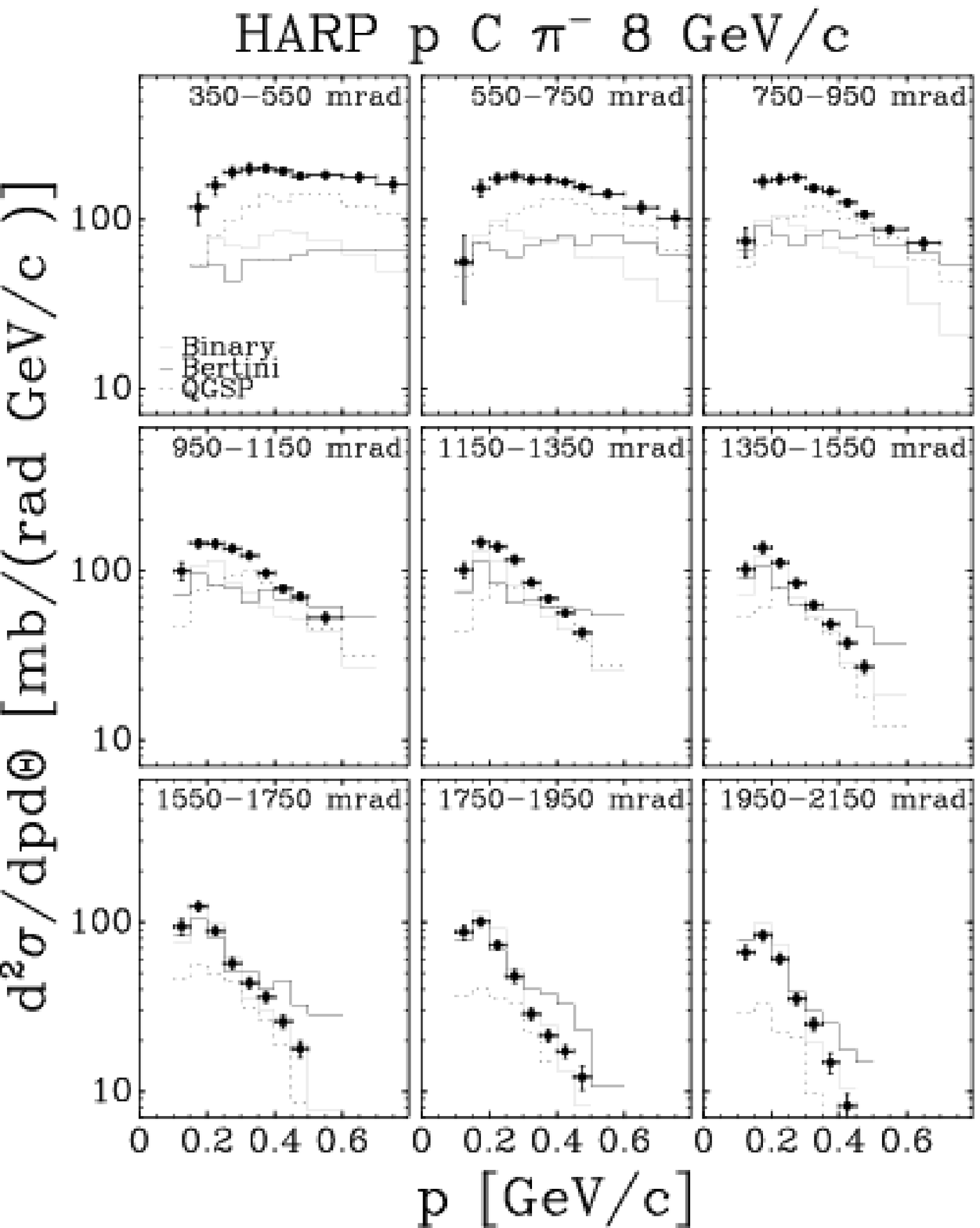}
  \includegraphics[width=0.47\textwidth]{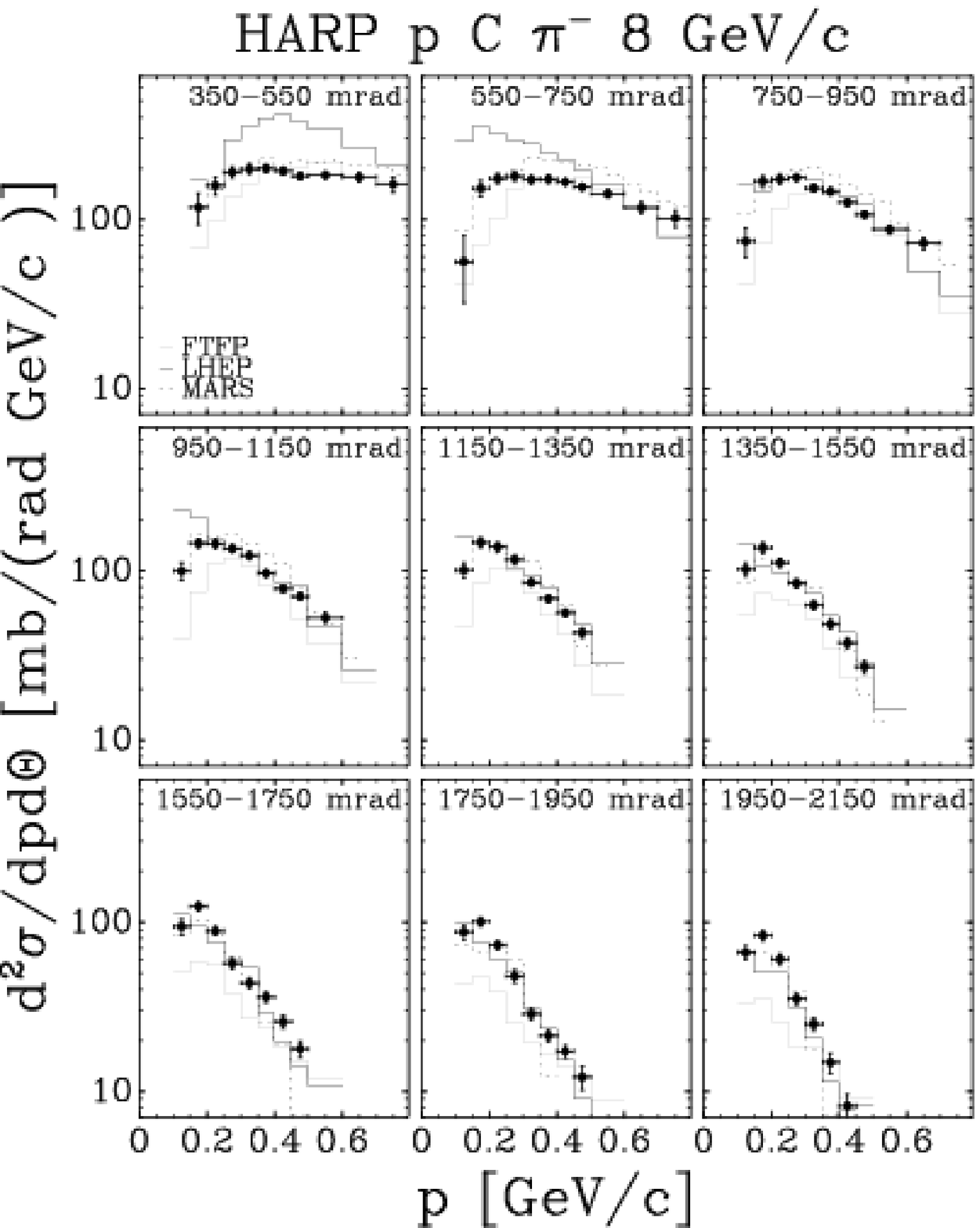}
\end{center}
\caption{
 Comparison of HARP double-differential $\pi^{-}$ cross sections for p--C at 8 GeV/c with
 GEANT4 and MARS MC predictions, using several generator models (see text for details).
}
\label{fig:G45b}
\end{figure*}

\begin{figure*}[htbp]
\begin{center}
  \includegraphics[width=0.47\textwidth]{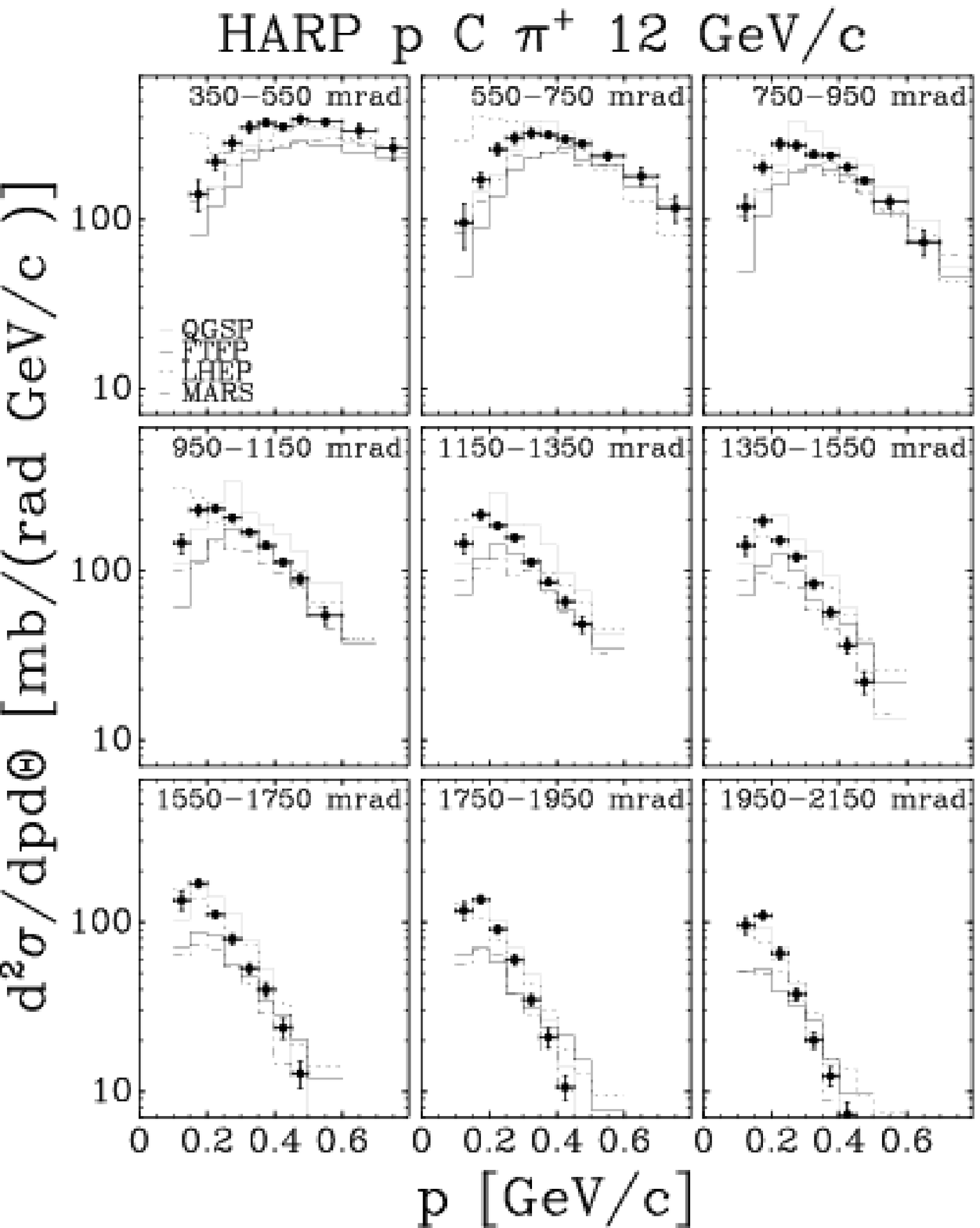}
  \includegraphics[width=0.47\textwidth]{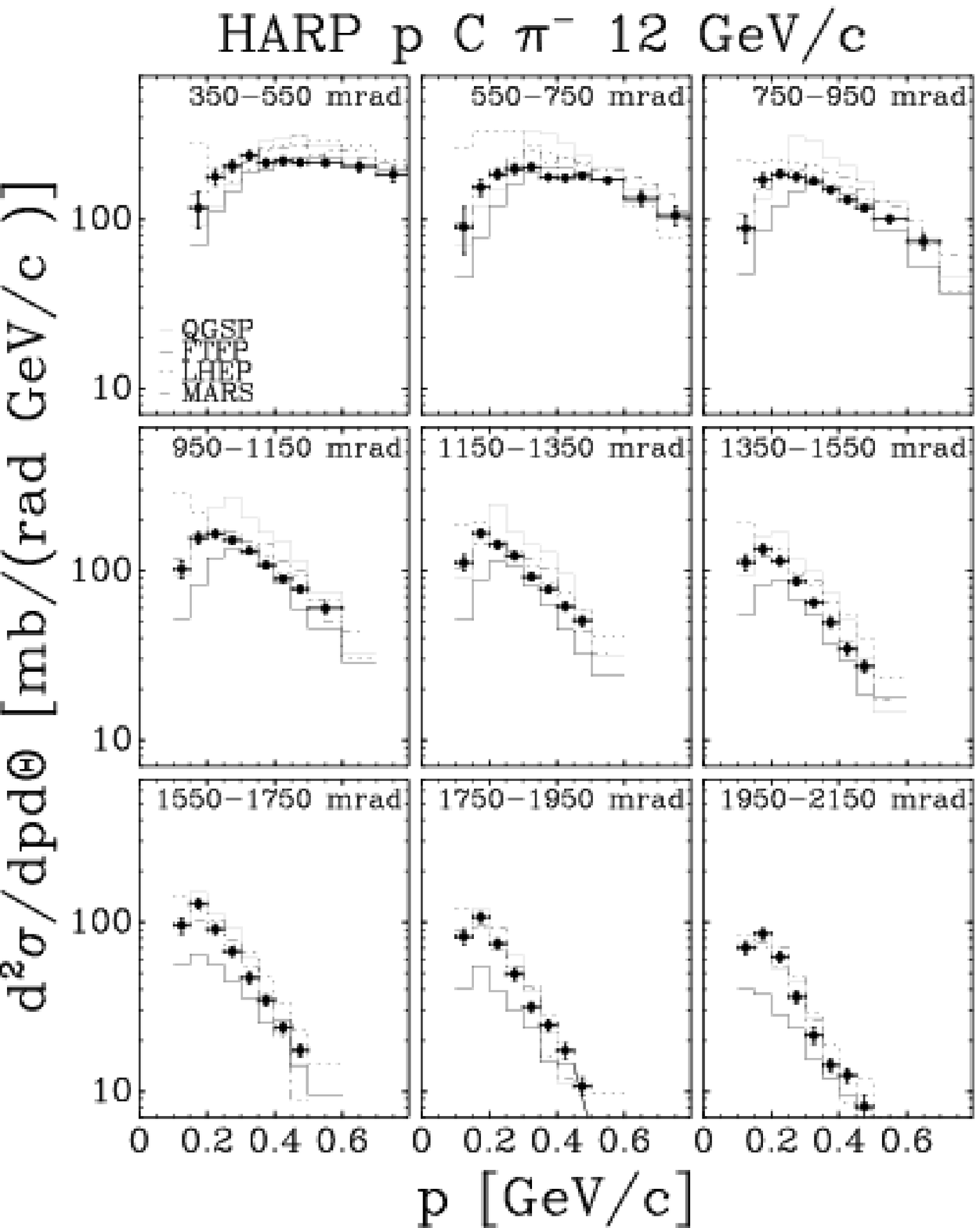}
\end{center}
\caption{
 Comparison of HARP double-differential $\pi^{\pm}$ cross sections for p--C at 12 GeV/c with
 GEANT4 and MARS MC predictions, using several generator models (see text for details).
}
\label{fig:G46}
\end{figure*}

\begin{figure*}[htbp]
\begin{center}
  \includegraphics[width=0.47\textwidth]{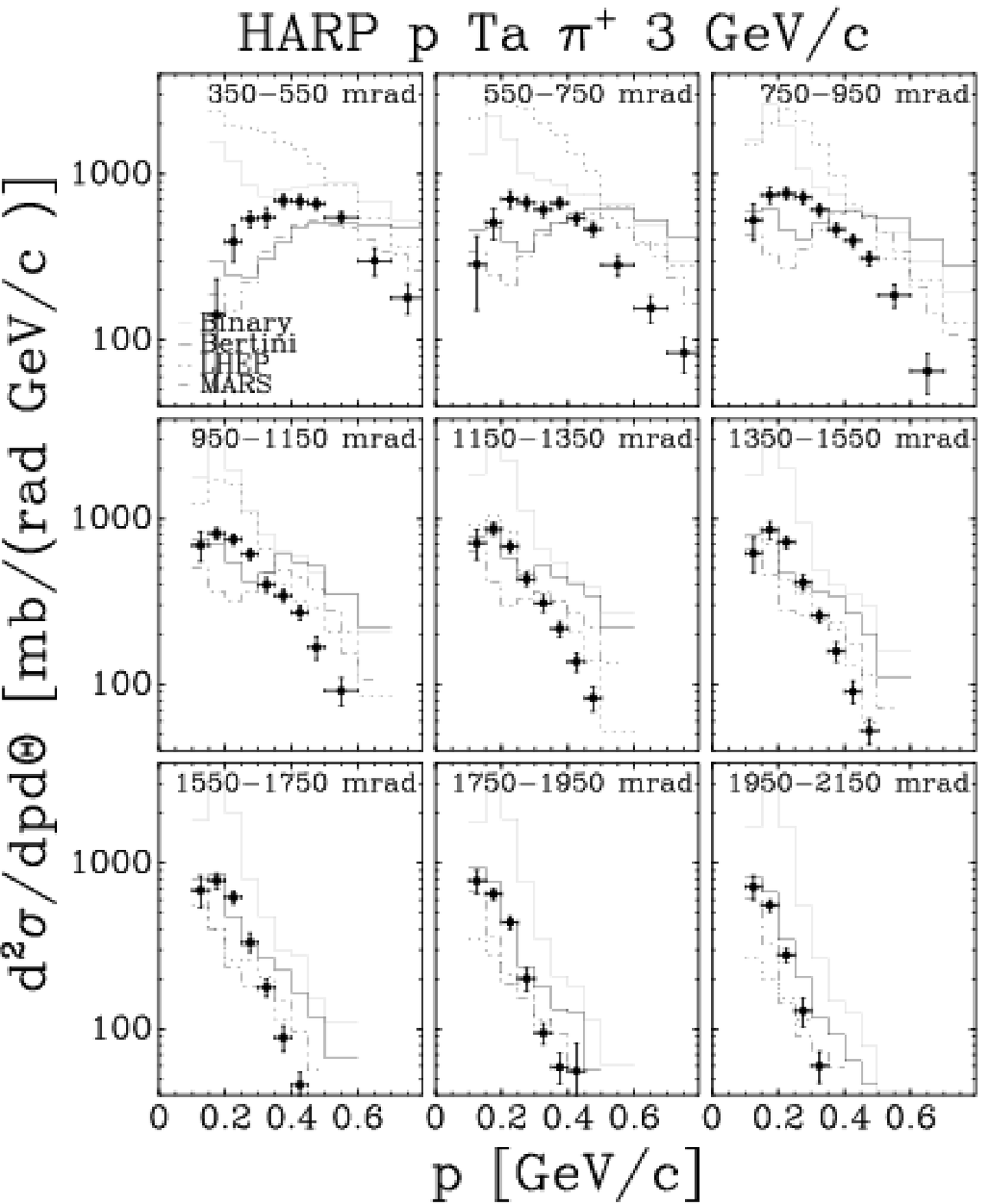}
  \includegraphics[width=0.47\textwidth]{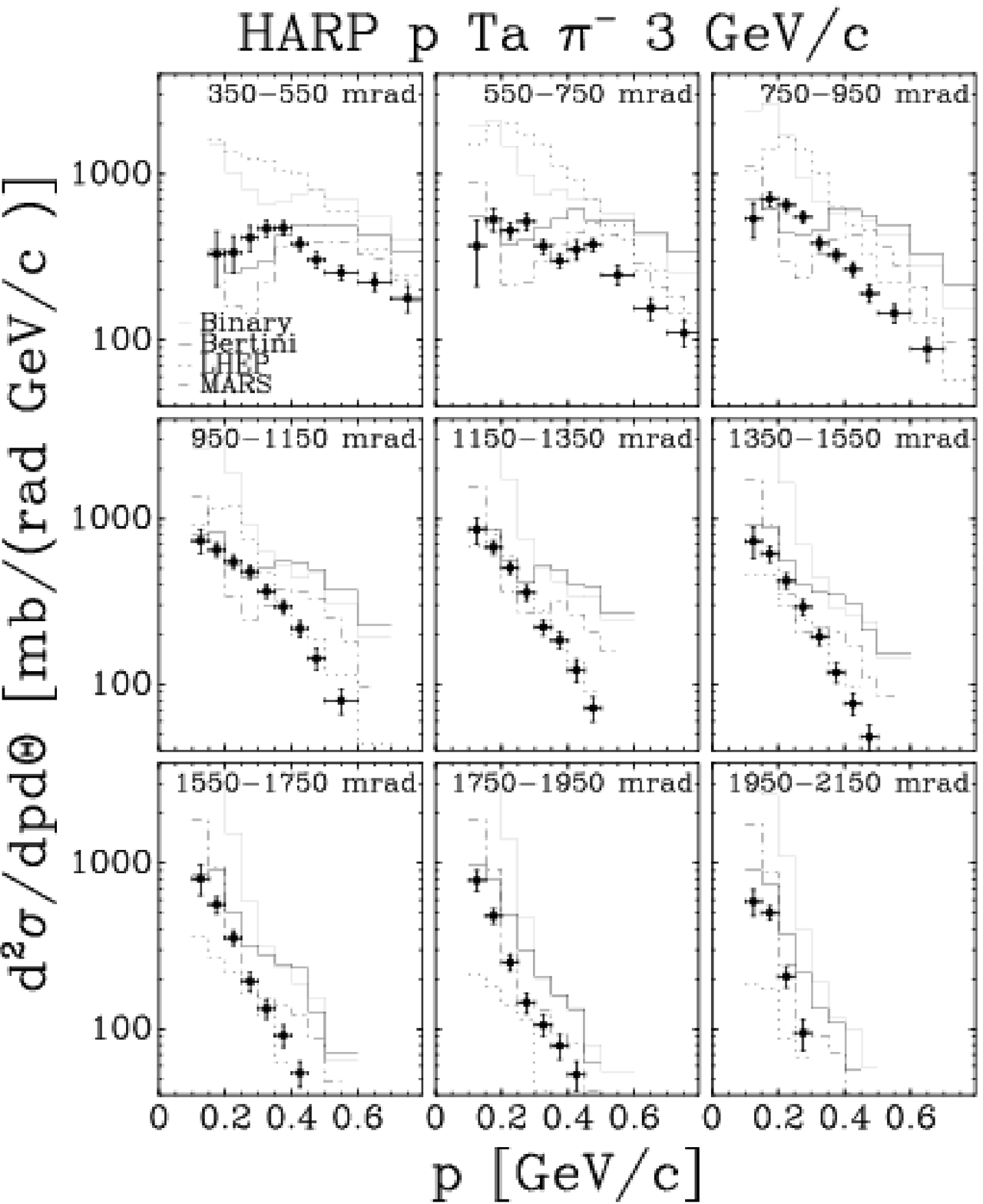}
\end{center}
\caption{
 Comparison of HARP double-differential $\pi^{\pm}$ cross sections for p--Ta at 3 GeV/c with
 GEANT4 and MARS MC predictions, using several generator models (see text for details).
}
\label{fig:G53}
\end{figure*}

\begin{figure*}[htbp]
\begin{center}
  \includegraphics[width=0.47\textwidth]{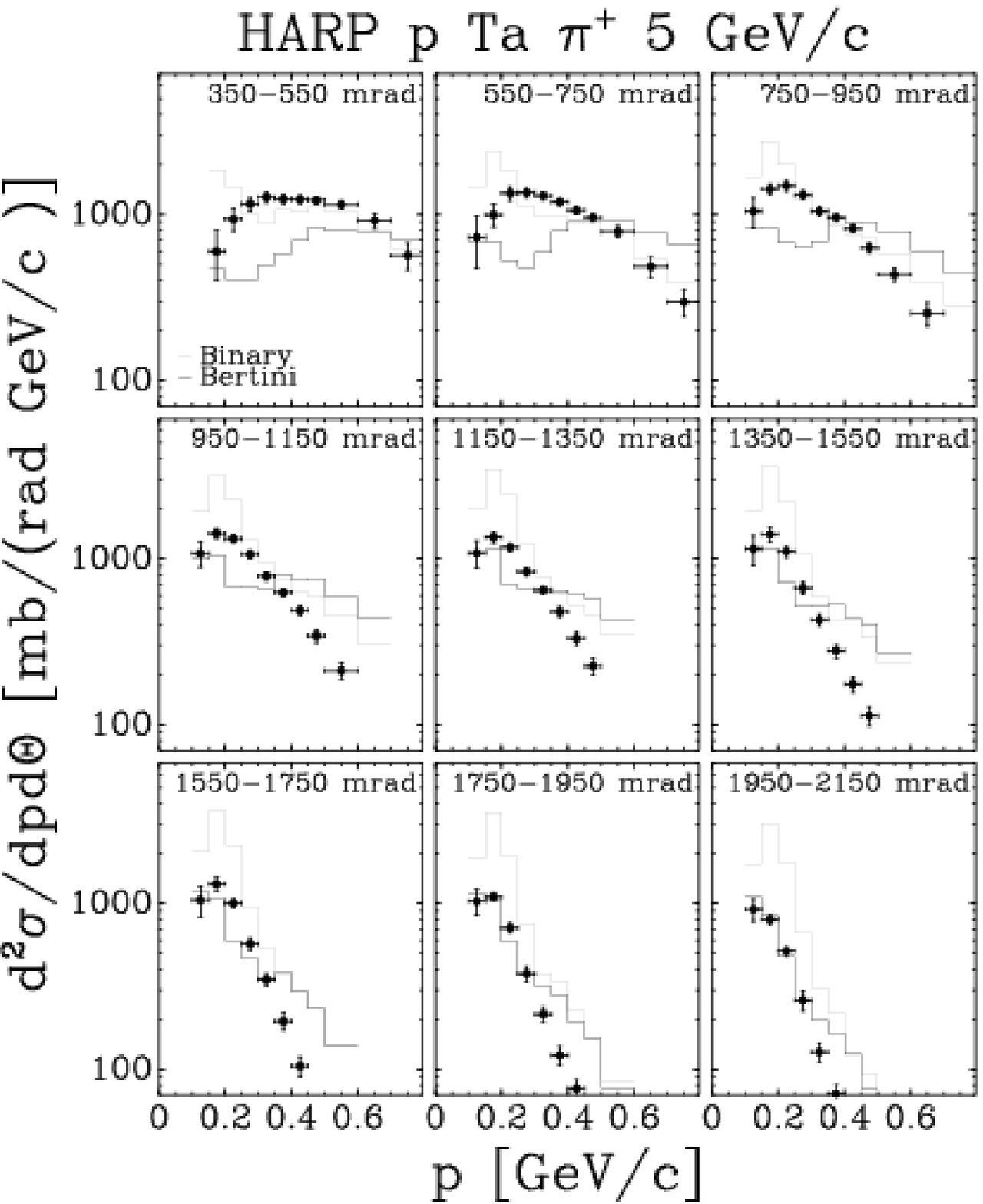}
  \includegraphics[width=0.47\textwidth]{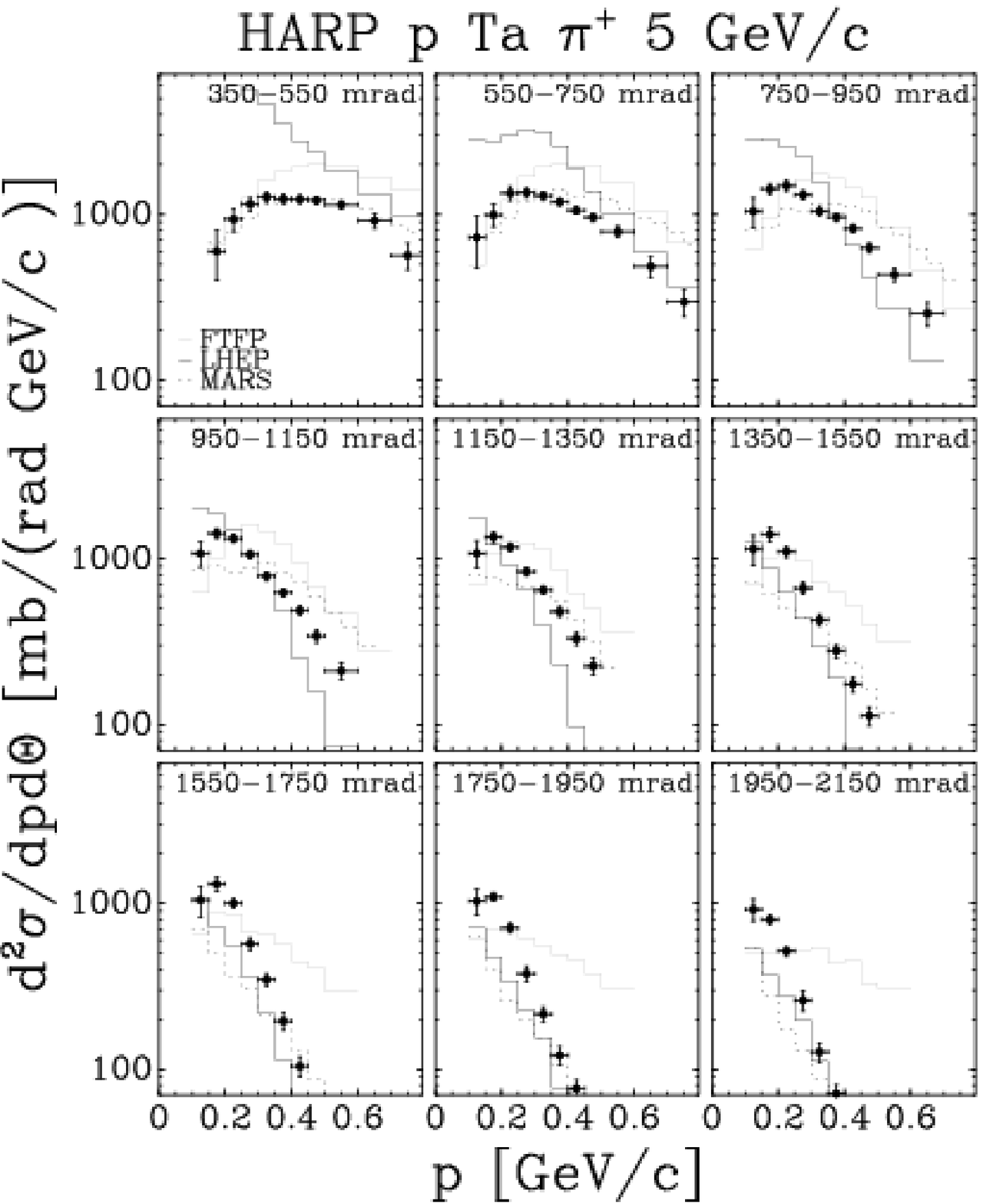}
\end{center}
\caption{
 Comparison of HARP double-differential $\pi^{+}$ cross sections for p--Ta at 5 GeV/c with
 GEANT4 and MARS MC predictions, using several generator models (see text for details).
}
\label{fig:G54a}
\end{figure*}

\begin{figure*}[htbp]
\begin{center}
  \includegraphics[width=0.47\textwidth]{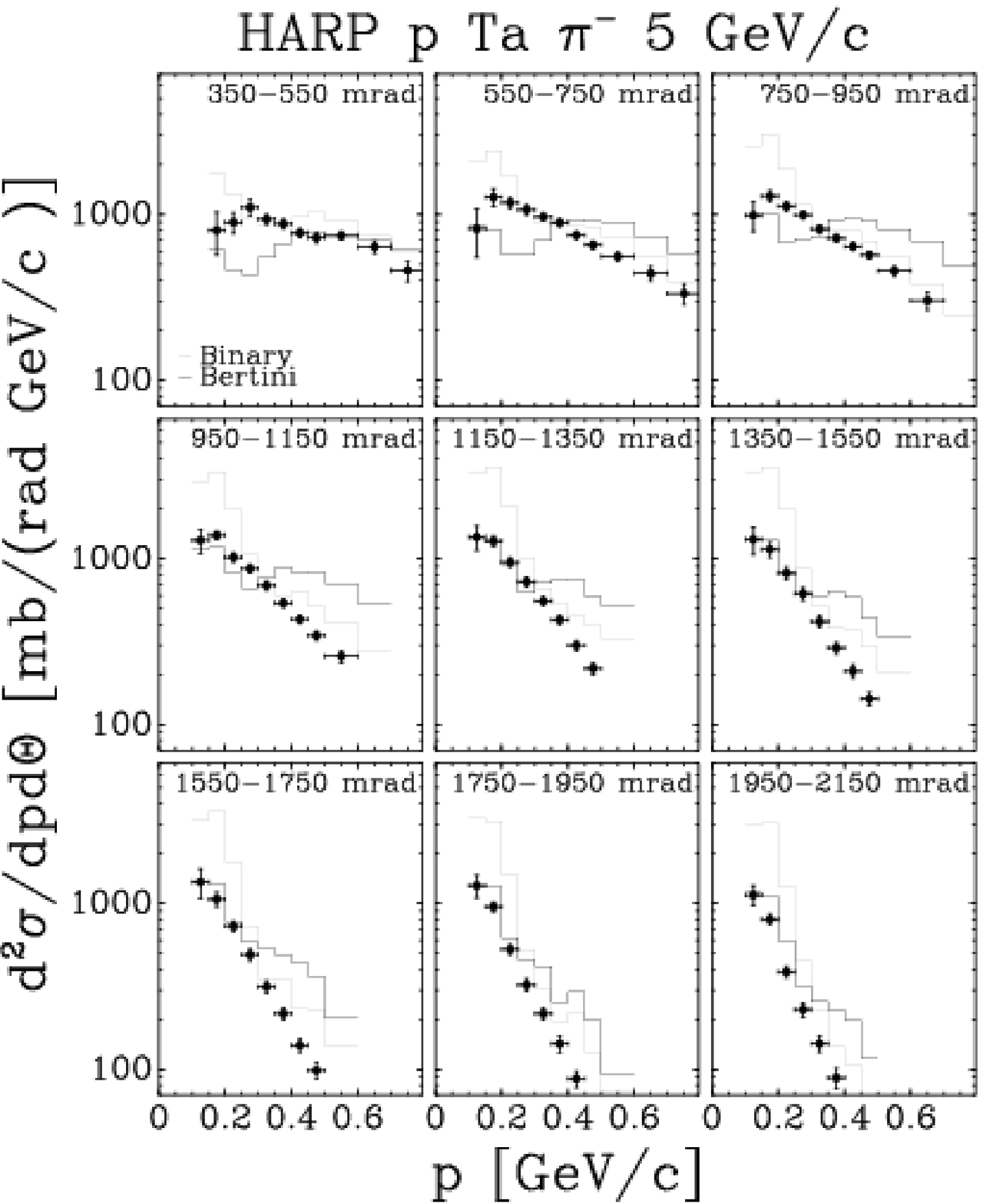}
  \includegraphics[width=0.47\textwidth]{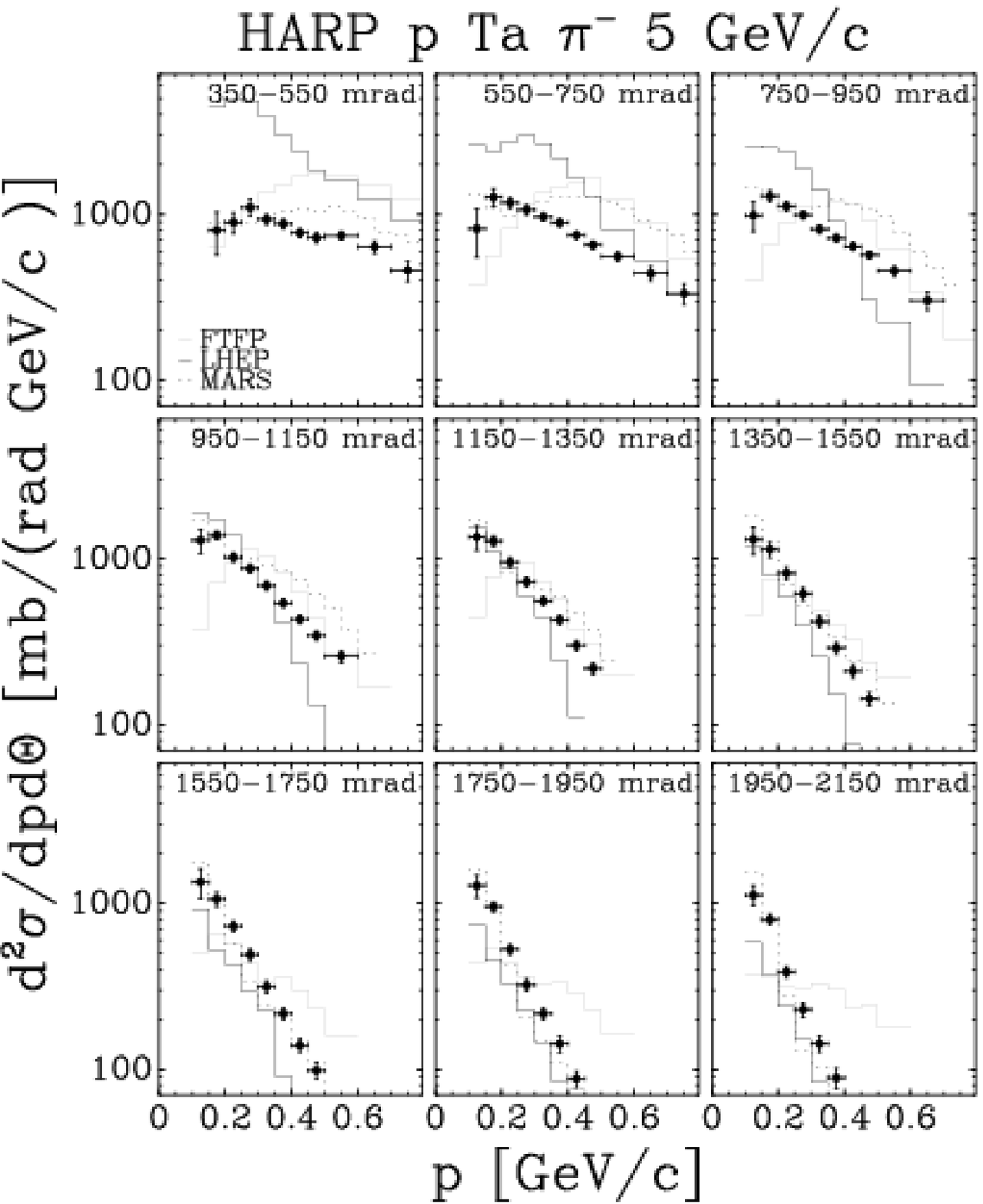}
\end{center}
\caption{
 Comparison of HARP double-differential $\pi^{-}$ cross sections for p--Ta at 5 GeV/c with
 GEANT4 and MARS MC predictions, using several generator models (see text for details).
}
\label{fig:G54b}
\end{figure*}

\begin{figure*}[htbp]
\begin{center}
  \includegraphics[width=0.47\textwidth]{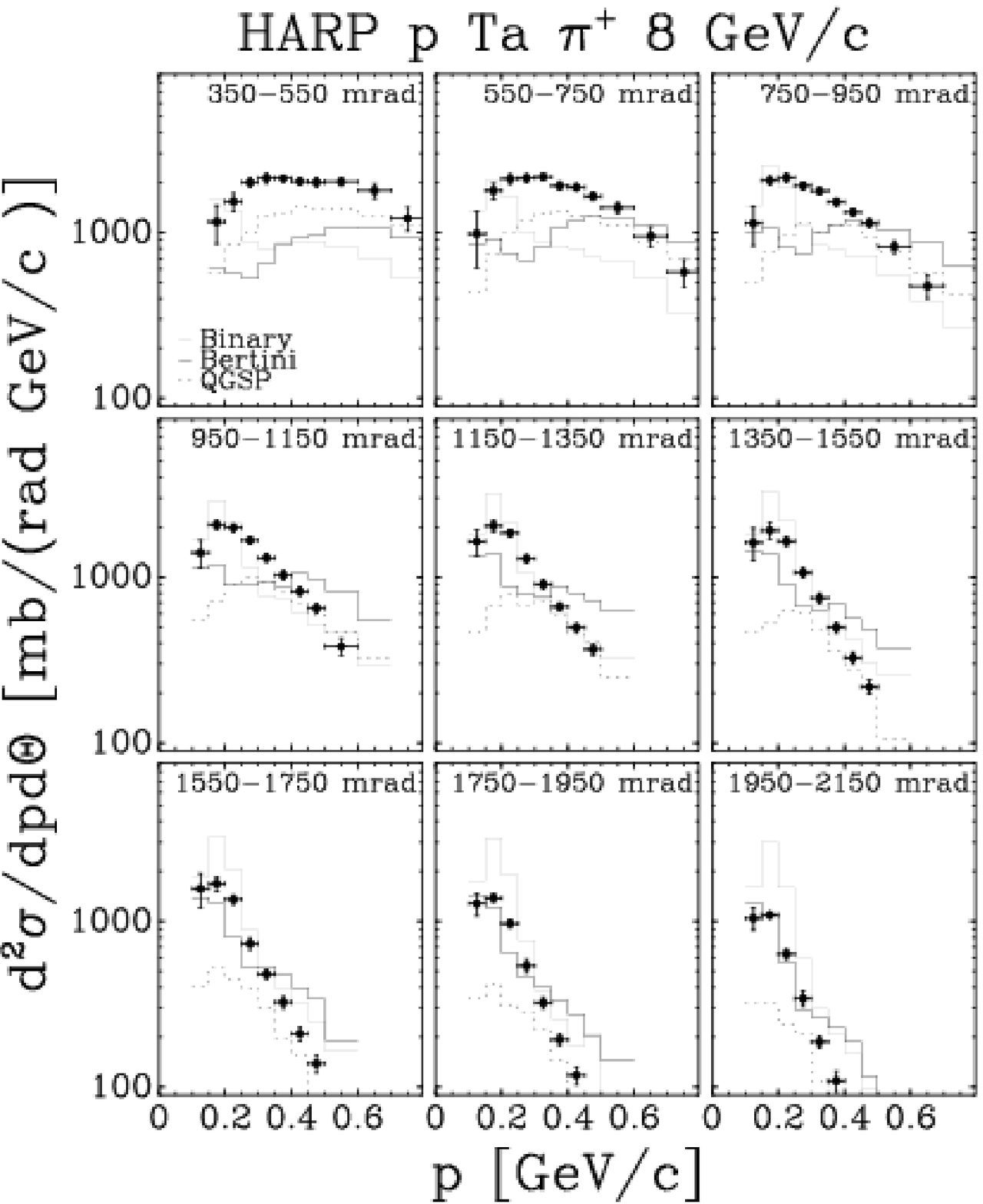}
  \includegraphics[width=0.47\textwidth]{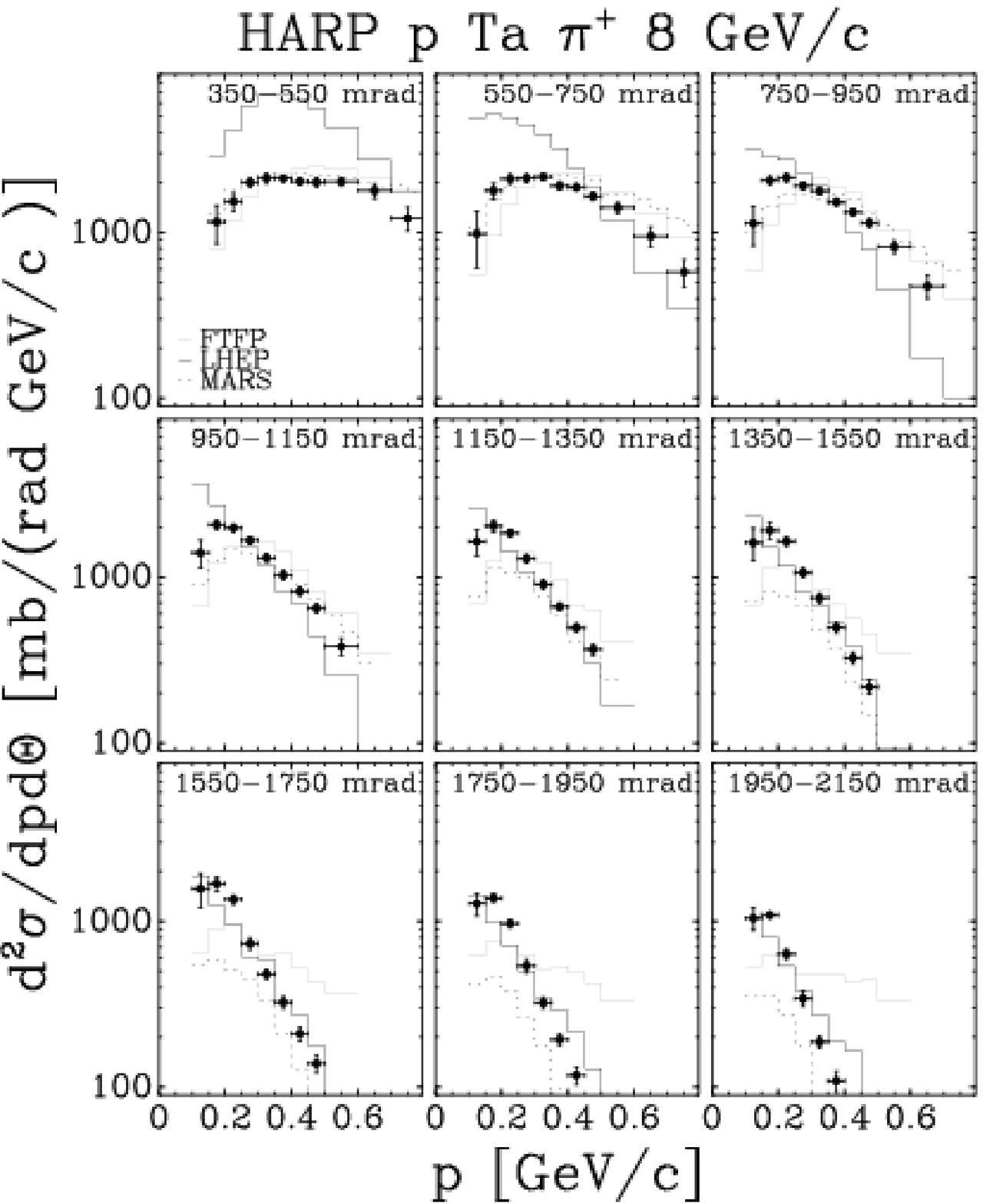}
\end{center}
\caption{
 Comparison of HARP double-differential $\pi^{+}$ cross sections for p--Ta at 8 GeV/c with
 GEANT4 and MARS MC predictions, using several generator models (see text for details).
}
\label{fig:G55a}
\end{figure*}

\begin{figure*}[htbp]
\begin{center}
  \includegraphics[width=0.47\textwidth]{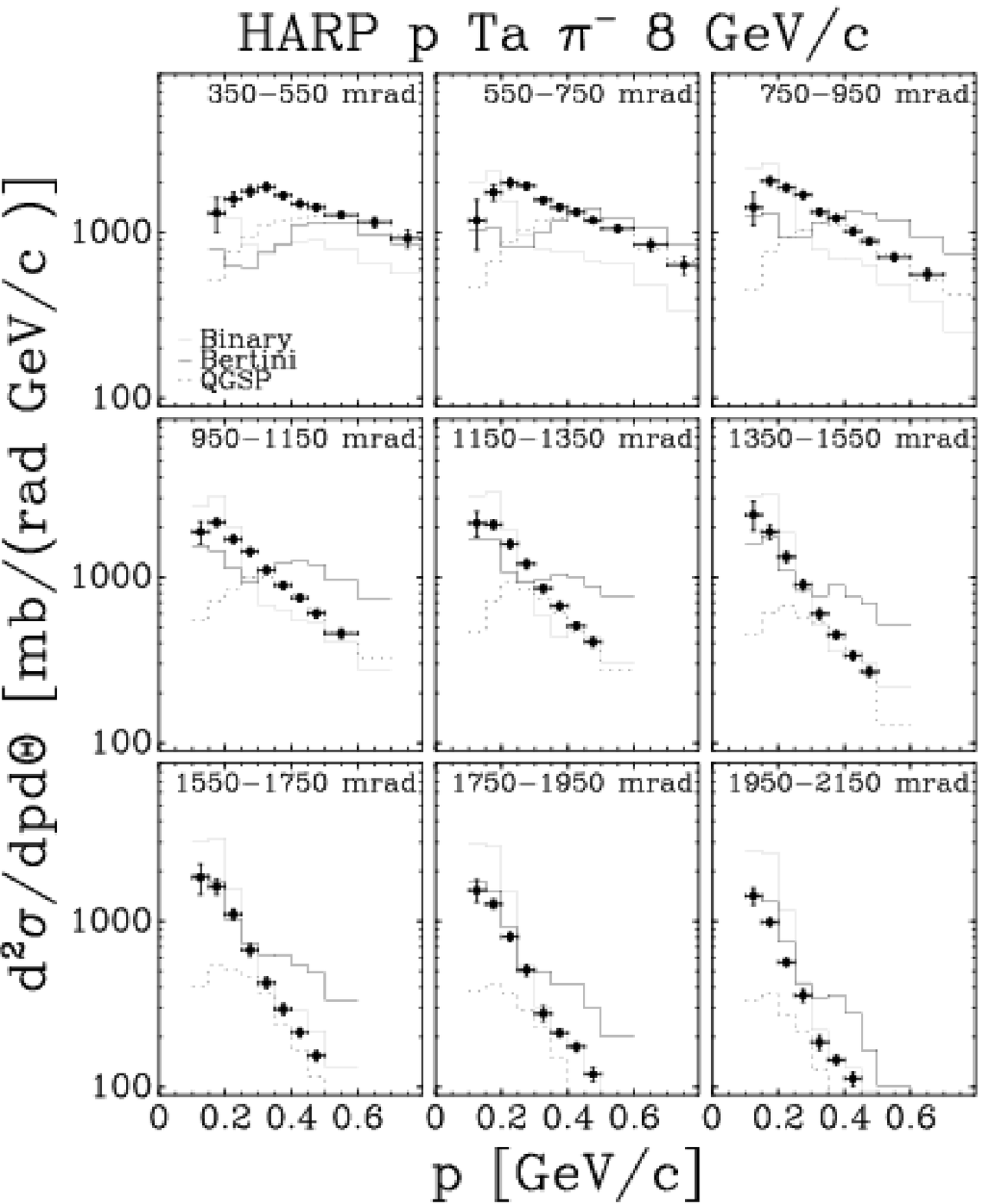}
  \includegraphics[width=0.47\textwidth]{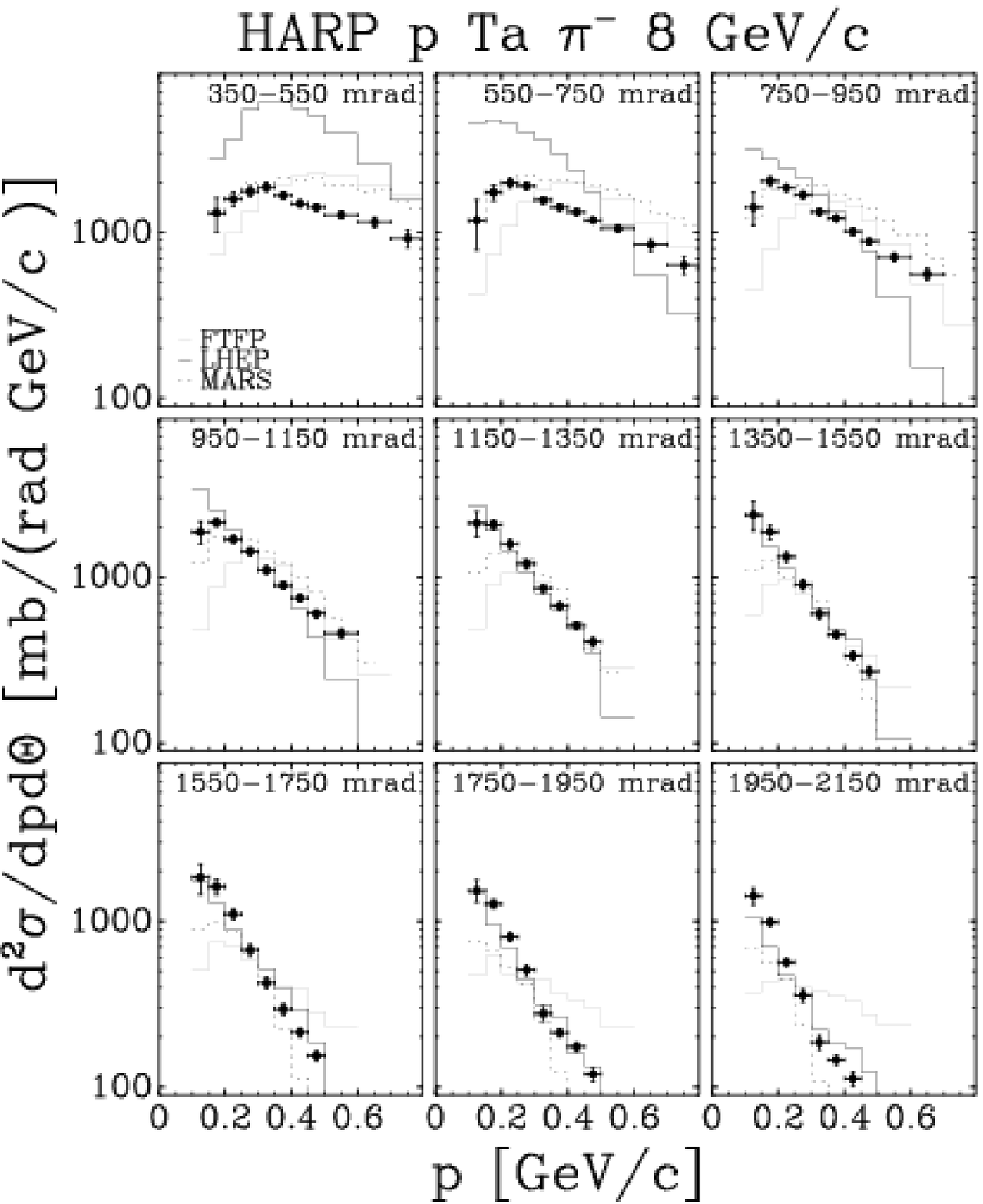}
\end{center}
\caption{
 Comparison of HARP double-differential $\pi^{-}$ cross sections for p--Ta at 8 GeV/c with
 GEANT4 and MARS MC predictions, using several generator models (see text for details).
}
\label{fig:G55b}
\end{figure*}

\begin{figure*}[htbp]
\begin{center}
  \includegraphics[width=0.47\textwidth]{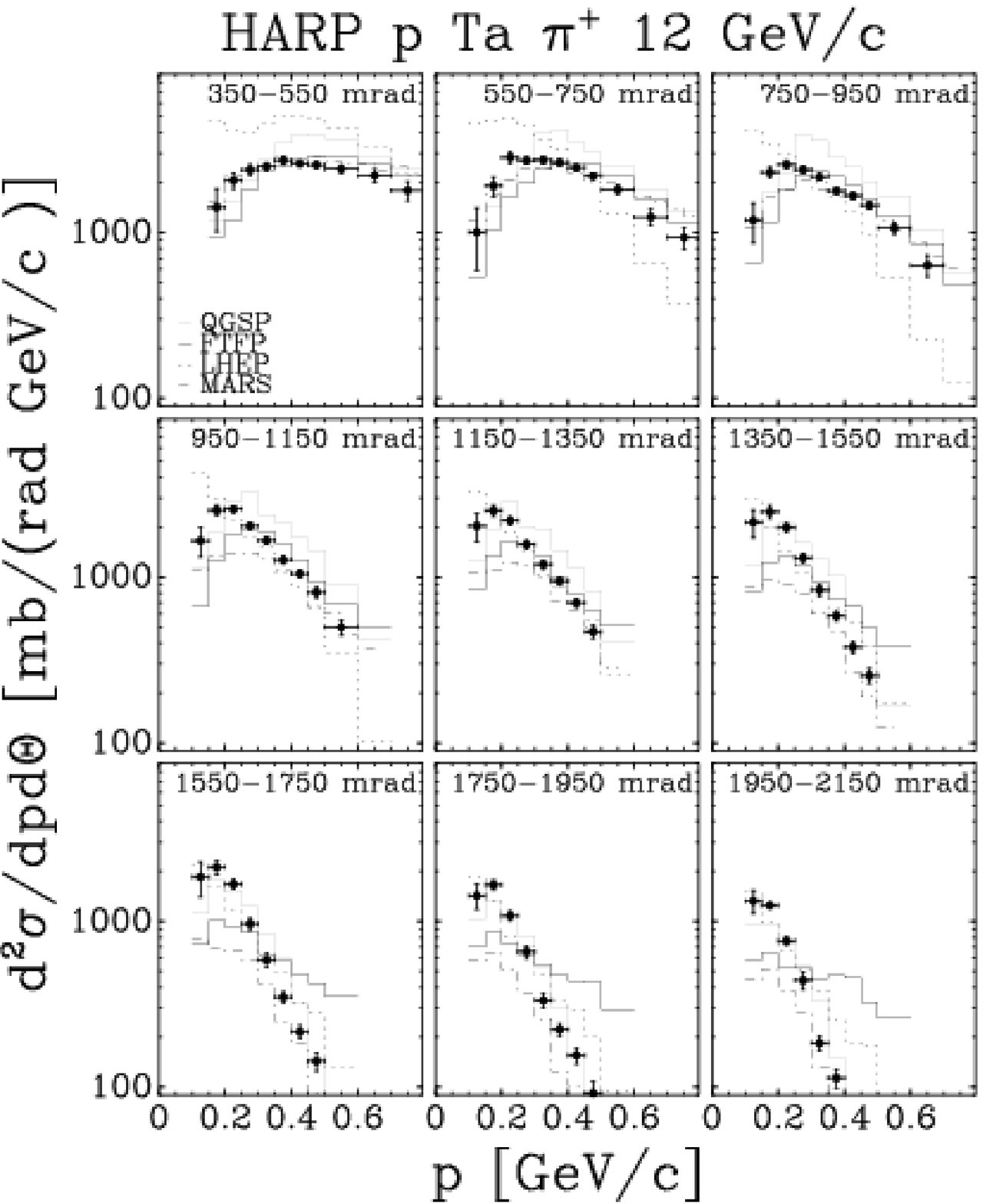}
  \includegraphics[width=0.47\textwidth]{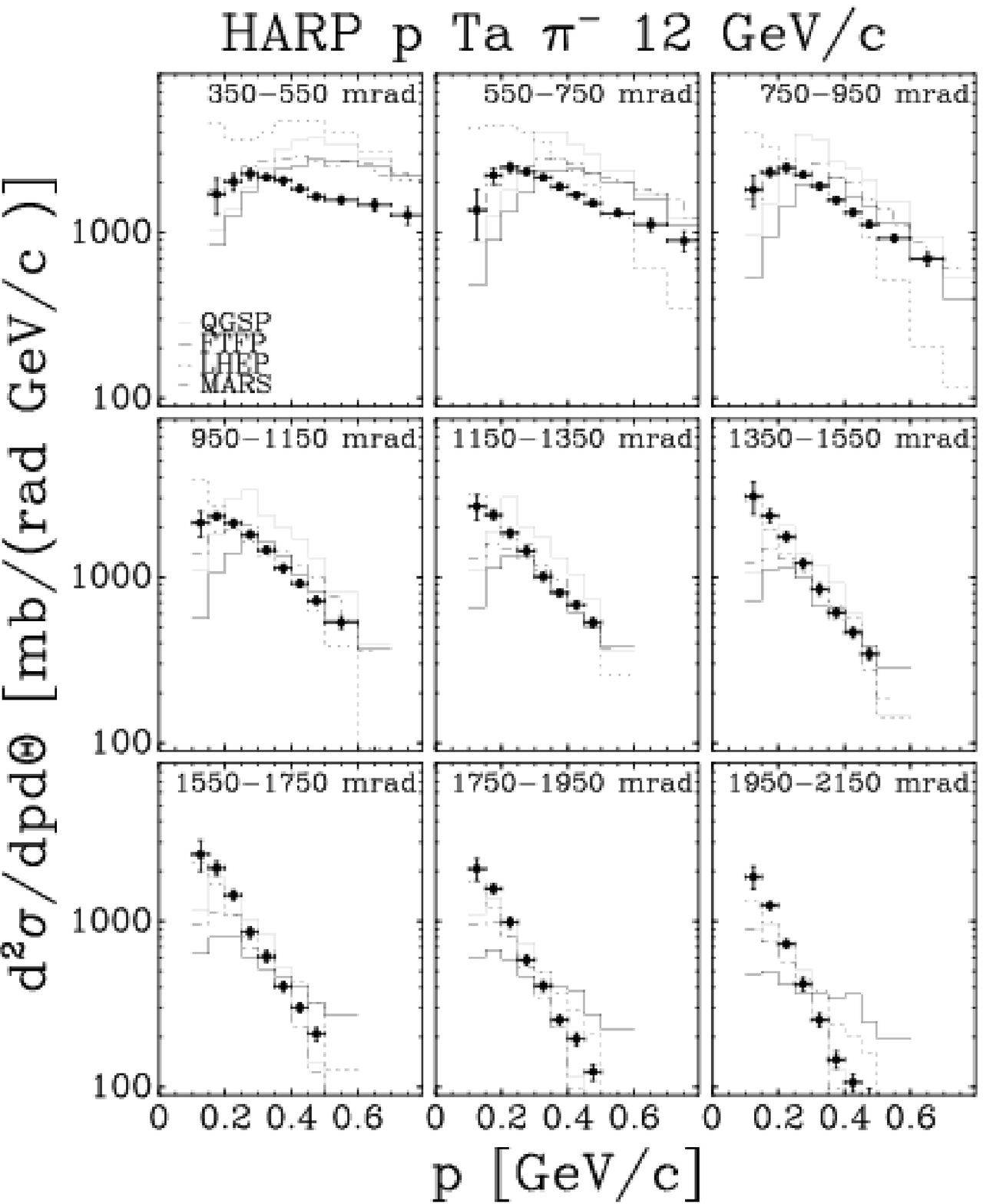}
\end{center}
\caption{
 Comparison of HARP double-differential $\pi^{\pm}$ cross sections for p--Ta at 12 GeV/c with
 GEANT4 and MARS MC predictions, using several generator models (see text for details).
}
\label{fig:G56}
\end{figure*}

In the following we will show only some comparisons with  publicly
available MonteCarlo simulations: GEANT4~\cite{ref:geant4} 
and MARS~\cite{ref:mars}, using different 
models.
We stress that no tuning to our data has been done by the 
GEANT4 or MARS teams. The comparison will be shown for a limited set of plots
and only for the C and Ta targets, as examples of a light and a heavy target
nucleus,
in figures \ref{fig:G43a} to \ref{fig:G56}.
 
At intermediate energies (up to 5-10 GeV), 
GEANT4 uses two types of intra-nuclear cascade models: the Bertini 
model~\cite{ref:bert,ref:bert1} (valid up to $\sim 10$ GeV) and the binary
model~\cite{ref:bin} (valid up to $\sim 3$ GeV). Both models treat the target
nucleus in detail, taking into account density variations and tracking in the
nuclear field. 
The binary model is based on hadron collisions with nucleons, giving 
resonances that decay according to their quantum numbers. The Bertini
model is based on the cascade code reported in \cite{ref:bert2}
and hadron collisions are assumed to proceed according to free-space partial
cross sections and final state distributions measured for the incident 
particle types.
 
At higher energies, instead, two parton string models, 
the quark-gluon string (QGS)  model~\cite{ref:bert,ref:QGSP} and the Fritiof
(FTP) model~\cite{ref:QGSP} are used, in addition to a High Energy 
Parametrized model (HEP)
derived from the high energy part of the Gheisha code used inside 
GEANT3~\cite{ref:gheisha}.

The parametrized models of GEANT4 (HEP and LEP) are intended to be fast,
but conserve energy and momentum on average and not event by 
event. 

A realistic GEANT4 simulation is built by combining models and physics processes
into what is called a ``physics list''. In high energy calorimetry the two
most commonly used are the QGSP physics list, based on the QGS model, 
the pre-compound
nucleus model and some of the Low Energy Parametrized (LEP) model (that 
has its root in the GHEISHA code inside
GEANT3) and the LHEP physics list~\cite{ref:lhep} based on the parametrized 
LEP model and HEP models. 

The MARS code system~\cite{ref:mars} uses as basic model an inclusive
approach multiparticle production originated by R.~Feynman. Above 3~\GeVc
phenomenological particle production models are used. Below 5~\GeVc
a cascade-exciton model~\cite{ref:casca} combined with the Fermi break-up model, the coalescence
model, an evaporation model and a multifragmentation extension is used
instead.  

None of the considered models describe fully our data. However, backward
or central region production seems to be described better than 
relatively more forward
production, especially at higher incident momenta.
In our data, the lowest angular bin corresponds to a
transition region from forward to central production,
that is more difficult to describe by MC models.
 
In general, $\pi^{+}$ production is better described than
$\pi^{-}$ production. 
At higher energies the FTP model (from GEANT4) and the MARS model 
describe better the data, while at the lowest energies the Bertini
and binary cascade models (from GEANT4) seem more appropriate.
Parametrized models, as LHEP from GEANT4, show relevant discrepancies:
up to a factor three in the forward region at low energies. 

The comparison, just outlined in our paper,  between data and models 
shows that the full set of HARP data, taken with targets 
spanning the full periodic table of elements, with small total errors and large
coverage of the solid angle with a single detector, may greatly help the 
tuning  of models used in hadronic simulations in the difficult 
energy range between
3~\GeVc and 15~\GeVc of incident momentum. 






\FloatBarrier

\section{Summary and Conclusions}
\label{sec:summary}

An analysis of the production of pions at
large angles with respect to the beam direction for protons of
3~\GeVc, 5~\GeVc, 8~\GeVc, 8.9~\GeVc (Be only), 12~\GeVc  and 
12.9~\GeVc (Al only) beam momentum impinging on  thin
(5\% interaction length) beryllium, carbon, aluminium, copper, tin, tantalum   and lead  targets is described. 
The secondary pion yield is measured in a large angular and momentum
range and double-differential cross-sections are obtained.
A detailed error estimation has been discussed.
Results on the dependence from atomic number A of pion production
are also presented.

The data taken with the lead and tantalum targets are relevant 
for the optimization of the targetry of a Neutrino Factory.
The pion yield increases with momentum and in our kinematic 
range the optimum is between 5 GeV/c and 8 GeV/c.

The use of a single detector for a range of beam momenta makes it
possible to measure the dependence of the pion yield on the secondary
particle momentum and emission angle $\theta$ with high precision.
The $A$--dependence of the cross-section can be studied,
using data from a single experiment. 
Very few pion production measurements in this energy range are reported
in the literature.
The only comparable results found in the literature agrees with the
analysis described in this paper.
Hadronic production models describing this energy range have now been
compared with our new results. 

\section{Acknowledgements}

We gratefully acknowledge the help and support of the PS beam staff
and of the numerous technical collaborators who contributed to the
detector design, construction, commissioning and operation.  
In particular, we would like to thank
G.~Barichello,
R.~Brocard,
K.~Burin,
V.~Carassiti,
F.~Chignoli,
D.~Conventi,
G.~Decreuse,
M.~Delattre,
C.~Detraz,  
A.~Domeniconi,
M.~Dwuznik,   
F.~Evangelisti,
B.~Friend,
A.~Iaciofano,
I.~Krasin, 
D.~Lacroix,
J.-C.~Legrand,
M.~Lobello, 
M.~Lollo,
J.~Loquet,
F.~Marinilli,
R.~Mazza,
J.~Mulon,
L.~Musa,
R.~Nicholson,
A.~Pepato,
P.~Petev, 
X.~Pons,
I.~Rusinov,
M.~Scandurra,
E.~Usenko,
and
R.~van der Vlugt,
for their support in the construction of the detector.
The collaboration acknowledges the major contributions and advice of
M.~Baldo-Ceolin, 
L.~Linssen, 
M.T.~Muciaccia and A. Pullia
during the construction of the experiment.
The collaboration is indebted to 
V.~Ableev,
P.~Arce,   
F.~Bergsma,
P.~Binko,
E.~Boter,
C.~Buttar,  
M.~Calvi, 
M.~Campanelli, 
C.~Cavion, 
A.~Chukanov, 
A.~De~Min,    
M.~Doucet,
D.~D\"{u}llmann,
R.~Engel,   
V.~Ermilova, 
W.~Flegel,
P.~Gruber,   
Y.~Hayato,
P.~Hodgson,  
A.~Ichikawa,
I.~Kato,  
O.~Klimov,
T.~Kobayashi,
D.~Kustov,
M.~Laveder,  
M.~Mass,
H.~Meinhard,
T.~Nakaya,
K.~Nishikawa,
M.~Paganoni,     
F.~Paleari,  
M.~Pasquali,
J.~Pasternak,   
C.~Pattison,    
M.~Placentino,
S.~Robbins,   
G.~Santin,  
S.~Simone,
A.~Tornero,   
S.~Troquereau,
S.~Ueda, 
A.~Valassi,
F.~Vannucci   
and
K.~Zuber   
for their contributions to the experiment
and to P. Dini for help in MC production. 

We acknowledge the contributions of 
V.~Ammosov,
G.~Chelkov,
D.~Dedovich,
F.~Dydak,
M.~Gostkin,
A.~Guskov, 
D.~Khartchenko, 
V.~Koreshev,
Z.~Kroumchtein,
I.~Nefedov,
A.~Semak, 
J.~Wotschack,
V.~Zaets and
A.~Zhemchugov
to the work described in this paper.

 The experiment was made possible by grants from
the Institut Interuniversitaire des Sciences Nucl\'eair\-es and the
Interuniversitair Instituut voor Kernwetenschappen (Belgium), 
Ministerio de Educacion y Ciencia, Grant FPA2003-06921-c02-02 and
Generalitat Valenciana, grant GV00-054-1,
CERN (Geneva, Switzerland), 
the German Bundesministerium f\"ur Bildung und Forschung (Germany), 
the Istituto Na\-zio\-na\-le di Fisica Nucleare (Italy), 
INR RAS (Moscow) and the Particle Physics and Astronomy Research Council (UK).
We gratefully acknowledge their support.
This work was supported in part by the Swiss National Science Foundation
and the Swiss Agency for Development and Cooperation in the framework of
the programme SCOPES - Scientific co-operation between Eastern Europe
and Switzerland. 

\begin{appendix}
\section{Cross-section data}\label{app:data}
Results on double differential cross section for protons impinging on
thin beryllium,  carbon, aluminium, copper, tin, tantalum and lead 
targets are reported here.
%
\begin{table*}[hp!]
\begin{center}
\small{
  \caption{\label{tab:xsec-p-be}
    HARP results for the double-differential $\pi^+$ production
    cross-section in the laboratory system,
    $d^2\sigma^{\pi^+}/(dpd\theta)$ for p--Be interactions. Each row refers to a
    different $(p_{\hbox{\small min}} \le p<p_{\hbox{\small max}},
    \theta_{\hbox{\small min}} \le \theta<\theta_{\hbox{\small max}})$ bin,
    where $p$ and $\theta$ are the pion momentum and polar angle, respectively.
    The central value as well as the square-root of the diagonal elements
    of the covariance matrix are given.}
\vspace{2mm}
\begin{tabular}{rrrr|r@{$\pm$}lr@{$\pm$}lr@{$\pm$}lr@{$\pm$}lr@{$\pm$}l}
\hline
$\theta_{\hbox{\small min}}$ &
$\theta_{\hbox{\small max}}$ &
$p_{\hbox{\small min}}$ &
$p_{\hbox{\small max}}$ &
\multicolumn{10}{c}{$d^2\sigma^{\pi^+}/(dpd\theta)$}
\\
(rad) & (rad) & (\GeVc) & (\GeVc) &
\multicolumn{10}{c}{\bgr}
\\
  &  &  &
&\multicolumn{2}{c}{$ \bf{3 \ \GeVc}$}
&\multicolumn{2}{c}{$ \bf{5 \ \GeVc}$}
&\multicolumn{2}{c}{$ \bf{8 \ \GeVc}$}
&\multicolumn{2}{c}{$ \bf{8.9 \ \GeVc}$}
&\multicolumn{2}{c}{$ \bf{12 \ \GeVc}$}
\\
\hline 
 0.35 & 0.55 & 0.15 & 0.20& 0.062 &  0.011& 0.107 &  0.015& 0.122 &  0.017& 0.140 &  0.017& 0.137 &  0.020\\ 
      &      & 0.20 & 0.25& 0.079 &  0.010& 0.138 &  0.013& 0.162 &  0.013& 0.184 &  0.013& 0.199 &  0.013\\ 
      &      & 0.25 & 0.30& 0.117 &  0.014& 0.187 &  0.018& 0.217 &  0.020& 0.231 &  0.017& 0.240 &  0.020\\ 
      &      & 0.30 & 0.35& 0.156 &  0.017& 0.204 &  0.017& 0.241 &  0.017& 0.249 &  0.020& 0.269 &  0.018\\ 
      &      & 0.35 & 0.40& 0.164 &  0.014& 0.238 &  0.017& 0.258 &  0.021& 0.269 &  0.016& 0.251 &  0.014\\ 
      &      & 0.40 & 0.45& 0.165 &  0.011& 0.223 &  0.012& 0.264 &  0.012& 0.284 &  0.012& 0.297 &  0.022\\ 
      &      & 0.45 & 0.50& 0.164 &  0.011& 0.230 &  0.014& 0.271 &  0.014& 0.300 &  0.016& 0.319 &  0.016\\ 
      &      & 0.50 & 0.60& 0.162 &  0.013& 0.230 &  0.013& 0.284 &  0.016& 0.292 &  0.017& 0.324 &  0.016\\ 
      &      & 0.60 & 0.70& 0.121 &  0.018& 0.214 &  0.021& 0.282 &  0.027& 0.250 &  0.025& 0.271 &  0.029\\ 
      &      & 0.70 & 0.80& 0.068 &  0.016& 0.157 &  0.028& 0.210 &  0.032& 0.182 &  0.029& 0.211 &  0.033\\ 
\hline  
 0.55 & 0.75 & 0.10 & 0.15& 0.047 &  0.013& 0.080 &  0.017& 0.082 &  0.019& 0.088 &  0.016& 0.085 &  0.019\\ 
      &      & 0.15 & 0.20& 0.111 &  0.012& 0.136 &  0.013& 0.145 &  0.012& 0.141 &  0.010& 0.145 &  0.010\\ 
      &      & 0.20 & 0.25& 0.146 &  0.015& 0.182 &  0.014& 0.218 &  0.016& 0.205 &  0.015& 0.201 &  0.018\\ 
      &      & 0.25 & 0.30& 0.161 &  0.013& 0.198 &  0.016& 0.220 &  0.016& 0.229 &  0.015& 0.237 &  0.015\\ 
      &      & 0.30 & 0.35& 0.171 &  0.017& 0.211 &  0.017& 0.226 &  0.013& 0.239 &  0.013& 0.252 &  0.017\\ 
      &      & 0.35 & 0.40& 0.173 &  0.011& 0.213 &  0.011& 0.219 &  0.011& 0.231 &  0.010& 0.258 &  0.012\\ 
      &      & 0.40 & 0.45& 0.145 &  0.011& 0.172 &  0.009& 0.224 &  0.011& 0.218 &  0.008& 0.237 &  0.010\\ 
      &      & 0.45 & 0.50& 0.130 &  0.010& 0.171 &  0.009& 0.206 &  0.010& 0.211 &  0.009& 0.210 &  0.009\\ 
      &      & 0.50 & 0.60& 0.089 &  0.012& 0.150 &  0.012& 0.187 &  0.013& 0.179 &  0.013& 0.186 &  0.012\\ 
      &      & 0.60 & 0.70& 0.051 &  0.010& 0.099 &  0.015& 0.129 &  0.019& 0.128 &  0.018& 0.138 &  0.019\\ 
      &      & 0.70 & 0.80& 0.029 &  0.006& 0.057 &  0.012& 0.083 &  0.017& 0.080 &  0.016& 0.086 &  0.020\\ 
\hline  
 0.75 & 0.95 & 0.10 & 0.15& 0.082 &  0.013& 0.073 &  0.014& 0.093 &  0.014& 0.092 &  0.013& 0.101 &  0.015\\ 
      &      & 0.15 & 0.20& 0.138 &  0.015& 0.161 &  0.013& 0.158 &  0.011& 0.163 &  0.011& 0.163 &  0.013\\ 
      &      & 0.20 & 0.25& 0.156 &  0.014& 0.175 &  0.012& 0.181 &  0.014& 0.195 &  0.012& 0.197 &  0.012\\ 
      &      & 0.25 & 0.30& 0.154 &  0.012& 0.169 &  0.010& 0.185 &  0.011& 0.214 &  0.011& 0.197 &  0.012\\ 
      &      & 0.30 & 0.35& 0.126 &  0.010& 0.148 &  0.010& 0.192 &  0.012& 0.189 &  0.008& 0.197 &  0.010\\ 
      &      & 0.35 & 0.40& 0.120 &  0.009& 0.139 &  0.008& 0.165 &  0.008& 0.165 &  0.006& 0.172 &  0.007\\ 
      &      & 0.40 & 0.45& 0.098 &  0.009& 0.125 &  0.007& 0.136 &  0.007& 0.140 &  0.005& 0.143 &  0.006\\ 
      &      & 0.45 & 0.50& 0.071 &  0.008& 0.109 &  0.008& 0.118 &  0.006& 0.123 &  0.006& 0.128 &  0.007\\ 
      &      & 0.50 & 0.60& 0.047 &  0.006& 0.072 &  0.009& 0.099 &  0.008& 0.092 &  0.009& 0.096 &  0.009\\ 
      &      & 0.60 & 0.70& 0.025 &  0.006& 0.039 &  0.008& 0.063 &  0.010& 0.055 &  0.009& 0.057 &  0.010\\ 
\hline  
 0.95 & 1.15 & 0.10 & 0.15& 0.097 &  0.012& 0.087 &  0.013& 0.089 &  0.013& 0.094 &  0.012& 0.094 &  0.013\\ 
      &      & 0.15 & 0.20& 0.127 &  0.013& 0.144 &  0.011& 0.153 &  0.011& 0.154 &  0.009& 0.175 &  0.012\\ 
      &      & 0.20 & 0.25& 0.145 &  0.011& 0.180 &  0.014& 0.160 &  0.009& 0.168 &  0.008& 0.177 &  0.010\\ 
      &      & 0.25 & 0.30& 0.120 &  0.010& 0.154 &  0.011& 0.148 &  0.008& 0.159 &  0.008& 0.157 &  0.007\\ 
      &      & 0.30 & 0.35& 0.097 &  0.009& 0.107 &  0.006& 0.125 &  0.007& 0.123 &  0.006& 0.126 &  0.006\\ 
      &      & 0.35 & 0.40& 0.058 &  0.006& 0.089 &  0.006& 0.102 &  0.006& 0.107 &  0.005& 0.108 &  0.005\\ 
      &      & 0.40 & 0.45& 0.045 &  0.004& 0.071 &  0.005& 0.080 &  0.005& 0.081 &  0.005& 0.090 &  0.005\\ 
      &      & 0.45 & 0.50& 0.033 &  0.004& 0.055 &  0.005& 0.062 &  0.005& 0.062 &  0.005& 0.069 &  0.006\\ 
      &      & 0.50 & 0.60& 0.018 &  0.003& 0.034 &  0.005& 0.040 &  0.005& 0.041 &  0.005& 0.043 &  0.006\\ 
\hline \\ 
\end{tabular}
}
\end{center}
\end{table*}

\begin{table*}[hp!]
\begin{center}
\small{
\begin{tabular}{rrrr|r@{$\pm$}lr@{$\pm$}lr@{$\pm$}lr@{$\pm$}lr@{$\pm$}l}
\hline
$\theta_{\hbox{\small min}}$ &
$\theta_{\hbox{\small max}}$ &
$p_{\hbox{\small min}}$ &
$p_{\hbox{\small max}}$ &
\multicolumn{10}{c}{$d^2\sigma^{\pi^+}/(dpd\theta)$}
\\
(rad) & (rad) & (\GeVc) & (\GeVc) &
\multicolumn{10}{c}{\bgr}
\\
  &  &  &
&\multicolumn{2}{c}{$ \bf{3 \ \GeVc}$}
&\multicolumn{2}{c}{$ \bf{5 \ \GeVc}$}
&\multicolumn{2}{c}{$ \bf{8 \ \GeVc}$}
&\multicolumn{2}{c}{$ \bf{8.9 \ \GeVc}$}
&\multicolumn{2}{c}{$ \bf{12 \ \GeVc}$}
\\
\hline 
 1.15 & 1.35 & 0.10 & 0.15& 0.093 &  0.012& 0.084 &  0.011& 0.099 &  0.012& 0.101 &  0.013& 0.099 &  0.013\\ 
      &      & 0.15 & 0.20& 0.121 &  0.013& 0.131 &  0.010& 0.157 &  0.010& 0.151 &  0.008& 0.134 &  0.009\\ 
      &      & 0.20 & 0.25& 0.120 &  0.010& 0.126 &  0.008& 0.133 &  0.007& 0.141 &  0.006& 0.137 &  0.006\\ 
      &      & 0.25 & 0.30& 0.078 &  0.008& 0.106 &  0.008& 0.102 &  0.006& 0.106 &  0.005& 0.115 &  0.006\\ 
      &      & 0.30 & 0.35& 0.048 &  0.007& 0.071 &  0.005& 0.081 &  0.006& 0.078 &  0.003& 0.078 &  0.005\\ 
      &      & 0.35 & 0.40& 0.031 &  0.003& 0.051 &  0.004& 0.067 &  0.004& 0.065 &  0.003& 0.066 &  0.003\\ 
      &      & 0.40 & 0.45& 0.025 &  0.003& 0.038 &  0.004& 0.050 &  0.004& 0.046 &  0.004& 0.049 &  0.004\\ 
      &      & 0.45 & 0.50& 0.016 &  0.003& 0.025 &  0.003& 0.035 &  0.004& 0.032 &  0.003& 0.035 &  0.004\\ 
\hline  
1.35 & 1.55 & 0.10 & 0.15& 0.086 &  0.011& 0.102 &  0.012& 0.082 &  0.011& 0.101 &  0.011& 0.096 &  0.012\\ 
      &      & 0.15 & 0.20& 0.118 &  0.012& 0.119 &  0.009& 0.139 &  0.011& 0.135 &  0.007& 0.136 &  0.009\\ 
      &      & 0.20 & 0.25& 0.084 &  0.009& 0.120 &  0.008& 0.116 &  0.007& 0.110 &  0.006& 0.108 &  0.006\\ 
      &      & 0.25 & 0.30& 0.062 &  0.007& 0.078 &  0.007& 0.081 &  0.007& 0.081 &  0.005& 0.082 &  0.005\\ 
      &      & 0.30 & 0.35& 0.036 &  0.005& 0.048 &  0.005& 0.055 &  0.004& 0.054 &  0.003& 0.056 &  0.004\\ 
      &      & 0.35 & 0.40& 0.021 &  0.003& 0.034 &  0.003& 0.040 &  0.003& 0.039 &  0.003& 0.040 &  0.003\\ 
      &      & 0.40 & 0.45& 0.012 &  0.002& 0.022 &  0.003& 0.026 &  0.003& 0.026 &  0.002& 0.028 &  0.003\\ 
      &      & 0.45 & 0.50& 0.008 &  0.002& 0.013 &  0.002& 0.015 &  0.003& 0.019 &  0.003& 0.018 &  0.003\\ 
\hline  
 1.55 & 1.75 & 0.10 & 0.15& 0.070 &  0.012& 0.080 &  0.010& 0.085 &  0.012& 0.085 &  0.011& 0.095 &  0.011\\ 
      &      & 0.15 & 0.20& 0.093 &  0.009& 0.118 &  0.009& 0.116 &  0.007& 0.115 &  0.005& 0.106 &  0.006\\ 
      &      & 0.20 & 0.25& 0.054 &  0.007& 0.076 &  0.007& 0.076 &  0.005& 0.086 &  0.005& 0.084 &  0.005\\ 
      &      & 0.25 & 0.30& 0.035 &  0.005& 0.050 &  0.005& 0.057 &  0.004& 0.056 &  0.003& 0.051 &  0.005\\ 
      &      & 0.30 & 0.35& 0.020 &  0.004& 0.029 &  0.003& 0.037 &  0.003& 0.037 &  0.003& 0.030 &  0.002\\ 
      &      & 0.35 & 0.40& 0.010 &  0.002& 0.019 &  0.002& 0.026 &  0.003& 0.023 &  0.002& 0.023 &  0.002\\ 
      &      & 0.40 & 0.45& 0.008 &  0.002& 0.011 &  0.002& 0.015 &  0.002& 0.014 &  0.002& 0.014 &  0.002\\ 
      &      & 0.45 & 0.50& 0.004 &  0.001& 0.006 &  0.001& 0.008 &  0.002& 0.009 &  0.001& 0.008 &  0.002\\ 
\hline  
 1.75 & 1.95 & 0.10 & 0.15& 0.060 &  0.009& 0.090 &  0.011& 0.086 &  0.011& 0.081 &  0.009& 0.073 &  0.010\\ 
      &      & 0.15 & 0.20& 0.096 &  0.011& 0.096 &  0.007& 0.096 &  0.006& 0.099 &  0.004& 0.099 &  0.006\\ 
      &      & 0.20 & 0.25& 0.060 &  0.009& 0.047 &  0.005& 0.065 &  0.005& 0.062 &  0.004& 0.056 &  0.005\\ 
      &      & 0.25 & 0.30& 0.020 &  0.005& 0.036 &  0.004& 0.036 &  0.004& 0.041 &  0.003& 0.033 &  0.002\\ 
      &      & 0.30 & 0.35& 0.011 &  0.002& 0.018 &  0.003& 0.023 &  0.002& 0.026 &  0.002& 0.024 &  0.002\\ 
      &      & 0.35 & 0.40& 0.006 &  0.002& 0.011 &  0.002& 0.013 &  0.002& 0.015 &  0.002& 0.015 &  0.002\\ 
      &      & 0.40 & 0.45& 0.003 &  0.001& 0.005 &  0.001& 0.006 &  0.001& 0.007 &  0.001& 0.007 &  0.002\\ 
      &      & 0.45 & 0.50& 0.001 &  0.001& 0.003 &  0.001& 0.003 &  0.001& 0.004 &  0.001& 0.004 &  0.001\\ 
\hline  
 1.95 & 2.15 & 0.10 & 0.15& 0.063 &  0.011& 0.060 &  0.008& 0.063 &  0.008& 0.068 &  0.007& 0.055 &  0.007\\ 
      &      & 0.15 & 0.20& 0.063 &  0.008& 0.082 &  0.006& 0.080 &  0.006& 0.070 &  0.004& 0.073 &  0.005\\ 
      &      & 0.20 & 0.25& 0.033 &  0.005& 0.041 &  0.005& 0.044 &  0.004& 0.047 &  0.003& 0.047 &  0.005\\ 
      &      & 0.25 & 0.30& 0.013 &  0.003& 0.020 &  0.003& 0.027 &  0.003& 0.027 &  0.002& 0.026 &  0.003\\ 
      &      & 0.30 & 0.35& 0.008 &  0.002& 0.009 &  0.001& 0.014 &  0.002& 0.016 &  0.001& 0.013 &  0.002\\ 
      &      & 0.35 & 0.40& 0.004 &  0.002& 0.009 &  0.002& 0.007 &  0.001& 0.008 &  0.001& 0.008 &  0.001\\ 
      &      & 0.40 & 0.45& 0.002 &  0.001& 0.004 &  0.001& 0.004 &  0.001& 0.004 &  0.001& 0.004 &  0.001\\ 
      &      & 0.45 & 0.50& \multicolumn{2}{c}{}& 0.002 &  0.001& 0.002 &  0.001& 0.002 &  0.001& 0.002 &  0.001\\ 
\hline
\end{tabular} 
}
\end{center}
\end{table*}

\begin{table*}[hp!]
\begin{center}
\small{
  \caption{\label{tab:xsec-n-be}
    HARP results for the double-differential $\pi^-$ production
    cross-section in the laboratory system,
    $d^2\sigma^{\pi^-}/(dpd\theta)$ for p--Be interactions. Each row refers to a
    different $(p_{\hbox{\small min}} \le p<p_{\hbox{\small max}},
    \theta_{\hbox{\small min}} \le \theta<\theta_{\hbox{\small max}})$ bin,
    where $p$ and $\theta$ are the pion momentum and polar angle, respectively.
    The central value as well as the square-root of the diagonal elements
    of the covariance matrix are given.}
\vspace{2mm}
\begin{tabular}{rrrr|r@{$\pm$}lr@{$\pm$}lr@{$\pm$}lr@{$\pm$}lr@{$\pm$}l}
\hline
$\theta_{\hbox{\small min}}$ &
$\theta_{\hbox{\small max}}$ &
$p_{\hbox{\small min}}$ &
$p_{\hbox{\small max}}$ &
\multicolumn{10}{c}{$d^2\sigma^{\pi^-}/(dpd\theta)$}
\\
(rad) & (rad) & (\GeVc) & (\GeVc) &
\multicolumn{10}{c}{\bgr}
\\
  &  &  &
&\multicolumn{2}{c}{$ \bf{3 \ \GeVc}$}
&\multicolumn{2}{c}{$ \bf{5 \ \GeVc}$}
&\multicolumn{2}{c}{$ \bf{8 \ \GeVc}$}
&\multicolumn{2}{c}{$ \bf{8.9 \ \GeVc}$}
&\multicolumn{2}{c}{$ \bf{12 \ \GeVc}$}
\\
\hline

\hline  
 0.35 & 0.55 & 0.15 & 0.20& 0.048 &  0.009& 0.084 &  0.013& 0.114 &  0.015& 0.124 &  0.015& 0.124 &  0.020\\ 
      &      & 0.20 & 0.25& 0.068 &  0.009& 0.095 &  0.008& 0.139 &  0.011& 0.147 &  0.009& 0.170 &  0.011\\ 
      &      & 0.25 & 0.30& 0.060 &  0.008& 0.134 &  0.013& 0.160 &  0.012& 0.170 &  0.011& 0.182 &  0.012\\ 
      &      & 0.30 & 0.35& 0.079 &  0.010& 0.129 &  0.009& 0.154 &  0.011& 0.171 &  0.009& 0.186 &  0.013\\ 
      &      & 0.35 & 0.40& 0.069 &  0.007& 0.111 &  0.007& 0.162 &  0.010& 0.163 &  0.008& 0.184 &  0.010\\ 
      &      & 0.40 & 0.45& 0.066 &  0.007& 0.116 &  0.010& 0.155 &  0.008& 0.165 &  0.009& 0.168 &  0.008\\ 
      &      & 0.45 & 0.50& 0.057 &  0.006& 0.121 &  0.008& 0.148 &  0.008& 0.164 &  0.008& 0.162 &  0.008\\ 
      &      & 0.50 & 0.60& 0.053 &  0.005& 0.129 &  0.008& 0.159 &  0.010& 0.160 &  0.009& 0.172 &  0.010\\ 
      &      & 0.60 & 0.70& 0.051 &  0.006& 0.101 &  0.012& 0.157 &  0.013& 0.152 &  0.013& 0.170 &  0.013\\ 
      &      & 0.70 & 0.80& 0.039 &  0.007& 0.084 &  0.011& 0.143 &  0.018& 0.128 &  0.014& 0.155 &  0.018\\ 
\hline  
 0.55 & 0.75 & 0.10 & 0.15& 0.026 &  0.008& 0.069 &  0.015& 0.070 &  0.016& 0.073 &  0.015& 0.092 &  0.020\\ 
      &      & 0.15 & 0.20& 0.058 &  0.009& 0.093 &  0.009& 0.111 &  0.010& 0.123 &  0.006& 0.131 &  0.008\\ 
      &      & 0.20 & 0.25& 0.057 &  0.007& 0.126 &  0.009& 0.161 &  0.012& 0.148 &  0.010& 0.135 &  0.011\\ 
      &      & 0.25 & 0.30& 0.062 &  0.008& 0.119 &  0.009& 0.152 &  0.008& 0.153 &  0.009& 0.160 &  0.009\\ 
      &      & 0.30 & 0.35& 0.077 &  0.008& 0.113 &  0.009& 0.134 &  0.008& 0.150 &  0.008& 0.159 &  0.009\\ 
      &      & 0.35 & 0.40& 0.076 &  0.007& 0.111 &  0.007& 0.144 &  0.010& 0.146 &  0.007& 0.151 &  0.007\\ 
      &      & 0.40 & 0.45& 0.059 &  0.006& 0.107 &  0.006& 0.146 &  0.008& 0.137 &  0.006& 0.148 &  0.007\\ 
      &      & 0.45 & 0.50& 0.052 &  0.005& 0.105 &  0.007& 0.127 &  0.006& 0.129 &  0.005& 0.140 &  0.006\\ 
      &      & 0.50 & 0.60& 0.049 &  0.005& 0.095 &  0.007& 0.111 &  0.006& 0.120 &  0.006& 0.131 &  0.007\\ 
      &      & 0.60 & 0.70& 0.041 &  0.006& 0.076 &  0.007& 0.100 &  0.008& 0.098 &  0.010& 0.111 &  0.011\\ 
      &      & 0.70 & 0.80& 0.027 &  0.006& 0.057 &  0.010& 0.086 &  0.013& 0.075 &  0.011& 0.096 &  0.012\\ 
\hline  
 0.75 & 0.95 & 0.10 & 0.15& 0.044 &  0.008& 0.056 &  0.009& 0.070 &  0.010& 0.069 &  0.009& 0.079 &  0.011\\ 
      &      & 0.15 & 0.20& 0.091 &  0.011& 0.107 &  0.010& 0.120 &  0.010& 0.130 &  0.008& 0.134 &  0.010\\ 
      &      & 0.20 & 0.25& 0.078 &  0.008& 0.114 &  0.008& 0.144 &  0.010& 0.139 &  0.007& 0.136 &  0.008\\ 
      &      & 0.25 & 0.30& 0.073 &  0.008& 0.111 &  0.007& 0.119 &  0.006& 0.135 &  0.007& 0.141 &  0.008\\ 
      &      & 0.30 & 0.35& 0.063 &  0.006& 0.096 &  0.006& 0.110 &  0.007& 0.123 &  0.005& 0.132 &  0.008\\ 
      &      & 0.35 & 0.40& 0.060 &  0.006& 0.084 &  0.005& 0.109 &  0.006& 0.108 &  0.004& 0.115 &  0.005\\ 
      &      & 0.40 & 0.45& 0.048 &  0.005& 0.080 &  0.005& 0.091 &  0.004& 0.099 &  0.003& 0.099 &  0.004\\ 
      &      & 0.45 & 0.50& 0.036 &  0.004& 0.074 &  0.004& 0.087 &  0.005& 0.091 &  0.003& 0.091 &  0.004\\ 
      &      & 0.50 & 0.60& 0.033 &  0.004& 0.060 &  0.005& 0.075 &  0.005& 0.078 &  0.004& 0.080 &  0.005\\ 
      &      & 0.60 & 0.70& 0.023 &  0.004& 0.042 &  0.006& 0.060 &  0.006& 0.058 &  0.006& 0.062 &  0.006\\ 
\hline  
 0.95 & 1.15 & 0.10 & 0.15& 0.038 &  0.007& 0.066 &  0.009& 0.070 &  0.009& 0.077 &  0.008& 0.072 &  0.009\\ 
      &      & 0.15 & 0.20& 0.069 &  0.008& 0.120 &  0.011& 0.109 &  0.009& 0.131 &  0.008& 0.119 &  0.008\\ 
      &      & 0.20 & 0.25& 0.065 &  0.008& 0.107 &  0.008& 0.117 &  0.008& 0.126 &  0.006& 0.122 &  0.007\\ 
      &      & 0.25 & 0.30& 0.061 &  0.006& 0.090 &  0.008& 0.116 &  0.007& 0.114 &  0.005& 0.114 &  0.006\\ 
      &      & 0.30 & 0.35& 0.065 &  0.007& 0.076 &  0.005& 0.097 &  0.005& 0.098 &  0.004& 0.096 &  0.004\\ 
      &      & 0.35 & 0.40& 0.049 &  0.006& 0.062 &  0.004& 0.088 &  0.004& 0.079 &  0.003& 0.084 &  0.004\\ 
      &      & 0.40 & 0.45& 0.034 &  0.004& 0.055 &  0.004& 0.068 &  0.005& 0.068 &  0.003& 0.070 &  0.004\\ 
      &      & 0.45 & 0.50& 0.025 &  0.004& 0.043 &  0.004& 0.054 &  0.003& 0.058 &  0.002& 0.059 &  0.003\\ 
      &      & 0.50 & 0.60& 0.018 &  0.003& 0.029 &  0.003& 0.044 &  0.003& 0.044 &  0.003& 0.046 &  0.003\\ 
\hline
\end{tabular}
}
\end{center}
\end{table*}

\begin{table*}[hp!]
\begin{center}
\small{
\begin{tabular}{rrrr|r@{$\pm$}lr@{$\pm$}lr@{$\pm$}lr@{$\pm$}lr@{$\pm$}l}
\hline
$\theta_{\hbox{\small min}}$ &
$\theta_{\hbox{\small max}}$ &
$p_{\hbox{\small min}}$ &
$p_{\hbox{\small max}}$ &
\multicolumn{10}{c}{$d^2\sigma^{\pi^-}/(dpd\theta)$}
\\
(rad) & (rad) & (\GeVc) & (\GeVc) &
\multicolumn{10}{c}{\bgr}
\\
  &  &  &
&\multicolumn{2}{c}{$ \bf{3 \ \GeVc}$}
&\multicolumn{2}{c}{$ \bf{5 \ \GeVc}$}
&\multicolumn{2}{c}{$ \bf{8 \ \GeVc}$}
&\multicolumn{2}{c}{$ \bf{8.9 \ \GeVc}$}
&\multicolumn{2}{c}{$ \bf{12 \ \GeVc}$}
\\
\hline
 1.15 & 1.35 & 0.10 & 0.15& 0.039 &  0.007& 0.073 &  0.010& 0.085 &  0.008& 0.075 &  0.007& 0.085 &  0.009\\ 
      &      & 0.15 & 0.20& 0.067 &  0.009& 0.101 &  0.007& 0.105 &  0.007& 0.122 &  0.008& 0.117 &  0.008\\ 
      &      & 0.20 & 0.25& 0.088 &  0.010& 0.089 &  0.006& 0.100 &  0.006& 0.109 &  0.004& 0.112 &  0.006\\ 
      &      & 0.25 & 0.30& 0.059 &  0.008& 0.077 &  0.005& 0.087 &  0.005& 0.091 &  0.004& 0.092 &  0.005\\ 
      &      & 0.30 & 0.35& 0.034 &  0.005& 0.050 &  0.005& 0.072 &  0.004& 0.069 &  0.003& 0.075 &  0.004\\ 
      &      & 0.35 & 0.40& 0.022 &  0.003& 0.036 &  0.003& 0.055 &  0.004& 0.055 &  0.002& 0.057 &  0.003\\ 
      &      & 0.40 & 0.45& 0.015 &  0.002& 0.034 &  0.003& 0.043 &  0.003& 0.044 &  0.002& 0.043 &  0.003\\ 
      &      & 0.45 & 0.50& 0.012 &  0.002& 0.031 &  0.003& 0.036 &  0.002& 0.035 &  0.002& 0.034 &  0.002\\ 
\hline  
 1.35 & 1.55 & 0.10 & 0.15& 0.057 &  0.009& 0.073 &  0.009& 0.061 &  0.007& 0.080 &  0.008& 0.078 &  0.009\\ 
      &      & 0.15 & 0.20& 0.061 &  0.007& 0.098 &  0.008& 0.106 &  0.009& 0.109 &  0.006& 0.110 &  0.009\\ 
      &      & 0.20 & 0.25& 0.063 &  0.007& 0.077 &  0.006& 0.102 &  0.006& 0.095 &  0.004& 0.097 &  0.006\\ 
      &      & 0.25 & 0.30& 0.034 &  0.005& 0.060 &  0.005& 0.077 &  0.005& 0.070 &  0.004& 0.074 &  0.005\\ 
      &      & 0.30 & 0.35& 0.027 &  0.004& 0.043 &  0.005& 0.051 &  0.004& 0.050 &  0.002& 0.049 &  0.003\\ 
      &      & 0.35 & 0.40& 0.019 &  0.003& 0.027 &  0.003& 0.036 &  0.003& 0.037 &  0.002& 0.038 &  0.002\\ 
      &      & 0.40 & 0.45& 0.015 &  0.002& 0.022 &  0.002& 0.026 &  0.002& 0.029 &  0.002& 0.029 &  0.003\\ 
      &      & 0.45 & 0.50& 0.010 &  0.002& 0.019 &  0.002& 0.020 &  0.002& 0.021 &  0.002& 0.021 &  0.002\\ 
\hline  
 1.55 & 1.75 & 0.10 & 0.15& 0.042 &  0.007& 0.068 &  0.008& 0.072 &  0.009& 0.073 &  0.008& 0.084 &  0.010\\ 
      &      & 0.15 & 0.20& 0.067 &  0.009& 0.085 &  0.007& 0.090 &  0.006& 0.095 &  0.005& 0.090 &  0.006\\ 
      &      & 0.20 & 0.25& 0.052 &  0.009& 0.073 &  0.006& 0.072 &  0.005& 0.078 &  0.004& 0.078 &  0.005\\ 
      &      & 0.25 & 0.30& 0.016 &  0.003& 0.057 &  0.005& 0.055 &  0.005& 0.055 &  0.003& 0.054 &  0.004\\ 
      &      & 0.30 & 0.35& 0.012 &  0.002& 0.035 &  0.005& 0.038 &  0.003& 0.037 &  0.002& 0.041 &  0.003\\ 
      &      & 0.35 & 0.40& 0.014 &  0.003& 0.020 &  0.003& 0.032 &  0.003& 0.027 &  0.002& 0.027 &  0.003\\ 
      &      & 0.40 & 0.45& 0.009 &  0.002& 0.012 &  0.002& 0.021 &  0.002& 0.020 &  0.001& 0.018 &  0.002\\ 
      &      & 0.45 & 0.50& 0.006 &  0.002& 0.008 &  0.001& 0.015 &  0.002& 0.013 &  0.001& 0.014 &  0.001\\ 
\hline  
 1.75 & 1.95 & 0.10 & 0.15& 0.038 &  0.007& 0.053 &  0.006& 0.063 &  0.008& 0.063 &  0.006& 0.060 &  0.008\\ 
      &      & 0.15 & 0.20& 0.052 &  0.007& 0.067 &  0.006& 0.081 &  0.006& 0.086 &  0.004& 0.085 &  0.006\\ 
      &      & 0.20 & 0.25& 0.029 &  0.006& 0.053 &  0.005& 0.054 &  0.004& 0.061 &  0.003& 0.063 &  0.004\\ 
      &      & 0.25 & 0.30& 0.020 &  0.004& 0.037 &  0.004& 0.038 &  0.003& 0.039 &  0.003& 0.041 &  0.003\\ 
      &      & 0.30 & 0.35& 0.013 &  0.004& 0.019 &  0.003& 0.027 &  0.002& 0.024 &  0.002& 0.031 &  0.002\\ 
      &      & 0.35 & 0.40& 0.006 &  0.002& 0.013 &  0.002& 0.018 &  0.002& 0.017 &  0.001& 0.019 &  0.003\\ 
      &      & 0.40 & 0.45& 0.004 &  0.001& 0.008 &  0.001& 0.012 &  0.001& 0.012 &  0.001& 0.010 &  0.002\\ 
      &      & 0.45 & 0.50& 0.003 &  0.001& 0.005 &  0.001& 0.008 &  0.001& 0.009 &  0.001& 0.007 &  0.001\\ 
\hline  
 1.95 & 2.15 & 0.10 & 0.15& 0.049 &  0.008& 0.047 &  0.007& 0.062 &  0.007& 0.055 &  0.006& 0.047 &  0.005\\ 
      &      & 0.15 & 0.20& 0.038 &  0.006& 0.068 &  0.006& 0.061 &  0.005& 0.067 &  0.003& 0.068 &  0.006\\ 
      &      & 0.20 & 0.25& 0.023 &  0.005& 0.048 &  0.005& 0.046 &  0.004& 0.049 &  0.002& 0.044 &  0.005\\ 
      &      & 0.25 & 0.30& 0.009 &  0.003& 0.026 &  0.003& 0.025 &  0.003& 0.029 &  0.002& 0.023 &  0.002\\ 
      &      & 0.30 & 0.35& 0.005 &  0.002& 0.015 &  0.002& 0.017 &  0.002& 0.017 &  0.001& 0.019 &  0.002\\ 
      &      & 0.35 & 0.40& 0.001 &  0.001& 0.009 &  0.001& 0.012 &  0.001& 0.013 &  0.001& 0.012 &  0.001\\ 
      &      & 0.40 & 0.45& \multicolumn{2}{c}{}& 0.006 &  0.001& 0.009 &  0.002& 0.008 &  0.001& 0.009 &  0.001\\ 
      &      & 0.45 & 0.50& \multicolumn{2}{c}{}& 0.003 &  0.001& 0.005 &  0.001& 0.005 &  0.001& 0.006 &  0.001\\ 

\hline
\end{tabular}
}
\end{center}
\end{table*}
%
%
\begin{table*}[hp!]
\begin{center}
\small{
  \caption{\label{tab:xsec-p:c}
    HARP results for the double-differential $\pi^+$ production
    cross-section in the laboratory system,
    $d^2\sigma^{\pi^+}/(dpd\theta)$ for p--C interactions. Each row refers to a
    different $(p_{\hbox{\small min}} \le p<p_{\hbox{\small max}},
    \theta_{\hbox{\small min}} \le \theta<\theta_{\hbox{\small max}})$ bin,
    where $p$ and $\theta$ are the pion momentum and polar angle, respectively.
    The central value as well as the square-root of the diagonal elements
    of the covariance matrix are given.}
\vspace{2mm}
\begin{tabular}{rrrr|r@{$\pm$}lr@{$\pm$}lr@{$\pm$}lr@{$\pm$}l}
\hline
$\theta_{\hbox{\small min}}$ &
$\theta_{\hbox{\small max}}$ &
$p_{\hbox{\small min}}$ &
$p_{\hbox{\small max}}$ &
\multicolumn{8}{c}{$d^2\sigma^{\pi^+}/(dpd\theta)$}
\\
(rad) & (rad) & (\GeVc) & (\GeVc) &
\multicolumn{8}{c}{\bgr}
\\
  &  &  &
&\multicolumn{2}{c}{$ \bf{3 \ \GeVc}$}
&\multicolumn{2}{c}{$ \bf{5 \ \GeVc}$}
&\multicolumn{2}{c}{$ \bf{8 \ \GeVc}$}
&\multicolumn{2}{c}{$ \bf{12 \ \GeVc}$}
\\
\hline 
 0.35 & 0.55 & 0.15 & 0.20& 0.05 &  0.02& 0.12 &  0.03& 0.15 &  0.03& 0.14 &  0.03\\ 
      &      & 0.20 & 0.25& 0.10 &  0.02& 0.18 &  0.02& 0.20 &  0.02& 0.22 &  0.02\\ 
      &      & 0.25 & 0.30& 0.15 &  0.02& 0.23 &  0.02& 0.28 &  0.03& 0.28 &  0.02\\ 
      &      & 0.30 & 0.35& 0.16 &  0.02& 0.25 &  0.03& 0.30 &  0.02& 0.35 &  0.03\\ 
      &      & 0.35 & 0.40& 0.18 &  0.02& 0.29 &  0.02& 0.33 &  0.03& 0.37 &  0.02\\ 
      &      & 0.40 & 0.45& 0.19 &  0.01& 0.29 &  0.02& 0.35 &  0.02& 0.35 &  0.02\\ 
      &      & 0.45 & 0.50& 0.19 &  0.01& 0.29 &  0.01& 0.36 &  0.02& 0.39 &  0.02\\ 
      &      & 0.50 & 0.60& 0.18 &  0.01& 0.28 &  0.01& 0.36 &  0.02& 0.38 &  0.02\\ 
      &      & 0.60 & 0.70& 0.13 &  0.02& 0.25 &  0.02& 0.33 &  0.03& 0.33 &  0.03\\ 
      &      & 0.70 & 0.80& 0.07 &  0.02& 0.18 &  0.03& 0.26 &  0.04& 0.26 &  0.04\\ 
\hline  
 0.55 & 0.75 & 0.10 & 0.15& 0.09 &  0.03& 0.09 &  0.03& 0.10 &  0.03& 0.10 &  0.03\\ 
      &      & 0.15 & 0.20& 0.13 &  0.02& 0.18 &  0.02& 0.19 &  0.02& 0.17 &  0.02\\ 
      &      & 0.20 & 0.25& 0.20 &  0.02& 0.22 &  0.02& 0.26 &  0.02& 0.26 &  0.02\\ 
      &      & 0.25 & 0.30& 0.22 &  0.02& 0.27 &  0.02& 0.31 &  0.03& 0.30 &  0.02\\ 
      &      & 0.30 & 0.35& 0.24 &  0.02& 0.26 &  0.01& 0.29 &  0.02& 0.32 &  0.02\\ 
      &      & 0.35 & 0.40& 0.21 &  0.02& 0.26 &  0.01& 0.30 &  0.02& 0.32 &  0.01\\ 
      &      & 0.40 & 0.45& 0.18 &  0.01& 0.23 &  0.01& 0.28 &  0.01& 0.30 &  0.01\\ 
      &      & 0.45 & 0.50& 0.18 &  0.01& 0.21 &  0.01& 0.26 &  0.01& 0.28 &  0.01\\ 
      &      & 0.50 & 0.60& 0.11 &  0.02& 0.17 &  0.01& 0.22 &  0.01& 0.24 &  0.01\\ 
      &      & 0.60 & 0.70& 0.08 &  0.01& 0.12 &  0.01& 0.17 &  0.02& 0.18 &  0.02\\ 
      &      & 0.70 & 0.80& 0.04 &  0.01& 0.07 &  0.01& 0.12 &  0.02& 0.12 &  0.02\\ 
\hline  
 0.75 & 0.95 & 0.10 & 0.15& 0.09 &  0.02& 0.12 &  0.02& 0.12 &  0.02& 0.12 &  0.02\\ 
      &      & 0.15 & 0.20& 0.17 &  0.02& 0.22 &  0.02& 0.21 &  0.02& 0.20 &  0.02\\ 
      &      & 0.20 & 0.25& 0.21 &  0.02& 0.25 &  0.02& 0.24 &  0.02& 0.28 &  0.02\\ 
      &      & 0.25 & 0.30& 0.19 &  0.02& 0.23 &  0.01& 0.25 &  0.02& 0.27 &  0.01\\ 
      &      & 0.30 & 0.35& 0.20 &  0.02& 0.21 &  0.01& 0.24 &  0.01& 0.24 &  0.01\\ 
      &      & 0.35 & 0.40& 0.14 &  0.02& 0.17 &  0.01& 0.22 &  0.01& 0.24 &  0.01\\ 
      &      & 0.40 & 0.45& 0.10 &  0.01& 0.15 &  0.01& 0.18 &  0.01& 0.20 &  0.01\\ 
      &      & 0.45 & 0.50& 0.08 &  0.01& 0.13 &  0.01& 0.16 &  0.01& 0.17 &  0.01\\ 
      &      & 0.50 & 0.60& 0.05 &  0.01& 0.09 &  0.01& 0.12 &  0.01& 0.13 &  0.01\\ 
      &      & 0.60 & 0.70& 0.03 &  0.01& 0.05 &  0.01& 0.07 &  0.01& 0.07 &  0.01\\ 
\hline  
 0.95 & 1.15 & 0.10 & 0.15& 0.14 &  0.02& 0.13 &  0.02& 0.12 &  0.02& 0.15 &  0.02\\ 
      &      & 0.15 & 0.20& 0.22 &  0.02& 0.20 &  0.02& 0.23 &  0.02& 0.23 &  0.02\\ 
      &      & 0.20 & 0.25& 0.19 &  0.01& 0.21 &  0.01& 0.24 &  0.01& 0.23 &  0.01\\ 
      &      & 0.25 & 0.30& 0.16 &  0.01& 0.18 &  0.01& 0.22 &  0.01& 0.21 &  0.01\\ 
      &      & 0.30 & 0.35& 0.10 &  0.01& 0.15 &  0.01& 0.16 &  0.01& 0.17 &  0.01\\ 
      &      & 0.35 & 0.40& 0.07 &  0.01& 0.12 &  0.01& 0.13 &  0.01& 0.14 &  0.01\\ 
      &      & 0.40 & 0.45& 0.06 &  0.01& 0.09 &  0.01& 0.11 &  0.01& 0.11 &  0.01\\ 
      &      & 0.45 & 0.50& 0.04 &  0.01& 0.07 &  0.01& 0.09 &  0.01& 0.09 &  0.01\\ 
      &      & 0.50 & 0.60& 0.02 &  0.01& 0.04 &  0.01& 0.06 &  0.01& 0.05 &  0.01\\ 

\hline
\end{tabular}
}
\end{center}
\end{table*}

\begin{table*}[hp!]
\begin{center}
\small{

\begin{tabular}{rrrr|r@{$\pm$}lr@{$\pm$}lr@{$\pm$}lr@{$\pm$}l}
\hline
$\theta_{\hbox{\small min}}$ &
$\theta_{\hbox{\small max}}$ &
$p_{\hbox{\small min}}$ &
$p_{\hbox{\small max}}$ &
\multicolumn{8}{c}{$d^2\sigma^{\pi^+}/(dpd\theta)$}
\\
(rad) & (rad) & (\GeVc) & (\GeVc) &
\multicolumn{8}{c}{\bgr}
\\
  &  &  &
&\multicolumn{2}{c}{$ \bf{3 \ \GeVc}$}
&\multicolumn{2}{c}{$ \bf{5 \ \GeVc}$}
&\multicolumn{2}{c}{$ \bf{8 \ \GeVc}$}
&\multicolumn{2}{c}{$ \bf{12 \ \GeVc}$}
\\
\hline 
 1.15 & 1.35 & 0.10 & 0.15& 0.14 &  0.02& 0.15 &  0.02& 0.14 &  0.02& 0.14 &  0.02\\ 
      &      & 0.15 & 0.20& 0.18 &  0.02& 0.20 &  0.01& 0.21 &  0.01& 0.21 &  0.01\\ 
      &      & 0.20 & 0.25& 0.16 &  0.01& 0.17 &  0.01& 0.20 &  0.01& 0.18 &  0.01\\ 
      &      & 0.25 & 0.30& 0.09 &  0.01& 0.13 &  0.01& 0.15 &  0.01& 0.16 &  0.01\\ 
      &      & 0.30 & 0.35& 0.07 &  0.01& 0.10 &  0.01& 0.12 &  0.01& 0.11 &  0.01\\ 
      &      & 0.35 & 0.40& 0.06 &  0.01& 0.07 &  0.01& 0.08 &  0.01& 0.09 &  0.01\\ 
      &      & 0.40 & 0.45& 0.03 &  0.01& 0.05 &  0.01& 0.06 &  0.01& 0.07 &  0.01\\ 
      &      & 0.45 & 0.50& 0.02 &  0.01& 0.03 &  0.01& 0.04 &  0.01& 0.05 &  0.01\\ 
\hline  
 1.35 & 1.55 & 0.10 & 0.15& 0.13 &  0.02& 0.14 &  0.02& 0.14 &  0.02& 0.14 &  0.02\\ 
      &      & 0.15 & 0.20& 0.19 &  0.02& 0.18 &  0.01& 0.20 &  0.01& 0.20 &  0.01\\ 
      &      & 0.20 & 0.25& 0.14 &  0.01& 0.15 &  0.01& 0.15 &  0.01& 0.15 &  0.01\\ 
      &      & 0.25 & 0.30& 0.10 &  0.01& 0.12 &  0.01& 0.10 &  0.01& 0.12 &  0.01\\ 
      &      & 0.30 & 0.35& 0.05 &  0.01& 0.07 &  0.01& 0.08 &  0.01& 0.08 &  0.01\\ 
      &      & 0.35 & 0.40& 0.03 &  0.01& 0.04 &  0.01& 0.06 &  0.01& 0.06 &  0.01\\ 
      &      & 0.40 & 0.45& 0.02 &  0.01& 0.03 &  0.01& 0.04 &  0.01& 0.04 &  0.01\\ 
      &      & 0.45 & 0.50& 0.01 &  0.01& 0.02 &  0.01& 0.02 &  0.01& 0.02 &  0.01\\ 
\hline  
 1.55 & 1.75 & 0.10 & 0.15& 0.14 &  0.02& 0.13 &  0.02& 0.13 &  0.02& 0.13 &  0.02\\ 
      &      & 0.15 & 0.20& 0.18 &  0.02& 0.15 &  0.01& 0.16 &  0.01& 0.17 &  0.01\\ 
      &      & 0.20 & 0.25& 0.09 &  0.01& 0.11 &  0.01& 0.12 &  0.01& 0.11 &  0.01\\ 
      &      & 0.25 & 0.30& 0.05 &  0.01& 0.07 &  0.01& 0.07 &  0.01& 0.08 &  0.01\\ 
      &      & 0.30 & 0.35& 0.04 &  0.01& 0.04 &  0.01& 0.05 &  0.01& 0.05 &  0.01\\ 
      &      & 0.35 & 0.40& 0.02 &  0.01& 0.03 &  0.01& 0.03 &  0.01& 0.04 &  0.01\\ 
      &      & 0.40 & 0.45& 0.01 &  0.01& 0.01 &  0.01& 0.02 &  0.01& 0.02 &  0.01\\ 
      &      & 0.45 & 0.50& \multicolumn{2}{c}{}& \multicolumn{2}{c}{}& 0.01 &  0.01& 0.01 &  0.01\\ 
\hline  
 1.75 & 1.95 & 0.10 & 0.15& 0.12 &  0.02& 0.13 &  0.01& 0.12 &  0.02& 0.12 &  0.01\\ 
      &      & 0.15 & 0.20& 0.13 &  0.01& 0.13 &  0.01& 0.14 &  0.01& 0.14 &  0.01\\ 
      &      & 0.20 & 0.25& 0.08 &  0.01& 0.09 &  0.01& 0.10 &  0.01& 0.09 &  0.01\\ 
      &      & 0.25 & 0.30& 0.03 &  0.01& 0.05 &  0.01& 0.06 &  0.01& 0.06 &  0.01\\ 
      &      & 0.30 & 0.35& 0.02 &  0.01& 0.03 &  0.01& 0.03 &  0.01& 0.03 &  0.01\\ 
      &      & 0.35 & 0.40& \multicolumn{2}{c}{}& 0.01 &  0.01& 0.02 &  0.01& 0.02 &  0.01\\ 
      &      & 0.40 & 0.45& \multicolumn{2}{c}{}& \multicolumn{2}{c}{}& 0.01 &  0.01& 0.01 &  0.01\\ 
      &      & 0.45 & 0.50& \multicolumn{2}{c}{}& \multicolumn{2}{c}{}& \multicolumn{2}{c}{}& \multicolumn{2}{c}{}\\ 
\hline  
 1.95 & 2.15 & 0.10 & 0.15& 0.10 &  0.01& 0.10 &  0.01& 0.11 &  0.01& 0.10 &  0.01\\ 
      &      & 0.15 & 0.20& 0.10 &  0.01& 0.11 &  0.01& 0.11 &  0.01& 0.11 &  0.01\\ 
      &      & 0.20 & 0.25& 0.06 &  0.01& 0.06 &  0.01& 0.06 &  0.01& 0.07 &  0.01\\ 
      &      & 0.25 & 0.30& 0.02 &  0.01& 0.03 &  0.01& 0.04 &  0.01& 0.04 &  0.01\\ 
      &      & 0.30 & 0.35& 0.01 &  0.01& 0.02 &  0.01& 0.02 &  0.01& 0.02 &  0.01\\ 
      &      & 0.35 & 0.40& \multicolumn{2}{c}{}& 0.01 &  0.01& 0.01 &  0.01& 0.01 &  0.01\\ 
      &      & 0.40 & 0.45& \multicolumn{2}{c}{}& \multicolumn{2}{c}{}& \multicolumn{2}{c}{}& \multicolumn{2}{c}{}\\ 
      &      & 0.45 & 0.50& \multicolumn{2}{c}{}& \multicolumn{2}{c}{}& \multicolumn{2}{c}{}& \multicolumn{2}{c}{}\\ 

\hline

\end{tabular}
}
\end{center}
\end{table*}

\begin{table*}[hp!]
\begin{center}
\small{
 
 \caption{\label{tab:xsec-n:c}
    HARP results for the double-differential $\pi^-$ production
    cross-section in the laboratory system,
    $d^2\sigma^{\pi^-}/(dpd\theta)$ for p--C interactions. Each row refers to a
    different $(p_{\hbox{\small min}} \le p<p_{\hbox{\small max}},
    \theta_{\hbox{\small min}} \le \theta<\theta_{\hbox{\small max}})$ bin,
    where $p$ and $\theta$ are the pion momentum and polar angle, respectively.
    The central value as well as the square-root of the diagonal elements
    of the covariance matrix are given.}
\vspace{2mm}
\begin{tabular}{rrrr|r@{$\pm$}lr@{$\pm$}lr@{$\pm$}lr@{$\pm$}l}
\hline
$\theta_{\hbox{\small min}}$ &
$\theta_{\hbox{\small max}}$ &
$p_{\hbox{\small min}}$ &
$p_{\hbox{\small max}}$ &
\multicolumn{8}{c}{$d^2\sigma^{\pi^-}/(dpd\theta)$}
\\
(rad) & (rad) & (\GeVc) & (\GeVc) &
\multicolumn{8}{c}{\bgr}
\\
  &  &  &
&\multicolumn{2}{c}{$ \bf{3 \ \GeVc}$}
&\multicolumn{2}{c}{$ \bf{5 \ \GeVc}$}
&\multicolumn{2}{c}{$ \bf{8 \ \GeVc}$}
&\multicolumn{2}{c}{$ \bf{12 \ \GeVc}$}
\\
\hline 
 0.35 & 0.55 & 0.15 & 0.20& 0.04 &  0.02& 0.08 &  0.03& 0.12 &  0.03& 0.12 &  0.03\\ 
      &      & 0.20 & 0.25& 0.05 &  0.02& 0.12 &  0.02& 0.16 &  0.02& 0.18 &  0.02\\ 
      &      & 0.25 & 0.30& 0.06 &  0.01& 0.13 &  0.01& 0.19 &  0.01& 0.21 &  0.02\\ 
      &      & 0.30 & 0.35& 0.09 &  0.01& 0.13 &  0.01& 0.20 &  0.01& 0.24 &  0.01\\ 
      &      & 0.35 & 0.40& 0.08 &  0.01& 0.14 &  0.01& 0.20 &  0.01& 0.22 &  0.01\\ 
      &      & 0.40 & 0.45& 0.07 &  0.01& 0.13 &  0.01& 0.19 &  0.01& 0.22 &  0.01\\ 
      &      & 0.45 & 0.50& 0.07 &  0.01& 0.12 &  0.01& 0.18 &  0.01& 0.22 &  0.01\\ 
      &      & 0.50 & 0.60& 0.06 &  0.01& 0.13 &  0.01& 0.18 &  0.01& 0.22 &  0.01\\ 
      &      & 0.60 & 0.70& 0.05 &  0.01& 0.12 &  0.01& 0.18 &  0.01& 0.20 &  0.01\\ 
      &      & 0.70 & 0.80& 0.04 &  0.01& 0.10 &  0.01& 0.16 &  0.02& 0.18 &  0.02\\ 
\hline  
 0.55 & 0.75 & 0.10 & 0.15& 0.03 &  0.02& 0.06 &  0.02& 0.06 &  0.02& 0.09 &  0.03\\ 
      &      & 0.15 & 0.20& 0.08 &  0.02& 0.10 &  0.02& 0.15 &  0.02& 0.15 &  0.02\\ 
      &      & 0.20 & 0.25& 0.10 &  0.01& 0.15 &  0.01& 0.17 &  0.01& 0.18 &  0.01\\ 
      &      & 0.25 & 0.30& 0.09 &  0.01& 0.15 &  0.01& 0.18 &  0.01& 0.20 &  0.01\\ 
      &      & 0.30 & 0.35& 0.09 &  0.01& 0.13 &  0.01& 0.17 &  0.01& 0.20 &  0.01\\ 
      &      & 0.35 & 0.40& 0.08 &  0.01& 0.13 &  0.01& 0.17 &  0.01& 0.18 &  0.01\\ 
      &      & 0.40 & 0.45& 0.08 &  0.01& 0.11 &  0.01& 0.17 &  0.01& 0.18 &  0.01\\ 
      &      & 0.45 & 0.50& 0.07 &  0.01& 0.10 &  0.01& 0.15 &  0.01& 0.18 &  0.01\\ 
      &      & 0.50 & 0.60& 0.05 &  0.01& 0.10 &  0.01& 0.14 &  0.01& 0.17 &  0.01\\ 
      &      & 0.60 & 0.70& 0.04 &  0.01& 0.08 &  0.01& 0.12 &  0.01& 0.13 &  0.01\\ 
      &      & 0.70 & 0.80& 0.03 &  0.01& 0.06 &  0.01& 0.10 &  0.01& 0.11 &  0.01\\ 
\hline  
 0.75 & 0.95 & 0.10 & 0.15& 0.04 &  0.02& 0.05 &  0.01& 0.07 &  0.01& 0.09 &  0.02\\ 
      &      & 0.15 & 0.20& 0.08 &  0.01& 0.12 &  0.01& 0.17 &  0.01& 0.17 &  0.02\\ 
      &      & 0.20 & 0.25& 0.09 &  0.01& 0.13 &  0.01& 0.17 &  0.01& 0.18 &  0.01\\ 
      &      & 0.25 & 0.30& 0.08 &  0.01& 0.12 &  0.01& 0.18 &  0.01& 0.18 &  0.01\\ 
      &      & 0.30 & 0.35& 0.07 &  0.01& 0.12 &  0.01& 0.15 &  0.01& 0.17 &  0.01\\ 
      &      & 0.35 & 0.40& 0.07 &  0.01& 0.11 &  0.01& 0.15 &  0.01& 0.15 &  0.01\\ 
      &      & 0.40 & 0.45& 0.05 &  0.01& 0.10 &  0.01& 0.13 &  0.01& 0.13 &  0.01\\ 
      &      & 0.45 & 0.50& 0.04 &  0.01& 0.08 &  0.01& 0.11 &  0.01& 0.12 &  0.01\\ 
      &      & 0.50 & 0.60& 0.03 &  0.01& 0.06 &  0.01& 0.09 &  0.01& 0.10 &  0.01\\ 
      &      & 0.60 & 0.70& 0.02 &  0.01& 0.05 &  0.01& 0.07 &  0.01& 0.07 &  0.01\\ 
\hline  
 0.95 & 1.15 & 0.10 & 0.15& 0.05 &  0.01& 0.07 &  0.01& 0.10 &  0.01& 0.10 &  0.01\\ 
      &      & 0.15 & 0.20& 0.09 &  0.01& 0.14 &  0.01& 0.15 &  0.01& 0.16 &  0.01\\ 
      &      & 0.20 & 0.25& 0.09 &  0.01& 0.12 &  0.01& 0.14 &  0.01& 0.16 &  0.01\\ 
      &      & 0.25 & 0.30& 0.09 &  0.01& 0.12 &  0.01& 0.13 &  0.01& 0.15 &  0.01\\ 
      &      & 0.30 & 0.35& 0.06 &  0.01& 0.09 &  0.01& 0.12 &  0.01& 0.13 &  0.01\\ 
      &      & 0.35 & 0.40& 0.05 &  0.01& 0.07 &  0.01& 0.10 &  0.01& 0.11 &  0.01\\ 
      &      & 0.40 & 0.45& 0.03 &  0.01& 0.06 &  0.01& 0.08 &  0.01& 0.09 &  0.01\\ 
      &      & 0.45 & 0.50& 0.03 &  0.01& 0.05 &  0.01& 0.07 &  0.01& 0.08 &  0.01\\ 
      &      & 0.50 & 0.60& 0.02 &  0.01& 0.04 &  0.01& 0.05 &  0.01& 0.06 &  0.01\\ 
\hline
\end{tabular}
}
\end{center}
\end{table*}

\begin{table*}[hp!]
\begin{center}
\small{
\begin{tabular}{rrrr|r@{$\pm$}lr@{$\pm$}lr@{$\pm$}lr@{$\pm$}l}
\hline
$\theta_{\hbox{\small min}}$ &
$\theta_{\hbox{\small max}}$ &
$p_{\hbox{\small min}}$ &
$p_{\hbox{\small max}}$ &
\multicolumn{8}{c}{$d^2\sigma^{\pi^-}/(dpd\theta)$}
\\
(rad) & (rad) & (\GeVc) & (\GeVc) &
\multicolumn{8}{c}{\bgr}
\\
  &  &  &
&\multicolumn{2}{c}{$ \bf{3 \ \GeVc}$}
&\multicolumn{2}{c}{$ \bf{5 \ \GeVc}$}
&\multicolumn{2}{c}{$ \bf{8 \ \GeVc}$}
&\multicolumn{2}{c}{$ \bf{12 \ \GeVc}$}
\\
\hline 
 1.15 & 1.35 & 0.10 & 0.15& 0.07 &  0.02& 0.08 &  0.01& 0.10 &  0.01& 0.11 &  0.01\\ 
      &      & 0.15 & 0.20& 0.11 &  0.01& 0.13 &  0.01& 0.15 &  0.01& 0.17 &  0.01\\ 
      &      & 0.20 & 0.25& 0.07 &  0.01& 0.12 &  0.01& 0.14 &  0.01& 0.14 &  0.01\\ 
      &      & 0.25 & 0.30& 0.06 &  0.01& 0.10 &  0.01& 0.12 &  0.01& 0.12 &  0.01\\ 
      &      & 0.30 & 0.35& 0.04 &  0.01& 0.07 &  0.01& 0.09 &  0.01& 0.09 &  0.01\\ 
      &      & 0.35 & 0.40& 0.02 &  0.01& 0.05 &  0.01& 0.07 &  0.01& 0.08 &  0.01\\ 
      &      & 0.40 & 0.45& 0.02 &  0.01& 0.04 &  0.01& 0.06 &  0.01& 0.06 &  0.01\\ 
      &      & 0.45 & 0.50& 0.02 &  0.01& 0.03 &  0.01& 0.04 &  0.01& 0.05 &  0.01\\ 
\hline  
 1.35 & 1.55 & 0.10 & 0.15& 0.06 &  0.01& 0.09 &  0.01& 0.10 &  0.01& 0.11 &  0.01\\ 
      &      & 0.15 & 0.20& 0.08 &  0.01& 0.12 &  0.01& 0.14 &  0.01& 0.13 &  0.01\\ 
      &      & 0.20 & 0.25& 0.05 &  0.01& 0.10 &  0.01& 0.11 &  0.01& 0.11 &  0.01\\ 
      &      & 0.25 & 0.30& 0.05 &  0.01& 0.08 &  0.01& 0.08 &  0.01& 0.09 &  0.01\\ 
      &      & 0.30 & 0.35& 0.04 &  0.01& 0.05 &  0.01& 0.06 &  0.01& 0.06 &  0.01\\ 
      &      & 0.35 & 0.40& 0.03 &  0.01& 0.04 &  0.01& 0.05 &  0.01& 0.05 &  0.01\\ 
      &      & 0.40 & 0.45& 0.02 &  0.01& 0.03 &  0.01& 0.04 &  0.01& 0.03 &  0.01\\ 
      &      & 0.45 & 0.50& 0.01 &  0.01& 0.02 &  0.01& 0.03 &  0.01& 0.03 &  0.01\\ 
\hline  
 1.55 & 1.75 & 0.10 & 0.15& 0.07 &  0.01& 0.08 &  0.01& 0.09 &  0.01& 0.10 &  0.01\\ 
      &      & 0.15 & 0.20& 0.09 &  0.01& 0.12 &  0.01& 0.12 &  0.01& 0.13 &  0.01\\ 
      &      & 0.20 & 0.25& 0.06 &  0.01& 0.07 &  0.01& 0.09 &  0.01& 0.09 &  0.01\\ 
      &      & 0.25 & 0.30& 0.03 &  0.01& 0.05 &  0.01& 0.06 &  0.01& 0.07 &  0.01\\ 
      &      & 0.30 & 0.35& 0.01 &  0.01& 0.04 &  0.01& 0.04 &  0.01& 0.05 &  0.01\\ 
      &      & 0.35 & 0.40& \multicolumn{2}{c}{}& 0.03 &  0.01& 0.04 &  0.01& 0.03 &  0.01\\ 
      &      & 0.40 & 0.45& \multicolumn{2}{c}{}& 0.02 &  0.01& 0.03 &  0.01& 0.02 &  0.01\\ 
      &      & 0.45 & 0.50& \multicolumn{2}{c}{}& 0.01 &  0.01& 0.02 &  0.01& 0.02 &  0.01\\ 
\hline  
 1.75 & 1.95 & 0.10 & 0.15& 0.07 &  0.01& 0.08 &  0.01& 0.09 &  0.01& 0.08 &  0.01\\ 
      &      & 0.15 & 0.20& 0.08 &  0.01& 0.10 &  0.01& 0.10 &  0.01& 0.11 &  0.01\\ 
      &      & 0.20 & 0.25& 0.04 &  0.01& 0.07 &  0.01& 0.07 &  0.01& 0.07 &  0.01\\ 
      &      & 0.25 & 0.30& 0.03 &  0.01& 0.04 &  0.01& 0.05 &  0.01& 0.05 &  0.01\\ 
      &      & 0.30 & 0.35& 0.02 &  0.01& 0.03 &  0.01& 0.03 &  0.01& 0.03 &  0.01\\ 
      &      & 0.35 & 0.40& 0.01 &  0.01& 0.02 &  0.01& 0.02 &  0.01& 0.02 &  0.01\\ 
      &      & 0.40 & 0.45& \multicolumn{2}{c}{}& 0.01 &  0.01& 0.02 &  0.01& 0.02 &  0.01\\ 
      &      & 0.45 & 0.50& \multicolumn{2}{c}{}& \multicolumn{2}{c}{}& 0.01 &  0.01& 0.01 &  0.01\\ 
\hline  
 1.95 & 2.15 & 0.10 & 0.15& 0.04 &  0.01& 0.06 &  0.01& 0.07 &  0.01& 0.07 &  0.01\\ 
      &      & 0.15 & 0.20& 0.07 &  0.01& 0.09 &  0.01& 0.08 &  0.01& 0.09 &  0.01\\ 
      &      & 0.20 & 0.25& 0.04 &  0.01& 0.05 &  0.01& 0.06 &  0.01& 0.06 &  0.01\\ 
      &      & 0.25 & 0.30& 0.01 &  0.01& 0.03 &  0.01& 0.04 &  0.01& 0.04 &  0.01\\ 
      &      & 0.30 & 0.35& \multicolumn{2}{c}{}& 0.02 &  0.01& 0.02 &  0.01& 0.02 &  0.01\\ 
      &      & 0.35 & 0.40& \multicolumn{2}{c}{}& 0.01 &  0.01& 0.01 &  0.01& 0.01 &  0.01\\ 
      &      & 0.40 & 0.45& \multicolumn{2}{c}{}& \multicolumn{2}{c}{}& \multicolumn{2}{c}{}& 0.01 &  0.01\\ 
      &      & 0.45 & 0.50& \multicolumn{2}{c}{}& \multicolumn{2}{c}{}& \multicolumn{2}{c}{}& \multicolumn{2}{c}{}\\ 

\hline
\end{tabular}
}
\end{center}
\end{table*}
%


%
\begin{table*}[hp!]
\begin{center}
\small{

  \caption{\label{tab:xsec-p-al}
    HARP results for the double-differential $\pi^+$ production
    cross-section in the laboratory system,
    $d^2\sigma^{\pi^+}/(dpd\theta)$ for p--Al interactions. Each row refers to a
    different $(p_{\hbox{\small min}} \le p<p_{\hbox{\small max}},
    \theta_{\hbox{\small min}} \le \theta<\theta_{\hbox{\small max}})$ bin,
    where $p$ and $\theta$ are the pion momentum and polar angle, respectively.
    The central value as well as the square-root of the diagonal elements
    of the covariance matrix are given.}
\vspace{2mm}
\begin{tabular}{rrrr|r@{$\pm$}lr@{$\pm$}lr@{$\pm$}lr@{$\pm$}lr@{$\pm$}l}
\hline
$\theta_{\hbox{\small min}}$ &
$\theta_{\hbox{\small max}}$ &
$p_{\hbox{\small min}}$ &
$p_{\hbox{\small max}}$ &
\multicolumn{10}{c}{$d^2\sigma^{\pi^+}/(dpd\theta)$}
\\
(rad) & (rad) & (\GeVc) & (\GeVc) &
\multicolumn{10}{c}{\bgr}
\\
  &  &  &
&\multicolumn{2}{c}{$ \bf{3 \ \GeVc}$}
&\multicolumn{2}{c}{$ \bf{5 \ \GeVc}$}
&\multicolumn{2}{c}{$ \bf{8 \ \GeVc}$}
&\multicolumn{2}{c}{$ \bf{12 \ \GeVc}$}
&\multicolumn{2}{c}{$ \bf{12.9 \ \GeVc}$}
\\
\hline
 0.35 & 0.55 & 0.15 & 0.20& 0.156 &  0.027& 0.296 &  0.045& 0.351 &  0.048& 0.407 &  0.066& 0.424 &  0.064\\ 
      &      & 0.20 & 0.25& 0.203 &  0.025& 0.354 &  0.030& 0.488 &  0.044& 0.559 &  0.038& 0.573 &  0.038\\ 
      &      & 0.25 & 0.30& 0.226 &  0.025& 0.454 &  0.038& 0.549 &  0.037& 0.667 &  0.053& 0.635 &  0.042\\ 
      &      & 0.30 & 0.35& 0.282 &  0.034& 0.503 &  0.036& 0.621 &  0.043& 0.662 &  0.038& 0.716 &  0.051\\ 
      &      & 0.35 & 0.40& 0.297 &  0.024& 0.469 &  0.027& 0.615 &  0.033& 0.723 &  0.047& 0.753 &  0.043\\ 
      &      & 0.40 & 0.45& 0.308 &  0.025& 0.484 &  0.027& 0.678 &  0.045& 0.758 &  0.046& 0.784 &  0.034\\ 
      &      & 0.45 & 0.50& 0.314 &  0.024& 0.515 &  0.030& 0.696 &  0.033& 0.779 &  0.038& 0.747 &  0.029\\ 
      &      & 0.50 & 0.60& 0.297 &  0.027& 0.546 &  0.032& 0.644 &  0.035& 0.761 &  0.044& 0.760 &  0.036\\ 
      &      & 0.60 & 0.70& 0.199 &  0.030& 0.439 &  0.048& 0.616 &  0.058& 0.739 &  0.076& 0.708 &  0.072\\ 
      &      & 0.70 & 0.80& 0.112 &  0.025& 0.315 &  0.054& 0.503 &  0.074& 0.569 &  0.091& 0.555 &  0.095\\ 
\hline  
 0.55 & 0.75 & 0.10 & 0.15& 0.189 &  0.046& 0.261 &  0.061& 0.225 &  0.054& 0.298 &  0.074& 0.277 &  0.070\\ 
      &      & 0.15 & 0.20& 0.253 &  0.029& 0.372 &  0.026& 0.395 &  0.036& 0.419 &  0.035& 0.447 &  0.037\\ 
      &      & 0.20 & 0.25& 0.318 &  0.031& 0.477 &  0.046& 0.570 &  0.045& 0.579 &  0.041& 0.634 &  0.037\\ 
      &      & 0.25 & 0.30& 0.377 &  0.037& 0.476 &  0.027& 0.615 &  0.041& 0.621 &  0.039& 0.647 &  0.040\\ 
      &      & 0.30 & 0.35& 0.285 &  0.025& 0.465 &  0.026& 0.583 &  0.027& 0.607 &  0.028& 0.639 &  0.031\\ 
      &      & 0.35 & 0.40& 0.307 &  0.032& 0.432 &  0.027& 0.580 &  0.028& 0.622 &  0.038& 0.679 &  0.032\\ 
      &      & 0.40 & 0.45& 0.305 &  0.022& 0.424 &  0.021& 0.523 &  0.025& 0.646 &  0.029& 0.625 &  0.021\\ 
      &      & 0.45 & 0.50& 0.268 &  0.021& 0.407 &  0.020& 0.504 &  0.024& 0.587 &  0.028& 0.577 &  0.019\\ 
      &      & 0.50 & 0.60& 0.186 &  0.024& 0.333 &  0.025& 0.485 &  0.029& 0.501 &  0.035& 0.496 &  0.030\\ 
      &      & 0.60 & 0.70& 0.116 &  0.021& 0.227 &  0.030& 0.340 &  0.049& 0.361 &  0.053& 0.374 &  0.049\\ 
      &      & 0.70 & 0.80& 0.054 &  0.019& 0.147 &  0.028& 0.233 &  0.044& 0.227 &  0.043& 0.252 &  0.047\\ 
\hline  
 0.75 & 0.95 & 0.10 & 0.15& 0.245 &  0.038& 0.302 &  0.048& 0.276 &  0.048& 0.292 &  0.047& 0.317 &  0.051\\ 
      &      & 0.15 & 0.20& 0.337 &  0.032& 0.410 &  0.028& 0.479 &  0.032& 0.469 &  0.041& 0.534 &  0.035\\ 
      &      & 0.20 & 0.25& 0.367 &  0.032& 0.443 &  0.024& 0.554 &  0.035& 0.548 &  0.034& 0.620 &  0.039\\ 
      &      & 0.25 & 0.30& 0.355 &  0.027& 0.437 &  0.029& 0.542 &  0.028& 0.598 &  0.038& 0.573 &  0.024\\ 
      &      & 0.30 & 0.35& 0.253 &  0.021& 0.381 &  0.021& 0.460 &  0.022& 0.519 &  0.022& 0.532 &  0.019\\ 
      &      & 0.35 & 0.40& 0.216 &  0.017& 0.347 &  0.017& 0.424 &  0.020& 0.439 &  0.018& 0.489 &  0.017\\ 
      &      & 0.40 & 0.45& 0.180 &  0.017& 0.280 &  0.017& 0.362 &  0.017& 0.392 &  0.018& 0.413 &  0.015\\ 
      &      & 0.45 & 0.50& 0.137 &  0.014& 0.242 &  0.014& 0.313 &  0.017& 0.346 &  0.016& 0.358 &  0.017\\ 
      &      & 0.50 & 0.60& 0.094 &  0.012& 0.180 &  0.018& 0.253 &  0.021& 0.269 &  0.022& 0.287 &  0.025\\ 
      &      & 0.60 & 0.70& 0.052 &  0.010& 0.104 &  0.019& 0.165 &  0.028& 0.178 &  0.029& 0.167 &  0.027\\ 
\hline  
 0.95 & 1.15 & 0.10 & 0.15& 0.229 &  0.036& 0.319 &  0.041& 0.283 &  0.037& 0.291 &  0.046& 0.323 &  0.045\\ 
      &      & 0.15 & 0.20& 0.385 &  0.033& 0.412 &  0.021& 0.494 &  0.037& 0.521 &  0.043& 0.506 &  0.028\\ 
      &      & 0.20 & 0.25& 0.324 &  0.027& 0.420 &  0.031& 0.504 &  0.023& 0.519 &  0.024& 0.538 &  0.028\\ 
      &      & 0.25 & 0.30& 0.244 &  0.022& 0.354 &  0.022& 0.390 &  0.020& 0.404 &  0.019& 0.456 &  0.021\\ 
      &      & 0.30 & 0.35& 0.169 &  0.019& 0.271 &  0.015& 0.346 &  0.020& 0.341 &  0.017& 0.380 &  0.013\\ 
      &      & 0.35 & 0.40& 0.128 &  0.012& 0.221 &  0.012& 0.298 &  0.016& 0.288 &  0.014& 0.320 &  0.010\\ 
      &      & 0.40 & 0.45& 0.084 &  0.011& 0.171 &  0.012& 0.217 &  0.015& 0.245 &  0.012& 0.259 &  0.014\\ 
      &      & 0.45 & 0.50& 0.059 &  0.009& 0.114 &  0.013& 0.173 &  0.011& 0.201 &  0.014& 0.194 &  0.017\\ 
      &      & 0.50 & 0.60& 0.037 &  0.006& 0.073 &  0.010& 0.128 &  0.013& 0.137 &  0.017& 0.125 &  0.014\\ 
\hline
\end{tabular}
}
\end{center}
\end{table*}

\begin{table*}[hp!]
\begin{center}
\small{

\begin{tabular}{rrrr|r@{$\pm$}lr@{$\pm$}lr@{$\pm$}lr@{$\pm$}lr@{$\pm$}l}
\hline
$\theta_{\hbox{\small min}}$ &
$\theta_{\hbox{\small max}}$ &
$p_{\hbox{\small min}}$ &
$p_{\hbox{\small max}}$ &
\multicolumn{10}{c}{$d^2\sigma^{\pi^+}/(dpd\theta)$}
\\
(rad) & (rad) & (\GeVc) & (\GeVc) &
\multicolumn{10}{c}{\bgr}
\\
  &  &  &
&\multicolumn{2}{c}{$ \bf{3 \ \GeVc}$}
&\multicolumn{2}{c}{$ \bf{5 \ \GeVc}$}
&\multicolumn{2}{c}{$ \bf{8 \ \GeVc}$}
&\multicolumn{2}{c}{$ \bf{8.9 \ \GeVc}$}
&\multicolumn{2}{c}{$ \bf{12 \ \GeVc}$}
\\
\hline 
1.15 & 1.35 & 0.10 & 0.15& 0.239 &  0.034& 0.320 &  0.042& 0.320 &  0.044& 0.298 &  0.041& 0.355 &  0.050\\ 
      &      & 0.15 & 0.20& 0.328 &  0.030& 0.428 &  0.024& 0.460 &  0.027& 0.474 &  0.031& 0.483 &  0.022\\ 
      &      & 0.20 & 0.25& 0.269 &  0.023& 0.344 &  0.021& 0.406 &  0.019& 0.438 &  0.022& 0.438 &  0.018\\ 
      &      & 0.25 & 0.30& 0.197 &  0.020& 0.245 &  0.016& 0.332 &  0.018& 0.332 &  0.017& 0.347 &  0.015\\ 
      &      & 0.30 & 0.35& 0.128 &  0.017& 0.198 &  0.011& 0.247 &  0.013& 0.260 &  0.015& 0.258 &  0.012\\ 
      &      & 0.35 & 0.40& 0.073 &  0.009& 0.140 &  0.010& 0.195 &  0.012& 0.173 &  0.015& 0.199 &  0.010\\ 
      &      & 0.40 & 0.45& 0.051 &  0.006& 0.103 &  0.011& 0.142 &  0.011& 0.139 &  0.009& 0.140 &  0.009\\ 
      &      & 0.45 & 0.50& 0.033 &  0.006& 0.072 &  0.008& 0.101 &  0.010& 0.104 &  0.009& 0.105 &  0.010\\ 
\hline 
 1.35 & 1.55 & 0.10 & 0.15& 0.214 &  0.030& 0.324 &  0.041& 0.357 &  0.040& 0.354 &  0.049& 0.365 &  0.051\\ 
      &      & 0.15 & 0.20& 0.334 &  0.034& 0.385 &  0.024& 0.422 &  0.024& 0.433 &  0.025& 0.436 &  0.020\\ 
      &      & 0.20 & 0.25& 0.255 &  0.032& 0.283 &  0.018& 0.299 &  0.016& 0.356 &  0.019& 0.364 &  0.015\\ 
      &      & 0.25 & 0.30& 0.112 &  0.013& 0.179 &  0.011& 0.232 &  0.014& 0.241 &  0.014& 0.254 &  0.014\\ 
      &      & 0.30 & 0.35& 0.076 &  0.009& 0.150 &  0.010& 0.171 &  0.012& 0.178 &  0.011& 0.179 &  0.009\\ 
      &      & 0.35 & 0.40& 0.054 &  0.008& 0.087 &  0.009& 0.119 &  0.009& 0.137 &  0.011& 0.129 &  0.008\\ 
      &      & 0.40 & 0.45& 0.035 &  0.007& 0.058 &  0.008& 0.078 &  0.008& 0.089 &  0.010& 0.087 &  0.009\\ 
      &      & 0.45 & 0.50& 0.021 &  0.005& 0.036 &  0.006& 0.049 &  0.006& 0.058 &  0.009& 0.056 &  0.008\\ 
\hline  
 1.55 & 1.75 & 0.10 & 0.15& 0.224 &  0.033& 0.311 &  0.038& 0.310 &  0.037& 0.301 &  0.039& 0.341 &  0.041\\ 
      &      & 0.15 & 0.20& 0.304 &  0.026& 0.312 &  0.019& 0.360 &  0.018& 0.370 &  0.023& 0.391 &  0.017\\ 
      &      & 0.20 & 0.25& 0.176 &  0.019& 0.234 &  0.016& 0.228 &  0.013& 0.274 &  0.020& 0.260 &  0.012\\ 
      &      & 0.25 & 0.30& 0.113 &  0.013& 0.142 &  0.012& 0.171 &  0.011& 0.190 &  0.013& 0.173 &  0.010\\ 
      &      & 0.30 & 0.35& 0.052 &  0.009& 0.101 &  0.008& 0.118 &  0.009& 0.101 &  0.011& 0.115 &  0.006\\ 
      &      & 0.35 & 0.40& 0.026 &  0.006& 0.063 &  0.006& 0.075 &  0.007& 0.062 &  0.005& 0.086 &  0.007\\ 
      &      & 0.40 & 0.45& 0.012 &  0.004& 0.038 &  0.008& 0.047 &  0.006& 0.046 &  0.005& 0.053 &  0.006\\ 
      &      & 0.45 & 0.50& 0.005 &  0.002& 0.021 &  0.004& 0.028 &  0.005& 0.029 &  0.005& 0.032 &  0.005\\ 
\hline  
 1.75 & 1.95 & 0.10 & 0.15& 0.230 &  0.032& 0.281 &  0.032& 0.306 &  0.034& 0.292 &  0.036& 0.296 &  0.033\\ 
      &      & 0.15 & 0.20& 0.235 &  0.021& 0.292 &  0.017& 0.309 &  0.022& 0.332 &  0.019& 0.315 &  0.014\\ 
      &      & 0.20 & 0.25& 0.147 &  0.017& 0.193 &  0.013& 0.211 &  0.014& 0.192 &  0.016& 0.207 &  0.009\\ 
      &      & 0.25 & 0.30& 0.094 &  0.012& 0.093 &  0.011& 0.117 &  0.011& 0.116 &  0.010& 0.126 &  0.009\\ 
      &      & 0.30 & 0.35& 0.046 &  0.011& 0.051 &  0.006& 0.065 &  0.006& 0.063 &  0.008& 0.078 &  0.006\\ 
      &      & 0.35 & 0.40& 0.014 &  0.005& 0.034 &  0.004& 0.041 &  0.005& 0.041 &  0.004& 0.049 &  0.005\\ 
      &      & 0.40 & 0.45& 0.006 &  0.002& 0.020 &  0.004& 0.024 &  0.004& 0.026 &  0.004& 0.031 &  0.004\\ 
      &      & 0.45 & 0.50& 0.004 &  0.002& 0.012 &  0.003& 0.013 &  0.003& 0.016 &  0.003& 0.017 &  0.002\\ 
\hline  
 1.95 & 2.15 & 0.10 & 0.15& 0.228 &  0.027& 0.205 &  0.025& 0.268 &  0.031& 0.223 &  0.028& 0.234 &  0.029\\ 
      &      & 0.15 & 0.20& 0.189 &  0.020& 0.253 &  0.016& 0.260 &  0.014& 0.246 &  0.016& 0.258 &  0.011\\ 
      &      & 0.20 & 0.25& 0.084 &  0.016& 0.121 &  0.012& 0.158 &  0.018& 0.140 &  0.015& 0.156 &  0.009\\ 
      &      & 0.25 & 0.30& 0.040 &  0.008& 0.061 &  0.008& 0.072 &  0.007& 0.073 &  0.006& 0.085 &  0.007\\ 
      &      & 0.30 & 0.35& 0.015 &  0.005& 0.032 &  0.004& 0.045 &  0.005& 0.046 &  0.006& 0.048 &  0.004\\ 
      &      & 0.35 & 0.40& 0.008 &  0.003& 0.022 &  0.003& 0.021 &  0.004& 0.031 &  0.004& 0.028 &  0.003\\ 
      &      & 0.40 & 0.45& 0.006 &  0.003& 0.010 &  0.002& 0.012 &  0.002& 0.015 &  0.004& 0.015 &  0.002\\ 
      &      & 0.45 & 0.50& 0.003 &  0.002& 0.005 &  0.002& 0.006 &  0.001& 0.007 &  0.002& 0.010 &  0.002\\ 
\hline
\end{tabular}
}
\end{center}
\end{table*}

%
\begin{table*}[hp!]
\begin{center}
\small{

  \caption{\label{tab:xsec-n-al}
    HARP results for the double-differential $\pi^-$ production
    cross-section in the laboratory system,
    $d^2\sigma^{\pi^-}/(dpd\theta)$ for p--Al interactions. Each row refers to a
    different $(p_{\hbox{\small min}} \le p<p_{\hbox{\small max}},
    \theta_{\hbox{\small min}} \le \theta<\theta_{\hbox{\small max}})$ bin,
    where $p$ and $\theta$ are the pion momentum and polar angle, respectively.
    The central value as well as the square-root of the diagonal elements
    of the covariance matrix are given.}
\vspace{2mm}
\begin{tabular}{rrrr|r@{$\pm$}lr@{$\pm$}lr@{$\pm$}lr@{$\pm$}lr@{$\pm$}l}
\hline
$\theta_{\hbox{\small min}}$ &
$\theta_{\hbox{\small max}}$ &
$p_{\hbox{\small min}}$ &
$p_{\hbox{\small max}}$ &
\multicolumn{10}{c}{$d^2\sigma^{\pi^-}/(dpd\theta)$}
\\
(rad) & (rad) & (\GeVc) & (\GeVc) &
\multicolumn{10}{c}{\bgr}
\\
  &  &  &
&\multicolumn{2}{c}{$ \bf{3 \ \GeVc}$}
&\multicolumn{2}{c}{$ \bf{5 \ \GeVc}$}
&\multicolumn{2}{c}{$ \bf{8 \ \GeVc}$}
&\multicolumn{2}{c}{$ \bf{12 \ \GeVc}$}
&\multicolumn{2}{c}{$ \bf{12.9 \ \GeVc}$}
\\
\hline 
 0.35 & 0.55 & 0.15 & 0.20& 0.131 &  0.030& 0.243 &  0.042& 0.338 &  0.051& 0.342 &  0.053& 0.397 &  0.065\\ 
      &      & 0.20 & 0.25& 0.124 &  0.021& 0.258 &  0.022& 0.411 &  0.037& 0.442 &  0.037& 0.492 &  0.035\\ 
      &      & 0.25 & 0.30& 0.126 &  0.022& 0.289 &  0.024& 0.479 &  0.027& 0.516 &  0.032& 0.548 &  0.030\\ 
      &      & 0.30 & 0.35& 0.170 &  0.020& 0.312 &  0.024& 0.387 &  0.023& 0.470 &  0.031& 0.529 &  0.020\\ 
      &      & 0.35 & 0.40& 0.186 &  0.021& 0.293 &  0.018& 0.428 &  0.034& 0.485 &  0.027& 0.468 &  0.018\\ 
      &      & 0.40 & 0.45& 0.132 &  0.017& 0.262 &  0.015& 0.463 &  0.026& 0.485 &  0.025& 0.475 &  0.024\\ 
      &      & 0.45 & 0.50& 0.112 &  0.012& 0.257 &  0.015& 0.428 &  0.022& 0.440 &  0.022& 0.473 &  0.020\\ 
      &      & 0.50 & 0.60& 0.131 &  0.014& 0.246 &  0.017& 0.383 &  0.022& 0.431 &  0.025& 0.458 &  0.024\\ 
      &      & 0.60 & 0.70& 0.087 &  0.013& 0.217 &  0.019& 0.377 &  0.029& 0.415 &  0.034& 0.454 &  0.035\\ 
      &      & 0.70 & 0.80& 0.061 &  0.013& 0.166 &  0.025& 0.334 &  0.040& 0.398 &  0.041& 0.388 &  0.049\\ 
\hline  
 0.55 & 0.75 & 0.10 & 0.15& 0.075 &  0.027& 0.176 &  0.047& 0.223 &  0.060& 0.275 &  0.075& 0.270 &  0.066\\ 
      &      & 0.15 & 0.20& 0.115 &  0.019& 0.290 &  0.024& 0.383 &  0.031& 0.353 &  0.028& 0.411 &  0.030\\ 
      &      & 0.20 & 0.25& 0.196 &  0.027& 0.273 &  0.020& 0.438 &  0.028& 0.389 &  0.029& 0.456 &  0.024\\ 
      &      & 0.25 & 0.30& 0.164 &  0.018& 0.279 &  0.018& 0.429 &  0.025& 0.443 &  0.030& 0.448 &  0.021\\ 
      &      & 0.30 & 0.35& 0.153 &  0.018& 0.265 &  0.016& 0.382 &  0.021& 0.447 &  0.028& 0.447 &  0.020\\ 
      &      & 0.35 & 0.40& 0.143 &  0.014& 0.231 &  0.012& 0.397 &  0.024& 0.371 &  0.015& 0.422 &  0.015\\ 
      &      & 0.40 & 0.45& 0.130 &  0.013& 0.219 &  0.015& 0.361 &  0.017& 0.374 &  0.020& 0.388 &  0.012\\ 
      &      & 0.45 & 0.50& 0.119 &  0.012& 0.228 &  0.014& 0.317 &  0.016& 0.374 &  0.017& 0.373 &  0.013\\ 
      &      & 0.50 & 0.60& 0.095 &  0.010& 0.201 &  0.013& 0.304 &  0.017& 0.337 &  0.020& 0.348 &  0.016\\ 
      &      & 0.60 & 0.70& 0.076 &  0.011& 0.156 &  0.017& 0.269 &  0.026& 0.287 &  0.027& 0.297 &  0.026\\ 
      &      & 0.70 & 0.80& 0.055 &  0.011& 0.126 &  0.020& 0.214 &  0.031& 0.240 &  0.034& 0.239 &  0.034\\ 
\hline  
 0.75 & 0.95 & 0.10 & 0.15& 0.099 &  0.025& 0.177 &  0.030& 0.243 &  0.043& 0.313 &  0.054& 0.292 &  0.050\\ 
      &      & 0.15 & 0.20& 0.200 &  0.023& 0.322 &  0.022& 0.394 &  0.023& 0.404 &  0.023& 0.463 &  0.023\\ 
      &      & 0.20 & 0.25& 0.170 &  0.019& 0.286 &  0.020& 0.384 &  0.025& 0.378 &  0.031& 0.412 &  0.019\\ 
      &      & 0.25 & 0.30& 0.145 &  0.015& 0.280 &  0.016& 0.350 &  0.021& 0.402 &  0.021& 0.411 &  0.020\\ 
      &      & 0.30 & 0.35& 0.134 &  0.015& 0.233 &  0.013& 0.327 &  0.016& 0.339 &  0.015& 0.371 &  0.014\\ 
      &      & 0.35 & 0.40& 0.125 &  0.012& 0.220 &  0.012& 0.304 &  0.015& 0.296 &  0.014& 0.330 &  0.011\\ 
      &      & 0.40 & 0.45& 0.115 &  0.012& 0.185 &  0.011& 0.267 &  0.012& 0.291 &  0.013& 0.274 &  0.008\\ 
      &      & 0.45 & 0.50& 0.094 &  0.011& 0.158 &  0.009& 0.241 &  0.011& 0.249 &  0.014& 0.259 &  0.009\\ 
      &      & 0.50 & 0.60& 0.059 &  0.010& 0.120 &  0.009& 0.205 &  0.013& 0.217 &  0.014& 0.219 &  0.012\\ 
      &      & 0.60 & 0.70& 0.034 &  0.006& 0.086 &  0.011& 0.158 &  0.017& 0.171 &  0.018& 0.172 &  0.019\\ 
\hline  
 0.95 & 1.15 & 0.10 & 0.15& 0.161 &  0.025& 0.190 &  0.024& 0.249 &  0.032& 0.253 &  0.037& 0.301 &  0.036\\ 
      &      & 0.15 & 0.20& 0.187 &  0.021& 0.305 &  0.027& 0.364 &  0.024& 0.356 &  0.026& 0.408 &  0.024\\ 
      &      & 0.20 & 0.25& 0.178 &  0.019& 0.300 &  0.017& 0.339 &  0.018& 0.373 &  0.021& 0.392 &  0.017\\ 
      &      & 0.25 & 0.30& 0.154 &  0.015& 0.227 &  0.013& 0.301 &  0.016& 0.321 &  0.018& 0.340 &  0.013\\ 
      &      & 0.30 & 0.35& 0.118 &  0.012& 0.186 &  0.011& 0.246 &  0.013& 0.265 &  0.013& 0.269 &  0.009\\ 
      &      & 0.35 & 0.40& 0.091 &  0.011& 0.147 &  0.009& 0.188 &  0.009& 0.214 &  0.010& 0.232 &  0.007\\ 
      &      & 0.40 & 0.45& 0.060 &  0.008& 0.128 &  0.008& 0.160 &  0.007& 0.187 &  0.010& 0.191 &  0.006\\ 
      &      & 0.45 & 0.50& 0.049 &  0.006& 0.097 &  0.007& 0.144 &  0.009& 0.157 &  0.009& 0.166 &  0.005\\ 
      &      & 0.50 & 0.60& 0.034 &  0.006& 0.072 &  0.007& 0.118 &  0.008& 0.126 &  0.009& 0.132 &  0.009\\ 
\hline  
\end{tabular}
}
\end{center}
\end{table*}

\begin{table*}[hp!]
\begin{center}
\small{

\begin{tabular}{rrrr|r@{$\pm$}lr@{$\pm$}lr@{$\pm$}lr@{$\pm$}lr@{$\pm$}l}
\hline
$\theta_{\hbox{\small min}}$ &
$\theta_{\hbox{\small max}}$ &
$p_{\hbox{\small min}}$ &
$p_{\hbox{\small max}}$ &
\multicolumn{10}{c}{$d^2\sigma^{\pi^-}/(dpd\theta)$}
\\
(rad) & (rad) & (\GeVc) & (\GeVc) &
\multicolumn{10}{c}{\bgr}
\\
  &  &  &
&\multicolumn{2}{c}{$ \bf{3 \ \GeVc}$}
&\multicolumn{2}{c}{$ \bf{5 \ \GeVc}$}
&\multicolumn{2}{c}{$ \bf{8 \ \GeVc}$}
&\multicolumn{2}{c}{$ \bf{8.9 \ \GeVc}$}
&\multicolumn{2}{c}{$ \bf{12 \ \GeVc}$}
\\
\hline

 1.15 & 1.35 & 0.10 & 0.15& 0.123 &  0.020& 0.211 &  0.025& 0.266 &  0.032& 0.266 &  0.037& 0.319 &  0.033\\ 
      &      & 0.15 & 0.20& 0.160 &  0.018& 0.288 &  0.019& 0.346 &  0.020& 0.366 &  0.023& 0.386 &  0.023\\ 
      &      & 0.20 & 0.25& 0.159 &  0.017& 0.239 &  0.015& 0.305 &  0.017& 0.321 &  0.018& 0.344 &  0.012\\ 
      &      & 0.25 & 0.30& 0.132 &  0.015& 0.185 &  0.013& 0.257 &  0.014& 0.243 &  0.014& 0.270 &  0.009\\ 
      &      & 0.30 & 0.35& 0.084 &  0.011& 0.164 &  0.010& 0.181 &  0.012& 0.208 &  0.012& 0.216 &  0.008\\ 
      &      & 0.35 & 0.40& 0.062 &  0.009& 0.114 &  0.008& 0.142 &  0.008& 0.156 &  0.010& 0.162 &  0.006\\ 
      &      & 0.40 & 0.45& 0.036 &  0.006& 0.087 &  0.007& 0.115 &  0.006& 0.126 &  0.008& 0.127 &  0.006\\ 
      &      & 0.45 & 0.50& 0.026 &  0.004& 0.064 &  0.006& 0.094 &  0.006& 0.100 &  0.007& 0.099 &  0.006\\ 
\hline  
 1.35 & 1.55 & 0.10 & 0.15& 0.187 &  0.027& 0.224 &  0.027& 0.251 &  0.027& 0.270 &  0.033& 0.305 &  0.035\\ 
      &      & 0.15 & 0.20& 0.198 &  0.020& 0.247 &  0.017& 0.289 &  0.017& 0.298 &  0.021& 0.341 &  0.017\\ 
      &      & 0.20 & 0.25& 0.130 &  0.016& 0.209 &  0.014& 0.251 &  0.013& 0.273 &  0.018& 0.284 &  0.012\\ 
      &      & 0.25 & 0.30& 0.083 &  0.011& 0.130 &  0.011& 0.190 &  0.012& 0.221 &  0.014& 0.221 &  0.010\\ 
      &      & 0.30 & 0.35& 0.067 &  0.010& 0.107 &  0.008& 0.147 &  0.009& 0.151 &  0.011& 0.160 &  0.008\\ 
      &      & 0.35 & 0.40& 0.035 &  0.006& 0.091 &  0.007& 0.110 &  0.007& 0.115 &  0.007& 0.117 &  0.007\\ 
      &      & 0.40 & 0.45& 0.025 &  0.005& 0.066 &  0.006& 0.089 &  0.006& 0.090 &  0.007& 0.083 &  0.005\\ 
      &      & 0.45 & 0.50& 0.019 &  0.004& 0.044 &  0.006& 0.067 &  0.006& 0.065 &  0.007& 0.062 &  0.005\\ 
\hline  
 1.55 & 1.75 & 0.10 & 0.15& 0.140 &  0.022& 0.224 &  0.025& 0.257 &  0.033& 0.267 &  0.036& 0.285 &  0.030\\ 
      &      & 0.15 & 0.20& 0.151 &  0.017& 0.241 &  0.015& 0.270 &  0.016& 0.300 &  0.019& 0.295 &  0.014\\ 
      &      & 0.20 & 0.25& 0.101 &  0.013& 0.171 &  0.013& 0.208 &  0.013& 0.229 &  0.014& 0.217 &  0.010\\ 
      &      & 0.25 & 0.30& 0.064 &  0.009& 0.102 &  0.009& 0.158 &  0.011& 0.171 &  0.013& 0.157 &  0.008\\ 
      &      & 0.30 & 0.35& 0.046 &  0.008& 0.079 &  0.006& 0.091 &  0.009& 0.095 &  0.011& 0.109 &  0.007\\ 
      &      & 0.35 & 0.40& 0.020 &  0.006& 0.057 &  0.005& 0.060 &  0.005& 0.068 &  0.005& 0.079 &  0.005\\ 
      &      & 0.40 & 0.45& 0.009 &  0.003& 0.040 &  0.005& 0.046 &  0.004& 0.049 &  0.004& 0.056 &  0.004\\ 
      &      & 0.45 & 0.50& 0.006 &  0.002& 0.026 &  0.004& 0.033 &  0.003& 0.040 &  0.005& 0.039 &  0.003\\ 
\hline  
 1.75 & 1.95 & 0.10 & 0.15& 0.118 &  0.019& 0.190 &  0.019& 0.227 &  0.023& 0.261 &  0.029& 0.247 &  0.026\\ 
      &      & 0.15 & 0.20& 0.163 &  0.018& 0.191 &  0.013& 0.235 &  0.013& 0.240 &  0.015& 0.244 &  0.010\\ 
      &      & 0.20 & 0.25& 0.080 &  0.012& 0.136 &  0.010& 0.168 &  0.010& 0.170 &  0.012& 0.181 &  0.006\\ 
      &      & 0.25 & 0.30& 0.053 &  0.009& 0.082 &  0.008& 0.129 &  0.009& 0.102 &  0.008& 0.119 &  0.007\\ 
      &      & 0.30 & 0.35& 0.035 &  0.007& 0.055 &  0.005& 0.070 &  0.008& 0.080 &  0.006& 0.075 &  0.005\\ 
      &      & 0.35 & 0.40& 0.025 &  0.006& 0.044 &  0.005& 0.047 &  0.004& 0.060 &  0.005& 0.054 &  0.002\\ 
      &      & 0.40 & 0.45& 0.013 &  0.004& 0.024 &  0.005& 0.037 &  0.003& 0.038 &  0.005& 0.041 &  0.003\\ 
      &      & 0.45 & 0.50& 0.007 &  0.003& 0.014 &  0.003& 0.026 &  0.003& 0.027 &  0.004& 0.027 &  0.003\\ 
\hline  
 1.95 & 2.15 & 0.10 & 0.15& 0.109 &  0.017& 0.132 &  0.015& 0.197 &  0.019& 0.187 &  0.020& 0.206 &  0.020\\ 
      &      & 0.15 & 0.20& 0.104 &  0.014& 0.173 &  0.012& 0.190 &  0.012& 0.196 &  0.015& 0.187 &  0.010\\ 
      &      & 0.20 & 0.25& 0.053 &  0.009& 0.103 &  0.009& 0.117 &  0.009& 0.130 &  0.012& 0.136 &  0.006\\ 
      &      & 0.25 & 0.30& 0.037 &  0.008& 0.066 &  0.007& 0.064 &  0.008& 0.078 &  0.007& 0.085 &  0.005\\ 
      &      & 0.30 & 0.35& 0.018 &  0.005& 0.038 &  0.005& 0.044 &  0.004& 0.052 &  0.006& 0.052 &  0.004\\ 
      &      & 0.35 & 0.40& 0.010 &  0.004& 0.024 &  0.004& 0.040 &  0.004& 0.029 &  0.004& 0.036 &  0.003\\ 
      &      & 0.40 & 0.45& 0.005 &  0.003& 0.012 &  0.003& 0.025 &  0.004& 0.019 &  0.003& 0.026 &  0.002\\ 
      &      & 0.45 & 0.50& 0.002 &  0.002& 0.008 &  0.002& 0.013 &  0.003& 0.015 &  0.003& 0.018 &  0.002\\ 

\hline
\end{tabular}
}
\end{center}
\end{table*}
%


\begin{table*}[hp!]
\begin{center}
\small{

  \caption{\label{tab:xsec-p:cu}
    HARP results for the double-differential $\pi^+$ production
    cross-section in the laboratory system,
    $d^2\sigma^{\pi^+}/(dpd\theta)$ for p--Cu interactions. Each row refers to a
    different $(p_{\hbox{\small min}} \le p<p_{\hbox{\small max}},
    \theta_{\hbox{\small min}} \le \theta<\theta_{\hbox{\small max}})$ bin,
    where $p$ and $\theta$ are the pion momentum and polar angle, respectively.
    The central value as well as the square-root of the diagonal elements
    of the covariance matrix are given.}
\vspace{2mm}
\begin{tabular}{rrrr|r@{$\pm$}lr@{$\pm$}lr@{$\pm$}lr@{$\pm$}l}
\hline
$\theta_{\hbox{\small min}}$ &
$\theta_{\hbox{\small max}}$ &
$p_{\hbox{\small min}}$ &
$p_{\hbox{\small max}}$ &
\multicolumn{8}{c}{$d^2\sigma^{\pi^+}/(dpd\theta)$}
\\
(rad) & (rad) & (\GeVc) & (\GeVc) &
\multicolumn{8}{c}{\bgr}
\\
  &  &  &
&\multicolumn{2}{c}{$ \bf{3 \ \GeVc}$}
&\multicolumn{2}{c}{$ \bf{5 \ \GeVc}$}
&\multicolumn{2}{c}{$ \bf{8 \ \GeVc}$}
&\multicolumn{2}{c}{$ \bf{12 \ \GeVc}$}
\\
\hline  
 0.35 & 0.55       & 0.15 & 0.20& 0.20 &  0.05& 0.54 &  0.09& 0.76 &  0.13& 0.87 &  0.15\\ 
      &      & 0.20 & 0.25& 0.22 &  0.05& 0.68 &  0.07& 0.94 &  0.08& 1.01 &  0.08\\ 
      &      & 0.25 & 0.30& 0.46 &  0.07& 0.81 &  0.06& 1.09 &  0.06& 1.32 &  0.10\\ 
      &      & 0.30 & 0.35& 0.45 &  0.05& 0.86 &  0.06& 1.11 &  0.08& 1.42 &  0.08\\ 
      &      & 0.35 & 0.40& 0.44 &  0.05& 0.82 &  0.04& 1.20 &  0.07& 1.40 &  0.07\\ 
      &      & 0.40 & 0.45& 0.37 &  0.04& 0.78 &  0.04& 1.25 &  0.06& 1.31 &  0.05\\ 
      &      & 0.45 & 0.50& 0.41 &  0.05& 0.78 &  0.04& 1.17 &  0.05& 1.35 &  0.08\\ 
      &      & 0.50 & 0.60& 0.41 &  0.04& 0.76 &  0.04& 1.13 &  0.05& 1.31 &  0.07\\ 
      &      & 0.60 & 0.70& 0.30 &  0.05& 0.61 &  0.07& 1.00 &  0.09& 1.19 &  0.11\\ 
      &      & 0.70 & 0.80& 0.15 &  0.04& 0.39 &  0.07& 0.77 &  0.11& 0.97 &  0.13\\ 
\hline  
 0.55 & 0.75 & 0.10 & 0.15& 0.40 &  0.11& 0.40 &  0.12& 0.51 &  0.15& 0.58 &  0.17\\ 
      &      & 0.15 & 0.20& 0.44 &  0.06& 0.71 &  0.07& 0.91 &  0.07& 0.92 &  0.08\\ 
      &      & 0.20 & 0.25& 0.46 &  0.06& 0.82 &  0.05& 1.08 &  0.08& 1.28 &  0.10\\ 
      &      & 0.25 & 0.30& 0.47 &  0.05& 0.81 &  0.06& 1.11 &  0.06& 1.33 &  0.06\\ 
      &      & 0.30 & 0.35& 0.53 &  0.08& 0.88 &  0.06& 1.12 &  0.06& 1.24 &  0.06\\ 
      &      & 0.35 & 0.40& 0.58 &  0.05& 0.82 &  0.04& 1.08 &  0.04& 1.25 &  0.05\\ 
      &      & 0.40 & 0.45& 0.44 &  0.04& 0.66 &  0.04& 0.99 &  0.04& 1.18 &  0.05\\ 
      &      & 0.45 & 0.50& 0.36 &  0.04& 0.61 &  0.03& 0.94 &  0.04& 1.07 &  0.04\\ 
      &      & 0.50 & 0.60& 0.26 &  0.03& 0.52 &  0.04& 0.78 &  0.05& 0.88 &  0.06\\ 
      &      & 0.60 & 0.70& 0.13 &  0.03& 0.35 &  0.04& 0.50 &  0.06& 0.64 &  0.08\\ 
      &      & 0.70 & 0.80& 0.06 &  0.02& 0.21 &  0.04& 0.33 &  0.06& 0.43 &  0.08\\ 
\hline  
 0.75 & 0.95 & 0.10 & 0.15& 0.41 &  0.10& 0.51 &  0.11& 0.56 &  0.13& 0.68 &  0.15\\ 
      &      & 0.15 & 0.20& 0.56 &  0.06& 0.83 &  0.06& 0.96 &  0.06& 1.19 &  0.08\\ 
      &      & 0.20 & 0.25& 0.49 &  0.05& 0.82 &  0.05& 1.13 &  0.08& 1.22 &  0.06\\ 
      &      & 0.25 & 0.30& 0.50 &  0.05& 0.71 &  0.04& 1.06 &  0.06& 1.15 &  0.07\\ 
      &      & 0.30 & 0.35& 0.37 &  0.04& 0.70 &  0.04& 0.94 &  0.04& 1.15 &  0.05\\ 
      &      & 0.35 & 0.40& 0.25 &  0.03& 0.58 &  0.03& 0.81 &  0.03& 0.95 &  0.04\\ 
      &      & 0.40 & 0.45& 0.20 &  0.02& 0.45 &  0.03& 0.67 &  0.03& 0.83 &  0.03\\ 
      &      & 0.45 & 0.50& 0.18 &  0.02& 0.38 &  0.02& 0.57 &  0.03& 0.67 &  0.04\\ 
      &      & 0.50 & 0.60& 0.10 &  0.02& 0.27 &  0.02& 0.42 &  0.03& 0.51 &  0.04\\ 
      &      & 0.60 & 0.70& 0.05 &  0.01& 0.16 &  0.03& 0.26 &  0.03& 0.30 &  0.05\\ 
\hline  
 0.95 & 1.15 & 0.10 & 0.15& 0.44 &  0.09& 0.65 &  0.11& 0.66 &  0.12& 0.82 &  0.16\\ 
      &      & 0.15 & 0.20& 0.59 &  0.06& 0.87 &  0.05& 1.08 &  0.07& 1.14 &  0.07\\ 
      &      & 0.20 & 0.25& 0.44 &  0.04& 0.66 &  0.04& 0.95 &  0.04& 1.11 &  0.06\\ 
      &      & 0.25 & 0.30& 0.36 &  0.04& 0.60 &  0.04& 0.80 &  0.04& 0.87 &  0.05\\ 
      &      & 0.30 & 0.35& 0.26 &  0.03& 0.52 &  0.03& 0.62 &  0.03& 0.76 &  0.04\\ 
      &      & 0.35 & 0.40& 0.18 &  0.02& 0.39 &  0.02& 0.52 &  0.03& 0.60 &  0.03\\ 
      &      & 0.40 & 0.45& 0.15 &  0.02& 0.30 &  0.02& 0.42 &  0.02& 0.47 &  0.03\\ 
      &      & 0.45 & 0.50& 0.11 &  0.02& 0.24 &  0.02& 0.32 &  0.02& 0.38 &  0.03\\ 
      &      & 0.50 & 0.60& 0.06 &  0.01& 0.15 &  0.02& 0.22 &  0.02& 0.24 &  0.03\\ 
\hline
\end{tabular}
}
\end{center}
\end{table*}

\begin{table*}[hp!]
\begin{center}
\small{

\begin{tabular}{rrrr|r@{$\pm$}lr@{$\pm$}lr@{$\pm$}lr@{$\pm$}l}
\hline
$\theta_{\hbox{\small min}}$ &
$\theta_{\hbox{\small max}}$ &
$p_{\hbox{\small min}}$ &
$p_{\hbox{\small max}}$ &
\multicolumn{8}{c}{$d^2\sigma^{\pi^+}/(dpd\theta)$}
\\
(rad) & (rad) & (\GeVc) & (\GeVc) &
\multicolumn{8}{c}{\bgr}
\\
  &  &  &
&\multicolumn{2}{c}{$ \bf{3 \ \GeVc}$}
&\multicolumn{2}{c}{$ \bf{5 \ \GeVc}$}
&\multicolumn{2}{c}{$ \bf{8 \ \GeVc}$}
&\multicolumn{2}{c}{$ \bf{12 \ \GeVc}$}
\\
\hline
 1.15 & 1.35 & 0.10 & 0.15& 0.51 &  0.10& 0.68 &  0.14& 0.72 &  0.15& 0.78 &  0.16\\ 
      &      & 0.15 & 0.20& 0.54 &  0.06& 0.81 &  0.05& 1.01 &  0.07& 1.07 &  0.08\\ 
      &      & 0.20 & 0.25& 0.41 &  0.04& 0.63 &  0.04& 0.87 &  0.04& 1.05 &  0.05\\ 
      &      & 0.25 & 0.30& 0.31 &  0.04& 0.48 &  0.03& 0.58 &  0.04& 0.73 &  0.04\\ 
      &      & 0.30 & 0.35& 0.18 &  0.02& 0.32 &  0.03& 0.41 &  0.02& 0.53 &  0.03\\ 
      &      & 0.35 & 0.40& 0.11 &  0.02& 0.23 &  0.02& 0.33 &  0.02& 0.36 &  0.02\\ 
      &      & 0.40 & 0.45& 0.08 &  0.01& 0.17 &  0.01& 0.25 &  0.01& 0.26 &  0.02\\ 
      &      & 0.45 & 0.50& 0.05 &  0.01& 0.12 &  0.01& 0.18 &  0.01& 0.19 &  0.02\\ 
\hline  
 1.35 & 1.55 & 0.10 & 0.15& 0.52 &  0.12& 0.70 &  0.15& 0.73 &  0.18& 0.83 &  0.18\\ 
      &      & 0.15 & 0.20& 0.47 &  0.05& 0.71 &  0.06& 0.90 &  0.07& 0.99 &  0.09\\ 
      &      & 0.20 & 0.25& 0.30 &  0.04& 0.52 &  0.04& 0.73 &  0.04& 0.84 &  0.05\\ 
      &      & 0.25 & 0.30& 0.18 &  0.02& 0.34 &  0.03& 0.48 &  0.04& 0.51 &  0.03\\ 
      &      & 0.30 & 0.35& 0.13 &  0.02& 0.23 &  0.02& 0.32 &  0.02& 0.38 &  0.02\\ 
      &      & 0.35 & 0.40& 0.07 &  0.01& 0.16 &  0.01& 0.22 &  0.02& 0.29 &  0.02\\ 
      &      & 0.40 & 0.45& 0.04 &  0.01& 0.11 &  0.01& 0.15 &  0.01& 0.20 &  0.02\\ 
      &      & 0.45 & 0.50& 0.03 &  0.01& 0.06 &  0.01& 0.10 &  0.01& 0.11 &  0.02\\ 
\hline  
 1.55 & 1.75 & 0.10 & 0.15& 0.53 &  0.13& 0.67 &  0.15& 0.83 &  0.19& 0.83 &  0.20\\ 
      &      & 0.15 & 0.20& 0.43 &  0.05& 0.72 &  0.05& 0.81 &  0.06& 0.87 &  0.07\\ 
      &      & 0.20 & 0.25& 0.27 &  0.03& 0.46 &  0.04& 0.61 &  0.04& 0.67 &  0.05\\ 
      &      & 0.25 & 0.30& 0.17 &  0.03& 0.24 &  0.02& 0.34 &  0.02& 0.36 &  0.03\\ 
      &      & 0.30 & 0.35& 0.13 &  0.02& 0.15 &  0.01& 0.22 &  0.02& 0.23 &  0.02\\ 
      &      & 0.35 & 0.40& 0.06 &  0.01& 0.12 &  0.01& 0.15 &  0.01& 0.18 &  0.01\\ 
      &      & 0.40 & 0.45& 0.03 &  0.01& 0.08 &  0.01& 0.09 &  0.01& 0.11 &  0.01\\ 
      &      & 0.45 & 0.50& 0.02 &  0.01& 0.04 &  0.01& 0.06 &  0.01& 0.06 &  0.01\\ 
\hline  
 1.75 & 1.95 & 0.10 & 0.15& 0.50 &  0.10& 0.61 &  0.10& 0.70 &  0.12& 0.70 &  0.12\\ 
      &      & 0.15 & 0.20& 0.51 &  0.05& 0.60 &  0.04& 0.69 &  0.04& 0.72 &  0.04\\ 
      &      & 0.20 & 0.25& 0.35 &  0.05& 0.34 &  0.03& 0.44 &  0.03& 0.46 &  0.03\\ 
      &      & 0.25 & 0.30& 0.08 &  0.02& 0.17 &  0.02& 0.21 &  0.02& 0.26 &  0.02\\ 
      &      & 0.30 & 0.35& 0.04 &  0.01& 0.10 &  0.01& 0.12 &  0.01& 0.16 &  0.02\\ 
      &      & 0.35 & 0.40& 0.02 &  0.01& 0.07 &  0.01& 0.09 &  0.01& 0.10 &  0.01\\ 
      &      & 0.40 & 0.45& 0.01 &  0.01& 0.05 &  0.01& 0.06 &  0.01& 0.06 &  0.01\\ 
      &      & 0.45 & 0.50& \multicolumn{2}{c}{}& 0.02 &  0.01& 0.03 &  0.01& 0.03 &  0.01\\ 
\hline  
 1.95 & 2.15 & 0.10 & 0.15& 0.38 &  0.06& 0.49 &  0.07& 0.56 &  0.08& 0.48 &  0.08\\ 
      &      & 0.15 & 0.20& 0.42 &  0.05& 0.44 &  0.03& 0.55 &  0.02& 0.56 &  0.03\\ 
      &      & 0.20 & 0.25& 0.18 &  0.03& 0.26 &  0.02& 0.29 &  0.02& 0.31 &  0.02\\ 
      &      & 0.25 & 0.30& 0.07 &  0.02& 0.11 &  0.01& 0.14 &  0.01& 0.19 &  0.02\\ 
      &      & 0.30 & 0.35& 0.04 &  0.01& 0.07 &  0.01& 0.08 &  0.01& 0.10 &  0.01\\ 
      &      & 0.35 & 0.40& 0.02 &  0.01& 0.04 &  0.01& 0.05 &  0.01& 0.06 &  0.01\\ 
      &      & 0.40 & 0.45& \multicolumn{2}{c}{}& 0.02 &  0.01& 0.03 &  0.01& 0.04 &  0.01\\ 
      &      & 0.45 & 0.50& \multicolumn{2}{c}{}& 0.01 &  0.01& 0.02 &  0.01& 0.02 &  0.01\\ 
\hline
\end{tabular}
}
\end{center}
\end{table*}

\begin{table*}[hp!]
\begin{center}
\small{

  \caption{\label{tab:xsec-n:cu}
    HARP results for the double-differential $\pi^-$ production
    cross-section in the laboratory system,
    $d^2\sigma^{\pi^-}/(dpd\theta)$ for p--Cu interactions. Each row refers to a
    different $(p_{\hbox{\small min}} \le p<p_{\hbox{\small max}},
    \theta_{\hbox{\small min}} \le \theta<\theta_{\hbox{\small max}})$ bin,
    where $p$ and $\theta$ are the pion momentum and polar angle, respectively.
    The central value as well as the square-root of the diagonal elements
    of the covariance matrix are given.}
\vspace{2mm}
\begin{tabular}{rrrr|r@{$\pm$}lr@{$\pm$}lr@{$\pm$}lr@{$\pm$}l}
\hline
$\theta_{\hbox{\small min}}$ &
$\theta_{\hbox{\small max}}$ &
$p_{\hbox{\small min}}$ &
$p_{\hbox{\small max}}$ &
\multicolumn{8}{c}{$d^2\sigma^{\pi^-}/(dpd\theta)$}
\\
(rad) & (rad) & (\GeVc) & (\GeVc) &
\multicolumn{8}{c}{\bgr}
\\
  &  &  &
&\multicolumn{2}{c}{$ \bf{3 \ \GeVc}$}
&\multicolumn{2}{c}{$ \bf{5 \ \GeVc}$}
&\multicolumn{2}{c}{$ \bf{8 \ \GeVc}$}
&\multicolumn{2}{c}{$ \bf{12 \ \GeVc}$}
\\
\hline  
 0.35 & 0.55        & 0.15 & 0.20& 0.15 &  0.06& 0.43 &  0.08& 0.70 &  0.11& 0.85 &  0.17\\ 
      &      & 0.20 & 0.25& 0.21 &  0.05& 0.48 &  0.05& 0.79 &  0.07& 1.06 &  0.07\\ 
      &      & 0.25 & 0.30& 0.13 &  0.04& 0.49 &  0.04& 0.89 &  0.06& 1.03 &  0.06\\ 
      &      & 0.30 & 0.35& 0.18 &  0.03& 0.49 &  0.03& 0.79 &  0.04& 0.99 &  0.06\\ 
      &      & 0.35 & 0.40& 0.17 &  0.03& 0.47 &  0.03& 0.78 &  0.04& 0.90 &  0.04\\ 
      &      & 0.40 & 0.45& 0.17 &  0.02& 0.43 &  0.03& 0.74 &  0.03& 0.81 &  0.03\\ 
      &      & 0.45 & 0.50& 0.19 &  0.03& 0.39 &  0.02& 0.70 &  0.03& 0.75 &  0.03\\ 
      &      & 0.50 & 0.60& 0.19 &  0.03& 0.41 &  0.02& 0.68 &  0.03& 0.78 &  0.04\\ 
      &      & 0.60 & 0.70& 0.13 &  0.02& 0.36 &  0.03& 0.63 &  0.04& 0.69 &  0.06\\ 
      &      & 0.70 & 0.80& 0.10 &  0.02& 0.29 &  0.03& 0.54 &  0.06& 0.63 &  0.07\\ 
\hline  
 0.55 & 0.75 & 0.10 & 0.15& 0.27 &  0.08& 0.49 &  0.13& 0.53 &  0.15& 0.57 &  0.18\\ 
      &      & 0.15 & 0.20& 0.28 &  0.06& 0.56 &  0.05& 0.74 &  0.07& 0.96 &  0.09\\ 
      &      & 0.20 & 0.25& 0.44 &  0.06& 0.64 &  0.05& 0.81 &  0.05& 1.06 &  0.06\\ 
      &      & 0.25 & 0.30& 0.20 &  0.03& 0.60 &  0.04& 0.85 &  0.06& 1.02 &  0.05\\ 
      &      & 0.30 & 0.35& 0.24 &  0.03& 0.52 &  0.03& 0.78 &  0.04& 0.86 &  0.04\\ 
      &      & 0.35 & 0.40& 0.24 &  0.03& 0.47 &  0.03& 0.67 &  0.03& 0.83 &  0.04\\ 
      &      & 0.40 & 0.45& 0.21 &  0.03& 0.42 &  0.02& 0.62 &  0.03& 0.78 &  0.03\\ 
      &      & 0.45 & 0.50& 0.17 &  0.02& 0.37 &  0.02& 0.61 &  0.02& 0.75 &  0.03\\ 
      &      & 0.50 & 0.60& 0.14 &  0.02& 0.29 &  0.02& 0.54 &  0.02& 0.63 &  0.04\\ 
      &      & 0.60 & 0.70& 0.09 &  0.02& 0.24 &  0.02& 0.44 &  0.04& 0.49 &  0.04\\ 
      &      & 0.70 & 0.80& 0.06 &  0.02& 0.19 &  0.02& 0.35 &  0.04& 0.42 &  0.05\\ 
\hline  
 0.75 & 0.95 & 0.10 & 0.15& 0.32 &  0.07& 0.48 &  0.09& 0.59 &  0.12& 0.63 &  0.13\\ 
      &      & 0.15 & 0.20& 0.33 &  0.05& 0.64 &  0.05& 0.86 &  0.05& 0.98 &  0.06\\ 
      &      & 0.20 & 0.25& 0.24 &  0.03& 0.59 &  0.04& 0.81 &  0.05& 0.92 &  0.05\\ 
      &      & 0.25 & 0.30& 0.22 &  0.03& 0.47 &  0.03& 0.73 &  0.04& 0.90 &  0.06\\ 
      &      & 0.30 & 0.35& 0.21 &  0.03& 0.42 &  0.02& 0.64 &  0.03& 0.82 &  0.04\\ 
      &      & 0.35 & 0.40& 0.23 &  0.03& 0.34 &  0.02& 0.61 &  0.03& 0.73 &  0.03\\ 
      &      & 0.40 & 0.45& 0.18 &  0.02& 0.28 &  0.02& 0.51 &  0.02& 0.59 &  0.03\\ 
      &      & 0.45 & 0.50& 0.14 &  0.02& 0.25 &  0.01& 0.42 &  0.02& 0.48 &  0.02\\ 
      &      & 0.50 & 0.60& 0.10 &  0.02& 0.21 &  0.01& 0.34 &  0.02& 0.40 &  0.02\\ 
      &      & 0.60 & 0.70& 0.06 &  0.01& 0.17 &  0.02& 0.27 &  0.02& 0.32 &  0.03\\ 
\hline  
 0.95 & 1.15 & 0.10 & 0.15& 0.25 &  0.06& 0.55 &  0.09& 0.68 &  0.11& 0.72 &  0.14\\ 
      &      & 0.15 & 0.20& 0.32 &  0.05& 0.65 &  0.04& 0.84 &  0.05& 0.89 &  0.06\\ 
      &      & 0.20 & 0.25& 0.32 &  0.04& 0.51 &  0.04& 0.77 &  0.04& 0.83 &  0.05\\ 
      &      & 0.25 & 0.30& 0.23 &  0.03& 0.43 &  0.03& 0.63 &  0.03& 0.72 &  0.04\\ 
      &      & 0.30 & 0.35& 0.13 &  0.02& 0.36 &  0.02& 0.51 &  0.03& 0.57 &  0.03\\ 
      &      & 0.35 & 0.40& 0.11 &  0.02& 0.27 &  0.02& 0.43 &  0.02& 0.49 &  0.02\\ 
      &      & 0.40 & 0.45& 0.12 &  0.02& 0.22 &  0.01& 0.35 &  0.02& 0.43 &  0.02\\ 
      &      & 0.45 & 0.50& 0.11 &  0.02& 0.19 &  0.01& 0.29 &  0.01& 0.36 &  0.02\\ 
      &      & 0.50 & 0.60& 0.07 &  0.01& 0.14 &  0.01& 0.22 &  0.02& 0.25 &  0.02\\ 
\hline \\
\end{tabular}
}
\end{center}
\end{table*}

\begin{table*}[hp!]
\begin{center}
\small{

\begin{tabular}{rrrr|r@{$\pm$}lr@{$\pm$}lr@{$\pm$}lr@{$\pm$}l}
\hline
$\theta_{\hbox{\small min}}$ &
$\theta_{\hbox{\small max}}$ &
$p_{\hbox{\small min}}$ &
$p_{\hbox{\small max}}$ &
\multicolumn{8}{c}{$d^2\sigma^{\pi^-}/(dpd\theta)$}
\\
(rad) & (rad) & (\GeVc) & (\GeVc) &
\multicolumn{8}{c}{\bgr}
\\
  &  &  &
&\multicolumn{2}{c}{$ \bf{3 \ \GeVc}$}
&\multicolumn{2}{c}{$ \bf{5 \ \GeVc}$}
&\multicolumn{2}{c}{$ \bf{8 \ \GeVc}$}
&\multicolumn{2}{c}{$ \bf{12 \ \GeVc}$}
\\
\hline
 
 1.15 & 1.35 & 0.10 & 0.15& 0.30 &  0.07& 0.63 &  0.11& 0.71 &  0.13& 0.79 &  0.17\\ 
      &      & 0.15 & 0.20& 0.38 &  0.05& 0.64 &  0.05& 0.75 &  0.06& 0.97 &  0.07\\ 
      &      & 0.20 & 0.25& 0.32 &  0.04& 0.48 &  0.03& 0.64 &  0.04& 0.79 &  0.05\\ 
      &      & 0.25 & 0.30& 0.16 &  0.03& 0.37 &  0.02& 0.51 &  0.03& 0.58 &  0.03\\ 
      &      & 0.30 & 0.35& 0.15 &  0.02& 0.29 &  0.02& 0.38 &  0.02& 0.45 &  0.02\\ 
      &      & 0.35 & 0.40& 0.10 &  0.02& 0.20 &  0.01& 0.30 &  0.01& 0.36 &  0.02\\ 
      &      & 0.40 & 0.45& 0.06 &  0.01& 0.15 &  0.01& 0.24 &  0.01& 0.28 &  0.02\\ 
      &      & 0.45 & 0.50& 0.04 &  0.01& 0.12 &  0.01& 0.19 &  0.01& 0.20 &  0.02\\ 
\hline  
 1.35 & 1.55 & 0.10 & 0.15& 0.39 &  0.09& 0.63 &  0.13& 0.72 &  0.16& 0.80 &  0.19\\ 
      &      & 0.15 & 0.20& 0.39 &  0.05& 0.57 &  0.05& 0.73 &  0.06& 0.85 &  0.07\\ 
      &      & 0.20 & 0.25& 0.27 &  0.04& 0.41 &  0.03& 0.53 &  0.03& 0.69 &  0.04\\ 
      &      & 0.25 & 0.30& 0.20 &  0.03& 0.30 &  0.02& 0.37 &  0.02& 0.47 &  0.03\\ 
      &      & 0.30 & 0.35& 0.15 &  0.02& 0.20 &  0.02& 0.29 &  0.02& 0.33 &  0.02\\ 
      &      & 0.35 & 0.40& 0.09 &  0.02& 0.13 &  0.01& 0.21 &  0.01& 0.24 &  0.02\\ 
      &      & 0.40 & 0.45& 0.07 &  0.01& 0.09 &  0.01& 0.16 &  0.01& 0.18 &  0.01\\ 
      &      & 0.45 & 0.50& 0.04 &  0.01& 0.06 &  0.01& 0.12 &  0.01& 0.13 &  0.01\\ 
\hline  
 1.55 & 1.75 & 0.10 & 0.15& 0.37 &  0.08& 0.56 &  0.12& 0.73 &  0.17& 0.84 &  0.20\\ 
      &      & 0.15 & 0.20& 0.27 &  0.04& 0.50 &  0.04& 0.66 &  0.05& 0.71 &  0.06\\ 
      &      & 0.20 & 0.25& 0.20 &  0.03& 0.33 &  0.03& 0.44 &  0.03& 0.50 &  0.04\\ 
      &      & 0.25 & 0.30& 0.12 &  0.02& 0.19 &  0.02& 0.31 &  0.02& 0.33 &  0.03\\ 
      &      & 0.30 & 0.35& 0.07 &  0.01& 0.13 &  0.01& 0.21 &  0.02& 0.23 &  0.02\\ 
      &      & 0.35 & 0.40& 0.04 &  0.01& 0.09 &  0.01& 0.14 &  0.01& 0.17 &  0.01\\ 
      &      & 0.40 & 0.45& 0.02 &  0.01& 0.06 &  0.01& 0.10 &  0.01& 0.11 &  0.01\\ 
      &      & 0.45 & 0.50& 0.01 &  0.01& 0.04 &  0.01& 0.07 &  0.01& 0.08 &  0.01\\ 
\hline  
 1.75 & 1.95 & 0.10 & 0.15& 0.26 &  0.05& 0.51 &  0.08& 0.63 &  0.10& 0.66 &  0.11\\ 
      &      & 0.15 & 0.20& 0.21 &  0.03& 0.42 &  0.03& 0.51 &  0.03& 0.64 &  0.04\\ 
      &      & 0.20 & 0.25& 0.19 &  0.03& 0.28 &  0.02& 0.35 &  0.02& 0.40 &  0.03\\ 
      &      & 0.25 & 0.30& 0.08 &  0.02& 0.15 &  0.02& 0.22 &  0.02& 0.24 &  0.02\\ 
      &      & 0.30 & 0.35& 0.06 &  0.01& 0.09 &  0.01& 0.14 &  0.01& 0.14 &  0.01\\ 
      &      & 0.35 & 0.40& 0.05 &  0.01& 0.06 &  0.01& 0.10 &  0.01& 0.11 &  0.01\\ 
      &      & 0.40 & 0.45& 0.02 &  0.01& 0.04 &  0.01& 0.07 &  0.01& 0.07 &  0.01\\ 
      &      & 0.45 & 0.50& 0.01 &  0.01& 0.03 &  0.01& 0.05 &  0.01& 0.05 &  0.01\\ 
\hline  
 1.95 & 2.15 & 0.10 & 0.15& 0.26 &  0.05& 0.42 &  0.06& 0.52 &  0.07& 0.55 &  0.08\\ 
      &      & 0.15 & 0.20& 0.20 &  0.03& 0.31 &  0.02& 0.42 &  0.02& 0.47 &  0.03\\ 
      &      & 0.20 & 0.25& 0.13 &  0.03& 0.17 &  0.01& 0.25 &  0.01& 0.28 &  0.02\\ 
      &      & 0.25 & 0.30& 0.06 &  0.02& 0.11 &  0.01& 0.16 &  0.01& 0.15 &  0.02\\ 
      &      & 0.30 & 0.35& 0.02 &  0.01& 0.07 &  0.01& 0.09 &  0.01& 0.10 &  0.01\\ 
      &      & 0.35 & 0.40& 0.02 &  0.01& 0.05 &  0.01& 0.05 &  0.01& 0.07 &  0.01\\ 
      &      & 0.40 & 0.45& 0.01 &  0.01& 0.03 &  0.01& 0.04 &  0.01& 0.06 &  0.01\\ 
      &      & 0.45 & 0.50& \multicolumn{2}{c}{}& 0.02 &  0.01& 0.03 &  0.01& 0.04 &  0.01\\ 
\hline \\
\end{tabular}
}
\end{center}
\end{table*}
%

\begin{table*}[hp!]
\begin{center}
\small{

  \caption{\label{tab:xsec-p:sn}
    HARP results for the double-differential $\pi^+$ production
    cross-section in the laboratory system,
    $d^2\sigma^{\pi^+}/(dpd\theta)$ for p--Sn interactions. Each row refers to a
    different $(p_{\hbox{\small min}} \le p<p_{\hbox{\small max}},
    \theta_{\hbox{\small min}} \le \theta<\theta_{\hbox{\small max}})$ bin,
    where $p$ and $\theta$ are the pion momentum and polar angle, respectively.
    The central value as well as the square-root of the diagonal elements
    of the covariance matrix are given.}
\vspace{2mm}
\begin{tabular}{rrrr|r@{$\pm$}lr@{$\pm$}lr@{$\pm$}lr@{$\pm$}l}
\hline
$\theta_{\hbox{\small min}}$ &
$\theta_{\hbox{\small max}}$ &
$p_{\hbox{\small min}}$ &
$p_{\hbox{\small max}}$ &
\multicolumn{8}{c}{$d^2\sigma^{\pi^+}/(dpd\theta)$}
\\
(rad) & (rad) & (\GeVc) & (\GeVc) &
\multicolumn{8}{c}{\bgr}
\\
  &  &  &
&\multicolumn{2}{c}{$ \bf{3 \ \GeVc}$}
&\multicolumn{2}{c}{$ \bf{5 \ \GeVc}$}
&\multicolumn{2}{c}{$ \bf{8 \ \GeVc}$}
&\multicolumn{2}{c}{$ \bf{12 \ \GeVc}$}

\\
\hline 
 0.35 & 0.55      & 0.15 & 0.20& 0.28 &  0.11& 0.74 &  0.14& 1.20 &  0.18& 1.55 &  0.25\\ 
      &      & 0.20 & 0.25& 0.41 &  0.10& 0.78 &  0.09& 1.42 &  0.12& 1.81 &  0.16\\ 
      &      & 0.25 & 0.30& 0.55 &  0.06& 1.05 &  0.10& 1.59 &  0.11& 2.18 &  0.14\\ 
      &      & 0.30 & 0.35& 0.53 &  0.05& 1.00 &  0.05& 1.67 &  0.11& 2.14 &  0.14\\ 
      &      & 0.35 & 0.40& 0.56 &  0.05& 1.05 &  0.08& 1.64 &  0.07& 2.14 &  0.07\\ 
      &      & 0.40 & 0.45& 0.50 &  0.04& 1.10 &  0.06& 1.55 &  0.08& 2.21 &  0.14\\ 
      &      & 0.45 & 0.50& 0.47 &  0.04& 1.10 &  0.05& 1.53 &  0.06& 2.15 &  0.07\\ 
      &      & 0.50 & 0.60& 0.44 &  0.04& 0.94 &  0.06& 1.45 &  0.07& 2.20 &  0.12\\ 
      &      & 0.60 & 0.70& 0.36 &  0.04& 0.80 &  0.08& 1.27 &  0.13& 2.12 &  0.18\\ 
      &      & 0.70 & 0.80& 0.26 &  0.04& 0.47 &  0.08& 0.93 &  0.16& 1.78 &  0.21\\ 
\hline  
 0.55 & 0.75 & 0.10 & 0.15& 0.33 &  0.12& 0.76 &  0.17& 0.87 &  0.21& 1.26 &  0.30\\ 
      &      & 0.15 & 0.20& 0.49 &  0.08& 1.02 &  0.10& 1.53 &  0.10& 1.77 &  0.14\\ 
      &      & 0.20 & 0.25& 0.59 &  0.09& 1.23 &  0.09& 1.72 &  0.12& 2.19 &  0.17\\ 
      &      & 0.25 & 0.30& 0.60 &  0.05& 1.11 &  0.07& 1.61 &  0.10& 2.15 &  0.17\\ 
      &      & 0.30 & 0.35& 0.56 &  0.05& 1.12 &  0.07& 1.54 &  0.07& 2.13 &  0.10\\ 
      &      & 0.35 & 0.40& 0.46 &  0.04& 1.03 &  0.05& 1.47 &  0.07& 2.08 &  0.07\\ 
      &      & 0.40 & 0.45& 0.46 &  0.04& 0.87 &  0.04& 1.27 &  0.05& 1.98 &  0.06\\ 
      &      & 0.45 & 0.50& 0.37 &  0.04& 0.80 &  0.04& 1.21 &  0.05& 1.76 &  0.07\\ 
      &      & 0.50 & 0.60& 0.27 &  0.03& 0.61 &  0.05& 1.03 &  0.06& 1.53 &  0.09\\ 
      &      & 0.60 & 0.70& 0.17 &  0.03& 0.38 &  0.05& 0.71 &  0.10& 1.21 &  0.11\\ 
      &      & 0.70 & 0.80& 0.11 &  0.02& 0.25 &  0.05& 0.43 &  0.08& 0.82 &  0.12\\ 
\hline  
 0.75 & 0.95 & 0.10 & 0.15& 0.47 &  0.10& 0.80 &  0.13& 1.07 &  0.17& 1.16 &  0.20\\ 
      &      & 0.15 & 0.20& 0.63 &  0.07& 1.17 &  0.07& 1.69 &  0.11& 1.95 &  0.12\\ 
      &      & 0.20 & 0.25& 0.76 &  0.06& 1.22 &  0.07& 1.64 &  0.09& 1.94 &  0.11\\ 
      &      & 0.25 & 0.30& 0.57 &  0.05& 0.95 &  0.05& 1.45 &  0.07& 1.92 &  0.11\\ 
      &      & 0.30 & 0.35& 0.57 &  0.04& 0.93 &  0.06& 1.31 &  0.06& 1.78 &  0.08\\ 
      &      & 0.35 & 0.40& 0.46 &  0.04& 0.73 &  0.04& 1.17 &  0.05& 1.40 &  0.05\\ 
      &      & 0.40 & 0.45& 0.36 &  0.03& 0.60 &  0.04& 0.99 &  0.04& 1.27 &  0.05\\ 
      &      & 0.45 & 0.50& 0.30 &  0.03& 0.50 &  0.03& 0.80 &  0.05& 1.11 &  0.05\\ 
      &      & 0.50 & 0.60& 0.19 &  0.03& 0.35 &  0.03& 0.56 &  0.05& 0.85 &  0.06\\ 
      &      & 0.60 & 0.70& 0.09 &  0.02& 0.20 &  0.03& 0.32 &  0.05& 0.55 &  0.06\\ 
\hline  
 0.95 & 1.15 & 0.10 & 0.15& 0.45 &  0.09& 0.80 &  0.11& 1.13 &  0.14& 1.14 &  0.17\\ 
      &      & 0.15 & 0.20& 0.68 &  0.07& 1.12 &  0.07& 1.55 &  0.09& 1.82 &  0.11\\ 
      &      & 0.20 & 0.25& 0.51 &  0.04& 1.01 &  0.06& 1.40 &  0.09& 1.72 &  0.08\\ 
      &      & 0.25 & 0.30& 0.43 &  0.04& 0.77 &  0.04& 1.16 &  0.05& 1.48 &  0.07\\ 
      &      & 0.30 & 0.35& 0.29 &  0.02& 0.59 &  0.04& 0.92 &  0.07& 1.20 &  0.06\\ 
      &      & 0.35 & 0.40& 0.27 &  0.03& 0.50 &  0.03& 0.72 &  0.04& 0.97 &  0.05\\ 
      &      & 0.40 & 0.45& 0.19 &  0.03& 0.41 &  0.03& 0.65 &  0.03& 0.79 &  0.05\\ 
      &      & 0.45 & 0.50& 0.14 &  0.02& 0.30 &  0.02& 0.49 &  0.04& 0.64 &  0.04\\ 
      &      & 0.50 & 0.60& 0.08 &  0.01& 0.20 &  0.02& 0.30 &  0.03& 0.46 &  0.04\\ 

\hline
\end{tabular}
}
\end{center}
\end{table*}

\begin{table*}[H]
\begin{center}
\small{

\begin{tabular}{rrrr|r@{$\pm$}lr@{$\pm$}lr@{$\pm$}lr@{$\pm$}l}
\hline
$\theta_{\hbox{\small min}}$ &
$\theta_{\hbox{\small max}}$ &
$p_{\hbox{\small min}}$ &
$p_{\hbox{\small max}}$ &
\multicolumn{8}{c}{$d^2\sigma^{\pi^+}/(dpd\theta)$}
\\
(rad) & (rad) & (\GeVc) & (\GeVc) &
\multicolumn{8}{c}{\bgr}
\\
  &  &  &
&\multicolumn{2}{c}{$ \bf{3 \ \GeVc}$}
&\multicolumn{2}{c}{$ \bf{5 \ \GeVc}$}
&\multicolumn{2}{c}{$ \bf{8 \ \GeVc}$}
&\multicolumn{2}{c}{$ \bf{12 \ \GeVc}$}
\\
\hline 
 1.15 & 1.35 & 0.10 & 0.15& 0.57 &  0.09& 0.76 &  0.10& 1.06 &  0.14& 1.10 &  0.18\\ 
      &      & 0.15 & 0.20& 0.64 &  0.05& 1.09 &  0.08& 1.39 &  0.10& 1.74 &  0.11\\ 
      &      & 0.20 & 0.25& 0.47 &  0.04& 0.85 &  0.05& 1.22 &  0.07& 1.43 &  0.07\\ 
      &      & 0.25 & 0.30& 0.40 &  0.04& 0.65 &  0.04& 0.93 &  0.05& 1.09 &  0.06\\ 
      &      & 0.30 & 0.35& 0.25 &  0.03& 0.46 &  0.03& 0.67 &  0.05& 0.86 &  0.05\\ 
      &      & 0.35 & 0.40& 0.15 &  0.02& 0.32 &  0.03& 0.48 &  0.02& 0.69 &  0.04\\ 
      &      & 0.40 & 0.45& 0.10 &  0.01& 0.23 &  0.02& 0.38 &  0.02& 0.51 &  0.04\\ 
      &      & 0.45 & 0.50& 0.08 &  0.01& 0.18 &  0.02& 0.28 &  0.03& 0.34 &  0.03\\ 
\hline  
 1.35 & 1.55 & 0.10 & 0.15& 0.72 &  0.12& 0.86 &  0.14& 1.26 &  0.21& 1.30 &  0.23\\ 
      &      & 0.15 & 0.20& 0.70 &  0.06& 1.11 &  0.09& 1.44 &  0.10& 1.77 &  0.16\\ 
      &      & 0.20 & 0.25& 0.44 &  0.04& 0.83 &  0.06& 1.11 &  0.09& 1.29 &  0.09\\ 
      &      & 0.25 & 0.30& 0.33 &  0.03& 0.52 &  0.04& 0.74 &  0.05& 0.92 &  0.06\\ 
      &      & 0.30 & 0.35& 0.19 &  0.03& 0.37 &  0.03& 0.52 &  0.03& 0.63 &  0.05\\ 
      &      & 0.35 & 0.40& 0.11 &  0.01& 0.26 &  0.02& 0.36 &  0.02& 0.48 &  0.03\\ 
      &      & 0.40 & 0.45& 0.07 &  0.01& 0.16 &  0.02& 0.24 &  0.02& 0.37 &  0.03\\ 
      &      & 0.45 & 0.50& 0.04 &  0.01& 0.09 &  0.01& 0.17 &  0.02& 0.21 &  0.03\\ 
\hline  
 1.55 & 1.75 & 0.10 & 0.15& 0.65 &  0.12& 0.99 &  0.16& 1.27 &  0.20& 1.29 &  0.23\\ 
      &      & 0.15 & 0.20& 0.71 &  0.06& 1.00 &  0.08& 1.40 &  0.10& 1.57 &  0.11\\ 
      &      & 0.20 & 0.25& 0.41 &  0.04& 0.68 &  0.05& 0.95 &  0.07& 1.04 &  0.08\\ 
      &      & 0.25 & 0.30& 0.24 &  0.03& 0.45 &  0.04& 0.57 &  0.04& 0.67 &  0.04\\ 
      &      & 0.30 & 0.35& 0.13 &  0.02& 0.23 &  0.02& 0.36 &  0.03& 0.46 &  0.03\\ 
      &      & 0.35 & 0.40& 0.08 &  0.01& 0.18 &  0.02& 0.24 &  0.02& 0.31 &  0.02\\ 
      &      & 0.40 & 0.45& 0.04 &  0.01& 0.09 &  0.01& 0.16 &  0.02& 0.20 &  0.02\\ 
      &      & 0.45 & 0.50& 0.03 &  0.01& 0.06 &  0.01& 0.10 &  0.02& 0.12 &  0.02\\ 
\hline  
 1.75 & 1.95 & 0.10 & 0.15& 0.60 &  0.09& 0.88 &  0.10& 0.99 &  0.11& 1.01 &  0.13\\ 
      &      & 0.15 & 0.20& 0.56 &  0.05& 0.77 &  0.05& 1.00 &  0.05& 1.13 &  0.06\\ 
      &      & 0.20 & 0.25& 0.28 &  0.03& 0.48 &  0.03& 0.62 &  0.05& 0.73 &  0.04\\ 
      &      & 0.25 & 0.30& 0.15 &  0.03& 0.26 &  0.03& 0.34 &  0.03& 0.40 &  0.03\\ 
      &      & 0.30 & 0.35& 0.07 &  0.01& 0.15 &  0.02& 0.22 &  0.02& 0.24 &  0.02\\ 
      &      & 0.35 & 0.40& 0.03 &  0.01& 0.08 &  0.01& 0.14 &  0.02& 0.15 &  0.02\\ 
      &      & 0.40 & 0.45& 0.01 &  0.01& 0.05 &  0.01& 0.08 &  0.01& 0.10 &  0.01\\ 
      &      & 0.45 & 0.50& \multicolumn{2}{c}{}& 0.02 &  0.01& 0.05 &  0.01& 0.06 &  0.01\\ 
\hline  
 1.95 & 2.15 & 0.10 & 0.15& 0.51 &  0.07& 0.63 &  0.07& 0.75 &  0.07& 0.73 &  0.08\\ 
      &      & 0.15 & 0.20& 0.45 &  0.04& 0.65 &  0.04& 0.72 &  0.04& 0.90 &  0.04\\ 
      &      & 0.20 & 0.25& 0.22 &  0.03& 0.36 &  0.03& 0.43 &  0.03& 0.52 &  0.04\\ 
      &      & 0.25 & 0.30& 0.11 &  0.02& 0.17 &  0.02& 0.24 &  0.02& 0.24 &  0.03\\ 
      &      & 0.30 & 0.35& 0.05 &  0.01& 0.10 &  0.01& 0.14 &  0.01& 0.17 &  0.01\\ 
      &      & 0.35 & 0.40& 0.02 &  0.01& 0.05 &  0.01& 0.09 &  0.01& 0.11 &  0.01\\ 
      &      & 0.40 & 0.45& 0.01 &  0.01& 0.03 &  0.01& 0.05 &  0.01& 0.06 &  0.01\\ 
      &      & 0.45 & 0.50& \multicolumn{2}{c}{}& 0.02 &  0.01& 0.02 &  0.01& 0.04 &  0.01\\ 

\hline
\end{tabular}
}
\end{center}
\end{table*}

\begin{table*}[hp!]
\begin{center}
\small{

  \caption{\label{tab:xsec-n:sn}
    HARP results for the double-differential $\pi^-$ production
    cross-section in the laboratory system,
    $d^2\sigma^{\pi^-}/(dpd\theta)$ for p--Sn interactions. Each row refers to a
    different $(p_{\hbox{\small min}} \le p<p_{\hbox{\small max}},
    \theta_{\hbox{\small min}} \le \theta<\theta_{\hbox{\small max}})$ bin,
    where $p$ and $\theta$ are the pion momentum and polar angle, respectively.
    The central value as well as the square-root of the diagonal elements
    of the covariance matrix are given.}
\vspace{2mm}
\begin{tabular}{rrrr|r@{$\pm$}lr@{$\pm$}lr@{$\pm$}lr@{$\pm$}l}
\hline
$\theta_{\hbox{\small min}}$ &
$\theta_{\hbox{\small max}}$ &
$p_{\hbox{\small min}}$ &
$p_{\hbox{\small max}}$ &
\multicolumn{8}{c}{$d^2\sigma^{\pi^-}/(dpd\theta)$}
\\
(rad) & (rad) & (\GeVc) & (\GeVc) &
\multicolumn{8}{c}{\bgr}
\\
  &  &  &
&\multicolumn{2}{c}{$ \bf{3 \ \GeVc}$}
&\multicolumn{2}{c}{$ \bf{5 \ \GeVc}$}
&\multicolumn{2}{c}{$ \bf{8 \ \GeVc}$}
&\multicolumn{2}{c}{$ \bf{12 \ \GeVc}$}
\\
\hline

0.35 & 0.55        & 0.15 & 0.20& 0.33 &  0.12& 0.74 &  0.13& 1.30 &  0.19& 1.62 &  0.24\\ 
      &      & 0.20 & 0.25& 0.39 &  0.08& 0.74 &  0.09& 1.41 &  0.11& 1.75 &  0.16\\ 
      &      & 0.25 & 0.30& 0.35 &  0.06& 0.76 &  0.07& 1.36 &  0.09& 1.88 &  0.12\\ 
      &      & 0.30 & 0.35& 0.36 &  0.04& 0.78 &  0.05& 1.35 &  0.07& 1.78 &  0.08\\ 
      &      & 0.35 & 0.40& 0.25 &  0.03& 0.66 &  0.04& 1.28 &  0.06& 1.65 &  0.06\\ 
      &      & 0.40 & 0.45& 0.34 &  0.04& 0.61 &  0.03& 1.16 &  0.05& 1.48 &  0.05\\ 
      &      & 0.45 & 0.50& 0.29 &  0.03& 0.56 &  0.03& 1.07 &  0.04& 1.42 &  0.06\\ 
      &      & 0.50 & 0.60& 0.23 &  0.02& 0.50 &  0.03& 0.96 &  0.05& 1.34 &  0.06\\ 
      &      & 0.60 & 0.70& 0.19 &  0.03& 0.46 &  0.04& 0.86 &  0.06& 1.31 &  0.09\\ 
      &      & 0.70 & 0.80& 0.13 &  0.03& 0.36 &  0.05& 0.68 &  0.08& 1.16 &  0.12\\ 
\hline  
 0.55 & 0.75 & 0.10 & 0.15& 0.30 &  0.12& 0.58 &  0.15& 1.07 &  0.22& 1.21 &  0.31\\ 
      &      & 0.15 & 0.20& 0.38 &  0.07& 0.97 &  0.10& 1.54 &  0.10& 1.67 &  0.14\\ 
      &      & 0.20 & 0.25& 0.43 &  0.06& 0.92 &  0.06& 1.35 &  0.08& 1.76 &  0.13\\ 
      &      & 0.25 & 0.30& 0.39 &  0.04& 0.78 &  0.05& 1.32 &  0.08& 1.66 &  0.07\\ 
      &      & 0.30 & 0.35& 0.32 &  0.03& 0.67 &  0.04& 1.28 &  0.06& 1.53 &  0.06\\ 
      &      & 0.35 & 0.40& 0.29 &  0.03& 0.63 &  0.03& 1.09 &  0.05& 1.38 &  0.06\\ 
      &      & 0.40 & 0.45& 0.27 &  0.03& 0.59 &  0.03& 1.00 &  0.04& 1.27 &  0.05\\ 
      &      & 0.45 & 0.50& 0.23 &  0.02& 0.52 &  0.03& 0.94 &  0.04& 1.15 &  0.04\\ 
      &      & 0.50 & 0.60& 0.17 &  0.02& 0.44 &  0.03& 0.77 &  0.04& 1.05 &  0.05\\ 
      &      & 0.60 & 0.70& 0.13 &  0.02& 0.37 &  0.03& 0.61 &  0.05& 0.88 &  0.07\\ 
      &      & 0.70 & 0.80& 0.10 &  0.02& 0.25 &  0.04& 0.46 &  0.07& 0.75 &  0.08\\ 
\hline  
 0.75 & 0.95 & 0.10 & 0.15& 0.30 &  0.08& 0.70 &  0.11& 1.07 &  0.14& 1.18 &  0.18\\ 
      &      & 0.15 & 0.20& 0.51 &  0.07& 1.00 &  0.07& 1.55 &  0.09& 1.83 &  0.12\\ 
      &      & 0.20 & 0.25& 0.45 &  0.05& 0.79 &  0.05& 1.36 &  0.07& 1.66 &  0.08\\ 
      &      & 0.25 & 0.30& 0.37 &  0.04& 0.78 &  0.05& 1.20 &  0.06& 1.51 &  0.07\\ 
      &      & 0.30 & 0.35& 0.31 &  0.03& 0.70 &  0.04& 1.05 &  0.05& 1.32 &  0.05\\ 
      &      & 0.35 & 0.40& 0.28 &  0.03& 0.56 &  0.03& 0.88 &  0.04& 1.15 &  0.04\\ 
      &      & 0.40 & 0.45& 0.20 &  0.03& 0.47 &  0.02& 0.77 &  0.03& 0.97 &  0.04\\ 
      &      & 0.45 & 0.50& 0.16 &  0.02& 0.39 &  0.02& 0.63 &  0.03& 0.88 &  0.04\\ 
      &      & 0.50 & 0.60& 0.13 &  0.02& 0.32 &  0.02& 0.54 &  0.03& 0.73 &  0.04\\ 
      &      & 0.60 & 0.70& 0.09 &  0.01& 0.24 &  0.03& 0.39 &  0.04& 0.56 &  0.05\\ 
\hline  
 0.95 & 1.15 & 0.10 & 0.15& 0.43 &  0.07& 0.69 &  0.09& 1.01 &  0.12& 1.16 &  0.14\\ 
      &      & 0.15 & 0.20& 0.50 &  0.05& 0.83 &  0.05& 1.30 &  0.07& 1.63 &  0.09\\ 
      &      & 0.20 & 0.25& 0.37 &  0.04& 0.75 &  0.05& 1.20 &  0.07& 1.39 &  0.08\\ 
      &      & 0.25 & 0.30& 0.32 &  0.03& 0.70 &  0.04& 1.07 &  0.05& 1.23 &  0.05\\ 
      &      & 0.30 & 0.35& 0.28 &  0.03& 0.55 &  0.03& 0.80 &  0.04& 1.04 &  0.05\\ 
      &      & 0.35 & 0.40& 0.21 &  0.02& 0.42 &  0.03& 0.66 &  0.03& 0.84 &  0.04\\ 
      &      & 0.40 & 0.45& 0.13 &  0.02& 0.31 &  0.02& 0.53 &  0.03& 0.69 &  0.03\\ 
      &      & 0.45 & 0.50& 0.08 &  0.01& 0.27 &  0.02& 0.44 &  0.02& 0.57 &  0.03\\ 
      &      & 0.50 & 0.60& 0.05 &  0.01& 0.21 &  0.02& 0.32 &  0.02& 0.44 &  0.03\\ 
\hline  
\end{tabular}
}
\end{center}
\end{table*}

\begin{table*}[hp!]
\begin{center}
\small{

\begin{tabular}{rrrr|r@{$\pm$}lr@{$\pm$}lr@{$\pm$}lr@{$\pm$}l}
\hline
$\theta_{\hbox{\small min}}$ &
$\theta_{\hbox{\small max}}$ &
$p_{\hbox{\small min}}$ &
$p_{\hbox{\small max}}$ &
\multicolumn{8}{c}{$d^2\sigma^{\pi^-}/(dpd\theta)$}
\\
(rad) & (rad) & (\GeVc) & (\GeVc) &
\multicolumn{8}{c}{\bgr}
\\
  &  &  &
&\multicolumn{2}{c}{$ \bf{3 \ \GeVc}$}
&\multicolumn{2}{c}{$ \bf{5 \ \GeVc}$}
&\multicolumn{2}{c}{$ \bf{8 \ \GeVc}$}
&\multicolumn{2}{c}{$ \bf{12 \ \GeVc}$}
\\
\hline
 1.15 & 1.35 & 0.10 & 0.15& 0.48 &  0.08& 0.68 &  0.09& 1.00 &  0.12& 1.21 &  0.19\\ 
      &      & 0.15 & 0.20& 0.37 &  0.04& 0.89 &  0.06& 1.26 &  0.07& 1.59 &  0.10\\ 
      &      & 0.20 & 0.25& 0.37 &  0.04& 0.76 &  0.04& 1.00 &  0.05& 1.23 &  0.06\\ 
      &      & 0.25 & 0.30& 0.29 &  0.03& 0.56 &  0.04& 0.83 &  0.05& 0.97 &  0.05\\ 
      &      & 0.30 & 0.35& 0.19 &  0.02& 0.39 &  0.03& 0.62 &  0.04& 0.76 &  0.04\\ 
      &      & 0.35 & 0.40& 0.16 &  0.02& 0.31 &  0.02& 0.48 &  0.03& 0.65 &  0.03\\ 
      &      & 0.40 & 0.45& 0.10 &  0.02& 0.24 &  0.02& 0.38 &  0.02& 0.49 &  0.03\\ 
      &      & 0.45 & 0.50& 0.07 &  0.01& 0.18 &  0.02& 0.28 &  0.02& 0.36 &  0.02\\ 
\hline
 1.35 & 1.55 & 0.10 & 0.15& 0.45 &  0.08& 0.81 &  0.13& 1.06 &  0.17& 1.45 &  0.26\\ 
      &      & 0.15 & 0.20& 0.50 &  0.05& 0.98 &  0.07& 1.33 &  0.09& 1.64 &  0.11\\ 
      &      & 0.20 & 0.25& 0.38 &  0.04& 0.68 &  0.05& 0.93 &  0.06& 1.20 &  0.07\\ 
      &      & 0.25 & 0.30& 0.24 &  0.03& 0.48 &  0.04& 0.76 &  0.05& 0.86 &  0.06\\ 
      &      & 0.30 & 0.35& 0.18 &  0.02& 0.32 &  0.02& 0.52 &  0.05& 0.60 &  0.04\\ 
      &      & 0.35 & 0.40& 0.12 &  0.02& 0.24 &  0.02& 0.34 &  0.02& 0.43 &  0.03\\ 
      &      & 0.40 & 0.45& 0.07 &  0.01& 0.15 &  0.01& 0.26 &  0.02& 0.32 &  0.02\\ 
      &      & 0.45 & 0.50& 0.04 &  0.01& 0.10 &  0.01& 0.19 &  0.02& 0.24 &  0.02\\ 
\hline  
 1.55 & 1.75 & 0.10 & 0.15& 0.46 &  0.08& 0.74 &  0.11& 0.91 &  0.14& 1.37 &  0.23\\ 
      &      & 0.15 & 0.20& 0.55 &  0.05& 0.87 &  0.06& 1.28 &  0.09& 1.43 &  0.10\\ 
      &      & 0.20 & 0.25& 0.33 &  0.03& 0.56 &  0.04& 0.77 &  0.06& 0.95 &  0.06\\ 
      &      & 0.25 & 0.30& 0.17 &  0.03& 0.35 &  0.03& 0.53 &  0.04& 0.64 &  0.05\\ 
      &      & 0.30 & 0.35& 0.10 &  0.01& 0.25 &  0.02& 0.37 &  0.03& 0.47 &  0.04\\ 
      &      & 0.35 & 0.40& 0.06 &  0.01& 0.16 &  0.02& 0.25 &  0.02& 0.31 &  0.02\\ 
      &      & 0.40 & 0.45& 0.04 &  0.01& 0.10 &  0.01& 0.17 &  0.01& 0.22 &  0.02\\ 
      &      & 0.45 & 0.50& 0.02 &  0.01& 0.08 &  0.01& 0.12 &  0.01& 0.16 &  0.01\\ 
\hline  
 1.75 & 1.95 & 0.10 & 0.15& 0.41 &  0.05& 0.60 &  0.07& 0.81 &  0.09& 0.94 &  0.12\\ 
      &      & 0.15 & 0.20& 0.35 &  0.03& 0.58 &  0.04& 0.89 &  0.04& 1.01 &  0.05\\ 
      &      & 0.20 & 0.25& 0.22 &  0.03& 0.40 &  0.03& 0.55 &  0.03& 0.62 &  0.04\\ 
      &      & 0.25 & 0.30& 0.14 &  0.02& 0.24 &  0.02& 0.36 &  0.03& 0.40 &  0.03\\ 
      &      & 0.30 & 0.35& 0.09 &  0.02& 0.16 &  0.01& 0.26 &  0.02& 0.27 &  0.02\\ 
      &      & 0.35 & 0.40& 0.06 &  0.01& 0.09 &  0.01& 0.18 &  0.01& 0.18 &  0.01\\ 
      &      & 0.40 & 0.45& 0.03 &  0.01& 0.07 &  0.01& 0.12 &  0.01& 0.15 &  0.01\\ 
      &      & 0.45 & 0.50& 0.02 &  0.01& 0.05 &  0.01& 0.08 &  0.01& 0.10 &  0.01\\ 
\hline  
 1.95 & 2.15 & 0.10 & 0.15& 0.33 &  0.04& 0.58 &  0.05& 0.72 &  0.05& 0.80 &  0.07\\ 
      &      & 0.15 & 0.20& 0.32 &  0.04& 0.50 &  0.04& 0.67 &  0.04& 0.73 &  0.04\\ 
      &      & 0.20 & 0.25& 0.16 &  0.02& 0.25 &  0.02& 0.39 &  0.02& 0.46 &  0.02\\ 
      &      & 0.25 & 0.30& 0.08 &  0.02& 0.15 &  0.02& 0.26 &  0.02& 0.31 &  0.02\\ 
      &      & 0.30 & 0.35& 0.04 &  0.01& 0.11 &  0.01& 0.15 &  0.02& 0.19 &  0.02\\ 
      &      & 0.35 & 0.40& 0.02 &  0.01& 0.07 &  0.01& 0.10 &  0.01& 0.11 &  0.01\\ 
      &      & 0.40 & 0.45& 0.01 &  0.01& 0.04 &  0.01& 0.08 &  0.01& 0.08 &  0.01\\ 
      &      & 0.45 & 0.50& \multicolumn{2}{c}{}& 0.03 &  0.01& 0.07 &  0.01& 0.05 &  0.01\\ 
\hline
\end{tabular}
}
\end{center}
\end{table*}
%


%
\begin{table*}[hp!]
\begin{center}
\small{

  \caption{\label{tab:xsec-p:ta}
    HARP results for the double-differential $\pi^+$ production
    cross-section in the laboratory system,
    $d^2\sigma^{\pi^+}/(dpd\theta)$ for p--Ta interactions. Each row refers to a
    different $(p_{\hbox{\small min}} \le p<p_{\hbox{\small max}},
    \theta_{\hbox{\small min}} \le \theta<\theta_{\hbox{\small max}})$ bin,
    where $p$ and $\theta$ are the pion momentum and polar angle, respectively.
    The central value as well as the square-root of the diagonal elements
    of the covariance matrix are given.}
\vspace{2mm}
\begin{tabular}{rrrr|r@{$\pm$}lr@{$\pm$}lr@{$\pm$}lr@{$\pm$}l}
\hline
$\theta_{\hbox{\small min}}$ &
$\theta_{\hbox{\small max}}$ &
$p_{\hbox{\small min}}$ &
$p_{\hbox{\small max}}$ &
\multicolumn{8}{c}{$d^2\sigma^{\pi^+}/(dpd\theta)$}
\\
(rad) & (rad) & (\GeVc) & (\GeVc) &
\multicolumn{8}{c}{\bgr}
\\
  &  &  &
&\multicolumn{2}{c}{$ \bf{3 \ \GeVc}$}
&\multicolumn{2}{c}{$ \bf{5 \ \GeVc}$}
&\multicolumn{2}{c}{$ \bf{8 \ \GeVc}$}
&\multicolumn{2}{c}{$ \bf{12 \ \GeVc}$}
\\
\hline 
 0.35 & 0.55 & 0.15 & 0.20& 0.14 &  0.09& 0.60 &  0.20& 1.16 &  0.30& 1.42 &  0.41\\ 
      &      & 0.20 & 0.25& 0.39 &  0.10& 0.93 &  0.15& 1.54 &  0.19& 2.06 &  0.23\\ 
      &      & 0.25 & 0.30& 0.54 &  0.06& 1.16 &  0.11& 2.00 &  0.14& 2.39 &  0.18\\ 
      &      & 0.30 & 0.35& 0.55 &  0.08& 1.27 &  0.09& 2.13 &  0.16& 2.50 &  0.13\\ 
      &      & 0.35 & 0.40& 0.69 &  0.06& 1.24 &  0.08& 2.12 &  0.10& 2.72 &  0.16\\ 
      &      & 0.40 & 0.45& 0.68 &  0.06& 1.23 &  0.07& 2.03 &  0.08& 2.60 &  0.10\\ 
      &      & 0.45 & 0.50& 0.66 &  0.05& 1.22 &  0.06& 2.01 &  0.13& 2.56 &  0.09\\ 
      &      & 0.50 & 0.60& 0.55 &  0.05& 1.15 &  0.06& 2.02 &  0.11& 2.42 &  0.13\\ 
      &      & 0.60 & 0.70& 0.30 &  0.06& 0.91 &  0.09& 1.78 &  0.18& 2.20 &  0.21\\ 
      &      & 0.70 & 0.80& 0.18 &  0.04& 0.57 &  0.10& 1.22 &  0.20& 1.79 &  0.24\\ 
\hline  
 0.55 & 0.75 & 0.10 & 0.15& 0.29 &  0.13& 0.73 &  0.26& 0.98 &  0.37& 1.00 &  0.41\\ 
      &      & 0.15 & 0.20& 0.51 &  0.11& 0.99 &  0.16& 1.79 &  0.19& 1.91 &  0.26\\ 
      &      & 0.20 & 0.25& 0.71 &  0.08& 1.34 &  0.13& 2.10 &  0.15& 2.84 &  0.21\\ 
      &      & 0.25 & 0.30& 0.67 &  0.07& 1.35 &  0.09& 2.13 &  0.13& 2.72 &  0.13\\ 
      &      & 0.30 & 0.35& 0.61 &  0.06& 1.30 &  0.08& 2.17 &  0.09& 2.74 &  0.13\\ 
      &      & 0.35 & 0.40& 0.67 &  0.05& 1.19 &  0.06& 1.91 &  0.11& 2.64 &  0.11\\ 
      &      & 0.40 & 0.45& 0.54 &  0.05& 1.06 &  0.05& 1.87 &  0.09& 2.46 &  0.09\\ 
      &      & 0.45 & 0.50& 0.47 &  0.04& 0.96 &  0.05& 1.65 &  0.09& 2.19 &  0.09\\ 
      &      & 0.50 & 0.60& 0.28 &  0.04& 0.78 &  0.06& 1.41 &  0.10& 1.82 &  0.12\\ 
      &      & 0.60 & 0.70& 0.16 &  0.03& 0.48 &  0.07& 0.95 &  0.13& 1.24 &  0.14\\ 
      &      & 0.70 & 0.80& 0.08 &  0.02& 0.30 &  0.05& 0.58 &  0.11& 0.93 &  0.14\\ 
\hline  
 0.75 & 0.95 & 0.10 & 0.15& 0.53 &  0.13& 1.05 &  0.22& 1.13 &  0.30& 1.19 &  0.32\\ 
      &      & 0.15 & 0.20& 0.75 &  0.08& 1.42 &  0.11& 2.06 &  0.12& 2.31 &  0.19\\ 
      &      & 0.20 & 0.25& 0.77 &  0.06& 1.49 &  0.11& 2.15 &  0.14& 2.57 &  0.14\\ 
      &      & 0.25 & 0.30& 0.72 &  0.06& 1.31 &  0.08& 1.92 &  0.09& 2.39 &  0.12\\ 
      &      & 0.30 & 0.35& 0.61 &  0.05& 1.05 &  0.06& 1.78 &  0.08& 2.16 &  0.09\\ 
      &      & 0.35 & 0.40& 0.46 &  0.04& 0.96 &  0.05& 1.53 &  0.08& 1.78 &  0.08\\ 
      &      & 0.40 & 0.45& 0.40 &  0.04& 0.82 &  0.05& 1.33 &  0.06& 1.66 &  0.07\\ 
      &      & 0.45 & 0.50& 0.31 &  0.03& 0.63 &  0.04& 1.14 &  0.06& 1.45 &  0.07\\ 
      &      & 0.50 & 0.60& 0.19 &  0.03& 0.43 &  0.04& 0.82 &  0.07& 1.06 &  0.08\\ 
      &      & 0.60 & 0.70& 0.06 &  0.02& 0.25 &  0.04& 0.48 &  0.07& 0.63 &  0.09\\ 
\hline  
 0.95 & 1.15 & 0.10 & 0.15& 0.69 &  0.14& 1.08 &  0.18& 1.40 &  0.27& 1.65 &  0.33\\ 
      &      & 0.15 & 0.20& 0.81 &  0.06& 1.42 &  0.10& 2.06 &  0.13& 2.53 &  0.18\\ 
      &      & 0.20 & 0.25& 0.75 &  0.06& 1.32 &  0.07& 1.98 &  0.11& 2.57 &  0.11\\ 
      &      & 0.25 & 0.30& 0.61 &  0.05& 1.06 &  0.06& 1.67 &  0.08& 2.03 &  0.09\\ 
      &      & 0.30 & 0.35& 0.40 &  0.04& 0.79 &  0.05& 1.30 &  0.08& 1.66 &  0.08\\ 
      &      & 0.35 & 0.40& 0.34 &  0.03& 0.63 &  0.03& 1.02 &  0.07& 1.27 &  0.07\\ 
      &      & 0.40 & 0.45& 0.27 &  0.03& 0.49 &  0.03& 0.82 &  0.05& 1.04 &  0.05\\ 
      &      & 0.45 & 0.50& 0.17 &  0.03& 0.34 &  0.03& 0.65 &  0.04& 0.81 &  0.06\\ 
      &      & 0.50 & 0.60& 0.09 &  0.02& 0.21 &  0.02& 0.38 &  0.04& 0.50 &  0.05\\ 
\hline
\end{tabular}
}
\end{center}
\end{table*}

\begin{table*}[hp!]
\begin{center}
\small{

\begin{tabular}{rrrr|r@{$\pm$}lr@{$\pm$}lr@{$\pm$}lr@{$\pm$}l}
\hline
$\theta_{\hbox{\small min}}$ &
$\theta_{\hbox{\small max}}$ &
$p_{\hbox{\small min}}$ &
$p_{\hbox{\small max}}$ &
\multicolumn{8}{c}{$d^2\sigma^{\pi^+}/(dpd\theta)$}
\\
(rad) & (rad) & (\GeVc) & (\GeVc) &
\multicolumn{8}{c}{\bgr}
\\
  &  &  &
&\multicolumn{2}{c}{$ \bf{3 \ \GeVc}$}
&\multicolumn{2}{c}{$ \bf{5 \ \GeVc}$}
&\multicolumn{2}{c}{$ \bf{8 \ \GeVc}$}
&\multicolumn{2}{c}{$ \bf{12 \ \GeVc}$}
\\
\hline 
 1.15 & 1.35 & 0.10 & 0.15& 0.71 &  0.15& 1.08 &  0.20& 1.64 &  0.29& 2.02 &  0.39\\ 
      &      & 0.15 & 0.20& 0.87 &  0.07& 1.35 &  0.10& 2.03 &  0.17& 2.51 &  0.20\\ 
      &      & 0.20 & 0.25& 0.68 &  0.05& 1.17 &  0.07& 1.84 &  0.10& 2.19 &  0.13\\ 
      &      & 0.25 & 0.30& 0.43 &  0.04& 0.84 &  0.05& 1.29 &  0.08& 1.57 &  0.10\\ 
      &      & 0.30 & 0.35& 0.31 &  0.04& 0.65 &  0.04& 0.90 &  0.06& 1.18 &  0.07\\ 
      &      & 0.35 & 0.40& 0.22 &  0.02& 0.48 &  0.03& 0.66 &  0.04& 0.94 &  0.05\\ 
      &      & 0.40 & 0.45& 0.14 &  0.02& 0.33 &  0.03& 0.49 &  0.03& 0.70 &  0.04\\ 
      &      & 0.45 & 0.50& 0.08 &  0.01& 0.23 &  0.03& 0.36 &  0.03& 0.47 &  0.04\\ 
\hline  
 1.35 & 1.55 & 0.10 & 0.15& 0.62 &  0.15& 1.15 &  0.24& 1.61 &  0.35& 2.13 &  0.41\\ 
      &      & 0.15 & 0.20& 0.86 &  0.10& 1.41 &  0.13& 1.91 &  0.21& 2.47 &  0.24\\ 
      &      & 0.20 & 0.25& 0.72 &  0.06& 1.11 &  0.09& 1.64 &  0.11& 1.98 &  0.13\\ 
      &      & 0.25 & 0.30& 0.41 &  0.04& 0.67 &  0.05& 1.06 &  0.07& 1.31 &  0.09\\ 
      &      & 0.30 & 0.35& 0.26 &  0.02& 0.43 &  0.03& 0.75 &  0.05& 0.83 &  0.07\\ 
      &      & 0.35 & 0.40& 0.16 &  0.02& 0.28 &  0.03& 0.50 &  0.04& 0.58 &  0.04\\ 
      &      & 0.40 & 0.45& 0.09 &  0.01& 0.18 &  0.02& 0.32 &  0.03& 0.38 &  0.03\\ 
      &      & 0.45 & 0.50& 0.05 &  0.01& 0.11 &  0.01& 0.22 &  0.02& 0.25 &  0.03\\ 
\hline  
 1.55 & 1.75 & 0.10 & 0.15& 0.68 &  0.15& 1.05 &  0.22& 1.57 &  0.37& 1.85 &  0.44\\ 
      &      & 0.15 & 0.20& 0.78 &  0.08& 1.31 &  0.12& 1.69 &  0.17& 2.12 &  0.21\\ 
      &      & 0.20 & 0.25& 0.62 &  0.06& 1.00 &  0.07& 1.36 &  0.09& 1.68 &  0.11\\ 
      &      & 0.25 & 0.30& 0.33 &  0.04& 0.57 &  0.05& 0.74 &  0.06& 0.97 &  0.08\\ 
      &      & 0.30 & 0.35& 0.18 &  0.02& 0.35 &  0.03& 0.48 &  0.03& 0.58 &  0.06\\ 
      &      & 0.35 & 0.40& 0.09 &  0.02& 0.19 &  0.02& 0.32 &  0.03& 0.35 &  0.03\\ 
      &      & 0.40 & 0.45& 0.05 &  0.01& 0.10 &  0.01& 0.21 &  0.02& 0.22 &  0.02\\ 
      &      & 0.45 & 0.50& 0.03 &  0.01& 0.06 &  0.01& 0.14 &  0.02& 0.14 &  0.02\\ 
\hline  
 1.75 & 1.95 & 0.10 & 0.15& 0.78 &  0.12& 1.03 &  0.19& 1.28 &  0.20& 1.43 &  0.25\\ 
      &      & 0.15 & 0.20& 0.65 &  0.05& 1.09 &  0.06& 1.39 &  0.08& 1.66 &  0.10\\ 
      &      & 0.20 & 0.25& 0.44 &  0.04& 0.71 &  0.05& 0.97 &  0.06& 1.08 &  0.07\\ 
      &      & 0.25 & 0.30& 0.20 &  0.03& 0.38 &  0.04& 0.54 &  0.05& 0.66 &  0.06\\ 
      &      & 0.30 & 0.35& 0.09 &  0.01& 0.21 &  0.02& 0.32 &  0.03& 0.33 &  0.03\\ 
      &      & 0.35 & 0.40& 0.06 &  0.01& 0.12 &  0.02& 0.19 &  0.02& 0.22 &  0.02\\ 
      &      & 0.40 & 0.45& 0.06 &  0.03& 0.08 &  0.01& 0.12 &  0.01& 0.16 &  0.02\\ 
      &      & 0.45 & 0.50& 0.01 &  0.01& 0.04 &  0.01& 0.07 &  0.01& 0.09 &  0.02\\ 
\hline  
 1.95 & 2.15 & 0.10 & 0.15& 0.72 &  0.12& 0.92 &  0.14& 1.04 &  0.16& 1.33 &  0.21\\ 
      &      & 0.15 & 0.20& 0.55 &  0.05& 0.80 &  0.05& 1.09 &  0.05& 1.25 &  0.06\\ 
      &      & 0.20 & 0.25& 0.28 &  0.03& 0.51 &  0.03& 0.64 &  0.05& 0.76 &  0.04\\ 
      &      & 0.25 & 0.30& 0.13 &  0.02& 0.26 &  0.03& 0.34 &  0.04& 0.44 &  0.05\\ 
      &      & 0.30 & 0.35& 0.06 &  0.01& 0.13 &  0.02& 0.19 &  0.02& 0.18 &  0.02\\ 
      &      & 0.35 & 0.40& 0.03 &  0.01& 0.07 &  0.01& 0.11 &  0.02& 0.11 &  0.01\\ 
      &      & 0.40 & 0.45& 0.02 &  0.01& 0.05 &  0.01& 0.05 &  0.01& 0.07 &  0.01\\ 
      &      & 0.45 & 0.50& 0.01 &  0.01& 0.02 &  0.01& 0.03 &  0.01& 0.04 &  0.01\\ 

\hline

\end{tabular}
}
\end{center}
\end{table*}

\begin{table*}[hp!]
\begin{center}
\small{

  \caption{\label{tab:xsec-n:ta}
    HARP results for the double-differential $\pi^-$ production
    cross-section in the laboratory system,
    $d^2\sigma^{\pi^-}/(dpd\theta)$ for p--Ta interactions. Each row refers to a
    different $(p_{\hbox{\small min}} \le p<p_{\hbox{\small max}},
    \theta_{\hbox{\small min}} \le \theta<\theta_{\hbox{\small max}})$ bin,
    where $p$ and $\theta$ are the pion momentum and polar angle, respectively.
    The central value as well as the square-root of the diagonal elements
    of the covariance matrix are given.}
\vspace{2mm}
\begin{tabular}{rrrr|r@{$\pm$}lr@{$\pm$}lr@{$\pm$}lr@{$\pm$}l}
\hline
$\theta_{\hbox{\small min}}$ &
$\theta_{\hbox{\small max}}$ &
$p_{\hbox{\small min}}$ &
$p_{\hbox{\small max}}$ &
\multicolumn{8}{c}{$d^2\sigma^{\pi^-}/(dpd\theta)$}
\\
(rad) & (rad) & (\GeVc) & (\GeVc) &
\multicolumn{8}{c}{\bgr}
\\
  &  &  &
&\multicolumn{2}{c}{$ \bf{3 \ \GeVc}$}
&\multicolumn{2}{c}{$ \bf{5 \ \GeVc}$}
&\multicolumn{2}{c}{$ \bf{8 \ \GeVc}$}
&\multicolumn{2}{c}{$ \bf{12 \ \GeVc}$}
\\
\hline
 0.35 & 0.55 & 0.15 & 0.20& 0.33 &  0.12& 0.80 &  0.23& 1.31 &  0.31& 1.70 &  0.41\\ 
      &      & 0.20 & 0.25& 0.34 &  0.08& 0.89 &  0.13& 1.59 &  0.16& 2.02 &  0.23\\ 
      &      & 0.25 & 0.30& 0.41 &  0.07& 1.10 &  0.12& 1.77 &  0.14& 2.26 &  0.17\\ 
      &      & 0.30 & 0.35& 0.47 &  0.05& 0.94 &  0.07& 1.88 &  0.11& 2.15 &  0.10\\ 
      &      & 0.35 & 0.40& 0.47 &  0.05& 0.88 &  0.06& 1.67 &  0.08& 2.05 &  0.10\\ 
      &      & 0.40 & 0.45& 0.38 &  0.04& 0.77 &  0.04& 1.49 &  0.07& 1.83 &  0.07\\ 
      &      & 0.45 & 0.50& 0.30 &  0.03& 0.72 &  0.05& 1.42 &  0.06& 1.64 &  0.08\\ 
      &      & 0.50 & 0.60& 0.25 &  0.03& 0.74 &  0.04& 1.28 &  0.07& 1.57 &  0.09\\ 
      &      & 0.60 & 0.70& 0.22 &  0.03& 0.64 &  0.05& 1.15 &  0.08& 1.47 &  0.13\\ 
      &      & 0.70 & 0.80& 0.18 &  0.03& 0.46 &  0.07& 0.92 &  0.11& 1.27 &  0.15\\ 
\hline  
 0.55 & 0.75 & 0.10 & 0.15& 0.37 &  0.16& 0.82 &  0.27& 1.18 &  0.39& 1.37 &  0.45\\ 
      &      & 0.15 & 0.20& 0.54 &  0.09& 1.27 &  0.16& 1.74 &  0.18& 2.21 &  0.25\\ 
      &      & 0.20 & 0.25& 0.46 &  0.05& 1.18 &  0.09& 1.99 &  0.14& 2.48 &  0.16\\ 
      &      & 0.25 & 0.30& 0.52 &  0.06& 1.07 &  0.08& 1.91 &  0.10& 2.33 &  0.09\\ 
      &      & 0.30 & 0.35& 0.37 &  0.04& 0.97 &  0.06& 1.57 &  0.08& 2.15 &  0.09\\ 
      &      & 0.35 & 0.40& 0.30 &  0.03& 0.89 &  0.05& 1.43 &  0.07& 1.89 &  0.07\\ 
      &      & 0.40 & 0.45& 0.35 &  0.05& 0.75 &  0.04& 1.33 &  0.06& 1.69 &  0.06\\ 
      &      & 0.45 & 0.50& 0.38 &  0.03& 0.65 &  0.03& 1.18 &  0.05& 1.50 &  0.06\\ 
      &      & 0.50 & 0.60& 0.25 &  0.03& 0.56 &  0.03& 1.05 &  0.05& 1.31 &  0.06\\ 
      &      & 0.60 & 0.70& 0.16 &  0.02& 0.44 &  0.04& 0.84 &  0.08& 1.12 &  0.10\\ 
      &      & 0.70 & 0.80& 0.11 &  0.02& 0.33 &  0.05& 0.64 &  0.08& 0.89 &  0.12\\ 
\hline  
 0.75 & 0.95 & 0.10 & 0.15& 0.54 &  0.13& 0.99 &  0.20& 1.42 &  0.32& 1.81 &  0.39\\ 
      &      & 0.15 & 0.20& 0.71 &  0.07& 1.29 &  0.10& 2.04 &  0.13& 2.32 &  0.16\\ 
      &      & 0.20 & 0.25& 0.65 &  0.06& 1.12 &  0.08& 1.87 &  0.10& 2.46 &  0.16\\ 
      &      & 0.25 & 0.30& 0.55 &  0.05& 0.99 &  0.05& 1.68 &  0.08& 2.22 &  0.10\\ 
      &      & 0.30 & 0.35& 0.38 &  0.04& 0.82 &  0.05& 1.33 &  0.06& 1.90 &  0.09\\ 
      &      & 0.35 & 0.40& 0.33 &  0.03& 0.72 &  0.04& 1.22 &  0.06& 1.57 &  0.07\\ 
      &      & 0.40 & 0.45& 0.27 &  0.03& 0.64 &  0.03& 1.01 &  0.05& 1.33 &  0.05\\ 
      &      & 0.45 & 0.50& 0.19 &  0.02& 0.57 &  0.03& 0.88 &  0.04& 1.12 &  0.05\\ 
      &      & 0.50 & 0.60& 0.14 &  0.02& 0.46 &  0.03& 0.71 &  0.04& 0.92 &  0.05\\ 
      &      & 0.60 & 0.70& 0.09 &  0.01& 0.30 &  0.04& 0.56 &  0.05& 0.69 &  0.07\\ 
\hline  
 0.95 & 1.15 & 0.10 & 0.15& 0.74 &  0.12& 1.30 &  0.21& 1.86 &  0.29& 2.12 &  0.36\\ 
      &      & 0.15 & 0.20& 0.65 &  0.06& 1.39 &  0.08& 2.13 &  0.13& 2.32 &  0.13\\ 
      &      & 0.20 & 0.25& 0.55 &  0.05& 1.02 &  0.07& 1.69 &  0.09& 2.11 &  0.12\\ 
      &      & 0.25 & 0.30& 0.48 &  0.04& 0.87 &  0.05& 1.42 &  0.08& 1.80 &  0.08\\ 
      &      & 0.30 & 0.35& 0.37 &  0.03& 0.69 &  0.04& 1.10 &  0.06& 1.45 &  0.07\\ 
      &      & 0.35 & 0.40& 0.29 &  0.03& 0.54 &  0.03& 0.89 &  0.04& 1.13 &  0.06\\ 
      &      & 0.40 & 0.45& 0.22 &  0.02& 0.43 &  0.02& 0.75 &  0.03& 0.91 &  0.05\\ 
      &      & 0.45 & 0.50& 0.14 &  0.02& 0.35 &  0.02& 0.60 &  0.03& 0.72 &  0.04\\ 
      &      & 0.50 & 0.60& 0.08 &  0.01& 0.26 &  0.02& 0.46 &  0.03& 0.53 &  0.04\\ 
\hline
\end{tabular}
}
\end{center}
\end{table*}

\begin{table*}[hp!]
\begin{center}
\small{

\begin{tabular}{rrrr|r@{$\pm$}lr@{$\pm$}lr@{$\pm$}lr@{$\pm$}l}
\hline
$\theta_{\hbox{\small min}}$ &
$\theta_{\hbox{\small max}}$ &
$p_{\hbox{\small min}}$ &
$p_{\hbox{\small max}}$ &
\multicolumn{8}{c}{$d^2\sigma^{\pi^-}/(dpd\theta)$}
\\
(rad) & (rad) & (\GeVc) & (\GeVc) &
\multicolumn{8}{c}{\bgr}
\\
  &  &  &
&\multicolumn{2}{c}{$ \bf{3 \ \GeVc}$}
&\multicolumn{2}{c}{$ \bf{5 \ \GeVc}$}
&\multicolumn{2}{c}{$ \bf{8 \ \GeVc}$}
&\multicolumn{2}{c}{$ \bf{12 \ \GeVc}$}
\\
\hline
 1.15 & 1.35 & 0.10 & 0.15& 0.86 &  0.15& 1.36 &  0.23& 2.11 &  0.36& 2.66 &  0.46\\ 
      &      & 0.15 & 0.20& 0.67 &  0.06& 1.28 &  0.09& 2.07 &  0.14& 2.37 &  0.17\\ 
      &      & 0.20 & 0.25& 0.50 &  0.04& 0.96 &  0.06& 1.58 &  0.09& 1.84 &  0.11\\ 
      &      & 0.25 & 0.30& 0.36 &  0.04& 0.73 &  0.05& 1.20 &  0.08& 1.44 &  0.08\\ 
      &      & 0.30 & 0.35& 0.22 &  0.02& 0.56 &  0.04& 0.85 &  0.05& 1.00 &  0.06\\ 
      &      & 0.35 & 0.40& 0.18 &  0.02& 0.43 &  0.03& 0.67 &  0.04& 0.80 &  0.04\\ 
      &      & 0.40 & 0.45& 0.12 &  0.02& 0.30 &  0.02& 0.51 &  0.03& 0.68 &  0.04\\ 
      &      & 0.45 & 0.50& 0.07 &  0.01& 0.22 &  0.02& 0.41 &  0.03& 0.53 &  0.04\\ 
\hline  
 1.35 & 1.55 & 0.10 & 0.15& 0.73 &  0.15& 1.31 &  0.25& 2.37 &  0.47& 3.06 &  0.66\\ 
      &      & 0.15 & 0.20& 0.61 &  0.07& 1.14 &  0.12& 1.87 &  0.18& 2.34 &  0.22\\ 
      &      & 0.20 & 0.25& 0.42 &  0.04& 0.82 &  0.07& 1.32 &  0.10& 1.74 &  0.12\\ 
      &      & 0.25 & 0.30& 0.29 &  0.03& 0.62 &  0.05& 0.90 &  0.07& 1.21 &  0.08\\ 
      &      & 0.30 & 0.35& 0.19 &  0.02& 0.42 &  0.04& 0.60 &  0.04& 0.85 &  0.07\\ 
      &      & 0.35 & 0.40& 0.12 &  0.02& 0.29 &  0.02& 0.45 &  0.03& 0.61 &  0.04\\ 
      &      & 0.40 & 0.45& 0.08 &  0.01& 0.21 &  0.02& 0.34 &  0.02& 0.46 &  0.03\\ 
      &      & 0.45 & 0.50& 0.05 &  0.01& 0.14 &  0.01& 0.27 &  0.02& 0.34 &  0.03\\ 
\hline  
 1.55 & 1.75 & 0.10 & 0.15& 0.80 &  0.16& 1.34 &  0.27& 1.84 &  0.38& 2.54 &  0.53\\ 
      &      & 0.15 & 0.20& 0.56 &  0.07& 1.06 &  0.10& 1.62 &  0.16& 2.10 &  0.21\\ 
      &      & 0.20 & 0.25& 0.35 &  0.04& 0.73 &  0.05& 1.11 &  0.08& 1.44 &  0.10\\ 
      &      & 0.25 & 0.30& 0.19 &  0.02& 0.49 &  0.04& 0.67 &  0.05& 0.86 &  0.07\\ 
      &      & 0.30 & 0.35& 0.13 &  0.02& 0.31 &  0.03& 0.43 &  0.04& 0.61 &  0.05\\ 
      &      & 0.35 & 0.40& 0.09 &  0.01& 0.22 &  0.02& 0.30 &  0.02& 0.41 &  0.03\\ 
      &      & 0.40 & 0.45& 0.05 &  0.01& 0.14 &  0.01& 0.21 &  0.01& 0.30 &  0.02\\ 
      &      & 0.45 & 0.50& 0.04 &  0.01& 0.10 &  0.01& 0.15 &  0.01& 0.21 &  0.02\\ 
\hline  
 1.75 & 1.95 & 0.10 & 0.15& 0.79 &  0.11& 1.28 &  0.20& 1.54 &  0.23& 2.07 &  0.31\\ 
      &      & 0.15 & 0.20& 0.48 &  0.05& 0.95 &  0.06& 1.27 &  0.08& 1.58 &  0.11\\ 
      &      & 0.20 & 0.25& 0.25 &  0.03& 0.53 &  0.04& 0.81 &  0.05& 0.99 &  0.07\\ 
      &      & 0.25 & 0.30& 0.14 &  0.02& 0.32 &  0.03& 0.51 &  0.04& 0.58 &  0.04\\ 
      &      & 0.30 & 0.35& 0.11 &  0.02& 0.22 &  0.02& 0.28 &  0.03& 0.41 &  0.03\\ 
      &      & 0.35 & 0.40& 0.08 &  0.01& 0.14 &  0.02& 0.21 &  0.01& 0.26 &  0.02\\ 
      &      & 0.40 & 0.45& 0.05 &  0.01& 0.09 &  0.01& 0.18 &  0.01& 0.20 &  0.02\\ 
      &      & 0.45 & 0.50& 0.03 &  0.01& 0.06 &  0.01& 0.12 &  0.01& 0.12 &  0.01\\ 
\hline  
 1.95 & 2.15 & 0.10 & 0.15& 0.59 &  0.11& 1.12 &  0.16& 1.43 &  0.17& 1.85 &  0.29\\ 
      &      & 0.15 & 0.20& 0.50 &  0.05& 0.80 &  0.06& 0.99 &  0.05& 1.24 &  0.07\\ 
      &      & 0.20 & 0.25& 0.21 &  0.03& 0.39 &  0.03& 0.57 &  0.03& 0.73 &  0.05\\ 
      &      & 0.25 & 0.30& 0.09 &  0.02& 0.23 &  0.02& 0.36 &  0.03& 0.42 &  0.04\\ 
      &      & 0.30 & 0.35& 0.04 &  0.01& 0.14 &  0.02& 0.19 &  0.02& 0.25 &  0.02\\ 
      &      & 0.35 & 0.40& 0.03 &  0.01& 0.09 &  0.01& 0.15 &  0.01& 0.15 &  0.02\\ 
      &      & 0.40 & 0.45& 0.02 &  0.01& 0.05 &  0.01& 0.11 &  0.01& 0.11 &  0.01\\ 
      &      & 0.45 & 0.50& 0.02 &  0.01& 0.03 &  0.01& 0.07 &  0.01& 0.09 &  0.01\\ 
\hline
\end{tabular}
}
\end{center}
\end{table*}
%



%
\begin{table*}[hp!]
\begin{center}
\small{

  \caption{\label{tab:xsec-p-pb}
    HARP results for the double-differential $\pi^+$ production
    cross-section in the laboratory system,
    $d^2\sigma^{\pi^+}/(dpd\theta)$ for p--Pb interactions. Each row refers to a
    different $(p_{\hbox{\small min}} \le p<p_{\hbox{\small max}},
    \theta_{\hbox{\small min}} \le \theta<\theta_{\hbox{\small max}})$ bin,
    where $p$ and $\theta$ are the pion momentum and polar angle, respectively.
    The central value as well as the square-root of the diagonal elements
    of the covariance matrix are given.}
\vspace{2mm}
\begin{tabular}{rrrr|r@{$\pm$}lr@{$\pm$}lr@{$\pm$}lr@{$\pm$}l}
\hline
$\theta_{\hbox{\small min}}$ &
$\theta_{\hbox{\small max}}$ &
$p_{\hbox{\small min}}$ &
$p_{\hbox{\small max}}$ &
\multicolumn{8}{c}{$d^2\sigma^{\pi^+}/(dpd\theta)$}
\\
(rad) & (rad) & (\GeVc) & (\GeVc) &
\multicolumn{8}{c}{\bgr}
\\
  &  &  &
&\multicolumn{2}{c}{$ \bf{3 \ \GeVc}$}
&\multicolumn{2}{c}{$ \bf{5 \ \GeVc}$}
&\multicolumn{2}{c}{$ \bf{8 \ \GeVc}$}
&\multicolumn{2}{c}{$ \bf{12 \ \GeVc}$}
\\
\hline
 0.35 & 0.55         & 0.15 & 0.20& 0.31 &  0.11& 0.85 &  0.22& 1.15 &  0.32& 1.38 &  0.41\\ 
      &      & 0.20 & 0.25& 0.29 &  0.08& 0.87 &  0.15& 1.75 &  0.20& 2.04 &  0.26\\ 
      &      & 0.25 & 0.30& 0.49 &  0.10& 1.18 &  0.13& 2.13 &  0.15& 2.66 &  0.25\\ 
      &      & 0.30 & 0.35& 0.76 &  0.08& 1.27 &  0.11& 2.19 &  0.13& 2.61 &  0.17\\ 
      &      & 0.35 & 0.40& 0.67 &  0.07& 1.34 &  0.09& 2.21 &  0.10& 2.84 &  0.18\\ 
      &      & 0.40 & 0.45& 0.76 &  0.07& 1.38 &  0.10& 2.20 &  0.09& 2.53 &  0.13\\ 
      &      & 0.45 & 0.50& 0.65 &  0.06& 1.42 &  0.08& 1.99 &  0.08& 2.40 &  0.12\\ 
      &      & 0.50 & 0.60& 0.48 &  0.06& 1.27 &  0.08& 1.96 &  0.09& 2.53 &  0.16\\ 
      &      & 0.60 & 0.70& 0.26 &  0.04& 0.99 &  0.10& 1.71 &  0.16& 2.19 &  0.22\\ 
      &      & 0.70 & 0.80& 0.18 &  0.03& 0.67 &  0.12& 1.12 &  0.18& 1.76 &  0.23\\ 
\hline  
 0.55 & 0.75 & 0.10 & 0.15& 0.35 &  0.15& 0.69 &  0.25& 0.84 &  0.35& 1.02 &  0.46\\ 
      &      & 0.15 & 0.20& 0.56 &  0.09& 1.31 &  0.17& 1.75 &  0.22& 2.19 &  0.29\\ 
      &      & 0.20 & 0.25& 0.70 &  0.09& 1.46 &  0.11& 2.31 &  0.15& 2.87 &  0.22\\ 
      &      & 0.25 & 0.30& 0.73 &  0.07& 1.46 &  0.11& 2.21 &  0.13& 3.13 &  0.20\\ 
      &      & 0.30 & 0.35& 0.60 &  0.05& 1.31 &  0.09& 2.16 &  0.12& 2.79 &  0.14\\ 
      &      & 0.35 & 0.40& 0.61 &  0.07& 1.08 &  0.08& 2.07 &  0.11& 2.65 &  0.13\\ 
      &      & 0.40 & 0.45& 0.62 &  0.06& 1.16 &  0.09& 1.87 &  0.08& 2.58 &  0.14\\ 
      &      & 0.45 & 0.50& 0.44 &  0.05& 1.17 &  0.07& 1.63 &  0.08& 2.41 &  0.12\\ 
      &      & 0.50 & 0.60& 0.31 &  0.04& 0.86 &  0.08& 1.33 &  0.08& 1.96 &  0.14\\ 
      &      & 0.60 & 0.70& 0.18 &  0.03& 0.48 &  0.08& 0.93 &  0.11& 1.39 &  0.14\\ 
      &      & 0.70 & 0.80& 0.10 &  0.02& 0.27 &  0.06& 0.59 &  0.11& 0.98 &  0.16\\ 
\hline  
 0.75 & 0.95 & 0.10 & 0.15& 0.52 &  0.13& 0.83 &  0.21& 1.20 &  0.31& 1.29 &  0.39\\ 
      &      & 0.15 & 0.20& 0.87 &  0.09& 1.54 &  0.14& 2.25 &  0.14& 2.78 &  0.20\\ 
      &      & 0.20 & 0.25& 0.86 &  0.08& 1.40 &  0.10& 2.25 &  0.14& 2.91 &  0.19\\ 
      &      & 0.25 & 0.30& 0.67 &  0.06& 1.20 &  0.12& 1.98 &  0.11& 2.69 &  0.15\\ 
      &      & 0.30 & 0.35& 0.59 &  0.05& 1.14 &  0.07& 1.83 &  0.08& 2.39 &  0.13\\ 
      &      & 0.35 & 0.40& 0.47 &  0.05& 0.97 &  0.06& 1.65 &  0.07& 2.06 &  0.11\\ 
      &      & 0.40 & 0.45& 0.37 &  0.04& 0.79 &  0.05& 1.38 &  0.06& 1.92 &  0.10\\ 
      &      & 0.45 & 0.50& 0.29 &  0.03& 0.66 &  0.04& 1.17 &  0.05& 1.61 &  0.10\\ 
      &      & 0.50 & 0.60& 0.19 &  0.03& 0.44 &  0.05& 0.81 &  0.07& 1.10 &  0.10\\ 
      &      & 0.60 & 0.70& 0.11 &  0.02& 0.25 &  0.04& 0.47 &  0.07& 0.70 &  0.09\\ 
\hline  
 0.95 & 1.15 & 0.10 & 0.15& 0.55 &  0.13& 1.02 &  0.20& 1.40 &  0.25& 1.55 &  0.36\\ 
      &      & 0.15 & 0.20& 0.92 &  0.08& 1.56 &  0.11& 2.31 &  0.17& 3.00 &  0.20\\ 
      &      & 0.20 & 0.25& 0.68 &  0.07& 1.43 &  0.09& 2.01 &  0.13& 2.68 &  0.16\\ 
      &      & 0.25 & 0.30& 0.71 &  0.07& 1.12 &  0.10& 1.56 &  0.10& 2.07 &  0.13\\ 
      &      & 0.30 & 0.35& 0.48 &  0.05& 0.84 &  0.07& 1.31 &  0.09& 1.74 &  0.11\\ 
      &      & 0.35 & 0.40& 0.32 &  0.04& 0.72 &  0.05& 1.07 &  0.06& 1.62 &  0.10\\ 
      &      & 0.40 & 0.45& 0.25 &  0.03& 0.55 &  0.04& 0.85 &  0.05& 1.33 &  0.08\\ 
      &      & 0.45 & 0.50& 0.16 &  0.02& 0.45 &  0.04& 0.65 &  0.04& 0.96 &  0.08\\ 
      &      & 0.50 & 0.60& 0.09 &  0.02& 0.24 &  0.03& 0.44 &  0.05& 0.59 &  0.07\\ 

\hline
\end{tabular}
}
\end{center}
\end{table*}

\begin{table*}[hp!]
\begin{center}
\small{

\begin{tabular}{rrrr|r@{$\pm$}lr@{$\pm$}lr@{$\pm$}lr@{$\pm$}l}
\hline
$\theta_{\hbox{\small min}}$ &
$\theta_{\hbox{\small max}}$ &
$p_{\hbox{\small min}}$ &
$p_{\hbox{\small max}}$ &
\multicolumn{8}{c}{$d^2\sigma^{\pi^+}/(dpd\theta)$}
\\
(rad) & (rad) & (\GeVc) & (\GeVc) &
\multicolumn{8}{c}{\bgr}
\\
  &  &  &
&\multicolumn{2}{c}{$ \bf{3 \ \GeVc}$}
&\multicolumn{2}{c}{$ \bf{5 \ \GeVc}$}
&\multicolumn{2}{c}{$ \bf{8 \ \GeVc}$}
&\multicolumn{2}{c}{$ \bf{12 \ \GeVc}$}
\\
\hline
 1.15 & 1.35 & 0.10 & 0.15& 0.67 &  0.14& 1.27 &  0.21& 1.68 &  0.28& 1.75 &  0.35\\ 
      &      & 0.15 & 0.20& 0.90 &  0.08& 1.49 &  0.10& 2.32 &  0.15& 2.88 &  0.23\\ 
      &      & 0.20 & 0.25& 0.64 &  0.06& 1.17 &  0.08& 1.96 &  0.10& 2.53 &  0.15\\ 
      &      & 0.25 & 0.30& 0.51 &  0.05& 0.92 &  0.06& 1.32 &  0.07& 1.87 &  0.13\\ 
      &      & 0.30 & 0.35& 0.32 &  0.04& 0.64 &  0.05& 0.91 &  0.07& 1.27 &  0.09\\ 
      &      & 0.35 & 0.40& 0.21 &  0.03& 0.45 &  0.04& 0.74 &  0.04& 0.89 &  0.06\\ 
      &      & 0.40 & 0.45& 0.15 &  0.02& 0.32 &  0.03& 0.55 &  0.04& 0.65 &  0.05\\ 
      &      & 0.45 & 0.50& 0.10 &  0.02& 0.22 &  0.03& 0.37 &  0.03& 0.43 &  0.05\\ 
\hline  
 1.35 & 1.55 & 0.10 & 0.15& 0.79 &  0.15& 1.39 &  0.27& 1.74 &  0.35& 1.83 &  0.40\\ 
      &      & 0.15 & 0.20& 0.90 &  0.08& 1.46 &  0.11& 2.08 &  0.17& 2.51 &  0.20\\ 
      &      & 0.20 & 0.25& 0.72 &  0.06& 1.02 &  0.07& 1.68 &  0.09& 2.01 &  0.14\\ 
      &      & 0.25 & 0.30& 0.36 &  0.05& 0.66 &  0.05& 1.01 &  0.06& 1.53 &  0.11\\ 
      &      & 0.30 & 0.35& 0.21 &  0.03& 0.47 &  0.05& 0.71 &  0.05& 0.89 &  0.08\\ 
      &      & 0.35 & 0.40& 0.13 &  0.02& 0.32 &  0.03& 0.53 &  0.04& 0.63 &  0.06\\ 
      &      & 0.40 & 0.45& 0.07 &  0.01& 0.21 &  0.02& 0.40 &  0.03& 0.43 &  0.04\\ 
      &      & 0.45 & 0.50& 0.05 &  0.01& 0.14 &  0.02& 0.23 &  0.03& 0.27 &  0.03\\ 
\hline  
 1.55 & 1.75 & 0.10 & 0.15& 0.80 &  0.17& 1.21 &  0.23& 1.62 &  0.32& 1.93 &  0.42\\ 
      &      & 0.15 & 0.20& 0.73 &  0.07& 1.38 &  0.11& 1.83 &  0.12& 2.28 &  0.16\\ 
      &      & 0.20 & 0.25& 0.52 &  0.05& 0.96 &  0.07& 1.38 &  0.09& 1.71 &  0.12\\ 
      &      & 0.25 & 0.30& 0.29 &  0.04& 0.56 &  0.05& 0.81 &  0.05& 1.00 &  0.09\\ 
      &      & 0.30 & 0.35& 0.17 &  0.02& 0.38 &  0.04& 0.54 &  0.04& 0.65 &  0.06\\ 
      &      & 0.35 & 0.40& 0.11 &  0.02& 0.25 &  0.02& 0.37 &  0.03& 0.46 &  0.04\\ 
      &      & 0.40 & 0.45& 0.06 &  0.01& 0.16 &  0.02& 0.23 &  0.02& 0.31 &  0.04\\ 
      &      & 0.45 & 0.50& 0.04 &  0.01& 0.11 &  0.01& 0.15 &  0.02& 0.18 &  0.03\\ 
\hline  
 1.75 & 1.95 & 0.10 & 0.15& 0.81 &  0.13& 1.02 &  0.17& 1.47 &  0.21& 1.64 &  0.28\\ 
      &      & 0.15 & 0.20& 0.67 &  0.06& 1.23 &  0.08& 1.60 &  0.08& 1.97 &  0.12\\ 
      &      & 0.20 & 0.25& 0.36 &  0.04& 0.75 &  0.07& 1.03 &  0.07& 1.29 &  0.10\\ 
      &      & 0.25 & 0.30& 0.16 &  0.03& 0.36 &  0.04& 0.56 &  0.04& 0.62 &  0.07\\ 
      &      & 0.30 & 0.35& 0.09 &  0.02& 0.22 &  0.03& 0.32 &  0.03& 0.40 &  0.04\\ 
      &      & 0.35 & 0.40& 0.07 &  0.02& 0.14 &  0.02& 0.20 &  0.02& 0.21 &  0.03\\ 
      &      & 0.40 & 0.45& 0.04 &  0.01& 0.09 &  0.01& 0.13 &  0.02& 0.12 &  0.02\\ 
      &      & 0.45 & 0.50& 0.02 &  0.01& 0.06 &  0.01& 0.06 &  0.01& 0.07 &  0.01\\ 
\hline  
 1.95 & 2.15 & 0.10 & 0.15& 0.88 &  0.13& 0.81 &  0.12& 1.13 &  0.17& 1.41 &  0.22\\ 
      &      & 0.15 & 0.20& 0.64 &  0.06& 0.93 &  0.06& 1.19 &  0.06& 1.40 &  0.11\\ 
      &      & 0.20 & 0.25& 0.36 &  0.05& 0.56 &  0.05& 0.74 &  0.06& 1.04 &  0.09\\ 
      &      & 0.25 & 0.30& 0.15 &  0.03& 0.25 &  0.05& 0.39 &  0.04& 0.44 &  0.06\\ 
      &      & 0.30 & 0.35& 0.07 &  0.02& 0.13 &  0.02& 0.19 &  0.02& 0.24 &  0.03\\ 
      &      & 0.35 & 0.40& 0.03 &  0.01& 0.10 &  0.01& 0.11 &  0.01& 0.16 &  0.02\\ 
      &      & 0.40 & 0.45& 0.02 &  0.01& 0.06 &  0.01& 0.06 &  0.01& 0.09 &  0.02\\ 
      &      & 0.45 & 0.50& 0.02 &  0.01& 0.03 &  0.01& 0.04 &  0.01& 0.06 &  0.02\\ 

\hline
\end{tabular}
}
\end{center}
\end{table*}

\begin{table*}[hp!]
\begin{center}
\small{

  \caption{\label{tab:xsec-n-pb}
    HARP results for the double-differential $\pi^-$ production
    cross-section in the laboratory system,
    $d^2\sigma^{\pi^-}/(dpd\theta)$ for p--Pb interactions. Each row refers to a
    different $(p_{\hbox{\small min}} \le p<p_{\hbox{\small max}},
    \theta_{\hbox{\small min}} \le \theta<\theta_{\hbox{\small max}})$ bin,
    where $p$ and $\theta$ are the pion momentum and polar angle, respectively.
    The central value as well as the square-root of the diagonal elements
    of the covariance matrix are given.}
\vspace{2mm}
\begin{tabular}{rrrr|r@{$\pm$}lr@{$\pm$}lr@{$\pm$}lr@{$\pm$}l}
\hline
$\theta_{\hbox{\small min}}$ &
$\theta_{\hbox{\small max}}$ &
$p_{\hbox{\small min}}$ &
$p_{\hbox{\small max}}$ &
\multicolumn{8}{c}{$d^2\sigma^{\pi^-}/(dpd\theta)$}
\\
(rad) & (rad) & (\GeVc) & (\GeVc) &
\multicolumn{8}{c}{\bgr}
\\
  &  &  &
&\multicolumn{2}{c}{$ \bf{3 \ \GeVc}$}
&\multicolumn{2}{c}{$ \bf{5 \ \GeVc}$}
&\multicolumn{2}{c}{$ \bf{8 \ \GeVc}$}
&\multicolumn{2}{c}{$ \bf{12 \ \GeVc}$}
\\
\hline
0.35 & 0.55       & 0.15 & 0.20& 0.20 &  0.10& 0.72 &  0.24& 1.47 &  0.32& 1.90 &  0.46\\ 
      &      & 0.20 & 0.25& 0.33 &  0.09& 0.78 &  0.16& 1.85 &  0.21& 2.36 &  0.26\\ 
      &      & 0.25 & 0.30& 0.37 &  0.06& 0.94 &  0.10& 2.02 &  0.12& 2.67 &  0.23\\ 
      &      & 0.30 & 0.35& 0.34 &  0.05& 1.01 &  0.11& 1.87 &  0.10& 2.40 &  0.14\\ 
      &      & 0.35 & 0.40& 0.39 &  0.05& 1.07 &  0.07& 1.67 &  0.08& 2.04 &  0.12\\ 
      &      & 0.40 & 0.45& 0.34 &  0.04& 0.80 &  0.06& 1.58 &  0.07& 1.74 &  0.09\\ 
      &      & 0.45 & 0.50& 0.35 &  0.04& 0.72 &  0.05& 1.42 &  0.06& 1.66 &  0.12\\ 
      &      & 0.50 & 0.60& 0.30 &  0.04& 0.75 &  0.05& 1.34 &  0.06& 1.58 &  0.11\\ 
      &      & 0.60 & 0.70& 0.18 &  0.03& 0.67 &  0.07& 1.22 &  0.09& 1.45 &  0.13\\ 
      &      & 0.70 & 0.80& 0.15 &  0.03& 0.46 &  0.08& 1.03 &  0.11& 1.51 &  0.17\\ 
\hline  
 0.55 & 0.75 & 0.10 & 0.15& 0.43 &  0.15& 0.93 &  0.32& 1.26 &  0.41& 1.72 &  0.54\\ 
      &      & 0.15 & 0.20& 0.39 &  0.09& 1.37 &  0.16& 1.91 &  0.20& 2.38 &  0.30\\ 
      &      & 0.20 & 0.25& 0.49 &  0.08& 1.19 &  0.10& 2.14 &  0.13& 2.36 &  0.15\\ 
      &      & 0.25 & 0.30& 0.53 &  0.06& 1.20 &  0.08& 2.06 &  0.11& 2.55 &  0.21\\ 
      &      & 0.30 & 0.35& 0.44 &  0.05& 1.00 &  0.08& 1.87 &  0.10& 2.26 &  0.11\\ 
      &      & 0.35 & 0.40& 0.42 &  0.05& 1.02 &  0.07& 1.52 &  0.08& 2.01 &  0.10\\ 
      &      & 0.40 & 0.45& 0.32 &  0.04& 0.84 &  0.06& 1.33 &  0.06& 1.81 &  0.09\\ 
      &      & 0.45 & 0.50& 0.30 &  0.03& 0.68 &  0.05& 1.25 &  0.06& 1.57 &  0.08\\ 
      &      & 0.50 & 0.60& 0.25 &  0.03& 0.58 &  0.04& 1.10 &  0.05& 1.42 &  0.08\\ 
      &      & 0.60 & 0.70& 0.16 &  0.03& 0.47 &  0.05& 0.88 &  0.07& 1.17 &  0.10\\ 
      &      & 0.70 & 0.80& 0.10 &  0.02& 0.35 &  0.06& 0.67 &  0.08& 0.92 &  0.12\\ 
\hline  
 0.75 & 0.95 & 0.10 & 0.15& 0.64 &  0.14& 1.19 &  0.23& 1.77 &  0.36& 1.78 &  0.40\\ 
      &      & 0.15 & 0.20& 0.62 &  0.07& 1.45 &  0.10& 2.24 &  0.13& 2.83 &  0.24\\ 
      &      & 0.20 & 0.25& 0.55 &  0.07& 1.15 &  0.08& 2.07 &  0.10& 2.69 &  0.15\\ 
      &      & 0.25 & 0.30& 0.47 &  0.05& 1.11 &  0.09& 1.79 &  0.09& 2.20 &  0.15\\ 
      &      & 0.30 & 0.35& 0.35 &  0.04& 0.98 &  0.07& 1.52 &  0.07& 2.14 &  0.11\\ 
      &      & 0.35 & 0.40& 0.34 &  0.04& 0.76 &  0.05& 1.21 &  0.06& 1.74 &  0.09\\ 
      &      & 0.40 & 0.45& 0.23 &  0.03& 0.59 &  0.05& 1.08 &  0.05& 1.48 &  0.08\\ 
      &      & 0.45 & 0.50& 0.19 &  0.02& 0.47 &  0.03& 0.93 &  0.04& 1.27 &  0.07\\ 
      &      & 0.50 & 0.60& 0.16 &  0.02& 0.40 &  0.03& 0.76 &  0.04& 0.99 &  0.07\\ 
      &      & 0.60 & 0.70& 0.11 &  0.02& 0.32 &  0.03& 0.59 &  0.06& 0.70 &  0.09\\ 
\hline  
 0.95 & 1.15 & 0.10 & 0.15& 0.62 &  0.12& 1.40 &  0.22& 2.14 &  0.32& 2.43 &  0.39\\ 
      &      & 0.15 & 0.20& 0.70 &  0.07& 1.46 &  0.11& 2.29 &  0.13& 3.02 &  0.20\\ 
      &      & 0.20 & 0.25& 0.56 &  0.06& 1.09 &  0.08& 1.85 &  0.10& 2.66 &  0.16\\ 
      &      & 0.25 & 0.30& 0.40 &  0.04& 0.96 &  0.08& 1.49 &  0.07& 2.10 &  0.12\\ 
      &      & 0.30 & 0.35& 0.30 &  0.03& 0.83 &  0.06& 1.15 &  0.06& 1.73 &  0.10\\ 
      &      & 0.35 & 0.40& 0.29 &  0.03& 0.63 &  0.05& 0.96 &  0.05& 1.38 &  0.08\\ 
      &      & 0.40 & 0.45& 0.28 &  0.03& 0.47 &  0.04& 0.78 &  0.04& 1.06 &  0.07\\ 
      &      & 0.45 & 0.50& 0.20 &  0.03& 0.36 &  0.03& 0.66 &  0.03& 0.81 &  0.06\\ 
      &      & 0.50 & 0.60& 0.10 &  0.02& 0.25 &  0.02& 0.50 &  0.03& 0.60 &  0.04\\ 
\hline
\end{tabular}
}
\end{center}
\end{table*}

\begin{table*}[hp!]
\begin{center}
\small{

\begin{tabular}{rrrr|r@{$\pm$}lr@{$\pm$}lr@{$\pm$}lr@{$\pm$}l}
\hline
$\theta_{\hbox{\small min}}$ &
$\theta_{\hbox{\small max}}$ &
$p_{\hbox{\small min}}$ &
$p_{\hbox{\small max}}$ &
\multicolumn{8}{c}{$d^2\sigma^{\pi^-}/(dpd\theta)$}
\\
(rad) & (rad) & (\GeVc) & (\GeVc) &
\multicolumn{8}{c}{\bgr}
\\
  &  &  &
&\multicolumn{2}{c}{$ \bf{3 \ \GeVc}$}
&\multicolumn{2}{c}{$ \bf{5 \ \GeVc}$}
&\multicolumn{2}{c}{$ \bf{8 \ \GeVc}$}
&\multicolumn{2}{c}{$ \bf{12 \ \GeVc}$}
\\
\hline
 1.15 & 1.35 & 0.10 & 0.15& 0.63 &  0.12& 1.51 &  0.24& 2.39 &  0.37& 3.18 &  0.51\\ 
      &      & 0.15 & 0.20& 0.68 &  0.07& 1.47 &  0.09& 2.20 &  0.15& 2.97 &  0.21\\ 
      &      & 0.20 & 0.25& 0.54 &  0.05& 1.01 &  0.08& 1.70 &  0.10& 2.28 &  0.14\\ 
      &      & 0.25 & 0.30& 0.33 &  0.04& 0.77 &  0.06& 1.25 &  0.07& 1.61 &  0.11\\ 
      &      & 0.30 & 0.35& 0.21 &  0.03& 0.58 &  0.05& 0.94 &  0.06& 1.12 &  0.08\\ 
      &      & 0.35 & 0.40& 0.15 &  0.02& 0.42 &  0.03& 0.70 &  0.04& 0.89 &  0.05\\ 
      &      & 0.40 & 0.45& 0.13 &  0.02& 0.33 &  0.03& 0.55 &  0.03& 0.74 &  0.05\\ 
      &      & 0.45 & 0.50& 0.09 &  0.02& 0.27 &  0.02& 0.45 &  0.03& 0.60 &  0.05\\ 
\hline  
 1.35 & 1.55 & 0.10 & 0.15& 0.95 &  0.20& 1.51 &  0.27& 2.36 &  0.46& 3.78 &  0.67\\ 
      &      & 0.15 & 0.20& 0.68 &  0.06& 1.29 &  0.10& 2.09 &  0.13& 2.84 &  0.18\\ 
      &      & 0.20 & 0.25& 0.44 &  0.05& 0.89 &  0.07& 1.50 &  0.09& 1.89 &  0.13\\ 
      &      & 0.25 & 0.30& 0.27 &  0.04& 0.54 &  0.05& 1.03 &  0.07& 1.27 &  0.09\\ 
      &      & 0.30 & 0.35& 0.15 &  0.02& 0.37 &  0.03& 0.68 &  0.05& 0.89 &  0.07\\ 
      &      & 0.35 & 0.40& 0.15 &  0.02& 0.30 &  0.02& 0.50 &  0.03& 0.72 &  0.05\\ 
      &      & 0.40 & 0.45& 0.09 &  0.02& 0.25 &  0.02& 0.37 &  0.03& 0.57 &  0.05\\ 
      &      & 0.45 & 0.50& 0.05 &  0.01& 0.19 &  0.02& 0.26 &  0.02& 0.40 &  0.04\\ 
\hline  
 1.55 & 1.75 & 0.10 & 0.15& 1.05 &  0.16& 1.36 &  0.25& 2.14 &  0.39& 3.43 &  0.65\\ 
      &      & 0.15 & 0.20& 0.57 &  0.06& 1.17 &  0.09& 1.78 &  0.10& 2.48 &  0.15\\ 
      &      & 0.20 & 0.25& 0.41 &  0.05& 0.75 &  0.06& 1.13 &  0.07& 1.38 &  0.10\\ 
      &      & 0.25 & 0.30& 0.28 &  0.04& 0.44 &  0.04& 0.81 &  0.06& 0.93 &  0.07\\ 
      &      & 0.30 & 0.35& 0.14 &  0.03& 0.34 &  0.03& 0.51 &  0.04& 0.67 &  0.05\\ 
      &      & 0.35 & 0.40& 0.09 &  0.01& 0.24 &  0.03& 0.35 &  0.03& 0.48 &  0.04\\ 
      &      & 0.40 & 0.45& 0.07 &  0.01& 0.15 &  0.02& 0.25 &  0.02& 0.33 &  0.03\\ 
      &      & 0.45 & 0.50& 0.05 &  0.01& 0.10 &  0.01& 0.19 &  0.01& 0.23 &  0.03\\ 
\hline  
 1.75 & 1.95 & 0.10 & 0.15& 0.85 &  0.13& 1.19 &  0.19& 1.78 &  0.25& 2.61 &  0.45\\ 
      &      & 0.15 & 0.20& 0.48 &  0.05& 1.06 &  0.06& 1.42 &  0.07& 1.89 &  0.12\\ 
      &      & 0.20 & 0.25& 0.32 &  0.04& 0.64 &  0.06& 0.83 &  0.05& 1.15 &  0.08\\ 
      &      & 0.25 & 0.30& 0.18 &  0.03& 0.30 &  0.03& 0.56 &  0.04& 0.67 &  0.06\\ 
      &      & 0.30 & 0.35& 0.09 &  0.02& 0.23 &  0.02& 0.33 &  0.03& 0.39 &  0.05\\ 
      &      & 0.35 & 0.40& 0.09 &  0.02& 0.17 &  0.02& 0.24 &  0.01& 0.28 &  0.03\\ 
      &      & 0.40 & 0.45& 0.06 &  0.02& 0.12 &  0.02& 0.20 &  0.01& 0.23 &  0.02\\ 
      &      & 0.45 & 0.50& 0.03 &  0.01& 0.08 &  0.01& 0.14 &  0.01& 0.17 &  0.02\\ 
\hline  
 1.95 & 2.15 & 0.10 & 0.15& 0.74 &  0.13& 0.97 &  0.17& 1.56 &  0.21& 2.02 &  0.31\\ 
      &      & 0.15 & 0.20& 0.44 &  0.05& 0.74 &  0.05& 1.16 &  0.05& 1.37 &  0.09\\ 
      &      & 0.20 & 0.25& 0.18 &  0.03& 0.40 &  0.05& 0.65 &  0.04& 0.93 &  0.08\\ 
      &      & 0.25 & 0.30& 0.10 &  0.02& 0.23 &  0.03& 0.39 &  0.03& 0.53 &  0.06\\ 
      &      & 0.30 & 0.35& 0.07 &  0.02& 0.14 &  0.03& 0.22 &  0.03& 0.30 &  0.04\\ 
      &      & 0.35 & 0.40& 0.03 &  0.01& 0.08 &  0.01& 0.14 &  0.01& 0.23 &  0.03\\ 
      &      & 0.40 & 0.45& 0.02 &  0.01& 0.09 &  0.01& 0.10 &  0.01& 0.15 &  0.02\\ 
      &      & 0.45 & 0.50& 0.01 &  0.01& 0.08 &  0.01& 0.08 &  0.01& 0.09 &  0.02\\ 
\hline
\end{tabular}
}
\end{center}
\end{table*}
%

\end{appendix}
\FloatBarrier



\end{document}